\documentclass[aps,prb,superscriptaddress,twocolumn,preprintnumbers,amsmath,amssymb]{revtex4-1}
\usepackage{color}
\usepackage{array}
\usepackage{amsmath}
\usepackage{amssymb}
\usepackage{graphicx}
\usepackage{epstopdf}
\usepackage{hyperref}
\usepackage{float}
\usepackage{bibentry}

\newcommand{\ee}{\mathrm{e}}
\newcommand{\ii}{\mathrm{i}}
\newcommand{\UU}{\mathrm{U}}
\long\def\/*#1*/{}

\begin{document}
\title{Symmetric tensor networks and practical simulation algorithms to sharply identify classes of quantum phases distinguishable by short-range physics}
\author{Shenghan Jiang, Ying Ran}
\affiliation{Department of Physics, Boston College, Chestnut Hill, MA 02467}
\date{\today}

\begin{abstract}
Phases of matter are sharply defined in the thermodynamic limit. One major challenge of accurately simulating quantum phase diagrams of interacting quantum systems is due to the fact that numerical simulations usually deal with the energy density, a local property of quantum wavefunctions, while identifying different quantum phases generally relies on long-range physics. In this paper we construct generic fully symmetric quantum wavefunctions under certain assumptions using a type of tensor networks: projected entangled pair states, and provide practical simulation algorithms based on them. We find that quantum phases can be organized into crude classes distinguished by short-range physics, which is related to the fractionalization of both on-site symmetries and space-group symmetries. Consequently, our simulation algorithms, which are useful to study long-range physics as well, are expected to be able to sharply determine crude classes in interacting quantum systems efficiently. Examples of these crude 
classes are demonstrated in half-integer quantum spin systems on the kagome lattice. Limitations and generalizations of our results are discussed.
\end{abstract}

\maketitle
\tableofcontents

\section{Introduction}
Reliably simulating quantum phase diagrams of realistic interacting systems has been one of the central issues in condensed matter physics. A number of numerical methods have been developed in the past decades, including exact diagonalization, quantum Monte Carlo(for a review, see Ref.\onlinecite{Foulkes:2001p33}), variational Monte Carlo\cite{Foulkes:2001p33,Gros:1989p53}, the density matrix renormalization group method (DMRG)\cite{White:1992p2863,Schollwoeck:2011p96}, and methods based on tensor network representations of quantum wavefunctions\cite{Verstraete:2004p,Vidal:2007p220405,Vidal:2008p110501,Corboz:2009p165129,Kraus:2010p52338}. Although with advantages and disadvantages, these methods have been demonstrated to be able to successfully simulate various interacting quantum models. For instance, an exotic quantum spin liquid phase have been recently identified in the spin-1/2 Heisenberg model on the kagome lattice\cite{Yan:2011p1173,Jiang:2012p902,Depenbrock:2012p67201} using DMRG methods. 

One major source of the challenges of accurately simulating realistic quantum models is the following fact. The full many-body quantum Hamiltonian cannot be exactly diagonalized as long as the sample size is not very small. Therefore even for intermediate sample size, except for systems that do not suffer from the sign problem in quantum Monte Carlo, one has to come up with variational wavefunctions, using which to search for the true ground states of quantum systems. The guiding principle of all variational simulations is simply to minimize the energy density, a local property of quantum states, of a given sample. On the other hand, generally distinguishing different quantum phases relies on the long range physics. Consequently in these variational methods we are trying to determine long range physics based on local physics. However, competing quantum phases could have similar energy densities. In fact, it can be shown that different quantum phases could give arbitrarily close energy densities\cite{Chen:2010p165119,Balents:2014p245116}. 

To concretely demonstrate this challenge let's consider frustrated quantum spin systems, for instance, nearest neighbor spin-1/2 Heisenberg models on the triangular lattice and the kagome lattice. In the triangular lattice case a good understanding of the ground state is known, based on results from various numerical simulations\cite{Bernu:1994p10048,Capriotti:1999p3899,Zheng:2006p224420,White:2007p127004} which show that the system has a long range 120$^\circ$ coplanar magnetic order. To establish this long range magnetic order, a statement about the long-range physics, it is important to perform finite-size scaling since most numerical simulations study samples with small to intermediate sizes. The successful identification of the long range order in the triangular lattice model, to a large extent, is a consequence of the fact that the magnetic ordering in this system is quite strong\footnote{For instance, the magnetic moment is found to be $0.205(10)$ in Ref.\onlinecite{Capriotti:1999p3899}} Even the 
finite size scalings performed quite some time ago\cite{Capriotti:1999p3899} on small to intermediate sized samples give clear evidences of the order.

The situation for the kagome lattice model is drastically different. In the past it was known that even if a long range order does exist in this system, it is very weak. Thus in order to identify the presence or absence of a long range order, which represent two different quantum  phases: a symmetry-breaking phase and a quantum spin liquid phase, one needs to perform finite size scaling in samples with larger sizes. The simulations on these samples become possible only recently due to the progresses in DMRG methods.

Generally speaking, in order to fully determine the quantum phase diagrams of correlated systems in numerical simulations, one cannot avoid studying samples with large sizes, simply because general quantum phases are sharply defined by the long range physics. But practically the larger the system size is, the more challenging the simulation is. 

But are all quantum phases only distinguished by long-range physics? Before we provide an answer to this question, it is better to elaborate the question in a slightly sharper way. First we emphasize that the a phase is defined only when the global symmetry of the system is specified, which may or may not be \emph{spontaneously} broken. When limited to finite size samples, the ground state wavefunctions necessarily form (generally, irreducible) representations of the global symmetry which is usually a combination of on-site symmetries like spin rotations and space-group symmetries like translations. This statement is true even when the global symmetry is spontaneously broken in the long range physics.

We again demonstrate the above statement in the context of frustrated spin-1/2 models. In this context, quantum spin liquids (QSL) are states of matter that do not break translation and spin rotation symmetries. In particular, evidences of a fully gapped $Z_2$ QSL were reported in the kagome lattice model mentioned above. Recent theoretical work\cite{Zaletel:2014p} supports that this $Z_2$ QSL has a topological order which can be described as a usual $Z_2$ gauge theory (i.e., the same topological order as in Kitaev's toric code model\cite{Kitaev:2003p2}). In such a $Z_2$ QSL, quasiparticle excitations include bosonic spin-1/2 spinon-$e$, bosonic vison-$m$, and their fermionic bound state $f=em$. Suppose that we can tune certain parameters in the spin model, it is possible that either the spinon $e$ boson condenses, which gives rise to certain long-range magnetically ordered (MO) phase, or the vison $m$ boson condenses, which gives rise to certain valence bond solid (VBS) phase since visons transform 
nontrivially under 
lattice symmetries. 

However, in this context, the boson condensations of $e$ or $m$ quasiparticles are only sharply defined in the long range physics. For instance, imagine one does a numerical simulation for a phase transition between the $Z_2$ QSL and a nearby MO phase (VBS phase) via $e$ ($m$) condensation. To avoid possible subtlety due to open boundary conditions, let's consider a finite size torus sample. The ground state wavefunctions on both sides of the phase transition must share the same quantum number in the vicinity of the phase transition. Basically the quantum phase transition in the long-range physics is not visible on a finite size sample unless a careful finite size scaling is performed in large system sizes. \footnote{There is a subtlety about the sharp meaning of the ground state quantum number in the $Z_2$ QSL due to the topological ground state degeneracy on a torus. We will carefully comment on this issue in Section.\ref{sec:torus_long_range_order} } For this reason, we say that the $Z_2$ QSL and the 
nearby MO phase 
(VBS phase) share the same short-range physics but are distinguished in the long-range physics.

On the other hand, there are lots of examples in which ground states of different candidate quantum phases give distinct symmetry representations on sequences of finite size samples, which persist all the way to the thermodynamic limit\cite{Kou:2009p224406,Yao:2010p166402,Jiang:2014p31040}. Trivial examples include ferromagnetic phases and paramagnetic phases in spin systems. As a somewhat nontrivial example, in a recent investigation of correlated electronic models on the honeycomb lattice, two candidate quantum phases: the chiral spin density wave phase and the d+id superconductor phase, are found to host distinct lattice quantum numbers on $4N\times 4N\times 2$ symmetric samples\cite{Jiang:2014p31040}. In these cases, at least on these sequences of samples, clearly these candidate phases really give completely different ground state wavefunctions which cannot be smoothly tuned from one to another. These quantum phases must be distinguished by short range physics. Note that the energy density, minimizing 
which is the guideline of all variational methods, is also a short-range property of the wavefunction. \emph{One should have the hope of generically being able to sharply identify candidate phases distinguishable by short range physics even on small or intermediate sized samples, without worrying about finite size scalings in larger system sizes.}

Note that we have made statements on ``short-range physics'' and ``long-range physics'' without sharply defining their meanings. Now it is a good moment to comment on the sharp meanings of these terms used in this paper. By ``long-range physics'', we really mean the long-range behavior of correlators measured in ground state wavefunctions. Such long-range correlators, e.g., spin-spin correlation functions in a spin model, can be interpreted as the conventional Ginzburg-Landau order parameters of quantum phases. 

The meaning of ``short-range physics'' in this paper is more unconventional, by which we really mean how global symmetries are implemented locally in a quantum wavefunction. We will provide a sharper definition of this term later since we firstly need to introduce some tools to diagnose a local patch of the whole quantum wavefunction. But it is important to mention that this ``short-range physics'' is directly related to the quantum numbers of ground state wavefunctions on finite size samples. In addition, both ``short-range physics'' and ``long-range physics'' in this paper are referred to properties of quantum wavefunctions, even in the absence of specific quantum Hamiltonians.

As an interesting example, let's consider candidate $Z_2$ QSL that may be realized in the kagome lattice Heisenberg model. Previous studies showed that there exist many time-reversal symmetric $Z_2$ QSL phases respecting the full space group symmetry of the kagome lattice \cite{Sachdev:1992p12377,Wang:2006p174423,Lu:2011p224413}. All these $Z_2$ QSL phases, by definition, are featureless in long-range correlators. So their distinctions completely lie in the short-range physics.

The above discussions lead to the following intuitive picture. Different quantum phases may be organized into crude classes according to short range physics. In each class, there may be multiple member phases that are distinguishable by long range physics. Although identifying a particular quantum phase in a correlated model generally requires careful and challenging finite size scaling, identifying a crude class should be easier, even without finite-size scaling in large samples. In addition, doing the latter is still very useful. First, it would give us sharp, although incomplete, information about the quantum phase diagram. Second, determining the crude class allows us to focus only on the candidate member phases within one class, which helps identifying the complete phase diagram significantly.

This picture motivates the us to separate the task of simulating the quantum phase diagrams into a short-range part and a long range part, and brings up the following questions. How to systematically, and hopefully completely, characterize these crude classes distinguishable by short-range physics? Can one construct generic variational wavefunctions for each given crude class and provide simulation algorithms based on them? The answers to these questions would lead to an efficient numerical method to completely solve the short-range part of the simulation task, which is very useful for the long-range part of the task as well. We will comment further on the sharp information on long-range physics (i.e. spontaneous symmetry breaking) that can be obtained from short-range physics in Sec.\ref{sec:long_range_order_torus}.

This paper is an attempt to address these questions to a certain level. Here we rely on a recently developed language to construct quite generic and physically relevant quantum wavefunctions: the projected entangled pair states (PEPS) \cite{Vidal:2008p110501,Verstraete:2004p60302,Cirac:2009p504004} that is a version of tensor networks. PEPS has been viewed as a powerful and efficient method to represent generic quantum states whose entanglement entropies do not violate the boundary law(see Ref.[\onlinecite{Eisert:2010p277}] for a recent review). In addition, in two spatial dimensions, PEPS provides a set of concrete numerical algorithms for practical simulations (for instance, Ref[\onlinecite{Lubasch:2014p33014}] discusses details of many PEPS algorithms). In this work we construct generic symmetric wavefunctions using PEPS under certain assumptions. We find that there are classes of symmetric PEPS which are sharply distinguished by short range physics. More precisely, the symmetry requirements on PEPS lead 
to discrete number of solutions. Each solution corresponds to one crude class mentioned above, and constrains a sub-Hilbert space that the tensors in the PEPS must live within.

We find that these classes are related to, but not limited to, fractionalizations of both the on-site symmetries and the space group symmetries of the system\cite{Wen:2002p165113,Wen:2003p65003,Kou:2008p155134,Kou:2009p224406,Yao:2010p,Essin:2013p104406,Hung:2013p195103,Lu:2013p,Mesaros:2013p155115,Barkeshli:2014p,Teo:2015p}. These classes are generally characterized by three sets of algebraic data, which are denoted as $\Theta$'s, $\chi$'s and $\eta$'s in this paper. The first set of data ($\Theta$'s) represents the direct contribution to the symmetry quantum numbers of quantum wavefunctions from each local tensor. The second set of data ($\chi$'s) is related to projective representations of the global symmetry, or the second cohomology group $H^2(SG, U(1))$ in mathematics, where $SG$ is the symmetry group of the system that is generally a combination of on-site symmetries and lattice symmetries. The third set of data ($\eta$'s) is related to the so-called projective symmetry group (PSG)\cite{Wen:2002p165113} characterizing symmetry 
fractionalizations of topological quasiparticles. Mathematically, $\eta$'s are related to the second cohomolgy group $H^2(SG,IGG)$, where $IGG$ is some invariant gauge group. We will explain the origin and constraints on $IGG$ in detail later. Different possible $IGG$ actually gives a hierarchical structure of the crude classes. As an example, half-integer spin systems on the square or kagome lattices have $IGG$'s which at least contain a $Z_2$ subgroup. 

Moreover, we provide concrete simulation algorithms based on these symmetric PEPS wavefunctions in two spatial dimensions (2d) and comment on possible algorithms in higher dimensions. We demonstrate the procedure of crude classifying and constructing symmetric PEPS wavefunctions for the half-integer spin system on the kagome lattice, in which case 32 distinct classes are found under the assumption that $IGG=Z_2$. Although we mainly consider 2d systems in this paper, the majority of our discussions can be easily generalized to other spatial dimensions except for the algorithms specific for 2d.

Not surprisingly, the choice of the kagome lattice spin system as the main example in this paper is motivated by the recent reports of a $Z_2$ QSL in the spin-$1/2$ Heisenberg model\cite{Yan:2011p1173,Jiang:2012p902}. It remains an open question that which one of many candidate QSL may be realized in the kagome lattice model\footnote{In fact, numerical evidences supporting a gapless U(1)-Dirac spin liquid have also been reported\cite{Ran:2007p117205,Iqbal:2013p60405}.}. And very recently there have been a number of works\cite{Zaletel:2015p,Qi:2015p100401} describing how to idenfify these distinct $Z_2$ QSL in numerical simulations, based on careful quantum number analysis. In our work, when $IGG=Z_2$ in a half-integer spin system, every crude class contains a distinct $Z_2$ QSL as a member 
phase. Therefore part of our 
results can be viewed as a classification and construction of $Z_2$ QSL for half-integer spin systems on the kagome lattice, which is somewhat finer than the previous classifications for the spin-$1/2$ case\cite{Wang:2006p174423,Lu:2011p224413}(see Sec.\ref{sec:discussion} for details), and is generally applicable for other half-integer spins. In addition, the simulation algorithms proposed here can be used to identify the nature of the $Z_2$ QSL realized in the kagome lattice spin-$1/2$ Heisenberg model efficiently.

For each given crude class, the other member phases can be viewed as ordered phases in the vicinity of the $Z_2$ QSL member phase, but with a  spontaneous symmetry breaking only sharply defined in the long range physics, e.g. MO phases (via $e$-condensations) or VBS phases (via $m$-condensations). \emph{The nonvanishing symmetry breaking long range order parameters in these phases are expected to be captured in the present symmetric PEPS contruction after performing a scaling with repect to both the virtual bond dimension $D$ (see Sec.\ref{sec:sym_gauge_peps} for definition) and system sizes. }

Note that the concepts of invariant gauge groups and projective symmetry groups(PSG) have been used to study and classify symmetry fractionalizations in topologically ordered phases. In this sense it is not surprising that we find many non-symmetry-breaking $Z_2$ QSL phases distinct by short-range physics.  But in a conventional symmetry breaking phase, such as the MO phases or VBS phases mentioned above, there is no topological order and the long-range gauge dynamics is confined. However, due to the generality of the PEPS language, this work suggests that the concepts of invariant gauge groups and projective symmetry groups are useful even in these conventional ordered phases.

This interesting question raised by the present work can be rephrased in the following way. \emph{Do the neighoring conventional symmetry-breaking phases still ``remember'' their parent non-symmetry-breaking liquid phase?} In many situations the answer to this question is known to be positive. For example, consider the two parent $Z_2$ QSL, Sachdev's $Q_1 = Q_2$ state and $Q_1 = -Q_2$ state\cite{Sachdev:1992p12377,Wang:2006p174423,Tay:2011p20404}. After the spinon-$e$ condensation, they lead distinct long-range MO phases, e.g. so-called $q=0$ MO (for the $Q_1 = Q_2$ QSL) and $\sqrt{3}\times\sqrt{3}$ MO (for the $Q_1 = -Q_2$ QSL). However, in some other situations, the answer to this question is expected to be negative. For instance, the vison-$m$ condensation in these two $Z_2$ QSL could lead to the same VBS phase\cite{Huh:2011p94419}. This phenomenon is related to following fact: for the $Q_1 = Q_2$ state and $Q_1 = -Q_2$ state, the PSGs for the spinon-$e$ are different, but the PSGs for the visons are the 
same. 

Therefore, within the framework proposed in this paper, although one non-symmetry-breaking phase only appears in a single crude class, we cannot rule out the situation that certain special symmetry-breaking phase appears as member phases in multiple crude classes. Namely, it seems possible that two distinct short-range implementations of global symmetry lead to the same symmetry breaking phase in the thermodynamic limit. We will come back to this issue in Sec.\ref{sec:long_range_order_torus}.

This work may be also useful regarding continuous quantum phase transitions. We have mentioned the phase transitions between member phases within one crude class, e.g. the transition between a $Z_2$ QSL and a nearby MO (VBS) phase. One may wonder whether it is possible to have a continuous phase transition between two phases belonging to distinct crude classes. We believe that this is possible and is related to the hierarchical structure of crude classes due to different $IGG$s. For instance, one may consider a parent crude class with $IGG=U(1)$ that has two distinct descendent $IGG=Z_2$ crude classes. Two phases belonging to these two distinct $IGG=Z_2$ crude classes, as a matter of principle, may be connected by a critical point described by the parent $IGG=U(1)$ crude class. We leave further discussions on this topic in Sec.\ref{sec:discussion}.

Before moving on to the main body of the paper, we comment on the limitations of this work. First, due to the fact that we use PEPS to construct ground state wavefunctions, the discussions in this paper is limited to those quantum phases whose entanglement entropies do not violate the boundary law. For instance quantum phases with Fermi surfaces are beyond the scope of the current work. Even within the PEPS language our work makes a nontrivial basic assumption: the on-site symmetries are implemented as representations or projective representations on the virtual degrees of freedom in PEPS. This assumption, although appears natural on the superficial level, is nontrivial and gives rise to limitations.

This problem is related to the recently developed understandings on symmetry protected topological (SPT) phases\cite{Chen:2012p1604}. SPT phases are gapped quantum phases without anyon excitations and protected by various global symmetries. They are generalizations of the topological insulators(see Ref[\onlinecite{Hasan:2010p3045,Qi:2011p1057}] for reviews) in non-interacting fermion systems. It is known that when attempting to represent SPT phases using PEPS, the constraint that the on-site symmetry transforms as representations or projective representations on virtual degrees of freedoms leads to problems, at least in the long range physics. For example a fermion state with nonzero Chern number constructed using PEPS with a fixed bond dimension $D$ under the above constraint is found to have power-law correlations in real space\cite{Dubail:2013p,Wahl:2013p236805,Hastings:2014p}.

This paper is organized as follows. In Sec.\ref{sec:sym_gauge_peps}, we introduce some basics of PEPS. In particular, We discuss gauge redundancy as well as the implementation of symmetries in PEPS. We introduce a special kind of gauge transformation named as invariant gauge group ($IGG$). In phases with no symmetry breaking, $IGG$ leads to low-energy gauge dynamics. Further, for fractional filled systems, there are minimal required nontrivial $IGG$s for any symmetric PEPS under our basic assumption. This phenomenon is consistent with the Hastings-Oshikawa-Lieb-Schultz-Mattis theorem\cite{Hastings:2005p824,Oshikawa:2000p1535,Lieb:1961p407}. In Sec.\ref{sec:algorithm_peps}, we classify symmetric PEPS according to their distinct short-range physics, which is characterized by algebraic data $\Theta$'s, $\chi$'s and $\eta$'s. Relations of the data $\chi$'s and $\eta$'s to second cohomology are discussed. And an introduction of relevant mathematics is given in Appendix \ref{app:proj_rep}. As a main example, we 
give 
the 
classification result for symmetric PEPS on the kagome lattice with a half-integer spin per site and $IGG=Z_2$, and obtain the constraints on the sub-Hilbert spaces for local tensors for each given class. The detailed calculation is presented in Appendix \ref{app:kagome_PEPS_Z2_PSG}. A simpler and pedagogical example on the square lattice can be found in Appendix \ref{app:square_C4}. We also give efficient algorithms for minimization of energy density for a given class of PEPS, which can be used to identify these crude classes in interacting quantum systems. We give the physical interpretation of the algebraic data in Sec.\ref{sec:lantern_operator_peps}. Particularly, we construct fractionalized symmetry operators to explicitly show that $\eta$'s are describing the symmetry fractionalization of spinons in the $Z_2$ QSL member phase. Detectable signatures of the data $\Theta$'s, $\chi$'s and $\eta$'s are discussed. In Sec.\ref{sec:vison_psg}, we construct a decorated version for symmetric PEPS, which serves 
as a more convenient tool to study properties of visons in the $Z_2$ QSL member phase and the properties 
of the vison-condensed phases. Algebraic methods to extract the information of the symmetry fractionalization on visons are given. In Sec. \ref{sec:torus_long_range_order} we discuss symmetry-breaking phases in the symmetric PEPS formulation, and study the effects of the symmetry-breaking orders in finite-size scaling on torus samples. In Sec.\ref{sec:discussion} we consider generalizations and limitations of our study, comment on relations with previous works, and conclude.

\section{Symmetry, Gauge and PEPS}\label{sec:sym_gauge_peps}
In this section, we will give a brief introduction to PEPS. As we will see later, even for the same many-body wavefunction, the PEPS representations are not unique, and different representations are connected by gauge transformations. Further, we will study the implementation of symmetry on PEPS as well as the gauge dynamics in the PEPS language. Particularly, for certain systems, gauge structures will naturally emerge.

\subsection{Introduction to PEPS}
\begin{figure}
\includegraphics[width=0.45\textwidth]{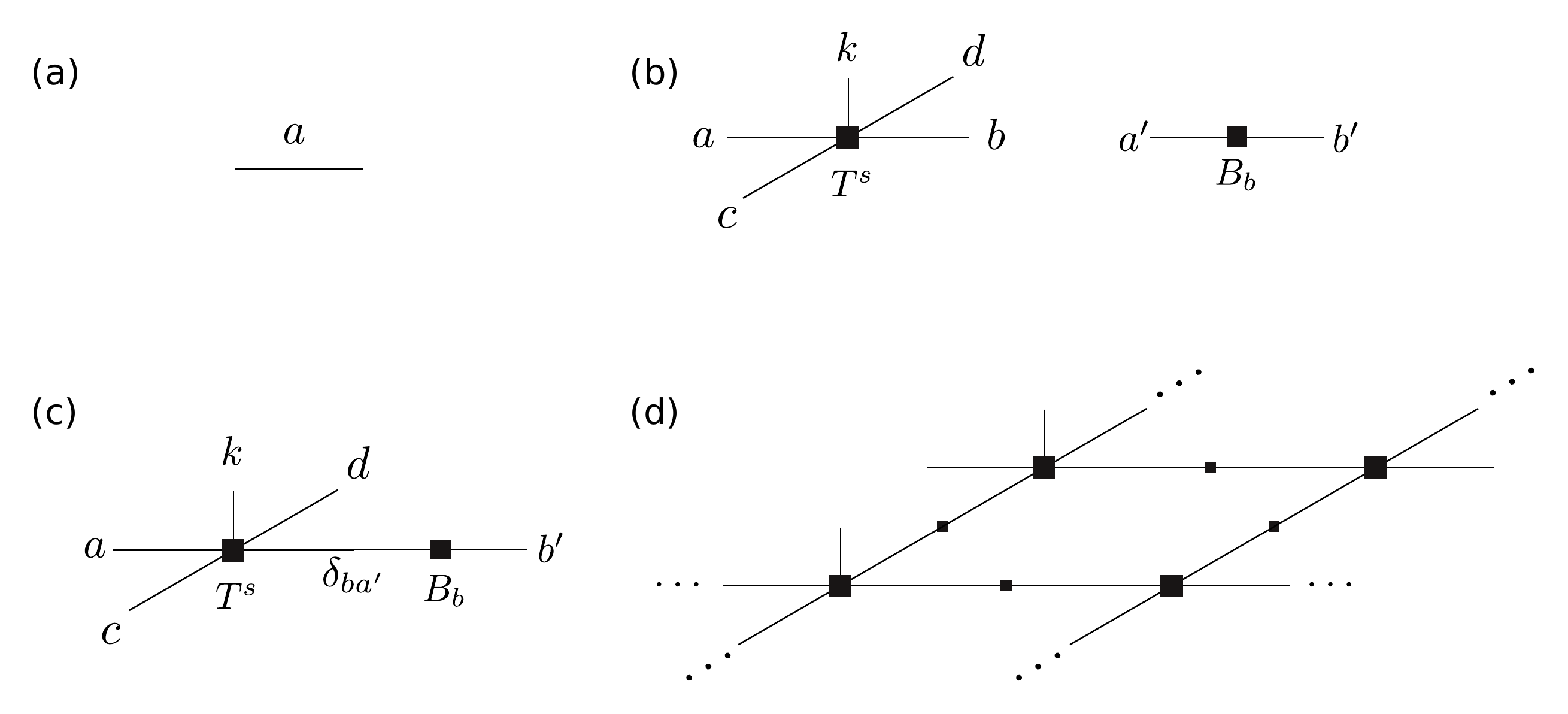}
\caption{(a): The leg $a$ is associated with the Hilbert space $\mathbb{V}^a$. (b): The site tensor (left) and the bond tensor (right) label quantum states on Hilbert spaces of tensor products of corresponding legs. (c): A new tensor can be obtained by contraction of the leg $b$ on $T^s$ and the leg $a'$ on $B_b$, which can be expressed as $(T^s)^k_{abcd}(B_b)_{a'b'}\delta_{ba'}$. Note that we require leg $b$ and leg $a'$ to be dual spaces. (d): The whole PEPS wavefunction is obtained by contracting all virtual legs of site tensors and bond tensors.}
  \label{fig:tensor_network}
\end{figure}

Projected Entangled Pair States (PEPS) is a type of tensor networks (TN). The basic ingredients of TN are ``legs'', and every leg is associated with a Hilbert space, as seen in Fig.(\ref{fig:tensor_network}a). In the following, we will use ``leg'' to denote the associated Hilbert space. As shown in Fig(\ref{fig:tensor_network}b), tensors formed by several legs simply describe quantum states living in the tensor product of these legs,
\begin{align}
  T^{abc\dots}\in\mathbb{V}^a\otimes\mathbb{V}^b\otimes\mathbb{V}^c\otimes\dots
  \label{}
\end{align}
where $\mathbb{V}^i$ labels Hilbert space associated with leg $i$. If two legs are the bra space and the ket space of the same set of quantum states, they are named as dual space to each other. New tensors can be obtained by contracting states in dual spaces, or by tracing out states in dual spaces, as shown in Fig.(\ref{fig:tensor_network}c).

A TN representation of many-body wavefunction can be viewed as a large tensor, which is obtained by contracting small building block tensors. Thus, a TN is formed by uncontracted legs (physical legs) and contracted legs (virtual legs). From another point of view, we can also treat a TN as a combination of a linear map from the virtual Hilbert space (the tensor product of all virtual legs) to the physical Hilbert space, together with an ``input'' virtual state.

Let us construct a PEPS on a two dimensional lattice. We first put tensors at both sites and bonds, named as site tensors ($T^{\mathrm{s}}$) and bond tensors ($B_{\mathrm{b}}$) respectively, see Fig.(\ref{fig:tensor_network}b). Every site tensor can be viewed as a linear map from several virtual legs to one physical leg, while a bond tensor, which is in fact a matrix, labels a quantum state (bond state) in the tensor product space of two virtual legs. Thus, as shown in Fig.(\ref{fig:tensor_network}d), by contracting virtual legs of site tensors with bond tensors, we get a PEPS as a combination of a linear map from the virtual Hilbert space to the physical Hilbert space together with an input virtual state, where the map is given by the tensor product of all site tensors and the input state is the tensor product of all bond states. We can express the PEPS representation of the wavefunction as
\begin{align}
  |\psi\rangle=\sum_{\{k_\mathrm{s}\}}\mathrm{tTr}\left((T^1)^{k_1}...(T^{N_\mathrm{s}})^{k_{N_\mathrm{s}}} B_{1}... B_{N_\mathrm{b}}\right)|k_1\dots k_{N_\mathrm{s}}\rangle,
  \label{eq:peps_wavefunction}
\end{align}
where $1,2,\dots N_\mathrm{s} (N_\mathrm{b})$ label sites (bonds), while $k_\mathrm{s}$ is the physical index. $\mathrm{tTr}$ means tensor trace, namely, contraction of all virtual legs. 

We define that a bond tensor (matrix) is a maximal entangled state, iff singular values of this matrix all equal some nonzero constant. By multiplying some constant, we can simply set singular values of maximal entangled states to be $1$. When performing numerical simulations, it is more convenient to use maximal entangled bond states, or even set bond tensors to be identity matrices. As we will see later, by using the gauge redundancy of PEPS, it is always possible to do so.

In the following, we will assume that all virtual legs label Hilbert spaces with the same dimension $D$, while a physical leg is associated with a $d-$dimensional Hilbert space.

\subsection{Gauge transformation on PEPS}
\begin{figure}
\includegraphics[width=0.45\textwidth]{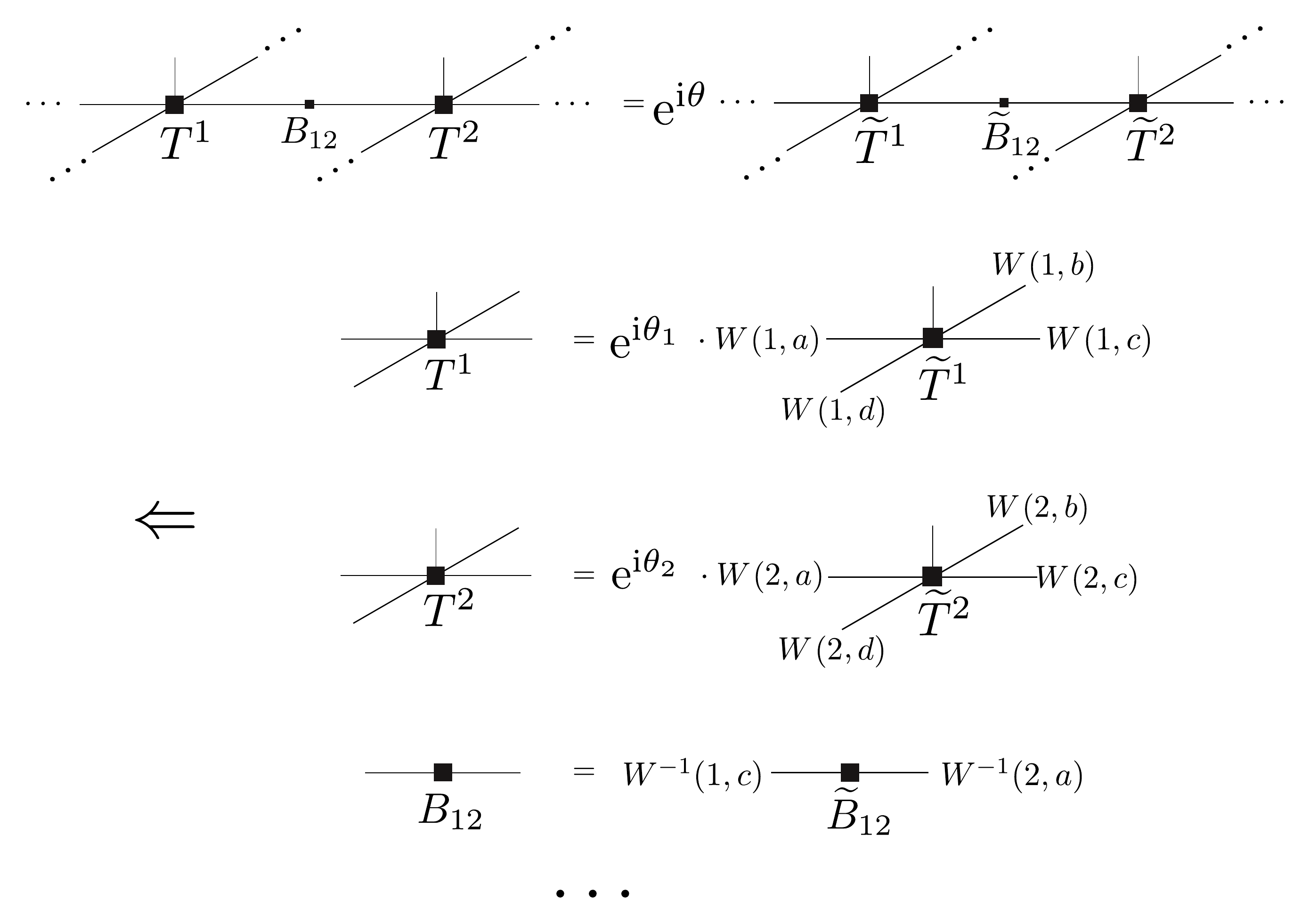}
\caption{Two PEPS describe the same quantum state, iff they are differ by gauge transformation together with $\UU(1)$ phase factor. The origin of the gauge transformation is that we can multiply identity matrix $\mathrm{I}=W\cdot W^{-1}$ between connected legs, which changes site tensors and bond tensors, but leave the whole wavefunction invariant. We can also view TN on the left as PEPS transformed by symmetry operation. Thus, this figure also express the condition for PEPS wavefunction to be symmetric.}
  \label{fig:gauge_freedom_peps}
\end{figure}

The representation of a many-body wavefunction on PEPS is far from unique. Particularly, as shown in Fig.(\ref{fig:gauge_freedom_peps}), we are always allowed to multiply $W$ and $W^{-1}$ to two connected virtual legs respectively. This action will change the connected small tensors while leaving the contracted tensor invariant,
\begin{align}
  (T^\mathrm{s})^k_{abcd}\delta_{ba'}(B_\mathrm{b})_{a'b'}=[(T^\mathrm{s})^k_{abcd}W_{bl}]\delta_{ll'}[(W^{-1})_{l'a'}(B_\mathrm{b})_{a'b'}]
  \label{}
\end{align}
Every contracted pair of virtual legs will contribute a gauge redundancy $\mathrm{GL}(D,\mathbb{C})$. All such gauge transformations form a group $[\mathrm{GL}(D,\mathbb{C})]^{2N_b}$ which we call the gauge transformation group of the PEPS ($N_b$ is the number of bond tensors in the TN). The meaning of the gauge transformation can be understood as a change of basis on virtual legs.

From another point of view, in general, for two PEPS whose tensors differ at most by gauge transformations defined above together with overall $\UU(1)$ phase factors, as shown in Fig.(\ref{fig:gauge_freedom_peps}), the two PEPS must describe the same physical state (up a $\UU(1)$ phase). In principal, these overall $\UU(1)$ phase factors can occur in gauge transformations on both site tensors and bond tensors. But it is straightforward to redefine the gauge transformations such that the phase factors only appear on site tensors. Mathematically, two PEPS denoted by $\{\widetilde T^\mathrm{s},\widetilde B_\mathrm{b}\}$ and $\{ T^\mathrm{s}, B_\mathrm{b}\}$ respectively describe the same physical state if there exist gauge transformations $\{W(\mathrm{s},i)\}$ and $\UU(1)$ phase factors $\{\ee^{\ii\theta(\mathrm{s})}\}$ ($\mathrm{s}$ labels a site and $i$ labels a virtual leg on the site.), such that
\begin{align}
  (T^\mathrm{s})^k_{\alpha\beta\dots}&=\ee^{\ii\theta(\mathrm{s})}\cdot[W(\mathrm{s},1)]_{\alpha\alpha'}[W(\mathrm{s},2)]_{\beta\beta'}\dots(\widetilde T^\mathrm{s})^k_{\alpha'\beta'\dots}\notag\\
  (B_{\mathrm{b}})_{\alpha\beta}&=[W(\mathrm{b},1)]_{\alpha\alpha'}[W(\mathrm{b},2)]_{\beta\beta'}(\widetilde B_\mathrm{b})_{\alpha'\beta'}.\notag\\
  \label{eq:gauge_equiv}
\end{align}
Here $W(\mathrm{b},j)$ represents a gauge transformation on the leg $j$ of the bond tensor $B_\mathrm{b}$, and if a site leg $(\mathrm{s},i)$ and a bond leg $(\mathrm{b},j)$ are connected, then $W(\mathrm{s},i)=[W(\mathrm{b},j)^{-1}]^\mathrm{t}$. (The superscript-$t$ stands for the matrix transpose.)

\subsection{Symmetric PEPS}\label{subsec:sym_peps}
The purpose of this section is to introduce a generic way to implement both on-site symmetries\cite{Perez-Garcia:2010p25010,Zhao:2010p174411,Singh:2010p50301,Singh:2011p115125,Bauer:2011p125106,Weichselbaum:2012p2972,Singh:2012p195114} and lattice space group symmetries\cite{Perez-Garcia:2010p25010} on PEPS. We firstly discuss the finite size symmetric quantum state that can be represented by a single PEPS; i.e., such a state would form a one-dimensional representation of the symmetry group. Then we define the symmetric PEPS on an infinite lattice, which is the main object to be (partially) classified in the current study.

\subsubsection{On-site unitary symmetries}
The action of a global on-site unitary symmetry $S$ on a finite size PEPS wavefunction is defined as
\begin{align}
  S|\psi\rangle=|\widetilde \psi\rangle=&\sum_{\{k_\mathrm{s}\}}\mathrm{tTr}\left((T^1)^{k_1}\dots (T^{N_\mathrm{s}})^{k_{N_\mathrm{s}}} B_{1}\dots B_{N_\mathrm{b}}\right)\notag\\
  &U_S\otimes U_S\dots|k_1k_2\dots k_{N_\mathrm{s}}\rangle,
  \label{eq:on-site_sym_peps}
\end{align}
$U_S$ is the representation of $S$ on Hilbert space of physical leg.  These local actions of an on-site symmetry give a new TN, with site tensors $\widetilde{T}^\mathrm{s}$ and bond tensors $\widetilde{B}_\mathrm{b}$ defined as,
\begin{align}
  \widetilde{T}^\mathrm{s}&=S\circ T^\mathrm{s}=\sum_l (U_S)_{kl}(T^\mathrm{s})^{l}\notag\\
  \widetilde{B}_\mathrm{b}&=S\circ B_\mathrm{b}=B_\mathrm{b}
  \label{eq:on-site_sym_site_bond}
\end{align}
We focus on those PEPS that are invariant under the global symmetry up to an overall $\UU(1)$ phase factor. Following the discussion in the previous section, we consider the PEPS $|\psi\rangle$ that differs from the transformed PEPS $|\widetilde{\psi}\rangle$ only by gauge transformations together with overall phase factors, as shown in Fig.(\ref{fig:gauge_freedom_peps}): 
\begin{align}
  T^\mathrm{s}&=\Theta_SW_SS\circ T^\mathrm{s}\notag\\
  B_\mathrm{b}&=W_SS\circ B_\mathrm{b}
  \label{eq:on_site_symmetric_tensor}
\end{align}
Here, gauge transformation $W_S$ and phase factor $\Theta_S$ associated with symmetry $S$ is defined as
\begin{align}
  \Theta_S\circ T^\mathrm{s}&=\ee^{\ii\theta_S(\mathrm{s})}(T^\mathrm{s})^k_{\alpha\beta\gamma\delta}\notag\\
  W_S\circ T^\mathrm{s}&=[W_S(\mathrm{s},1)]_{\alpha\alpha'}[W_S(\mathrm{s},2)]_{\beta\beta'}\dots(T^\mathrm{s})^k_{\alpha'\beta'\dots}\notag\\
  W_S\circ B_{\mathrm{b}}&=[W_S(\mathrm{b},1)]_{\alpha\alpha'}[W_S(\mathrm{b},2)]_{\beta\beta'}(B_\mathrm{b})_{\alpha'\beta'}.\notag\\
  \label{eq:on-site_sym_gauge_transf}
\end{align}
According to the definition of a gauge transformation, if site virtual leg $(\mathrm{s},i)$ and bond leg $(\mathrm{b},j)$ are connected, then $W_S(\mathrm{s},i)=[W_S(\mathrm{b},j)^{-1}]^\mathrm{t}$. Further, we always choose $W_S$ such that only site tensors transform with extra $\UU(1)$ phase factors.
Note that so far we do not require matrices on the leg $(\mathrm{s},i)$ $W_S(\mathrm{s},i)$ to form a representation of the on-site symmetry group when $S$ is tuned. We will come back to this shortly.

\subsubsection{Time reversal symmetry}
The representation of the global time reversal symmetry $\mathcal{T}$ on a many-body wavefunction is $U_{\mathcal{T}}\otimes U_{\mathcal{T}}\dots K$, where $K$ denotes the complex conjugation and $U_\mathcal{T}$ is a unitary matrix acting on local physical Hilbert space. Its action on PEPS is defined as 
\begin{align}
  \mathcal{T}|\psi\rangle=&\sum_{\{k_{\mathrm{s}}\}}\mathrm{tTr}\left((T^1)^{k_1}\dots (T^{N_\mathrm{s}})^{k_{N_\mathrm{s}}} B_1\dots B_{N_\mathrm{b}}\right)^*\notag\\
  &U_{\mathcal{T}}\otimes U_{\mathcal{T}}\dots|k_1k_2\dots k_{N_\mathrm{s}}\rangle,
  \label{eq:time_reversal_sym_peps}
\end{align}
Namely, the local actions on a single site or a bond tensor read
\begin{align}
  \widetilde{T}^\mathrm{s}=\mathcal{T}\circ T^\mathrm{s}&=\sum_l (U_{\mathcal{T}})_{kl}(T^\mathrm{s})^{*l}\notag\\
  \widetilde{B}_{\mathrm{b}}=\mathcal{T}\circ B_\mathrm{b}&=B_\mathrm{b}^*
  \label{eq:time_reversal_sym_site_bond}
\end{align}
We consider the PEPS that is symmetric under $\mathcal{T}$. Similar to the previous discussion, we consider a PEPS satisfying: 
\begin{align}
  T^\mathrm{s}&=\Theta_{\mathcal{T}}W_{\mathcal{T}}\mathcal{T}\circ T^\mathrm{s}\notag\\
  B_\mathrm{b}&=W_{\mathcal{T}}\mathcal{T}\circ B_\mathrm{b}\notag\\
  \label{eq:time_reversal_symmetric_tensor}
\end{align}
where $W_{\mathcal{T}}$ belongs to the gauge transformation group of the PEPS. 

\subsubsection{Lattice symmetry}
The definition of a lattice space group symmetry $R$ on PEPS is 
\begin{align}
  \widetilde{T}^\mathrm{s}&=R\circ (T^\mathrm{s})^k\equiv \sum_{\alpha\beta\dots}(T^{R^{-1}(\mathrm{s})})^k_{R^{-1}(\alpha\beta\dots)}\notag\\
  \widetilde{B}_{\mathrm{b}}&=R\circ B_{\mathrm{b}}\equiv \sum_{\alpha\beta}(B_{R^{-1}(\mathrm{b})})_{R^{-1}(\alpha\beta)}
  \label{eq:lattice_sym_site_bond}
\end{align}
The action of $R$ on site and bond tensor follows the natural definition of lattice symmetries. For instance, for a square lattice, after a translation along the right direction by one lattice spacing, the transformed site tensor at a given position equals the original site tensor on the left neighboring site. Note that the symmetry $R$ not only acts on site and bond indices; it may also act nontrivially on virtual legs. For example, the $90^{\circ}$ rotation of a site tensor on the square lattice permute the four virtual legs. Again, we consider those PEPS symmetric under $R$ satisfying the following conditions:
\begin{align}
  T^\mathrm{s}&=\Theta_{R}W_{R}R\circ T^\mathrm{s}\notag\\
  B_\mathrm{b}&=W_{R}R\circ B_\mathrm{b}\notag\\
  \label{eq:lattice_symmetric_tensor}
\end{align}
where $W_R$ belongs to the gauge transformation group of the PEPS.

\subsubsection{Symmetric PEPS on infinite lattices}
Space groups of lattices are usually defined for infinite lattices. This is because for a finite size sample, the lattice symmetry group is a finite group whose group structure is non-generic. In this paper, we will focus on PEPS on infinite lattices satisfying Eq.(\ref{eq:on_site_symmetric_tensor},\ref{eq:time_reversal_symmetric_tensor},\ref{eq:lattice_symmetric_tensor}) under symmetry transformations. And \emph{we define such PEPS as symmetric PEPS on infinite lattices, or simply as symmetric PEPS}. They form the main object to be (partially) classified in the current investigation.

A natural question that arises at this point is: are symmetric PEPS defined above general enough to capture ground states of quantum phases? Let us limit our discussion within those quantum phases whose entanglement entropies do not violate the boundary law so that in principle they may be represented as PEPS. 

Basically, \emph{we expect that the symmetric PEPS on infinite lattices defined above are capable to capture all non-symmetry-breaking liquid phases. After putting on finite lattices and performing a scaling with respect to both the bond dimension $D$ and lattice sizes, we expect the symmetric PEPS are also capable to capture the neighboring ordered phases of the liquid phases.} Here by ``neighboring'' (or ``in the vicinity below), we mean that the symmetry breaking in these phases is only sharply defined in the thermodynamic limit (namely, in the long-range physics). Note that we do not have a proof supporting the statement above. Nevertheless we are not aware of any counterexamples, so at least it is a reasonable conjecture.\footnote{On the other hand, this conjecture may be due to our current lack of understanding. For example, we are not aware how to construct a fully gapped (i.e., with correlators fall off exponentially) bosonic integer quantum hall liquid using a symmetric PEPS with a finite bond 
dimension $D$. But there is no known principle forbidding such a construction.}.

Sometimes one is forced to use more than one PEPS to represent ground state quantum wavefunctions. For instance, in a quantum spin system with $SU(2)$ spin rotation symmetry, this happens for the ferromagnetic phase, whose ground states form a large spin representation. However, such ferromagnetic phases are \emph{not} in the vicinity of any non-symmetry-breaking liquid phases. 

\emph{So far, we do not require the transformations matrices $W$'s on the virtual legs to form representations or even projective representations for the on-site unitary symmetries and the time-reversal symmetry.} As mentioned before, such a requirement leads to difficulties to represent SPT phases in two and higher spatial dimensions. (And SPT phases are non-symmetry-breaking liquid phases.) Indeed, if one translates the ground states of the exact solvable models of SPT phases with on-site symmetries into the language of PEPS, one still finds PEPS satisfying Eq.(\ref{eq:on_site_symmetric_tensor})\cite{Williamson:2014p}. But the transformation 
matrices $W$'s form neither representations nor projective representations of the on-site symmetry group.

Generally classifying symmetric PEPS defined here is a difficult task and we currently do not know how to solve. Next we will introduce the invariant gauge group for PEPS and will make further assumptions so that we could make progress on this difficult task.

\subsection{Invariant gauge group and gauge structure}\label{subsec:IGG_peps}
Among the gauge transformations, there is a special subgroup which we call the invariant gauge group ($IGG$). Note that generally a gauge transformation will leave the physical wavefunction invariant while transforming the site tensors and bond tensors nontrivially in a PEPS. However, by definition, the action of $IGG$ elements on PEPS even leaves all site tensors invariant up to overall U(1) phases and all bond tensors completely invariant\footnote{One could consider a gauge transformation leaving both site tensors and bond tensors up to overall U(1) phases. However one can always straightforwardly redefine the gauge transformation so that the bond tensors are completely invariant.}. So $IGG$ can be viewed as the ``symmetry'' of the building block tensors with actions only on virtual legs. In the following, we will see that $IGG$ is directly related to gauge dynamics\cite{Chen:2010p165119,Swingle:2010p,Schuch:2010p2153,Swingle:2010p,He:2014p205114}. We will also give examples where nontrivial $IGG$'s emerge 
naturally in fractional filled systems under a basic assumption. 

Note that the collection of all gauge transformations that leave all site tensors invariant up to overall U(1) phases and bond tensors completely invariant forms an infinite group, which we denote as $\overline{IGG}$. These gauge transformations satisfy Eq.(\ref{eq:gauge_equiv}) with $\widetilde T^\mathrm{s}=T^\mathrm{s},\widetilde B_\mathrm{b}=B_\mathrm{b}$. Namely, a gauge transformation $\{W(\mathrm{s},i)\}$ is in the $\overline{IGG}$ of a PEPS formed by $\{T^\mathrm{s},B_\mathrm{b}\}$ iff it satisfies:
\begin{align}
  (T^\mathrm{s})^k_{\alpha\beta\dots}&=\ee^{\ii\theta(\mathrm{s})}\cdot[W(\mathrm{s},1)]_{\alpha\alpha'}[W(\mathrm{s},2)]_{\beta\beta'}\dots( T^\mathrm{s})^k_{\alpha'\beta'\dots}\notag\\
  (B_{\mathrm{b}})_{\alpha\beta}&=[W(\mathrm{b},1)]_{\alpha\alpha'}[W(\mathrm{b},2)]_{\beta\beta'}( B_\mathrm{b})_{\alpha'\beta'},\notag\\
  \label{eq:IGG}
\end{align}
for certain U(1) phase factors $\{\ee^{\ii\theta(\mathrm{s})}\}$. Here $W(\mathrm{b},j)$ represents a gauge transformation on the leg $j$ of the bond tensor $B_\mathrm{b}$, and if a site leg $(\mathrm{s},i)$ and a bond leg $(\mathrm{b},j)$ are connected, then $W(\mathrm{s},i)=[W(\mathrm{b},j)^{-1}]^\mathrm{t}$.

Clearly, if certain gauge transformation $\{W(\mathrm{s},i)\}$ belongs to $\overline{IGG}$, then one can straightforwardly multiply U(1) phases $\chi(\mathrm{s},i)$ to the $W(\mathrm{s},i)$-matrices: $\{W(\mathrm{s},i)\}\rightarrow \{\widetilde W(\mathrm{s},i)=\chi(\mathrm{s},i)W(\mathrm{s},i)\}$ and obtain another element in $\overline{IGG}$, if $\chi(\mathrm{s},i)=\chi^*(\mathrm{s}',i')$ when $(\mathrm{s},i)$ and $(\mathrm{s}',i')$ are the two virtual legs connected by one bond tensor. If we view the U(1) phase factors $\{\chi(\mathrm{s},i)\}$ leaving the bond tensors completely invariant as a special kind of gauge transformations, they form an infinite abelian subgroup in the center of $\overline{IGG}$, which we denote as the $\chi-group$, since they commute with any gauge transformations. 

In general one should work with the infinite group $\overline{IGG}$. In this paper, for simplicity, we define $IGG$ as the quotient group:
\begin{align}
 IGG\equiv \frac{\overline{IGG}}{\chi-group}.
\end{align}
In addition, we will mainly focus on the cases in which $IGG$ is a simple finite abelian group $Z_n$. In this situation, it is straightforward to show that $\overline{IGG}=IGG\times \chi-group$, indicating $IGG$ is just a simpler way to express $\overline{IGG}$. This also means that we could equally view $IGG$ as a $Z_n$ subgroup of $\overline{IGG}$. In particular, there exist a generator $g\in \overline{IGG}$, but $g\not\in \chi-group$ and $g$ satisfies $g^n=\mathrm{I}$ where $\mathrm{I}$ is the identity gauge transformation --- the do-nothing gauge transformation. 

Note that if $IGG$ is a more complicated group, since the center extension with respect to $\chi-group$ can be nontrivial, it is possible that $\overline{IGG} \neq IGG\times \chi-group$. In this situation it is better to directly work with $\overline{IGG}$.

\subsubsection{$IGG$ and gauge dynamics}\label{sec:gauge_dynamics}

Here we will discuss the physical meaning of $IGG$. We use $IGG=Z_2$ as an example. The following discussion can be easily generalized to other $IGG$ groups. 

\begin{figure}
\includegraphics[width=0.45\textwidth]{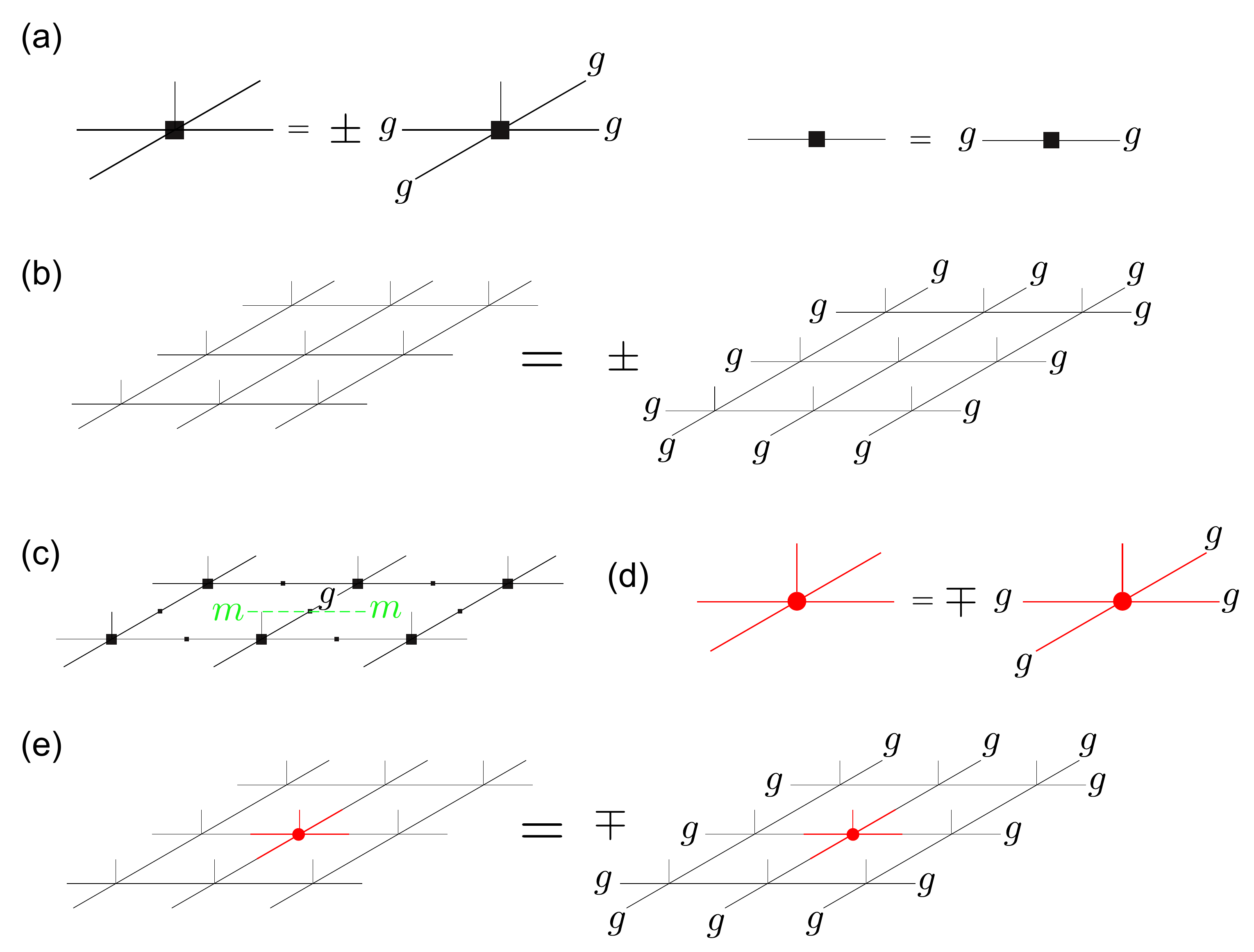}
\caption{(a): Site tensor and bond tensor are both invariant under $Z_2$ action on all virtual legs of tensors. (b): Tensors obtained by contracting $Z_2$ invariant tensors are also $Z_2$ invariant. (c): Acting $g$ on one virtual leg of single bond tensor creates two fluxons ($m$) in plaquettes sharing the bond. (d): $Z_2$ odd tensor indicates there sitting a chargon. (e): By applying $g$ (or creating fluxon loop) on the boundary of a region, we are able to determine chargon number is even or odd inside this region.}
  \label{fig:Z2_IGG}
\end{figure}

First, let us clarify the action of $Z_2$ $IGG$ on PEPS. Every virtual leg accommodates a representation of $Z_2=\{\mathrm{I},g\}$. Note that we do not require representations on different legs to be the same. However, we require two connected legs accommodate representations dual to each other, so that applying the $g$ actions on connected legs is just a special gauge transformation. The nontrivial $Z_2$ $IGG$ element is an action of $g$ on all virtual legs. Following the definition of $IGG$, all site tensors are invariant up to $\pm1$ and all bond tensors are completely invariant under this action, as shown in Fig.(\ref{fig:Z2_IGG}a). Further, it is straightforward to derive that any patch cut from PEPS is invariant up to $\pm1$ under the $g$ actions on boundary virtual legs, as shown in Fig.(\ref{fig:Z2_IGG}b).

The physical meaning of $IGG$ is related to the gauge dynamics. To see this, let us first review the $Z_2$ gauge theory. There are two phases in the $Z_2$ gauge theory: the deconfined phase and the confined phase. In the deconfined phase, the $Z_2$ gauge theory describes $Z_2$ topological order (toric code). The low-energy excitations include four types of quasiparticles: the trivial particle $1$, the chargon $e$, the fluxon $m$ and the bound state of chargon and fluxon $f=em$. $e,m$ and $f$ can only be created in pairs. Each particle is its own anti-particle, $e^2=m^2=f^2=1$. $e,m$ are bosons while $f$ is a fermion. The braiding statistics of the three nontrivial particles are mutually fermionic. In the confined phase, topologically nontrivial quasiparticles are confined.

To see the connection between $IGG$ and the gauge theory, let us create nontrivial excitations on PEPS with $Z_2$ $IGG$. We can define $e$ particles living on sites while $m$ particles living on plaquettes. As shown in Fig.(\ref{fig:Z2_IGG}c), to create two $m$ particles in neighboring plaquettes, we simply multiply the nontrivial $Z_2$ element $g$ on one of two contracted virtual legs shared by the two plaquettes. The insertion of $g$ only on one side of contracted legs is not a gauge transformation, and in general will change the wavefunction. One can also create a pair of $m$ particles spatially separated from each other by applying the single-sided $g$-actions over a string of bonds. The fluxons are located at the end of the string. Note that although the positions of fluxons are physical, the position of the string connecting them are not physical since one can perform $Z_2$ gauge transformations on site tensors to move the string around while leaving the physical wavefunction invariant.

Now, let us turn to $e$ particles. Let us first define $Z_2$ even/odd tensors. The action of $g$ on boundary virtual legs of a tensor generally gives a phase factor $\pm1$. If the phase factor is $+1$/$-1$, we call it $Z_2$ even/odd. The $Z_2$ parities of tensors depend on the representations of $g$ on virtual legs. If we do not worry about the lattice symmetry for the moment, for a $Z_2$ even/odd tensor, we can simply redefine $g$ on one virtual leg by $-1$, thus this tensor becomes $Z_2$ odd/even. So we can assume all tensors are $Z_2$ even for the remaining discussion in this subsection. Creating an $e$ particle on a single site corresponds to changing the site tensor from $Z_2$ even to $Z_2$ odd, as seen in Fig.(\ref{fig:Z2_IGG}d). To detect the number of chargons on a patch of PEPS, we simply apply $g$ on all boundary virtual legs; namely, we create an $m$ loop on the boundary. If there is an odd number of chargons on that patch, this patch tensor should be $Z_2$ odd and the $g$ action on the boundary 
picks 
up a $-1$, see Fig.(\ref{fig:Z2_IGG}e). This $-1$ can be understood as the Berry phase from braiding $e$ and $m$. One can easily convince oneself that an odd number of chargons cannot be created on a closed manifold.

If $IGG=Z_2$ PEPS describe deconfined phases, then separating topological quasiparticles is expected to cost zero tension. Consequently one can insert $m$ loops wrapping around torus holes to construct the four-fold degenerate ground states on a torus. However if $IGG=Z_2$ PEPS describe confined phases, which we expect to be possible after a scaling with both bond dimension $D$ and system sizes, this is no longer true. We will comment further on this in Sec.\ref{sec:torus_long_range_order}

As a final remark, there turns out to be two distinct types of $Z_2$ gauge theories: the toric code theory and the double-semion theory\cite{Dijkgraaf:1990p393,Kitaev:2003p2,Levin:2005p45110}. They have distinct topological orders; e.g., the topological spins (the exchange statistics phases) of quasiparticles are $[1,1,1,-1]$ ($[1,1,i,-i]$) for the $[1,e,m,em]$ particles in a toric code (double-semion) topological order. We emphasize that the $IGG=Z_2$ PEPS discussed here, when describing a deconfined phase, hosts the toric code topological order. The simplest way to see this is to realize the self braiding statistics phases of both the $e$ and the $m$ in the $IGG=Z_2$ PEPS are trivial, so they cannot be semions. 

Indeed, when moving an $e$ chargon around a loop by a sequence of hoppings, one realizes the Berry's phase is independent of whether there are other $e$ chargons inside the loop. Similarly, when moving an $m$ fluxon around a loop (giving rise to an $m$ loop), the topological Berry's phase is simply $\pm1$ depending on the $Z_2$ parity of the PEPS patch inside the loop, independent of whether there are other $m$ fluxons inside the loop. 

\subsubsection{Natural emergence of nontrivial $IGG$}
We will show that, under a basic assumption, the symmetric PEPS for certain quantum systems must have nontrivial $IGG$'s. \emph{This basic assumption is that the $W$ matrices on every virtual leg form (generally reducible) representations or projective representations for the on-site symmetries} (see Eq.\ref{eq:on_site_symmetric_tensor},\ref{eq:time_reversal_symmetric_tensor}). Under this assumption, the nontrivial $IGG$ in certain systems is a natural consequence of the global symmetry, even in the absence of specific Hamiltonians.

Consider a spin-$\frac{1}{2}$ system on a square lattice; i.e., the physical leg on every site tensor is a 2-dimensional spin-$\frac{1}{2}$ Hilbert space. For this system, we will show a symmetric PEPS under the basic assumption must feature an $IGG$ containing a $Z_2$ subgroup. Since $SU(2)$ spin rotation group has no projective representations, the basic assumption ensures that every virtual leg must form a representation of $SU(2)$, which generally is a direct sum of a number of half-integer spin representations and a number of integer spin representations. Eq.(\ref{eq:on_site_symmetric_tensor}) now has the following simple interpretation: the site tensors are spin singlets formed by the virtual spins and the physical spin-$\frac{1}{2}$, and the bond tensors are spin singlets formed the virtual spins only. 

Now we can consider the particular $2\pi$ $SU(2)$ rotation, and denote the corresponding $W(s,i)$ matrix on a virtual leg $(s,i)$ as $\mathrm{J}(s,i)$, which is simply a direct sum of the minus identity transformation in the half-integer spin subspace and the identity transformation in the integer spin subspace. Next, consider the combination of transformations $\{\mathrm{J}(s,i)\}$ \emph{acting on the virtual legs only} --- this is a particular gauge transformation. Since the physical spin-$\frac{1}{2}$ only picks up an overall $-1$ in the $2\pi$ $SU(2)$ rotation, and the bond tensors are spin singlets, we know that the gauge transformation $\{\mathrm{J}(s,i)\}$ is an element in $\overline{IGG}$.

To see this system featuring a nontrivial $IGG$, we only need to show $\overline{IGG}\neq \chi-group$. We will demonstrate that the gauge transformation $\{\mathrm{J}(s,i)\}\notin\chi-group$. To do this, we impose the $C_4$ rotational symmetry and the translation symmetry of the square lattice. Note that $\{\mathrm{J}(s,i)\}\in\chi-group$ if and only if for every virtual leg, the dimension of either the half-integer spin subspace or the integer spin subspace vanishes. However, this cannot be true. The site tensor is a spin singlet, which requires the virtual legs to combine into a spin-$\frac{1}{2}$ so that it can further combine with the physical spin-$\frac{1}{2}$ to form a singlet. Therefore, if $\{\mathrm{J}(s,i)\}\in\chi-group$, on a single site tensor, we must have an odd number of virtual legs which contain purely half-integer spins while the remaining virtual legs contain purely integer spins. This explicitly breaks the $C_4$ rotational symmetry. 

Consequently, there is at least one element $\mathrm{J}\equiv\{\mathrm{J}(s,i)\}$ in $\overline{IGG}$ but not in $\chi-group$, and $\mathrm{J}^2=e$. This tells us that $IGG$ at least contains a $Z_2$ subgroup $\{\mathrm{I},\mathrm{J}\}$. 

The above argument can be easily generalized to other symmetries, such as the time reversal symmetry. For the time reversal symmetry, consider a system with one Kramer doublet on every physical leg. To form a Kramer singlet PEPS, one must combine an odd number of Kramer doublets on virtual legs of every site tensor. However, for site tensors on a square lattice, there are even number (four) of virtual legs per site, and the $C_4$ symmetry dictates that the transformation $\mathcal{T}^2$ \emph{on virtual legs only} gives a nontrivial element of the $IGG$ which is at least $Z_2$.

We point out that translational symmetry itself is enough for the above argument and one does not necessarily consider $C_4$. This is because translational symmetry relates the left (down) virtual leg with the right (up) virtual leg connected to the same site tensor via the fact that the virtual legs connected by a bond need to form a spin singlet (or a Kramer singlet). What is really important for the above argument is the existence of a half-integer spin (or a Kramer doublet) per unit cell. One way to see this is to consider a honeycomb lattice with spin-$\frac{1}{2}$ per site, i.e., two spin-$\frac{1}{2}$'s per unit cell. In this case, every site has three virtual legs and it is possible to construct symmetric PEPS wavefunctions with purely half-integer spins on virtual legs, in which case the $2\pi$ spin rotation \emph{on the virtual legs only} becomes an element in the $\chi-group$.

Next, let us consider a system with fractional filled hard core bosons and see how a nontrivial $IGG$ naturally emerges. As an exercise, we can simply translate the previous discussions on spin-$\frac{1}{2}$ systems into $\frac{1}{2}$-filled hard-core boson systems on the square lattice. The physical leg for the hard-core bosons is two dimensional Hilbert space with basis labeled as $|0\rangle$ and $|1\rangle$. When mapped to a spin-$\frac{1}{2}$ system, $|0\rangle$($|1\rangle$) is identified as the down spin (up spin). The $\UU(1)$ charge transformation for the hard-core boson system can be written as $\exp[\ii\theta(S_z^i+\frac{1}{2})]$ using the spin operator on the leg-$i$. Note that spin-0 is identified as charge-$\frac{1}{2}$, a projective representation of the charge $\UU(1)$.  Since a bond tensor is a spin singlet formed by two virtual spins in the spin language, the representation of $\UU(1)$ group in the hard-core boson language on a bond tensor is
\begin{align}
  [\ee^{\ii\theta(S_z^a+\frac{1}{2})}\cdot\ee^{\ii\theta(S_z^b+\frac{1}{2})}]^*=\ee^{-\ii\theta}
  \label{}
\end{align}
where the complex conjugation comes from the fact that bond virtual legs transform as conjugate representation of site virtual legs, and we have used $S_z^a+S_z^b=0$ for the two virtual legs $a$ and $b$. So, every bond tensor carries charge $-1$. 

Further, since the site tensor is also a spin singlet, we require $\sum_{i=0}^5S_z^i=0$, where $i=0$ labels the physical leg and other $i\neq0$ label virtual legs. Therefore the representation of $\UU(1)$ symmetry on a site tensor reads
\begin{align}
  \prod_{i=0}^4\ee^{\ii\theta(S_z^i+\frac{1}{2})}=\ee^{\ii\frac{5}{2}\theta}
  \label{}
\end{align}
Namely, every site tensor carries charge-$\frac{5}{2}$. Consequently each unit cell carries charge-$\frac{1}{2}$.

Note that in this exercise, the bond tensor transform nontrivially under U(1), so the virtual leg transformation $W=\ee^{\ii\theta(S_z+\frac{1}{2})}$ does not satisfy Eq.(\ref{eq:on_site_symmetric_tensor}) in our definition of symmetric PEPS. But one could easily redefine the virtual leg transformation $W$'s so that the charge carried by the bond is absorbed to a neighboring site, and Eq.(\ref{eq:on_site_symmetric_tensor}) is satisfied using the redefined $W$'s.

The essential results from previous discussions on the spin-$\frac{1}{2}$ systems can now be translated as following statement: the virtual leg hosts both integer charges and half integer charges of $\UU(1)$, so $2\pi$ rotation of $\UU(1)$ symmetry on all virtual legs gives the nontrivial $Z_2$ $IGG$.

In the following, on the square lattice, we provide a general argument that a nontrivial minimal required $IGG$ emerges for a symmetric PEPS with fractional-filled bosons under our basic assumption. Further, this minimal required $IGG$ is given by the $2\pi$ rotation of the $\UU(1)$ symmetry on the virtual legs only.

Firstly, we have the physical legs carrying integer charges. And if the tensor network is symmetric under the $\UU(1)$ symmetry, for site tensors and bond tensors, we can rewrite Eq.(\ref{eq:on_site_symmetric_tensor}) as
\begin{align}
  W_SS\circ T^\mathrm{s}=\Theta_ST^\mathrm{s}\notag\\
  W_SS\circ B_\mathrm{b}=\Theta_SB_\mathrm{b}
  \label{}
\end{align}
where symmetry operation $S$ can be any $\UU(1)$ group element. Note that we put $\Theta_S$ operation on bond tensors as well to pick up the possible phase factors. As mentioned before, this phase factor on the bond can always be tuned away by redefining $W_S$. But for the moment, let us keep it since we want to include the previous exercise.  

We can view the left side as the $\UU(1)$ action on a site/bond tensor. Under the basic assumption, the above equation indicates every site/bond tensor carries a fixed $\UU(1)$ charge, which can be a fractional charge. In the presence of the lattice symmetry, we expect all virtual legs of site tensors share the same $\UU(1)$ reducible projective representation. (Virtual legs of bond tensors have the conjugate representation). Our plan is to assume the $2\pi$ rotation of $\UU(1)$ symmetry is trivial (only a phase factor) on the virtual leg, and then demonstrate a contradiction. This assumption dictates that the irreducible charges carried by a virtual leg can only differ by integer numbers. Namely, the basis for virtual 
legs of site tensors can be written as
\begin{align}
  \{|x\rangle,|x+n_1\rangle,|x+n_2\rangle,\dots\}
  \label{}
\end{align}
where $x$ can be any fractional number and $n_i$ are integers. Under symmetry operation $U_\theta$, state $|x+i\rangle$ transform as
\begin{align}
  U_\theta|x+n_i\rangle=\ee^{\ii\theta(x+n_i)}|x+n_i\rangle
  \label{}
\end{align}
So, $2\pi$ rotation on any state of the above Hilbert space will give the same phase factor $\ee^{\ii x\theta}$. Similarly, the basis for bond legs are
\begin{align}
  \{|-x\rangle,|-x-n_1\rangle,|-x-n_2\rangle,\dots\}
  \label{}
\end{align}
Recall that a single tensor should carry a fixed charge. Consequently a bond tensor should carry charge $-2x-n_b$, where $n_b$ is some integer. And a site tensor should carry charge $4x+n_s$. Since the physical leg only carries integer charges, $n_s$ should also be an integer. We then conclude that, for a single unit cell, the charge should be $n_s-n_b$, which must be an integer. This contradicts with the fact that the system is at a fractional filling. Therefore to construct a symmetric PEPS at a fractional filling under our basic assumption, the $2\pi$ rotation of $\UU(1)$ symmetry must be nontrivial on all virtual legs, and the nontrivial $IGG$ naturally emerges.

We discussed the naturally emerged $IGG$ in certain quantum systems. It is possible for the ground state symmetric PEPS to have a larger $IGG$ which contains the naturally emerged $IGG$ as a subgroup. We call the naturally emerged $IGG$ as the \emph{minimal required $IGG$}. A larger $IGG$ than the minimal required $IGG$ has important implications in both conceptual understandings and numerical simulations. We will come back to this point in Sec.(\ref{sec:algorithm_peps}) and Sec.(\ref{sec:discussion}).

The minimal required $IGG$'s in systems at fractional fillings are consistent with the Hastings-Oshikawa-Lieb-Schultz-Mattis (HOLSM) theorem.  Consider a 2+1D system with an odd number of spin-$\frac{1}{2}$ per unit cell, the HOLSM theorem states that it is impossible to have a featureless trivial insulator. In other words, the ground state must either be gapless, break the spin rotation or the lattice translation symmetry, or be topological ordered with a ground state degeneracy.

In our formalism, a half-integer spin per site on the square lattice (and similarly on the kagome lattice) enforces a minimal $Z_2$ $IGG$, consistent with the HOLSM theorem. For instance, if $IGG=Z_2$, the system could be in either a deconfined phase with a toric code topological order, or a confined phase. But the confined phase corresponds to either $e$ or $m$ condensation, which leads to spin rotation or lattice translation symmetry breaking.

For a honeycomb lattice spin-$\frac{1}{2}$ system, there are two spin-$\frac{1}{2}$ per unit cell and the HOLSM theorem does not apply. As mentioned above, we expect that symmetric PEPS on the honeycomb with a trivial $IGG$ can be constructed, which is consistent with the possible trivial symmetric insulator phase in this system as pointed out in Ref.[\onlinecite{Kimchi:2013p16378}].

\subsection{An example}\label{subsec:peps_rvb_example}
\begin{figure}
\includegraphics[width=0.5\textwidth]{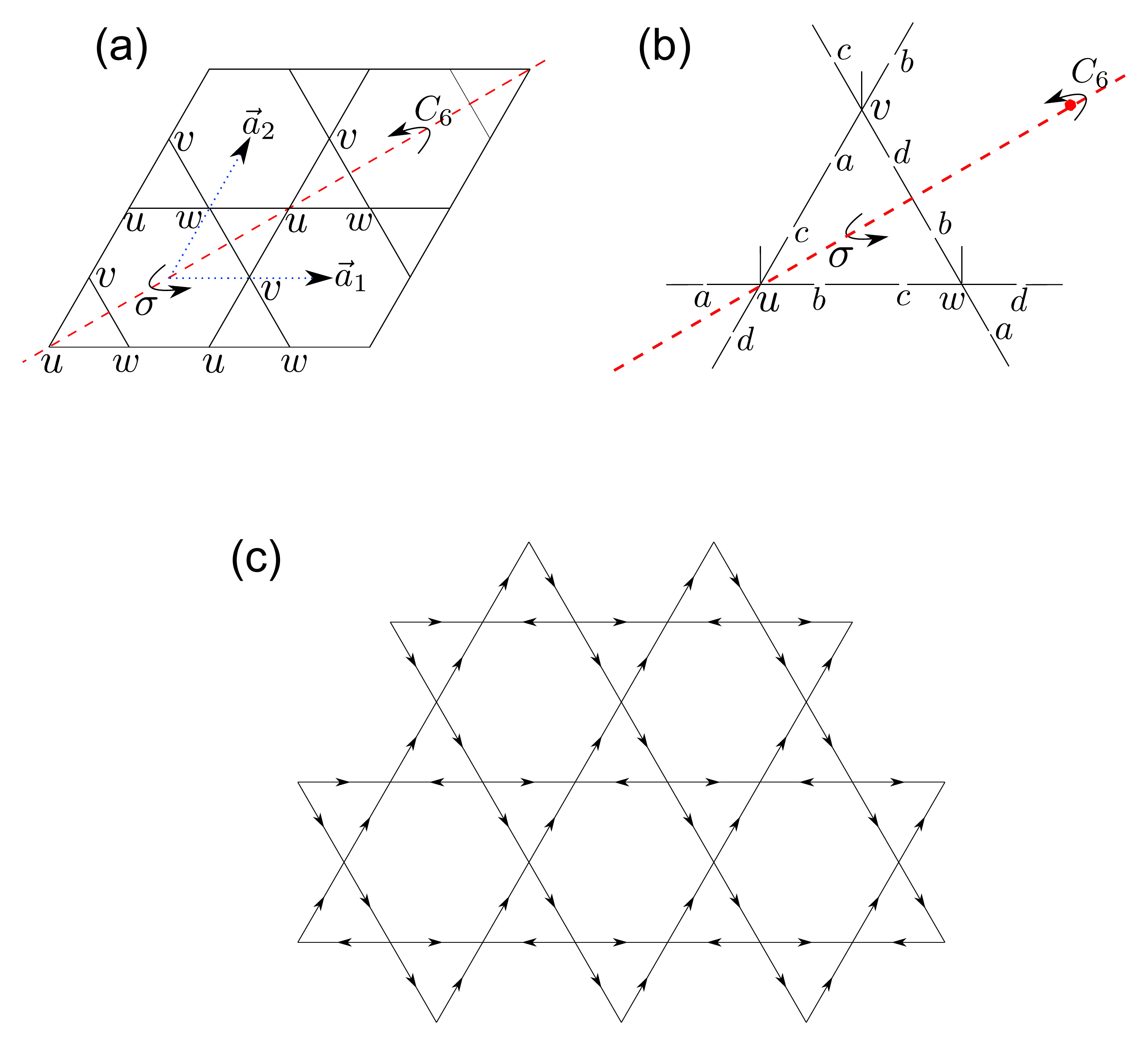}
\caption{(a): kagome lattice and the elements of its symmetry group. $\vec{a}_{1,2}$ are the translation unit vectors, $C_6$ denotes $\pi/3$ rotation around honeycomb center and $\sigma$ represents mirror reflection along the dashed red line. (b): Site tensor and bond tensor for kagome lattice in one unit cell. Virtual legs of site tensors are labeled as $(x,y,s,i)$, where $(x,y)$ denotes the position of unit cell, $s=u,v,w$ is the sublattice index and $i=a,b,c,d$ specifies one of four legs. (c): One possible orientation of kagome lattice. Particularly, for NN RVB state, the orientation of bonds denotes the direction of spin singlets.}
  \label{fig:kagome_lattice}
\end{figure}

Here, we will give a simple PEPS with $IGG=Z_2$ defined on the kagome lattice. In particular, we will write the PEPS description for a nearest neighboring (NN) resonating valence bond (RVB) state that preserves all lattice symmetry. The lattice symmetry generators for kagome lattice are shown in Fig.(\ref{fig:kagome_lattice}).

As shown in Ref.\cite{Yang:2012p147209}, there are four different kinds of symmetric NN RVB states defined on kagome lattice with spin-$\frac{1}{2}$ per site. Also, by solving projective symmetry group (PSG) equations for the Schwinger-boson mean field ansatz on the kagome lattice, one finds eight distinct PSG classes. And four of them can be realized by NN pairing terms\cite{Wang:2006p174423}. One can check that the four NN RVB states are exactly representative states for these four PSG classes. Here, we will focus on one particular PSG class, named as $Q_1=Q_2$ state in Ref.\cite{Sachdev:1992p12377,Wang:2006p174423}. This particular PSG class is a promising candidate phase\cite{Lu2011,Tay:2011p20404,Lu:2014p} for the $Z_2$ spin liquid reported in recent DMRG simulations\cite{Yan:2011p1173,Depenbrock:2012p67201,Jiang:2012p902}. Here, we will explicitly write down this NN RVB state in the PEPS language. 

In fact, this state has already been studied extensively in PEPS\cite{Schuch:2012p115108,Poilblanc:2012p14404}. Here, we will slightly modify the construction. Every physical leg is a spin-$\frac{1}{2}$ and virtual leg accommodates spin representation $0\oplus\frac{1}{2}$, with basis $\{|0\rangle,|\uparrow\rangle,|\downarrow\rangle\}$. Bond tensors are spin singlets, which can be written as a matrix in this basis,
\begin{align}
  B_{\mathrm{b}}=
\begin{pmatrix}
  1 & 0 & 0 \\
  0 & 0 & -\mathrm{i} \\
  0 & \mathrm{i} & 0 \\
\end{pmatrix}
\end{align}
where the direction of bond tensor is shown in Fig.(\ref{fig:kagome_lattice}c). A bond tensor with the inverse direction is transpose of the above matrix. Tensors for different sites are equal to each other, and can be written as,
\begin{align}
  T^\mathrm{s}=&|\uparrow\rangle\otimes(|\downarrow000\rangle+|0\downarrow00\rangle-\ii|00\downarrow0\rangle-\ii|000\downarrow\rangle)-\notag\\
  &|\downarrow\rangle\otimes(|\uparrow000\rangle+|0\uparrow00\rangle-\ii|00\uparrow0\rangle-\ii|000\uparrow\rangle)
  \label{eq:rvb_site_tensor}
\end{align}
where the order of site virtual legs is given in Fig.(\ref{fig:kagome_lattice}b). We can view site tensors as superposition of singlets formed by one physical leg and one of the four virtual legs, while the coefficient of singlets need to be carefully chosen to make PEPS symmetric under lattice symmetries. One can verify the state defined above is consistent with the PEPS representation of NN RVB given in Ref.\cite{Poilblanc:2012p14404} up to a gauge transformation.

As discussed before, the $Z_2$ $IGG$ here is generated by the $2\pi$ spin rotation of all virtual legs. Since all tensors are spin-singlet, they are invariant under this operation up to $-1$ factors on the site tensors. This NN RVB PEPS belongs to one of the crude classes proposed in this paper. Roughly speaking, according to global symmetry, we can find the generic sub-Hilbert space that the building block tensors must live within for each given crude class, which vastly generalize the one-dimensional sub-Hilbert space defined as in Eq.(\ref{eq:rvb_site_tensor}).

\section{Algorithm for Symmetric PEPS}\label{sec:algorithm_peps}
For a given quantum model with certain given symmetry groups, we propose a general simulation scheme to study its phase diagram as follows:
\begin{enumerate}
  \item One classifies symmetric PEPS according to their short-range physics. More precisely, crude classes are distinguished by ways of implementing symmetries on virtual legs.   
  \item For each class, by enforcing symmetry transformation rules, one finds constraint Hilbert spaces for the building block tensors in the PEPS representation. 
  \item One performs the energy density minimization for every class in the constrained Hilbert space, and determines the class which gives the lowest energy density. The quantum phase of the model will be a member phase of this crude class. This finishes the short-range part of the simulation task.
  \item At last, one could try to completely determine the quantum phase diagram by studying the long-range physics, e.g., by measuring correlation functions for the symmetric PEPS with the minimal energy density. With a careful scaling analysis, together with the sharp information on the long-range physics obtained from the short-range physics (see Sec.\ref{sec:torus_long_range_order} for details), possible long range symmetry breaking orders may be identified.
\end{enumerate}

As the main example, we will demonstrate this simulation scheme for a half-integer spin system on the kagome lattice. We will start with classifying and constructing generic symmetric PEPS with $IGG=Z_2$ that preserve the full lattice symmetry as well as the spin rotation and the time reversal symmetries. As we will show shortly, the condition $IGG=Z_2$ actually dictates that the virtual legs form (projective) representations of on-site symmetries. Therefore when we consider $IGG=Z_2$ symmetric PEPS, we already made our basic assumption in an implicit way. In addition,  although we focus on the minimal required $IGG$ under our basic assumption, the discussions can also be easily generalized to symmetric PEPS with a larger $IGG$.

\subsection{General framework for classification}\label{subsec:classification_general_framework}
From now on we assume $\overline{IGG}=IGG\times \chi-group$, which is always true if $IGG$ is a simple finite abelian group $Z_n$. 

Consider the gauge transformation associated with a symmetry $R$: $W_R$, and the corresponding phase on site tensors: $\Theta_R$. We have $T^\mathrm{s}=\Theta_RW_RR\circ T^\mathrm{s}$ and $B_\mathrm{b}=W_R\circ B_\mathrm{b}$, as shown in Sec.\ref{subsec:sym_peps}. However, since both site tensors and bond tensors are invariant under the $IGG$ action (up to phases for site tensors), we conclude that tensors are also invariant under a new symmetry operation defined as $W'_R\equiv\eta_R W_R$ and $\Theta'_R\equiv\mu_R\Theta_R$, 
\begin{align}
  &T^\mathrm{s}=\Theta'_R W'_RR\circ T^\mathrm{s}\notag\\
  &B_\mathrm{b}=W'_RR\circ B_\mathrm{b},
  \label{eq:group_extension_eta_ambiguity}
\end{align}
where $\eta_R \in IGG$ and $\mu_R\equiv\{\mu_R(\mathrm{s})\}$ is a set of phase factors on site tensors associated with $\eta_R$, such that $\mu_R\eta_R\circ T^s=T^s$. For instance, for a half-integer spin system described by PEPS with $IGG=\{\mathrm{I},\mathrm{J}\}$, if $\eta_R=\mathrm{J}$ corresponds to the $2\pi$ $SU(2)$ rotation on the virtual legs, then $\mu_R(s)=-1$ for all sites. 

Similarly one could modify $W_R$ and $\Theta_R$ with any element in the $\chi-group$, i.e., bond dependent phase factors $\{\varepsilon_R(\mathrm{s},i)\}$ as:
\begin{align}
  &W_R(\mathrm{s}/\mathrm{b},i)\rightarrow\varepsilon_R(\mathrm{s}/\mathrm{b},i)W_R(\mathrm{s}/\mathrm{b},i)\notag\\
  &\Theta_R(\mathrm{s})\rightarrow\prod_i\varepsilon_R^*(\mathrm{s},i)\Theta_R(\mathrm{s}),
  \label{eq:group_extension_epsilon_ambiguity}
\end{align}
where we have $\varepsilon_R(\mathrm{s},i)=\varepsilon_R(\mathrm{b},j)^*$ if $(\mathrm{s},i)$ and $(\mathrm{b},j)$ are connected. Further, $\varepsilon_R(\mathrm{b},1)=\varepsilon_R(\mathrm{b},2)^*$ for the two legs of the same bond tensor, as required in the definition of the $\chi-group$.

Basically, the symmetry transformation on the virtual legs $W_R$ is ambiguous since it can be combined with any element in $\overline{IGG}$. Mathematically, the representation of $R$ on the Hilbert space of PEPS (including both the virtual and physical Hilbert spaces) form a new group, which is the original symmetry group $SG$ extended by the $\overline{IGG}$. This extension is related to the 2-cohomology $H^2(SG,IGG)$ and $H^2(SG,U(1))$. (For details about projective representations and the 2-cohomology, see Appendix \ref{app:proj_rep}.) Particularly, we can view those $\overline{IGG}$ elements as ``representations'' of the identity element in the symmetry group on virtual legs.

Keeping these discussions in mind, let us consider a discrete symmetry group $SG$ as an example. $SG$ is always defined by a collection of group identities. For instance, elements $R_1,R_2,\dots,R_n\in SG$ satisfy the following relation:
\begin{align}
  R_1R_2\dots R_n=\mathrm{e}
  \label{eq:group_identity}
\end{align}
Then, acting $R_1R_2\dots R_n$ on a symmetric PEPS, one obtains a combined transformation sending every tensor back to the same tensor:
\begin{align}
  T^\mathrm{s}&=\Theta_{R_1}W_{R_1}R_1\Theta_{R_2}W_{R_2}R_2\dots \Theta_{R_n}W_{R_n}R_n\circ T^\mathrm{s}\notag\\
  B_\mathrm{b}&=W_{R_1}R_1W_{R_2}R_2\dots W_{R_n}R_n\circ B_\mathrm{b}
  \label{eq:psg_general_site_bond}
\end{align}
By definition, the transformation leaving all tensors invariant (up to phases on site tensors) can only be an element in $\overline{IGG}$. Explicitly writing down Eq.(\ref{eq:psg_general_site_bond}) on virtual legs of site tensors, we conclude that 
\begin{align}
  &W_{R_1}(\mathrm{s},i)W_{R_2}(R_1^{-1}(\mathrm{s},i))\dots \notag\\
  &W_{R_n}(R_{n-1}^{-1}\dots R_1^{-1}(\mathrm{s},i))=\eta(\mathrm{s},i)\chi(\mathrm{s},i)
  \label{eq:psg_equation_general}
\end{align}
where $\eta(\mathrm{s},i)$ is the action of $\eta\in IGG$ on the virtual leg $(\mathrm{s},i)$. Further, $\{\chi(\mathrm{s},i)\}$ is an element in the $\chi-group$. We point out that since $W_R(\mathrm{s},i)=[W^{-1}_R(\mathrm{b},j)]^{\mathrm{t}}$ if $(\mathrm{s},i)$ and $(\mathrm{b},j)$ are connected, $W_R$ on virtual legs of bond tensor gives us no extra equation. However, phase factors on site tensors will give an extra condition, which reads
\begin{align}
  &\Theta_{R_1}(\mathrm{s})\Theta_{R_2}(R_1^{-1}(\mathrm{s}))\dots 
  \Theta_{R_n}(R_{n-1}^{-1}\dots R_1^{-1}(\mathrm{s}))\notag\\
  &=\mu(\mathrm{s})\prod_i\chi^*(\mathrm{s},i)
  \label{eq:psg_equation_general_phase}
\end{align}
Here $\mu^*(\mathrm{s})$ is the phase factor obtained after applying $\eta$ on the $\mathrm{s}$-site tensor.

Our goal is to solve Eq.(\ref{eq:psg_equation_general}) and Eq.(\ref{eq:psg_equation_general_phase}) for all group identities and obtain the representations of symmetry operation on virtual legs ($W_R$) as well as phase factors on site tensors ($\Theta_R$). Recall that the same physical wavefunction can be represented by many PEPS which differ from each other by gauge transformations (note that these are general gauge transformations which may not be in $\overline{IGG}$.). One should really solve Eq.(\ref{eq:psg_equation_general}) and Eq.(\ref{eq:psg_equation_general_phase}) up to gauge equivalence.

Under a gauge transformation $V\equiv\{V(\mathrm{s},i)\}$ on virtual legs, $(T^\mathrm{s})'\equiv V\circ T^\mathrm{s}$ and $B'_\mathrm{b}\equiv V\circ B_\mathrm{b}$ satisfy the following conditions:
\begin{align}
  (T^\mathrm{s})'&=V\Theta_RW_RR\circ T^\mathrm{s}\notag\\
  &=(V\Theta_RV^{-1})(VW_RRV^{-1}R^{-1})R V\circ T^\mathrm{s}\notag\\
  &=\Theta_RW'_RR\circ(T^\mathrm{s})',\notag\\
  \label{eq:gauge_transf_on_psg_site}
\end{align}
and 
\begin{align}
  B'_\mathrm{b}&=VW_RR\circ B_\mathrm{b}\notag\\
  &=(VW_RRV^{-1}R^{-1})RV\circ B_\mathrm{b}\notag\\
  &={W}'_RR{B}'_\mathrm{b}
  \label{eq:gauge_transf_on_psg_bond}.
\end{align}
Here we use the fact that $V$ commutes with $\Theta_R$ in the last step of Eq.(\ref{eq:gauge_transf_on_psg_site}). Here, $W'_R\equiv VW_RRV^{-1}R^{-1}$. Writing the above expression explicitly on virtual leg $(\mathrm{s},i)$, we get
\begin{align}
  W_R(\mathrm{s},i)\rightarrow V(\mathrm{s},i)\cdot W_R(\mathrm{s},i)V^{-1}(R^{-1}(\mathrm{s},i))
  \label{eq:gauge_transf_on_psg_coord}
\end{align}
while $\Theta_R$ is invariant. Particularly, $\eta\in IGG$ changes as
\begin{align}
  \eta(\mathrm{s},i)\rightarrow V(\mathrm{s},i)\cdot \eta(\mathrm{s},i)V^{-1}(\mathrm{s},i)
  \label{eq:gauge_transf_on_igg}
\end{align}
And phase factors $\mu$ and $\chi$ in Eq.(\ref{eq:psg_equation_general_phase}) are invariant.

Apart from the above gauge transformation, one can change site tensors by phase factors, which do not affect physical observables. Note that one could also change bond tensors by phase factors, but such a modification is always equivalent to a gauge transformation together with a changing of phase factors on site tensors. Unlike gauge transformations, a modification of phase factors on site tensors may change the physical wavefunction up to an overall phase. When site tensors change as $T^\mathrm{s}\rightarrow \Phi\circ T^\mathrm{s}=\Phi(\mathrm{s})\cdot T^\mathrm{s}=\ee^{\ii\varphi(\mathrm{s})}T^\mathrm{s}$, $W_R$ associated with the symmetry $R$ is invariant, but $\Theta_R$ goes to $\Phi\Theta_RR\Phi^{-1} R^{-1}$. Namely, the phase factor $\Theta_R\equiv\{\ee^{\ii\theta_R(\mathrm{s})}\}$ will change as
\begin{align} 
  \Theta_R(\mathrm{s})\rightarrow\Theta_R(\mathrm{s})\Phi(\mathrm{s})\Phi^*(R^{-1}(\mathrm{s}))
  \label{eq:phase_factor_transf}
\end{align}

Basically, we should solve for the $W_R$ and $\Theta_R$ in Eq.(\ref{eq:psg_equation_general}) and Eq.(\ref{eq:psg_equation_general_phase}) up to two kinds of equivalences. First, if two sets of $W_R$ and $\Theta_R$ are related by Eq.(\ref{eq:gauge_transf_on_psg_coord}) and Eq.(\ref{eq:phase_factor_transf}), they are equivalent and we denote this situation as the \emph{gauge equivalence}. The gauge equivalence contains the $V$-ambiguity in Eq.(\ref{eq:gauge_transf_on_psg_coord}) and the $\Phi$-ambiguity in Eq.(\ref{eq:phase_factor_transf}).

Second, if two sets of $W_R$ and $\Theta_R$ are different by an $\overline{IGG}$ element, they are also equivalent and we denote this situation as the \emph{group extension equivalence}. Summarizing our discussion in Eq.(\ref{eq:group_extension_eta_ambiguity},\ref{eq:group_extension_epsilon_ambiguity}), it means that one could modify $W_R$ and $\Theta_R$ as $W_R\rightarrow {W}'_R=\eta_R\varepsilon_RW_R$ and $\Theta_R\rightarrow{\Theta}'_R=\mu_R\varepsilon_R\Theta_R$, where $\eta_R\in IGG$ and $\varepsilon_R\in \chi-group$ and
\begin{align}
  {W}'_R(\mathrm{s},i)=\eta_R(\mathrm{s},i)\varepsilon_R(\mathrm{s},i)W_R(\mathrm{s},i)\notag\\
  {\Theta}'_R(\mathrm{s})=\mu_R(\mathrm{s})\prod_i\varepsilon_R^*(\mathrm{s},i)\Theta(\mathrm{s}).
  \label{eq:epsilon_transf}
\end{align}
Note that to save notation, we define $\varepsilon_R\Theta_R$ as multiplying $\prod_i\varepsilon_R^*(\mathrm{s},i)$ on $\Theta(\mathrm{s})$. The group extension equivalence contains an $\eta$-ambiguity and an $\varepsilon$-ambiguity in Eq.(\ref{eq:epsilon_transf}). \emph{Note that different from the gauge equivalence, we have an $\eta$-ambiguity and an $\varepsilon$-ambiguity for each symmetry element $R$.}

We will solve Eq.(\ref{eq:psg_equation_general}) and Eq.(\ref{eq:psg_equation_general_phase}) for the whole symmetry group up to both the gauge equivalence and the group extension equivalence. Eventually we will obtain many classes of PEPS satisfying inequivalent $W_R$ and $\Theta_R$ transformation rules. Among all combinations of $W_R$ and $\Theta_R$ within the same equivalence class, we can choose a particular representative, and construct explicit forms of $W_R$ and $\Theta_R$ by fixing the $\eta$-ambiguity, the $\varepsilon$-ambiguity, the $V$-ambiguity and the $\Phi$-ambiguity. These $W_R$ and $\Theta_R$ specify the sub-Hilbert spaces for the building block tensors in each class. We sometimes call the whole procedure of fixing the four ambiguities as \emph{gauge fixing}. 

Practically, we  often firstly use the group extension equivalence to simplify Eq.(\ref{eq:psg_equation_general}) and Eq.(\ref{eq:psg_equation_general_phase}). For instance, one can use the $\varepsilon$-ambiguity to simplify $\{\chi(s,i)\}$ in Eq.(\ref{eq:psg_equation_general}) and Eq.(\ref{eq:psg_equation_general_phase}): under a transformation $W_{R_i}\rightarrow\varepsilon_{R_i}W_{R_i}$, according to Eq.(\ref{eq:psg_equation_general}), we find
\begin{align}
  \chi(\mathrm{s},i)\rightarrow \varepsilon_{R_1}(\mathrm{s},i)\dots\varepsilon_{R_n}(R_{n-1}^{-1}\dots R_1^{-1}(\mathrm{s},i))\chi(\mathrm{s},i).
  \label{eq:epsilon_transf_chi}
\end{align}
Moreover, one can use the $\eta$-ambiguity to simplify the $\{\eta(\mathrm{s},i)\}$ and $\{\mu(\mathrm{s})\}$ in Eq.(\ref{eq:psg_equation_general}) and Eq.(\ref{eq:psg_equation_general_phase}). For example, if some symmetry operation $R$ appears only once in the group identity $R_1R_2\dots R_n=\mathrm{e}$, one could use the $\eta$-ambiguity for $R$ to make sure $\{\eta(\mathrm{s},i)=\mathrm{I}\}$ and $\{\mu(\mathrm{s})=1\}$ for this group condition.

After the group extension equivalence is used, we will use the gauge equivalence (the $V$-ambiguity and the $\Phi$-ambiguity) to solve for explicit forms of $W_R$ and $\Theta_R$. Note that the group extension equivalence and the gauge equivalence are not completely independent. For example, after fixing the $V$-ambiguity and the $\Phi$-ambiguity, it is possible some part of the $\varepsilon$-ambiguity and the $\eta$-ambiguity are also fixed. In the following we demonstrate this procedure in an example: the half-integer spin systems on the kagome lattice.

\subsection{Classification of kagome PEPS}\label{subsec:classification_kagome_peps}
Here, we will classify symmetric kagome PEPS wavefunction with a half-integer spin-$S$ per site, which preserves all lattice symmetries, the time reversal symmetry as well as the spin rotation symmetry. We will only assume $IGG=Z_2=\{\mathrm{I},\mathrm{J}\}$ without specifying the physical meaning of $\mathrm{J}$. Later we will prove that $\mathrm{J}$ can always be chosen to be the $2\pi$ spin rotation on the virtual legs. Let us begin with setting up some useful facts.

First, we can use the $V$-ambiguity to diagonalize $\mathrm{J}(x,y,s,i)$ for every virtual leg $(x,y,s,i)$, where $(x,y,s)$ labels a site on the lattice by the coordinates of the unit cell $x,y$ and the sublattice index $s=u,v,w$, and $i=a,b,c,d$ labels one of the four virtual legs coming out of the site tensor. (see Fig.\ref{fig:kagome_lattice} for illustrations) In this gauge, $\forall (x,y,s,i)$, the matrix $\mathrm{J}(x,y,s,i)$ is a direct sum of an identity matrix and a minus identity matrix. Let us denote $\mathrm{J}(x_0,y_0,s_0,i_0)=\mathrm{I}_{D_1}\oplus(-\mathrm{I}_{D_2})$ for some given virtual leg $(x_0,y_0,s_0,i_0)$, where $D_1+D_2=D$. We will consider the generic case in which $D_1\neq D_2$. 

Using the lattice symmetry, it is straightforward to prove that one can always redefine $\{\mathrm{J}(x,y,s,i)\}$ by multiplying with an element $\varepsilon$ in the $\chi-group$: $\varepsilon(x,y,s,i)=\pm 1$ so that $\mathrm{J}(x,y,s,i)=\mathrm{I}_{D_1}\oplus(-\mathrm{I}_{D_2})$, $\forall (x,y,s,i)$. (Such a modification is allowed in our definition of $IGG$.) For example, consider a particular lattice symmetry operation $R$, which could be the $60^{\circ}$ degree rotation $C_6$ or the lattice translation $T_1$ or $T_2$ of the kagome lattice (see Appendix \ref{app:sym_group_kagome} for precise definitions), we always have a group relation $R^{-1}\cdot \mathrm{e}\cdot R=\mathrm{e}$. Using Eq.(\ref{eq:psg_equation_general}) for this group relation and choosing $\mathrm{J}$ to replace the $\mathrm{e}$ on the LHS:
\begin{align}
 &W_R^{-1}(R(x,y,s,i))\mathrm{J}(R(x,y,s,i))W_R(R(x,y,s,i))\notag\\
 =&\eta(x,y,s,i)\chi(x,y,s,i).\label{eq:trivial_group_id}
\end{align}
The $\eta$ on the RHS must be $\mathrm{J}$, otherwise we would find $\mathrm{J}$ to be an element in the $\chi-group$, violating $IGG=Z_2$. Therefore we know that $\mathrm{J}(R(x,y,s,i))$ and $\mathrm{J}(x,y,s,i)$, which are generally on two different virtual legs, are related by a similarity transformation $W_R(R(x,y,s,i))$ and an overall phase factor $\chi(x,y,s,i)$. But we are already in a gauge such that $\mathrm{J}(x,y,s,i)$ are all diagonal. We then conclude that $\mathrm{J}(R(x,y,s,i))=\pm \mathrm{J}(x,y,s,i)$. Since all virtual legs are related by lattice symmetries, we know $\mathrm{J}(x,y,s,i)=\varepsilon(x,y,s,i) \mathrm{J}(x_0,y_0,s_0,i_0)$, where $\varepsilon(x,y,s,i)=\pm1$ $\forall (x,y,s,i)$.

Next, we show $\{\varepsilon(x,y,s,i)\}\in \chi-group$. Namely, if $(x,y,s,i)$ and $(x',y',s',i')$ are connected by a bond tensor $B_b$, then $\varepsilon(x,y,s,i)=\varepsilon(x',y',s',i')$. This is because if $\varepsilon(x,y,s,i)=-\varepsilon(x',y',s',i')$, then the matrix $(B_b)_{\alpha\beta}$ satisfying Eq.(\ref{eq:IGG}) for $W=\mathrm{J}$ would not have a full rank, since $D_1\neq D_2$. This means that some singular value of $(B_b)$ vanishes, dictating an $IGG$ larger than $Z_2$. For instance, one can multiply an arbitrary $\UU(1)$ phase on the zero singular value eigenstate on one of the two virtual legs, leaving the bond tensor $B_b$ invariant.

Therefore $\{\varepsilon(x,y,s,i)\}\in \chi-group$ and we can always redefine $\mathrm{J}$ such that $\mathrm{J}(x,y,s,i)=\mathrm{I}_{D_1}\oplus(-\mathrm{I}_{D_2})$, $\forall (x,y,s,i)$. \emph{From now on we will work within this gauge} and denote the matrix $\mathrm{I}_{D_1}\oplus(-\mathrm{I}_{D_2})$ simply as $\mathrm{J}$.

This allows us to denote the $\eta(x,y,s,i)$ transformation in Eq.(\ref{eq:psg_equation_general}) simply as $\eta$ since it is site and virtual leg independent. In addition, according to Eq.(\ref{eq:gauge_transf_on_igg}), the remaining $V$-ambiguity: $V(x,y,s,i)$ must commute with $\mathrm{J}$. In other words, $V(x,y,s,i)$ are block diagonal with two blocks, and the sizes of blocks are $D_1$ and $D_2$ respectively.

Now we can consider an arbitrary symmetry transformation $R$, which could be either a lattice symmetry or an on-site symmetry. Eq.(\ref{eq:trivial_group_id}) still holds for $R$ and the $\eta$ on the RHS must be $J$. Consequently we have:
\begin{align}
  &W_R^{-1}(R(x,y,s,i))\cdot\mathrm{J}\cdot W_R(R(x,y,s,i))\notag\\
 =&\chi(x,y,s,i)\mathrm{J}.
\end{align}
Squaring this equation leads to $\chi(x,y,s,i)=\pm 1$. However only the $+$ sign is possible since otherwise the matrix $W_R(R(x,y,s,i))$ will not have a full rank, again due to $D_1\neq D_2$. Thus \emph{we have proved that $W_R(x,y,s,i)$ commutes with $\mathrm{J}$, $\forall (x,y,s,i)$ and $\forall R$}. Mathematically, this means that when we extend the symmetry group by $\overline{IGG}=IGG\times \chi-group$, $\overline{IGG}$ is in the center of the extended group.

Let us consider the phase factors $\mu_{\mathrm{J}}(x,y,s)$ on site tensors obtained when applying the nontrivial element $\mathrm{J}$ on the virtual legs. This determines whether the site tensor is $Z_2$ even or $Z_2$ odd. Now we are ready to show that \emph{$\mu_{\mathrm{J}}(x,y,s)$ is site independent in the current gauge}. Namely if one site tensor is $Z_2$ even (odd), the same is true for all site tensors. Consider a lattice symmetry $R$ which send a site $(x,y,s)$ to the site $(x',y',s')$, Eq.(\ref{eq:lattice_symmetric_tensor}) states that the two site tensors are related by a possible permutation of virtual indices (e.g. induced by a lattice rotation) together with multiplications of $W_R$ matrices on the virtual legs as well as a overall phase factor $\Theta_R(x,y,s)$. Because $W_R$ matrices all commute with $\mathrm{J}$, it is straightforward to see that the $\mu_{\mathrm{J}}(x,y,s)=\mu_{\mathrm{J}}(x',y',s')$. Because all sites are related to each other by lattice symmetries, $\mu_{\mathrm{J}}(x,y,
s)$ are identical for all sites. Thus in the discussion below we will simply denote the $\eta\in IGG$ associated phase factors $\mu(x,y,s)$ in Eq.(\ref{eq:psg_equation_general_phase}) as $\mu$, since it does not depend on the site.

By applying the condition $IGG=Z_2$ to the kagome lattice with the symmetry group described in Appendix \ref{app:sym_group_kagome}, we are able to solve the equations for symmetry operations, i.e. Eq.(\ref{eq:psg_equation_general},\ref{eq:psg_equation_general_phase}), by gauge fixing. For the purpose of presentation, here we only demonstrate the calculation for the translation symmetry, and list the full results of the classification. The calculation for other symmetries is in Appendix \ref{app:kagome_PEPS_Z2_PSG}. (We encourage interested readers to follow a simpler and pedagogical example on the square lattice in Appendix \ref{app:square_C4} before reading this more technical Appendix for the kagome lattice.)

Let us consider the translation symmetry group. This group is isomorphic to $Z\times Z$: the group is defined by its generators $T_1$, $T_2$ as well as the relation between generators,
\begin{align}
  T_2^{-1}T_1^{-1}T_2T_1=\mathrm{e}
  \label{}
\end{align}
As shown in Eq.(\ref{eq:lattice_symmetric_tensor}), for PEPS symmetric under $T_i$ $(i=1,2)$, we have
\begin{align}
  &T^{(x,y,s)}=\Theta_{T_i}W_{T_i}T_i\circ T^{(x,y,s)}\notag\\
  &B_{(xysi|x'y's'i')}=W_{T_i}T_i\circ B_{(xysi|x'y's'i')}
\end{align}

From the group relation $T_2^{-1}T_1^{-1}T_2T_1=\mathrm{e}$, we have
\begin{align}
  &W_{T_2}^{-1}(T_2(x,y,s,i))W_{T_1}^{-1}(T_1T_2(x,y,s,i))W_{T_2}(T_1T_2(x,y,s,i))\notag\\
  &W_{T_1}(T_1(x,y,s,i))=\eta_{12}\chi_{12}(x,y,s,i)
  \label{eq:eta_12}
\end{align}
as well as
\begin{align}
  &\Theta^*_{T_2}{(T_2(x,y,s))}\Theta^*_{T_1}{(T_1T_2(x,y,s))}\Theta_{T_2}{(T_1T_2(x,y,s))}\notag\\
  &\Theta_{T_1}{(T_1(x,y,s))}=\mu_{12}\prod_i\chi^*_{12}(x,y,s,i)
\end{align}
where $\eta_{12}\in\{\mathrm{I},\mathrm{J}\}$, and $\{\chi_{12}(x,y,s,i)\}\in \chi-group$.

Under transformations $W_{T_i}\rightarrow\varepsilon_{T_i}W_{T_i}$ and $\Theta_{T_i}\rightarrow\varepsilon_{T_i}\Theta_{T_i}$, we have
\begin{align}
  &\chi_{12}\rightarrow\varepsilon_{T_2}^*(x,y+1,s,i)\varepsilon_{T_1}^*(x+1,y+1,s,i)\cdot\notag\\
  &\varepsilon_{T_2}(x+1,y+1,s,i)\varepsilon_{T_1}(x+1,y,s,i)\chi_{12}(x,y,s,i)
  \label{}
\end{align}
Thus, we are able to set all $\chi_{12}(x,y,s,i)=1$ via the $\varepsilon_{T_i}$-ambiguity.

According to Eq.(\ref{eq:gauge_transf_on_psg_coord}) and Eq.(\ref{eq:phase_factor_transf}), by doing a gauge transformation $V(x,y,s,i)$ and multiply phase factors $\Phi(x,y,s)$:
\begin{align}
  W_{T_2}(x,y,s,i)&\rightarrow V(x,y,s,i)W_{T_2}(x,y,s,i)V^{-1}(x,y-1,s,i)\notag\\
  \Theta_{T_2}{(x,y,s)}&\rightarrow\Theta_{T_2}{(x,y,s)}\Phi{(x,y,s)}\Phi^*{(x,y-1,s)}
\end{align}
We are able to set $W_{T_2}(x,y,s,i)=\mathrm{I}$ as well as $\Theta_{T_2}{(x,y,s,i)}=1$. Thus we obtain $T^{(x,y,s)}=T^{(0,y,s)}$.  The remaining $V$-ambiguity preserving the form of $W_{T_2}$ should satisfy $V(x,y,s,i)=V(x,0,s,i)$, and the remaining $\Phi$-ambiguity preserving the form of $\Theta_{T_2}$ should satisfy $\Phi(x,y,s)=\Phi(x,0,s)$. In addition, any nontrivial $\varepsilon_{T_2}$ transformation will change the form of $W_{T_2}=\mathrm{I}$, so $\varepsilon_{T_2}$ is fixed to be $1$. Together with the condition $\chi_{12}(x,y,s,i)=1$, the remaining $\varepsilon_{T_1}$-ambiguity satisfies $\varepsilon_{T_1}(x,y,s,i)=\varepsilon_{T_1}(x,0,s,i)$.

Similarly, for $T_1$ transformation, using the remaining $V$-ambiguity and $\Phi$-ambiguity, we have
\begin{align}
  W_{T_1}(x,y,s,i)&\rightarrow V(x,0,s,i)W_{T_1}(x,y,s,i)V^{-1}(x-1,0,s,i)\notag\\
  \Theta_{T_1}{(x,y,s)}&\rightarrow\Theta_{T_1}{(x,y,s)}\Phi{(x,0,s)}\Phi^*{(x-1,0,s)}
\end{align}
Thus we can set $W_{T_1}(x,0,s,i)=\mathrm{I}$ and $\Theta_{T_1}(x,0,s)=1$. To maintain this form of $W_{T_1}$, we find that there is no remaining $\varepsilon_{T_1}$-ambiguity: $\varepsilon_{T_1}$ is fixed to be $1$. The remaining $V$-ambiguity and $\Phi$-ambiguity satisfy $V(x,y,s,i)=V(s,i)$ and $\Phi{(x,y,s)}=\Phi(s)$; namely they are only dependent on the sublattice index and the virtual leg index from a site, but are independent of the unit cell coordinates. Further, in this gauge, site tensors are translational invariant (but could be sublattice dependent),
\begin{align}
  T^{(x,y,s)}=T^{(x,0,s)}=T^{s}\doteq T^{(0,0,s)},\quad s=u,v,w
\end{align}

Thus, in the gauge that we choose so far, we can solve Eq.(\ref{eq:eta_12}), and get the implementation of translation symmetry on PEPS as
\begin{align}
  &W_{T_1}(x,y,s,i)=\eta_{12}^y\notag\\
  &W_{T_2}(x,y,s,i)=\mathrm{I}\notag\\
  &\Theta_{T_1}{(x,y,s)}=\mu^y_{12}\notag\\
  &\Theta_{T_2}{(x,y,s)}=1
  \label{}
\end{align}

So for systems with translational symmetries and $IGG=Z_2$, there are at least two distinct classes of wavefunction. In the context of quantum spin liquids, these two classes are known as {\it zero flux state} and {\it $\pi$ flux state}, corresponding to $\eta_{12}=\mathrm{I}$ and $\eta_{12}=\mathrm{J}$ respectively. Condensations of spinons in these two spin liquids lead to different types of magnetic orders\cite{Wang:2006p174423}. In the above gauge, although all site tensors related by the translation symmetry share the same form, bond states related by the translation symmetry are in general \emph{different} if $\eta_{12}$ is nontrivial.

The calculation for other symmetries is similar as the above procedure. The basic idea is to keep fixing gauge by the four ambiguities. And when we find certain algebraic data, such as the $\eta_{12}$ introduced above, that cannot be removed by the ambiguities, they describe different symmetric PEPS classes. We only list the result here, and put details in Appendix \ref{app:kagome_PEPS_Z2_PSG}. 

This classification scheme will always lead to three finite sets of algebraic indices $\eta$'s, $\chi$'s and $\Theta$'s and we will discuss their physical meanings in Sec.\ref{sec:lantern_operator_peps}. Although in general systems every set of indices is nonempty, for a half-integer spin  system on the kagome lattice described by PEPS with $IGG=Z_2$, we have:
\begin{itemize}
  \item $\eta_{12}$, $\eta_{C_6}$ and $\eta_\sigma$, where $\eta\in\{\mathrm{I},\mathrm{J}\}$. The corresponding $\mu_{12},\mu_{C_6},\mu_{\sigma}$ are determined by $\eta$'s.
  \item $\chi_\sigma$ and $\chi_{\mathcal{T}}$, where $\chi=\pm1$.
  \item There turns out to be no tunable $\Theta$ indices in this example.
\end{itemize}
So the number of classes equals to $2^5=32$. By choosing a gauge, the symmetry operations on PEPS can be solved as
\begin{align}
  &W_{T_1}(x,y,s,i)=\eta_{12}^y,\notag\\
  &W_{T_2}(x,y,s,i)=\mathrm{I}\notag,\\
  &W_{C_6}(x,y,u,i)=\eta_{12}^{xy+\frac{1}{2}x(x+1)+x+y}w_{C_6}(u,i),\notag\\
  &W_{C_6}(x,y,v,i)=\eta_{12}^{xy+\frac{1}{2}x(x+1)+x+y},\notag\\
  &W_{C_6}(x,y,w,i)=\eta_{12}^{xy+\frac{1}{2}x(x+1)},\notag\\
  &W_\sigma(x,y,s,i)=\eta_{12}^{x+y+xy}w_\sigma(s,i),\notag\\
  &W_{\mathcal{T}}(x,y,s,i)=w_{\mathcal{T}}(s,i),\notag\\
  &W_{\theta\vec{n}}(x,y,s,i)=\bigoplus_i(\mathrm{I}_{n_i}\otimes\ee^{\ii\theta\vec{n}\cdot\vec{S}_i}).
\end{align}
In this gauge all $W_R$ matrices are unitary. The last equation is for the $SU(2)$ spin rotation along $\vec n$ direction by an angle $\theta$. In addition, in this gauge we choose $\mathrm{J}=W_{2\pi}(x,y,s,i)=\bigoplus_i(\mathrm{I}_{n_i}\otimes\ee^{\ii2\pi\vec{n}\cdot\vec{S}_i})$; namely $\mathrm{J}$ is the direct sum of $\mathrm{I}_{D_1}$ for the integer spin subspace and $-\mathrm{I}_{D_2}$ for the half-integer spin subspace and $D_1+D_2=D$.

For the rotation transformation $w_{C_6}(u,i)$, we have
\begin{align}
  &w_{C_6}(u,a)=w_{C_6}(u,c)=\mathrm{I},\notag\\
  &w_{C_6}(u,b)=w_{C_6}(u,d)=\eta_{12}\eta_{C_6},
\end{align}
For the reflection transformation $w_{\sigma}(s,i)$, we have
\begin{align}
  &w_\sigma(u,a)=\mathrm{I},&\quad &w_\sigma(u,b)=\chi_\sigma\eta_{12}\eta_{C_6},\notag\\
  &w_\sigma(u,c)=\chi_\sigma\eta_{12}\eta_{C_6}\eta_\sigma,&\quad &w_\sigma(u,d)=\eta_\sigma;\notag\\
  &w_\sigma(v,a)=\eta_{12},&\quad &w_\sigma(v,b)=\chi_\sigma\eta_{12},\notag\\
  &w_\sigma(v,c)=\eta_{C_6}\eta_\sigma,&\quad &w_\sigma(v,d)=\chi_\sigma\eta_{C_6}\eta_\sigma;\notag\\
  &w_\sigma(w,a)=\chi_\sigma\eta_{C_6},&\quad &w_\sigma(w,b)=\eta_{C_6},\notag\\
  &w_\sigma(w,c)=\eta_{12}\eta_\sigma,&\quad &w_\sigma(w,d)=\chi_\sigma\eta_{12}\eta_\sigma;
\end{align}
And for the time reversal transformation $w_{\mathcal{T}}$, we have
\begin{align}
  &w_\mathcal{T}(u,a)=w_\mathcal{T},&\quad &w_\mathcal{T}(u,b)=\eta_{12}\eta_{C_6}w_\mathcal{T},\notag\\
  &w_\mathcal{T}(u,c)=\eta_{12}\eta_{C_6}\eta_{\sigma}w_\mathcal{T},&\quad &w_\mathcal{T}(u,d)=\eta_{\sigma}w_\mathcal{T};\notag\\
  &w_\mathcal{T}(v,a)=\eta_{12}\eta_{C_6}w_\mathcal{T},&\quad &w_\mathcal{T}(v,b)=w_\mathcal{T},\notag\\
  &w_\mathcal{T}(v,c)=\eta_{\sigma}w_\mathcal{T},&\quad &w_\mathcal{T}(v,d)=\eta_{12}\eta_{C_6}\eta_{\sigma}w_\mathcal{T};\notag\\
  &w_\mathcal{T}(w,a)=w_\mathcal{T},&\quad &w_\mathcal{T}(w,b)=\eta_{12}\eta_{C_6}w_\mathcal{T},\notag\\
  &w_\mathcal{T}(w,c)=\eta_{12}\eta_{C_6}\eta_{\sigma}w_\mathcal{T},&\quad &w_\mathcal{T}(w,d)=\eta_{\sigma}w_\mathcal{T};
\end{align}
where
\begin{align}
  w_\mathcal{T}=
  \left\{
    \begin{array}{l l}
      \bigoplus_i(\mathrm{I}_{n_i}\otimes\ee^{\ii\pi S_i^y})&\quad \text{if $\chi_\mathcal{T}=1$}\\
      \bigoplus_i(\Omega_{n_i}\otimes\ee^{\ii\pi S_i^y})&\quad \text{if $\chi_\mathcal{T}=-1$}
    \end{array}
    \right.
  \label{}
\end{align}
Here $n_i$ is dimension of the extra degeneracy associated with spin-$S_i$. Namely, the total degeneracy for spin-$S_i$ living on one virtual leg equals $n_i\times(2S_i+1)$. We have the virtual bond dimension
\begin{align}
  D=\sum_i n_i(2S_i+1)
  \label{}
\end{align}
And, $\Omega_{n_i}=\ii\sigma_y\otimes\mathrm{I}_{n_i/2}$ is a $n_i$ dimensional antisymmetric matrix.

For $\Theta_R$'s, we have
\begin{align}
  &\Theta_{T_1}(x,y,s)=\mu_{12}^y,\notag\\
  &\Theta_{T_2}(x,y,s)=1,\notag\\
  &\Theta_{C_6}(x,y,u)=\mu_{12}^{xy+\frac{1}{2}x(x+1)+x+y}\Theta_{C_6}(u),\notag\\
  &\Theta_{C_6}(x,y,v)=\mu_{12}^{xy+\frac{1}{2}x(x+1)+x+y},\notag\\
  &\Theta_{C_6}(x,y,w)=\mu_{12}^{xy+\frac{1}{2}x(x+1)},\notag\\
  &\Theta_\sigma(x,y,s)=\mu_{12}^{x+y+xy}\Theta_\sigma(s),\notag\\
  &\Theta_\mathcal{T}(x,y,u/w)=1,\notag\\
  &\Theta_\mathcal{T}(x,y,v)=\mu_{12}\mu_{C_6}\notag,\notag\\
  &\Theta_{\theta\vec{n}}=1,
\end{align}
where
\begin{align}
  &\Theta_{C_6}(u)=(\mu_{12}\mu_{C_6})^{\frac{1}{2}};\notag\\
  &\Theta_\sigma(u)=(\mu_\sigma)^{\frac{1}{2}};\notag\\
  &\Theta_\sigma(v)=\mu_{C_6}\Theta_{C_6}(u)\Theta_\sigma(u);\notag\\
  &\Theta_\sigma(w)=\mu_\sigma\mu_{C_6}(\Theta_{C_6}(u)\Theta_\sigma(u))^{-1}.\label{eq:kagome_theta}
\end{align}
Note that in Eq.(\ref{eq:kagome_theta}) $\Theta_{C_6}(u)$ and $\Theta_\sigma(u)$ contain square roots so there appear to be two possible values of each of them differing by a minus sign, giving rise to $\Theta$-indices. However, these minus signs can be tuned away using the $\eta$-ambiguities in the definition of $W_{C_6}$ and $W_{\sigma}$ since every site tensor is $Z_2$ odd. So one could simply fix an arbitrary choice for the square roots here. This is the reason why there turns out to be no tunable $\Theta$ indices in this example. 

Even after all these transformation rules are determined by gauge fixing, we still have some remaining $V$-ambiguity for each class. (Note that there is no remaining nontrivial $\eta$,$\varepsilon$ and $\Phi$ ambiguities.) To preserve the lattice symmetry, the remaining $V$-ambiguity is independent of sites and legs. To preserve the form of $W_{\theta\vec{n}}$, the remaining $V$-ambiguity must have the following form:
\begin{align}
      V=\bigoplus_i(\widetilde{V}_{S_i}\otimes\mathrm{I}_{2S_i+1}),
      \label{eq:remaining_V}
\end{align}
where $\widetilde{V}_{S_i}$ is a $n_i$ dimensional matrix.
In addition, the time-reversal transformation $W_{\mathcal{T}}$ further constrains the form of component matrices $\widetilde{V}_{S_i}$.  When $\chi_{\mathcal{T}}=1$, one can show that $\widetilde{V}_{S_i}$ must be a \emph{real} matrix. For the purpose of presentation we only consider $\chi_{\mathcal{T}}=1$ classes here. The $\chi_{\mathcal{T}}=-1$ cases involve quaternion matrices and we leave the general and detailed discussions in Appendix \ref{app:kagome_PEPS_Z2_PSG}.

Next, we are at the stage to construct the constrained sub-Hilbert spaces for building block tensors for all classes, according to the $W_R$ transformation rules. The basic idea is to determine the generic form of a single site/bond tensor using the $W_R$'s with $R$ leaving the site/bond invariant, and then generate all other site/bond tensors using all $W_R$'s. The generic forms of site tensors are straightforwardly determined in this fashion, with a set of real continuous variational parameters whose number basically equals the dimension of the constrained site sub-Hilbert space. However, for bond tensors, we will use the remaining $V$-ambiguity to bring them into canonical forms which are  maximal entangled bond states containing \emph{no} continuous variational parameters. 

To make sure a bond tensor $B_b$ to be invariant under the $SU(2)$ spin rotation, it must have the following form:
\begin{align}
  B_b=\bigoplus_{i=1}^{M}\left( \widetilde{B}_b^{S_i}\otimes K_{S_i} \right),
  \label{eq:bond_constrained_tensor}
\end{align}
where $\widetilde{B}_b^{S_i}$ is $n_i$ dimensional matrix, and $K_{S_i}$ is the fixed $(2S_i+1)$ dimensional matrix representing the spin singlet formed by two spin-$S_i$ on the two virtual legs shared by $B_b$. For example, we get $K_{S=0}=1$, $K_{S=\frac{1}{2}}=\ii\sigma_y$.

As shown in Appendix \ref{app:kagome_PEPS_Z2_PSG}, when $\chi_{\mathcal{T}}=1$ and a given $S_i$, depending on the four possible values of $\eta_{\sigma}$ and $\chi_{\sigma}$, the component matrix $\widetilde{B}_b^{S_i}$ must be a purely real/imaginary symmetric/antisymmetric matrix. Then we can use the remaining $V$-ambiguity in Eq.(\ref{eq:remaining_V}) to simplify $\widetilde{B}_b^{S_i}$, because under a $\widetilde{V}_{S_i}$ transformation, $\widetilde{B}_b^{S_i}$ transforms as:
 \begin{align}
      \widetilde{B}_b^{S_i}\rightarrow\widetilde{V}_{S_i}\cdot\widetilde{B}_b^{S_i}\cdot\widetilde{V}_{S_i}^\mathrm{t}
      \label{eq:remaining_V_transform_B}
 \end{align}
Clearly we can use a real orthogonal $\widetilde{V}_{S_i}$ to diagonalize (block diagonalize) $\widetilde{B}_b^{S_i}$ if $\widetilde{B}_b^{S_i}$ is a symmetric (antisymmetric) matrix. After this, the eigenvalues of $\widetilde{B}_b^{S_i}$ could have arbitrary norms. But then we can use another real diagonal $\widetilde{V}_{S_i}$ matrix to normalize the eigenvalues so that they are only $\pm 1$ (if $\widetilde{B}_b^{S_i}$ is purely real) or $\pm\ii$ (if $\widetilde{B}_b^{S_i}$ is purely imaginary). 

This procedure fixes $B_b$ to be maximal entangled states with no continuous variational parameters. However, the relative number of $+1$($+\ii$) eigenvalues and $-1$($-\ii$) eigenvalues cannot be further tuned away by gauge fixing and will serve as discrete variational parameters on the bond tensors.

The previous discussions in the subsection are general for any half-integer spin-$S$. \emph{Below we focus on the case with $S=\frac{1}{2}$.} For simplicity, we demonstrate the results for with $D=3$. The basis of virtual legs of site tensors are $\{|0\rangle,|\uparrow\rangle,|\downarrow\rangle\}$. Namely, virtual legs are formed by one spin singlet and one spin doublet. Note that virtual legs of bond tensors are dual to those of site tensors, so the basis are $\langle0|,\langle\uparrow|,\langle\downarrow|$.  Symmetric PEPS with larger $D$ are also conceptually straightforward but technically involved to obtain, and we leave the general construction in Appendix \ref{app:kagome_PEPS_Z2_PSG}

As discussed in Appendix \ref{app:kagome_PEPS_Z2_PSG}, only classes satisfying $\eta_\sigma=\mathrm{J}$, $\chi_\sigma=1$ and $\chi_\mathcal{T}=1$ can be realized with $D=3$. So the realizable classes reduce to $2^2=4$ with $D=3$. At such a small $D$, it turns out that each class has only two continuous variational parameters. (Note that for $D=6$, i.e. two spin singlet and two spin doublet on the virtual leg, we find that all the 32 classes can be realized. And each class has 47 continuous variational parameters.) Following the above procedure we can bring the bond tensor on a given bond $b_0$ into the canonical form:
\begin{align}
  B_{b_0}=
\begin{pmatrix}
  \pm1 & 0 & 0 \\
  0 & 0 & -\mathrm{i} \\
  0 & \mathrm{i} & 0 \\
\end{pmatrix}
\label{eq:bond_tensor_D3_main}
\end{align}
All other bond tensors are generated by combination of translation and rotation symmetries as:
\begin{align}
  B_{R(b)}=R^{-1}W_RR\circ B_{b_0}
  \label{}
\end{align}
where $R=T_1^{n_1}T_2^{n_2}C_6^{n_{C_6}}$ with $n_1,n_2,n_{C_6}\in\mathbb{Z}$.

One can view a bond tensor as a quantum state living in the Hilbert space formed by the tensor product of two virtual legs. Namely, we have
\begin{align}
  \hat{B}_{b_0}=\pm\langle0,0|-\ii\,\langle\uparrow,\downarrow|+\ii\,\langle\downarrow,\uparrow|
  \label{}
\end{align}
Here we use notation $\hat{B}_{b_0}$ as the quantum state representation while $B_{b_0}$ as the matrix (tensor) representation.

At a given site $s_0$, the generic form of the site tensor for all classes can be summarized as:
\begin{align}
  \hat{T}^{s_0}=&\{\hat{K}_0+\hat{K}_{12}(p_1,p_2)\}+\Theta_{C_6}(u)\{a\leftrightarrow b,c\leftrightarrow d\}+\Theta_{\sigma}(u)\cdot\notag\\
  &\{a\leftrightarrow d,b\leftrightarrow c\}+\mu_{12}\mu_{C_6}\Theta_{C_6}(u)\Theta_{\sigma}(u)\{a\leftrightarrow c,b\leftrightarrow d\}\notag\\
  \label{eq:site_tensor_D3_main}
\end{align}
with real continuous parameters $p_1,p_2$. Here $a,b,c,d$ denote virtual leg of sites, as shown in Fig.(\ref{fig:kagome_lattice}). $\hat{K}_0$ and $\hat{K}_{12}$ denote linear independent spin singlet states, which can be expressed as
\begin{align}
  \hat{K}_0=&|\uparrow\rangle\otimes|\downarrow000\rangle-|\downarrow\rangle\otimes|\uparrow000\rangle\notag\\
  \hat{K}_{12}=&p_1\cdot(|\uparrow\rangle\otimes|0\downarrow\uparrow\downarrow\rangle+|\downarrow\rangle\otimes|0\uparrow\downarrow\uparrow\rangle)+\notag\\
  &p_2\cdot(|\uparrow\rangle\otimes|0\downarrow\downarrow\uparrow\rangle+|\downarrow\rangle\otimes|0\uparrow\uparrow\downarrow\rangle)-\notag\\
  &(p_1+p_2)\cdot(|\uparrow\rangle\otimes|0\uparrow\downarrow\downarrow\rangle+|\downarrow\rangle\otimes|0\downarrow\uparrow\uparrow\rangle),
\end{align}
where the first spin lives on the physical leg, while the following four spins live on virtual legs $a,b,c,d$ respectively. Note that we have chosen a particular gauge such that all site tensors share the same form.

By direct comparison, the NN RVB state ($Q_1=Q_2$ state) given in Sec.(\ref{subsec:peps_rvb_example}) is represented as the PEPS defined in Eq.(\ref{eq:bond_tensor_D3_main}) and Eq.(\ref{eq:site_tensor_D3_main}), with $p_1=p_2=0$ and:
\begin{align}
  &\eta_{12}=\eta_{C_6}=\mathrm{I},\quad\eta_{\sigma}=\mathrm{J};\notag\\
  &\chi_{\sigma}=\chi_{\mathcal{T}}=1;\notag\\
\label{eq:Q1_Q2_RVB_class}
\end{align}

\subsection{Algorithm for minimization}\label{sec:minimization}
As demonstrated above, after the symmetry transformation rules are determined for each class, for a fixed bond dimension $D$ together with the specified on-site symmetry (projective) representation on this virtual Hilbert space, we can construct the generic symmetric PEPS. The strategy is to firstly construct one site tensor and one bond tensor, and to use spatial symmetries to generate all site tensors and bond tensors. For each class, in general, the maximal entangled bond tensor will be completely fixed up to a finite number of $\pm1$ signs. These signs are physical (they could modify the energy density) and should be treated as discrete variational parameters. The site tensor, however, will be fixed in a sub-Hilbert space for each give class, whose dimension (minus one if one considers normalized PEPS) represents the number of continuous variational parameters. Different classes give 
different sub-Hilbert spaces for the site tensors.

For efficient PEPS simulations, one generally chooses an open-boundary sample and introduces certain boundary conditions. In order to determine which symmetric PEPS class that the ground state of a model Hamiltonian belongs to, one needs to minimize the energy density \emph{near the center} of the sample by tuning all the variational parameters of each given class. And the ground state crude class is identified as the class which gives the lowest energy density. The effect of boundary conditions should decay quickly into the bulk of sample and the optimal energy density near sample center is expected to converge to a boundary condition independent value for intermediate sample sizes. 

By construction, PEPS belonging to different classes has distinct short-range physics (the site tensors and bond tensors transform according to inequivalent algebraic equations), therefore we expect that quite generically these optimal energy densities for different classes are significantly different. But, there are three scenarios that the optimal energy densities for distinct classes can be identical, at least in the thermodynamic limit. The first scenario is that the quantum Hamiltonian under investigation is right at a phase transition point, so that the optimal energy densities for the two symmetric PEPS on the two sides of the phase transition are the same. This scenario, however, is expect to occur only at phase transition points.

The second scenario has a deeper physical origin --- it is possible that the invariant gauge group of the PEPS representation of a quantum phase is larger than the assumed $IGG$ for the symmetric PEPS classification. We will comment more on this scenario at the end of the paper. For instance, in the example of kagome spin-1/2 systems, we classify all the symmetric PEPS whose $IGG=Z_2$ in the previous section. But it is possible that a given quantum phase actually require $IGG$ larger than $Z_2$. A PEPS with a larger $IGG$ indicates more invariance of the tensor network, and therefore it could be well approximated (with arbitrarily small wavefunction differences) by PEPS with the minimal required $Z_2$ $IGG$. Physically this is related to the Higgs mechanism breaking the large $IGG$ down to $Z_2$. In particular, it is possible that the same PEPS with a larger $IGG$ can be well approximated by two or more distinct $IGG=Z_2$ PEPS classes. This picture can actually be used as a feature of the algorithm proposed 
here. Namely,
 if 
numerically one 
finds that two distinct classes of $IGG=Z_2$ PEPS give the same optimal energy densities over a finite region in the phase diagram, it could be due to the fact that the true $IGG$ of the quantum phase is larger than $Z_2$.

The third scenario is the one that we currently do not fully understand. It seems possible that certain symmetry breaking conventional phases such as a valence bond solid phase that can be represented by distinct classes even with the minimal required $IGG$. It would be very interesting to perform numerically simulations on quantum models to understand whether this really occurs or not, and if this scenario occurs, how does it occur. We will comment further on this issue later in Sec.\ref{sec:torus_long_range_order}.

Note that although the bond tensors are position dependent in our construction, it is straightforward to absorb each bond tensor to a neighboring site tensor so that all the bond tensors become identity matrices. After this transformation the PEPS is in the conventional form used in numerical simulations. As a result, for a PEPS with given variational parameters, the measurement of the energy density near the sample center, which is one or few terms in the translational symmetric Hamiltonian, can be carried out in the same way as in the original PEPS algorithms\cite{Verstraete:2004p,Verstraete:2008p143}. The only new ingredient of the current minimization scheme is an algorithm to perform the energy density minimization by tuning the continuous variational parameters within each class. Below we propose such an algorithm based on the conjugate gradient method. Similar minimization algorithm was investigated in the context of one-dimensional matrix product states with translational symmetry\cite{Pirvu:2011p125104}.

\begin{figure}
 \includegraphics[width=0.5\textwidth]{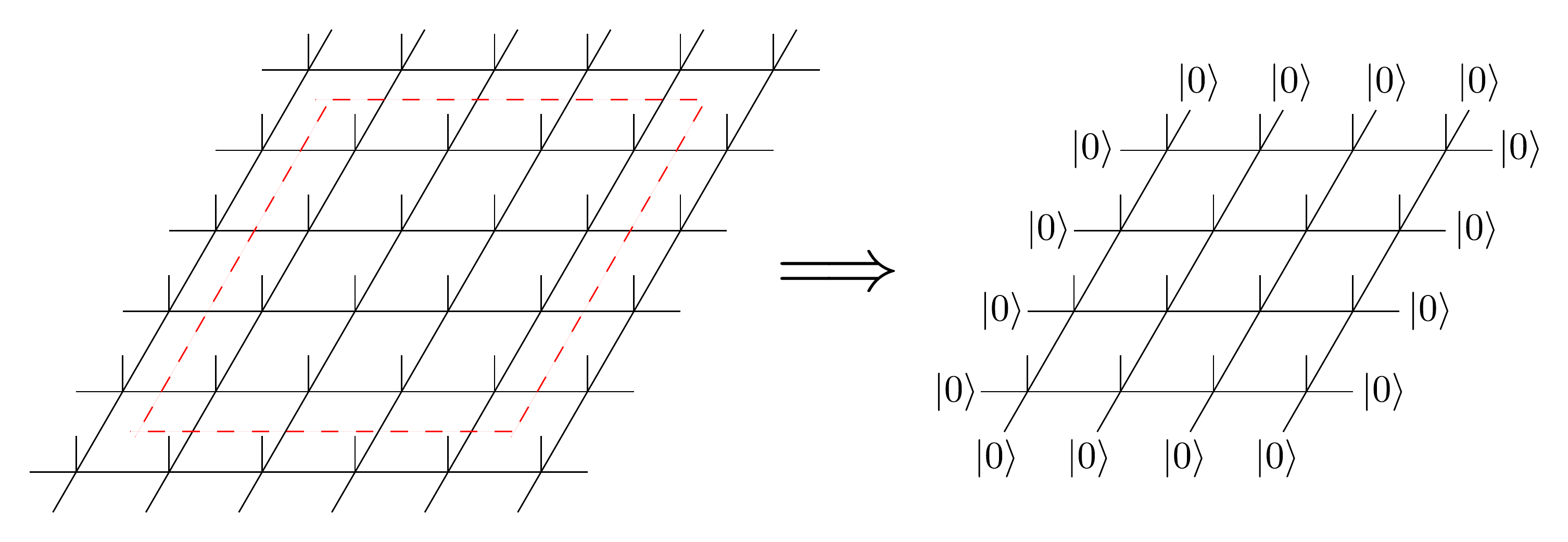}
 \caption{Illustration of truncating an infinite PEPS to a finite size PEPS with a choice of boundary condition.}
 \label{fig:truncation}
\end{figure}

Before moving towards the minimization algorithm, we introduce the well-established numerical methods in PEPS to perform energy density measurements\cite{Verstraete:2004p,Verstraete:2008p143}. As a demonstration of principles, let us consider a spin-1/2 symmetric PEPS on the square lattice; namely, for any site (bond) on the infinite square lattice, we have a well-defined site (bond) tensor. We firstly cut the infinite square lattice to a $L_x\times L_y$ finite sample. And the open boundary condition around the edge of the sample is specified by a certain choice of the virtual quantum states living on the dangling bonds. For instance, one can choose a simple boundary condition by requiring all the virtual states on the dangling bonds to be the spin singlet $|0\rangle$ in virtual Hilbert space (see Fig.\ref{fig:truncation} for illustration).
\begin{figure}
 \includegraphics[width=0.5\textwidth]{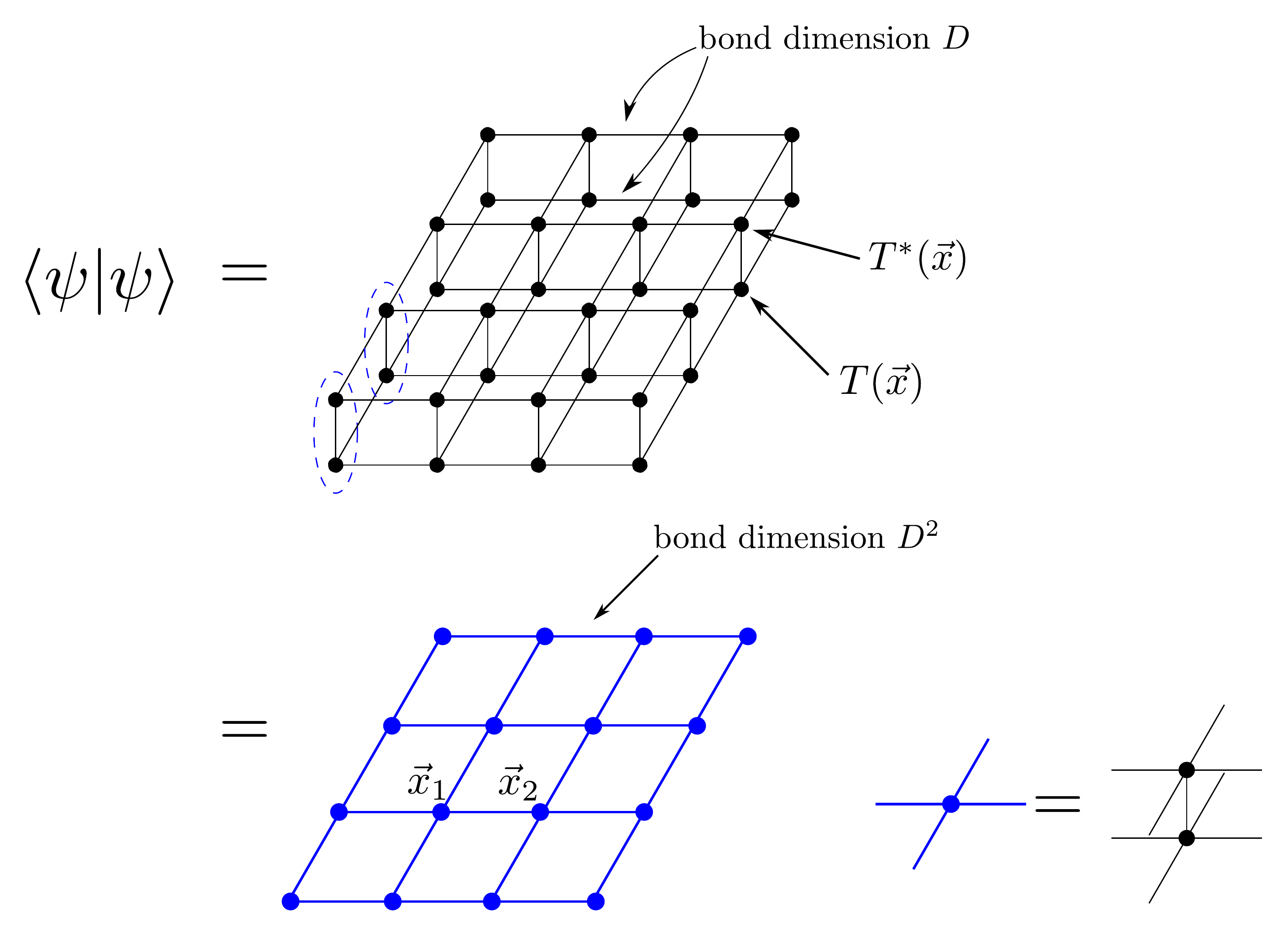}\\
 \includegraphics[width=0.5\textwidth]{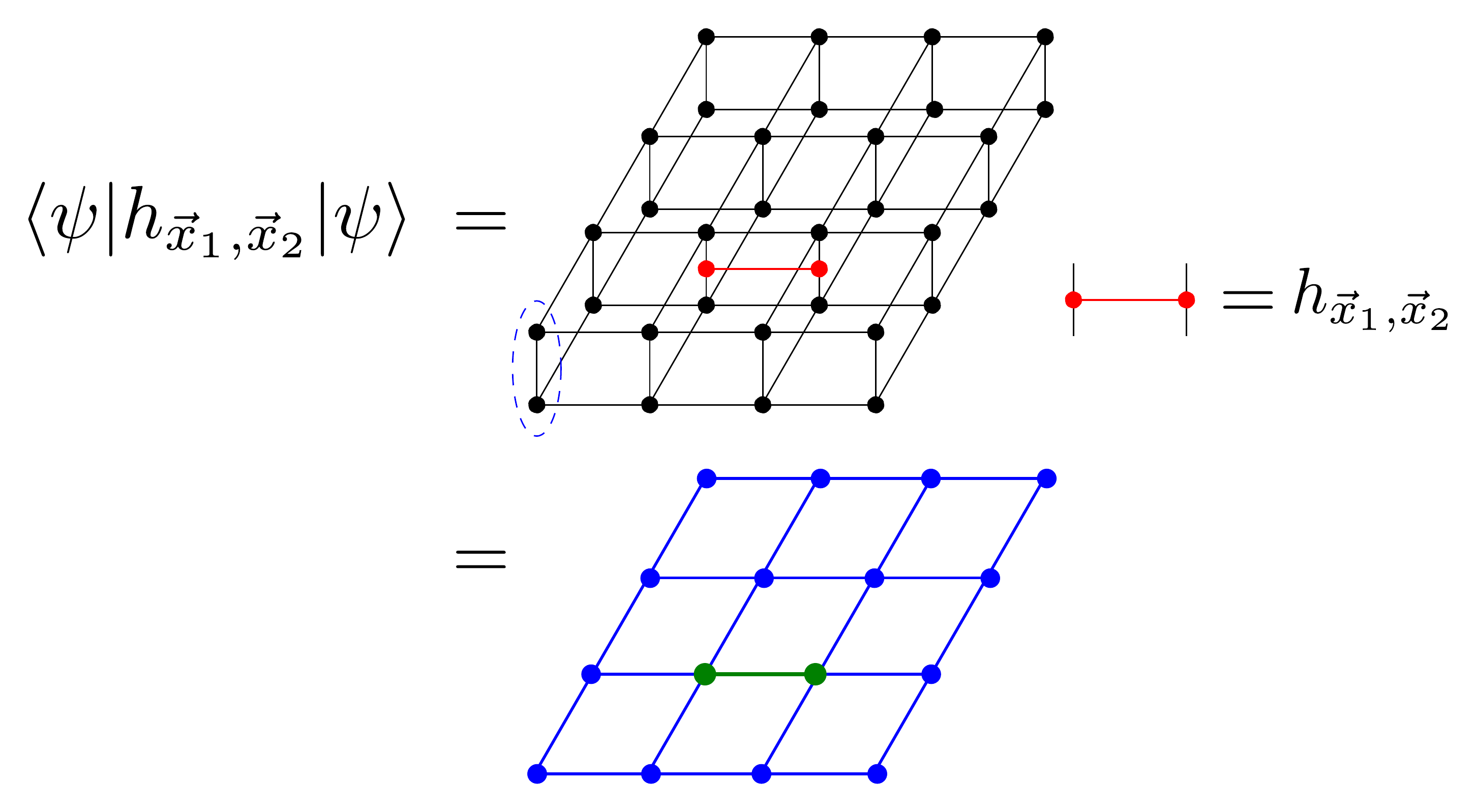}
 \caption{Illustration of measuring wavefunction norm and energy density of a PEPS.}
 \label{fig:measurement}
\end{figure}

With a well-defined finite size PEPS, the computation of wavefunction norm and the energy density near the center of the sample is illustrated in Fig.\ref{fig:measurement}. Note that the total Hamiltonian $H$ is a summation of many terms which are related to each other by lattice symmetry. For instance one can consider $H=\sum_{<\vec x \vec x'>}J\vec S_{\vec x}\cdot \vec S_{\vec x'}=\sum_{<\vec x \vec x'>}h_{\vec x\vec x'}$, where $<
\vec x\vec x'>$ represents the nearest neighbors and each energy density term is given by $h_{\vec x\vec x'}=J\vec S_{\vec x}\cdot \vec S_{\vec x'}$. Because by construction the PEPS is lattice space group symmetric, indicating that as system size increases, expectation values of all the energy density terms near the center of the sample would quickly converge to the same value. Therefore it is legitimate to optimize only a single energy density term as shown in Fig.\ref{fig:measurement}. 

These measurements become the standard problem in PEPS --- contracting a finite-size two-dimensional tensor network with the bond dimension $D^2$. Here the bond dimension is squared because we view a pair of a site tensor with its conjugate as a single tensor as shown in Fig.\ref{fig:measurement}. In general there is no way to exactly contract such a tensor network as long as the system size is not very small. However PEPS\cite{Verstraete:2004p} and other tensor network methods\cite{Levin:2007p120601,Gu:2008p205116,Xiang:2008p90603} provide efficient algorithms to approximately contract such tensor networks to high accuracies. The basic idea of the PEPS contraction algorithm is to view the first row of the 2d tensor network as a matrix product state (MPS), and view the other rows as matrix product operators acting on the MPS. Consequently the existing algorithm of compressing a MPS allows one to approximately contract the whole 2d tensor network in a row-by-row fashion. The details of the PEPS contraction 
algorithm can be found in review 
articles such as Ref.\onlinecite{
Verstraete:2008p143,Orus:2014p117}. 

Now we present an algorithm to perform the energy density minimization, based on the well-established conjugate gradient minimization algorithm. Namely, the quantity we need to minimize is:
\begin{align}
 E(\{p_i\})=\frac{\langle\psi|h_{\vec x\vec x'}|\psi\rangle}{\langle\psi|\psi\rangle},
\end{align}
where $\{p_i\}$ represents the finite collection of continuous variational parameters in a given symmetric PEPS class with a bond dimension $D$. When $\{p_i\}$ are tuned, every site tensor in the PEPS wavefunction $|\psi\rangle$ is modified. In order to apply the conjugate gradient minimization algorithm, one needs to compute the following derivatives:
\begin{align}
 \frac{\partial E}{\partial p_i}=\frac{1}{\langle\psi|\psi\rangle}\frac{\partial \langle\psi|h_{\vec x\vec x'}|\psi\rangle}{\partial p_i}-\frac{\langle\psi|h_{\vec x\vec x'}|\psi\rangle}{\langle\psi|\psi\rangle^2}\frac{\partial \langle\psi|\psi\rangle}{\partial p_i}.
\end{align}
The only new quantities that we need to compute are $\frac{\partial \langle\psi|\psi\rangle}{\partial p_i}$ and $\frac{\partial \langle\psi|h_{\vec x\vec x'}|\psi\rangle}{\partial p_i}$. But these quantities can also be efficiently computed using PEPS algorithms. 

\begin{figure}
 \includegraphics[width=0.4\textwidth]{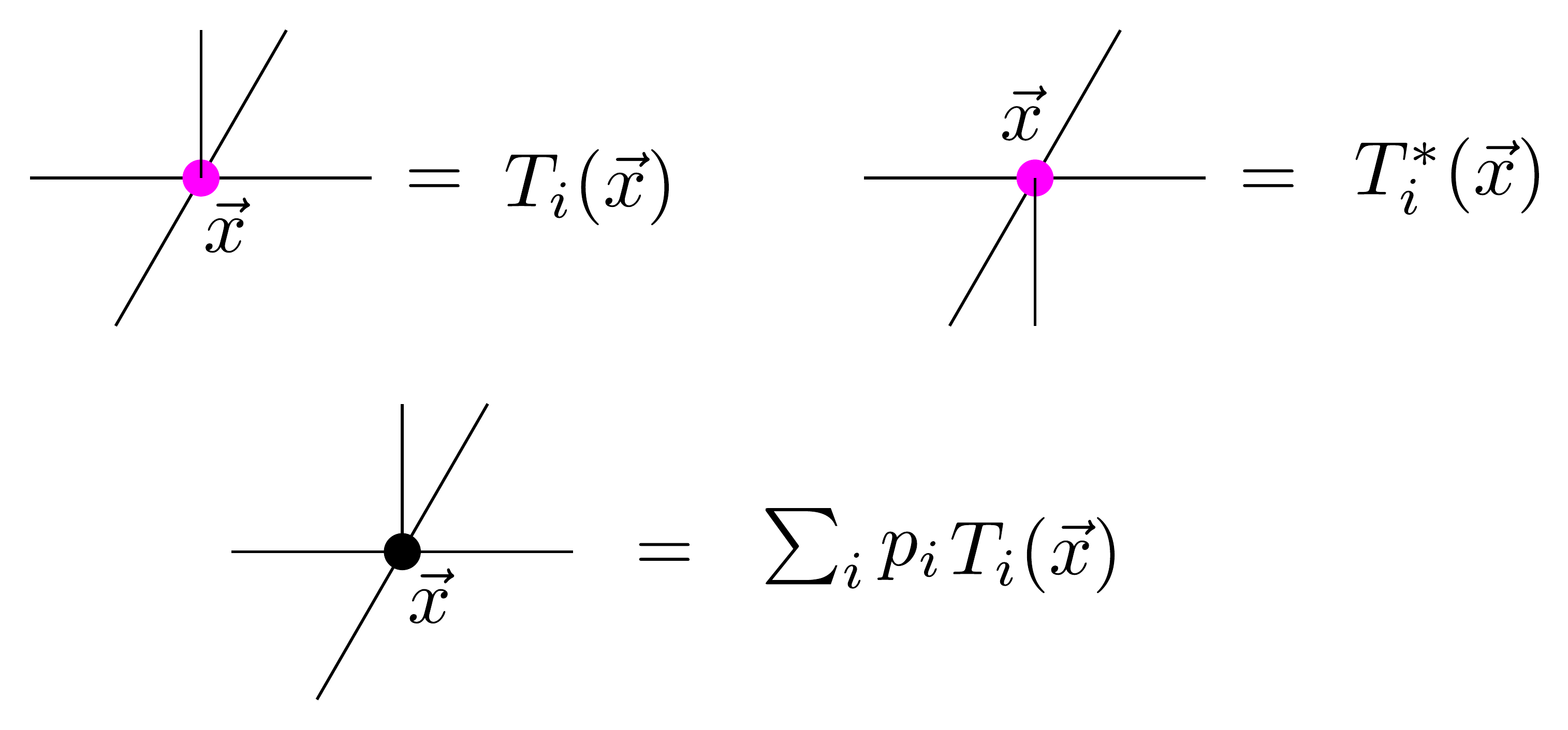}
 \caption{Each site tensor in a symmetric PEPS is a linear superposition of the states in a sub-Hilbert space.}
 \label{fig:site_tensor_variation}
\end{figure}

Due to the symmetric PEPS construction, we know that each site tensor is a linear superposition of the states in the symmetry constrained sub-Hilbert space (see Fig.\ref{fig:site_tensor_variation} for illustration):
\begin{align}
 T(\vec x)=\sum_i p_i T_i (\vec x),
\end{align}
where $\vec x$ labels the real space position of the site tensor. (Because the overall factor of all $p_i$'s does not change the normalized wavefunction, it may be convenient to set one of the $p_i$'s, say $p_0$ to be unity and only study the variation of other $p_i$'s.) Note that for different site $\vec x$, the form of $T_i (\vec x)$ are generally different due to the nontrivial symmetry transformation rules of tensors and the fact that we absorb the bond tensors into neighboring site tensors. 

\begin{figure}
 \includegraphics[width=0.4\textwidth]{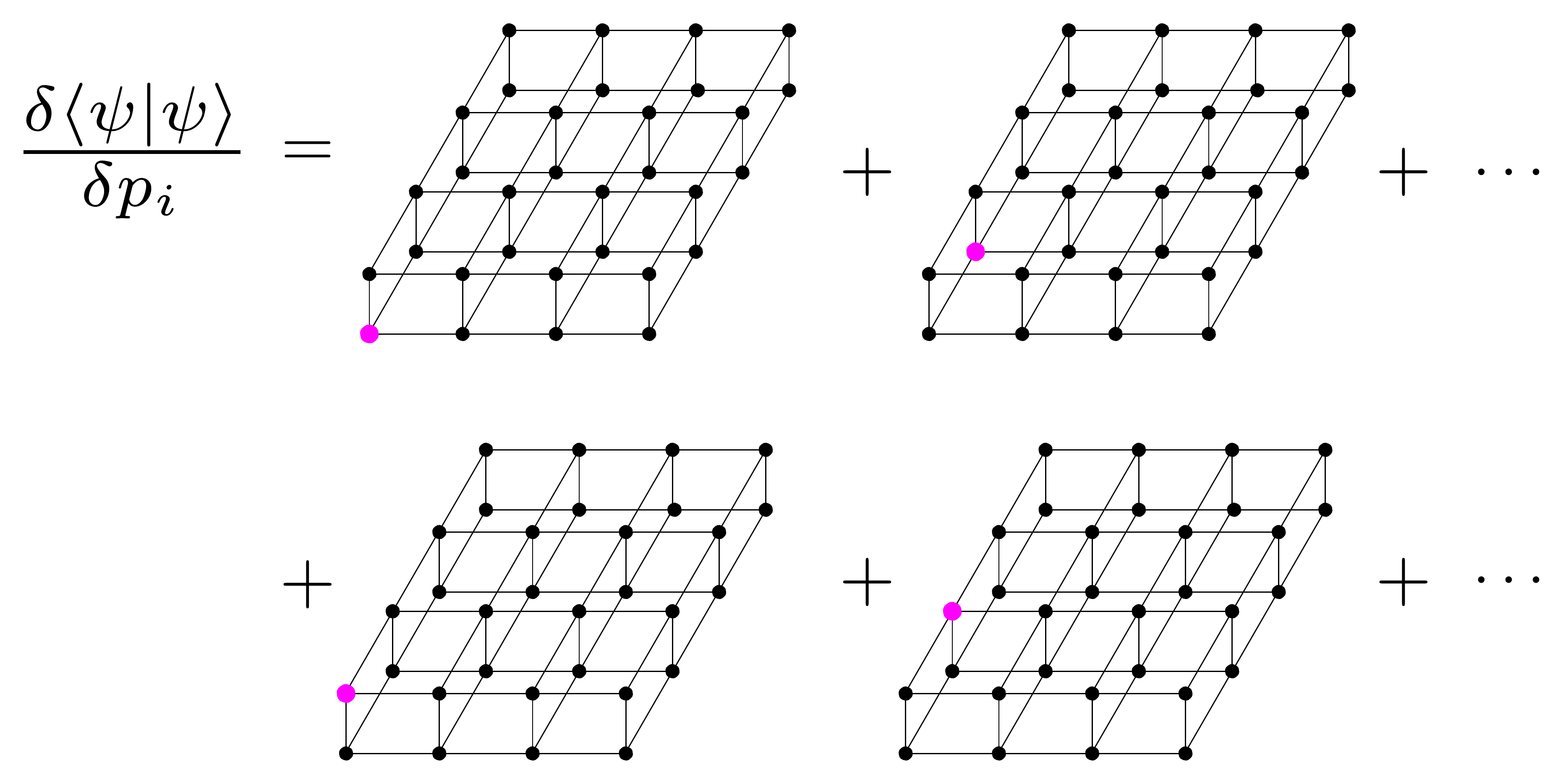}\\
 \includegraphics[width=0.5\textwidth]{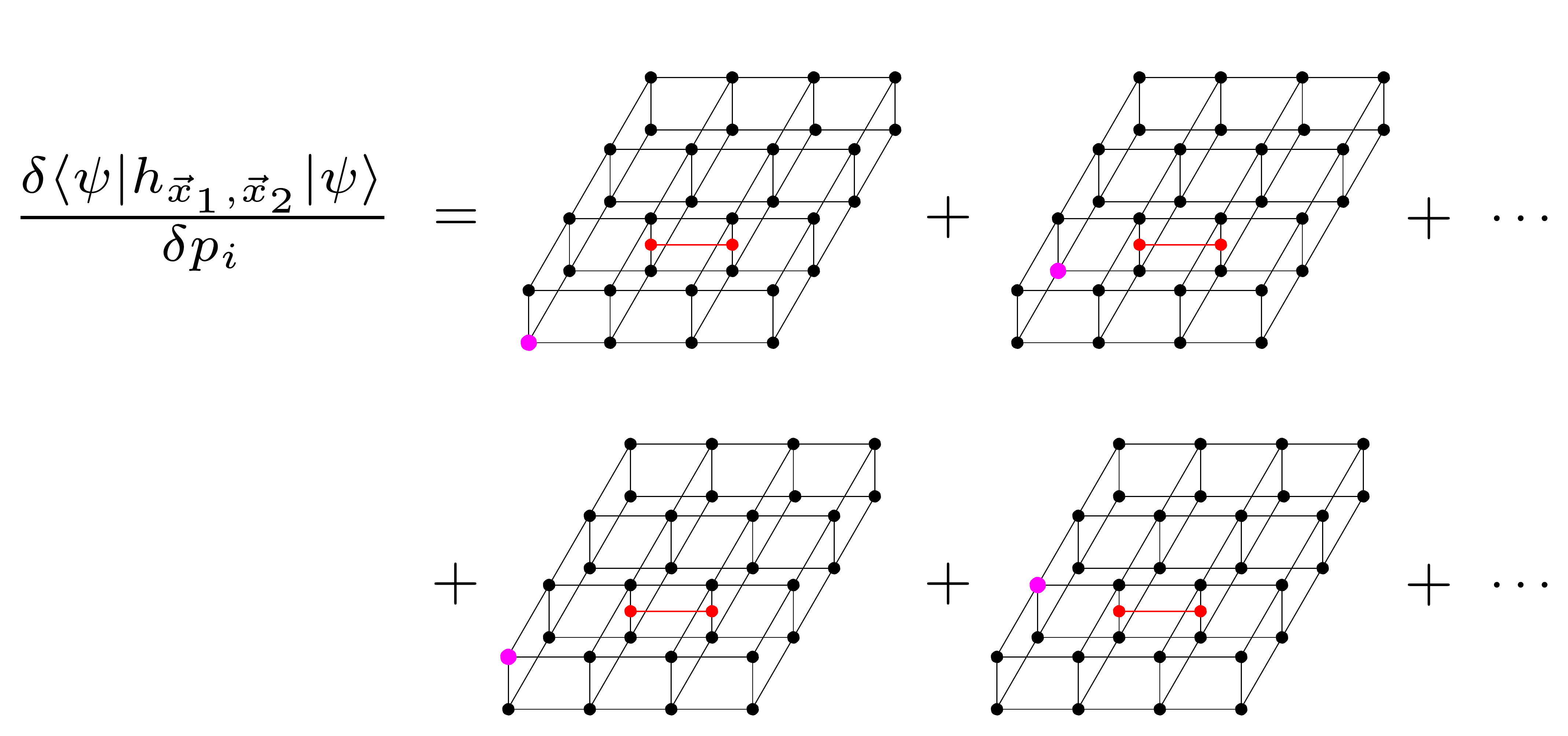}\\
 \caption{The linear order variations of $\partial \langle\psi|\psi\rangle$ and $\langle\psi|h_{\vec x\vec x'}|\psi\rangle$.}
 \label{fig:peps_variation}
\end{figure}

It is then straightforward to see that to the linear order, the variation of $ \langle\psi|\psi\rangle$ ($\langle\psi|h_{\vec x\vec x'}|\psi\rangle$) is a summation of $Lx\times L_y\times 2$ terms, as illustrated in Fig.\ref{fig:peps_variation}, where the factor of two is due to the fact that both the variation of $|\psi\rangle$ and the variation of $\langle\psi|$ contribute. Each term can be computed by the standard PEPS contraction algorithm.

Basically one needs to compute the derivatives $\frac{\partial \langle\psi|\psi\rangle}{\partial p_i}$ and $\frac{\partial \langle\psi|h_{\vec x\vec x'}|\psi\rangle}{\partial p_i}$ for each variational parameter $p_i$. For each derivative one needs to compute $\sim L_x\times L_y$ terms. (This number can be reduced by a factor if one chooses an open boundary sample respecting the point group symmetry.) 
However, one notes that every derivative term only involves one modified site tensor. So one could program the contraction sequence so that all the other tensors are contracted first, leaving an ``environment tensor'' $E(\vec{x})$ for the only remaining site tensor at $\vec{x}$, which is exactly the modified site tensor. Then, different $p_i$ derivatives on the given site $\vec{x}$ can be efficiently computed by contracting  the different $T_i(\vec{x})$ with the same $E(\vec{x})$.  We then end up with a factor of $\sim L_x\times L_y$ in the computation complexity due to the choices of modified sites. However this factor $\sim L_x\times L_y$ in the computation cost can be straightforwardly parallelized on $\sim L_x\times L_y$ computer nodes, simply because no communication is needed between different contractions. Consequently with a computing cluster, the minimization algorithm proposed here can be efficiently implemented.

Finally we comment on possible algorithms in spatial dimensions higher than two. Although our scheme of constructing symmetric PEPS classes still works in higher dimensions, the PEPS algorithm to contract the tensor no longer applies. However, it is still possible to approximately contract tensor networks in higher dimensions while giving up certain accuracy, efficiency and larger system sizes. For example, the tensor renormalization group techniques (TRG)\cite{Levin:2007p120601,Xiang:2008p90603,Gu:2008p205116} could be used to contract tensors in higher dimensions. It would be interesting to see whether the TRG algorithms, when combined with the symmetric PEPS construction studied here, can be used to efficiently study quantum phase diagrams in higher dimensional systems. We leave this as a subject of future investigations.

\section{Physical Interpretation of Classes}\label{sec:lantern_operator_peps}
We will discuss the physical meanings of different classes, which are labeled by $\Theta_R$, $\chi_R$ as well as $\eta_R$. In this section, we will focus on the non-symmetry-breaking liquid member phase in each crude class. We will comment on the meaning of these indices in the long-range ordered member phases in Sec.\ref{sec:torus_long_range_order}.

\subsection{Interpretation of $\Theta_R$ and $\chi_R$}
Although it happens to be true that the kagome half-integer spin example has no tunable $\Theta_R$ indices, $\Theta_R$ indices do appear in general quantum systems. In Appendix \ref{app:square_C4} we perform the crude classification for the half-integer spin systems on the square lattice with a space group generated by translation symmetries and the $C_4$ rotation symmetry only. Assuming $IGG$ being the minimal required $Z_2$, we find a tunable $\Theta_{C_4}$ index.

In fact, the $\Theta_R$ indices and the $\chi_R$ indices generally appear even when the $IGG$ is trivial. For instance, we could consider a system on the kagome lattice with \emph{no} on-site symmetry (i.e., remove the spin $SU(2)$ rotation and the time-reversal symmetry in our main example), and consequently the minimal required $IGG$ is trivial. Assuming $IGG$ being trivial in this system, we will not have the $\eta$ indices but still have the $\chi$ indices. The calculation procedure of transformation rules almost remains the same as before if we simply limit all the $\eta$'s to be identity. Eventually we will arrive at Eq.(\ref{eq:kagome_theta}) replacing all the $\mu_R$ by $+1$. Note that there is no $\eta$-ambiguities to tune away the signs for the square roots as in the half-integer spin case. In this system, apart from the $\chi$ indices, we do have two tunable $\Theta$ indices in the PEPS classification:  $\Theta_{C6}(u)=\pm1$ and $\Theta_{\sigma}(u)=\pm1$.

Different $\Theta_R$ indices can be viewed as different symmetry quantum numbers (for either on-site symmetries or space group symmetries) carried by each site tensor. These quantum numbers of the site tensors, generally speaking, directly contribute to the quantum numbers of a finite size sample. The physics of $\Theta_R$ indices is similar to the physics of the so-called ``fragile Mott insulator'' discussed by Yao and Kivelson\cite{Yao:2010p166402}. And similar indices in one-dimensional matrix product states have been investigated recently\cite{Fuji:2014p}. For instance, in the fragile Mott insulator example\cite{Yao:2010p166402}, a Mott insulator wavefunction is constructed on the checkerboard lattice which carries nontrivial point group quantum numbers on the odd-by-odd unit cell lattices. This distinguishes the fragile Mott insulator from trivial insulators which carries trivial quantum numbers on the same lattices. And such nontrivial quantum numbers can be traced 
back to the quantum numbers carried by the wavefunction on every square cluster on the checkerboard lattice. If one tries to use a site tensor in PEPS to represent the square cluster wavefunction, it is clear that this site tensor forms a nontrivial representation of the point group symmetry. 

The physical meaning of $\chi_R$ may be more well-known. These are generalizations of the symmetry fractionalizations in the 2d AKLT model\cite{Affleck:1987p799}. Let's firstly briefly describe the PEPS construction of the $SO(3)$ symmetric spin-2 AKLT state on the square lattice. In this construction, each virtual leg forms a spin-1/2 projective representation of the $SO(3)$ symmetry group of the spin-2 system. Each site tensor is given by the only singlet state formed by the physical spin-2 and the four virtual spin-1/2's, and each bond tensor is formed by the only spin singlet formed by the two spin-1/2's on the two ends of the bond. Such an AKLT wavefunction can be shown to be the unique gapped ground state of the AKLT Hamiltonian on the square lattice with periodic boundary conditions\cite{Garciz-Saez:2013p245118}. 

However, when the system has an open boundary, one needs to specify a symmetric boundary condition. But one encounters the following problem: each site tensor on the boundary has only three virtual spin-1/2's and it is impossible for form a spin-singlet with the physical spin-2. Basically each site on the boundary can be viewed as a half-integer spin --- which is a projective representation of the original $SO(3)$ group. One sometimes calls this phenomena as the symmetry fractionalization in 2d \emph{in the absence of topological orders}. When coupled together along a translational symmetric edge, the low energy dynamics of the edge states can be effectively described by a translational symmetric half-integer spin chain, which would give a gapless excitation spectrum assuming no spontaneous translational symmetry breaking. Clearly, in the PEPS construction, the origin of such symmetry fractionalization behavior is due to the fact that projective representations appear in the virtual legs.

For an on-site symmetry $R$, this is exactly the physics that $\chi_R$ captures. For instance, the $\chi_{\mathcal{T}}$ index appearing in the kagome example is really about the projective representations of the symmetry group $SU(2)\times \mathcal{T}$ on the virtual legs. As mentioned before, when $\chi_{\mathcal{T}}=1$, the half-integer (integer) spins on the virtual legs form Kramer doublet (singlet) under the time-reversal transformation. This is the usual representation of $SU(2)\times \mathcal{T}$. However when $\chi_{\mathcal{T}}=-1$, the half-integer (integer) spins on the virtual legs form Kramer singlet (doublet) under the time-reversal transformation. This is a nontrivial projective representation of $SU(2)\times \mathcal{T}$. We expect that $\chi_{\mathcal{T}}=-1$ would give rise to nontrivial signatures in entanglement spectra and physical edge states.

For a spatial symmetry $R$, the physical meaning of $\chi_R$ is less obvious. But it's one-dimensional analog has been investigated in the context of matrix product states\cite{Chen:2011p35107,Schuch:2011p165139,Pollmann:2010p64439,Pollmann:2012p75125}. In our example, the $\chi_{\sigma}$ is capturing similar physics in 2d kagome lattice, which basically describes how the tensor network forms possible projective representations of the spatial reflection. We speculate that nontrivial $\chi_{\sigma}$ would give rise to signatures in entanglement spectra when the partition of the system respects the $\sigma$ reflection.

In summary, $\Theta_R$ is capturing local contributions to symmetry group quantum numbers, and $\chi_R$ is capturing the symmetry fractionalizations \emph{not} due to topological orders.

\subsection{$\eta_R$ and symmetry fractionalization}\label{subsec:spinon_sym_frac}
Here, we will show that $\eta$'s are directly related to the symmetry fractionalization of spinon excitations (chargons). To see this, let us firstly introduce the concept of symmetry fractionalization in the presence of topological orders. We will use the unitary on-site symmetry as an example. Related discussions can be found in Ref.\onlinecite{Mesaros:2013p155115} and Ref.\onlinecite{Essin:2013p104406}.

\begin{figure}
  \includegraphics[width=0.35\textwidth]{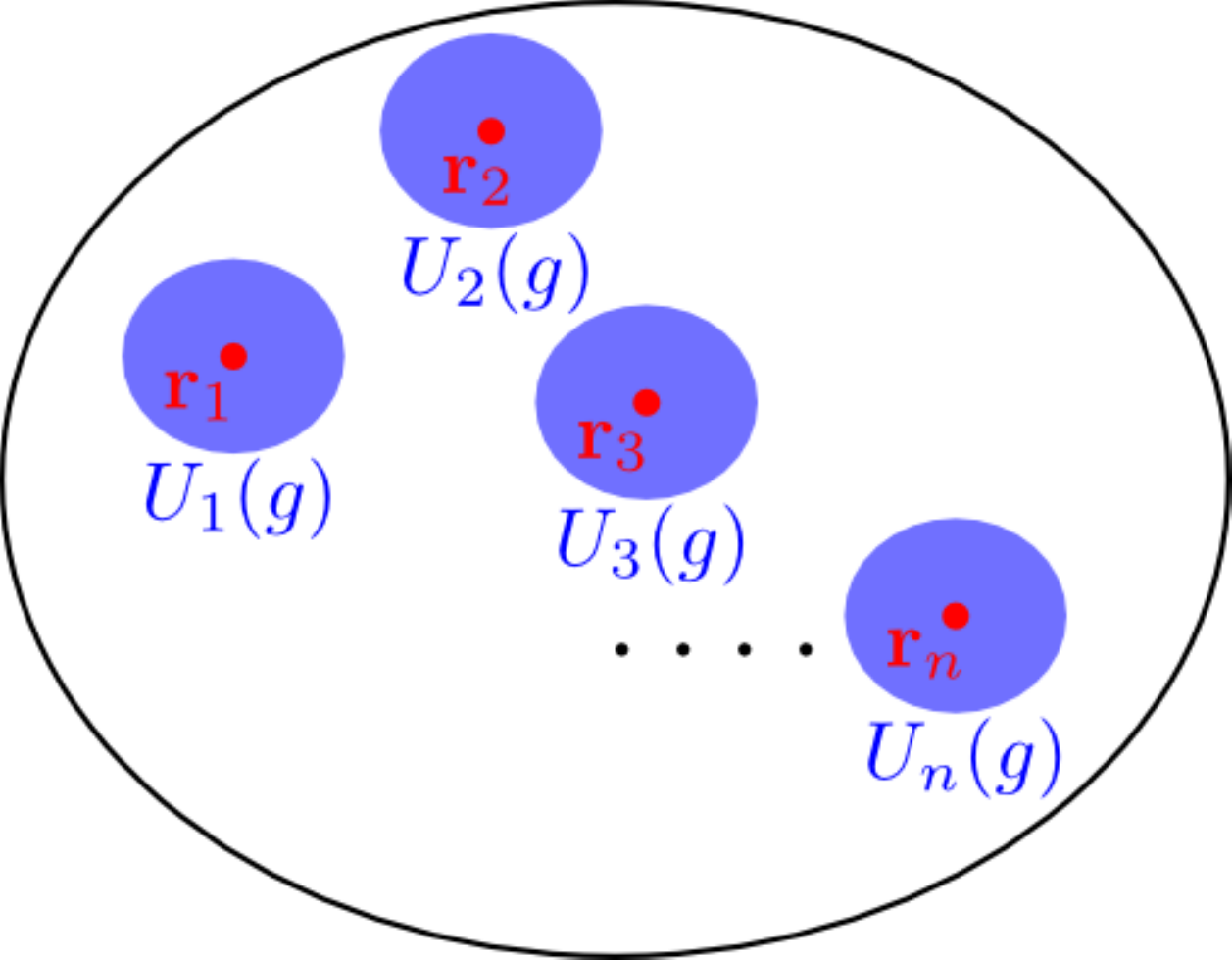}
  \caption{Illustration of symmetry fractionalization phenomena: Under a symmetry transformation $U(g)$ with $\forall g\in SG$, an excited state is transformed by the product of local transformation operators $U_i(g)$, with each operator only acting on one quasiparticle locally.}
  \label{fig:s_f_assumption}
\end{figure}

Starting from a topologically ordered ground state with a global symmetry group $SG$, consider an excited state, having $n-$quasiparticles (which do not have to be of the same type) spatially located at position $\mathbf{r}_1,\mathbf{r}_2,\dots,\mathbf{r}_n$, far apart from one another. Let's denote this state by $|\psi(\mathbf{r}_1,\mathbf{r}_2,\dots,\mathbf{r}_n)\rangle$. For any symmetry transformation $U(g)$ by a group element $g\in SG$, $U(g)$ will generally transform this state to another state:
\begin{align}
  U(g)\circ|\psi(\mathbf{r}_1,\mathbf{r}_2,\dots,\mathbf{r}_n)\rangle\rightarrow|\widetilde{\psi}(\mathbf{r}_1,\mathbf{r}_2,\dots,\mathbf{r}_n)\rangle
  \label{}
\end{align}
One way to describe the symmetry fractionalization on quasiparticles is the following condition: there exist local operators $U_1(g),U_2(g),\dots,U_n(g)$, such that $U_i(g)$ is a local operator acting only in a finite region around the spatial position $\mathbf{r}_i$, and does not touch the other quasiparticles; in addition, $U_1(g),U_2(g),\dots,U_n(g)$ satisfy:
\begin{align}
  &U_1(g)\cdot U_2(g)\cdots U_n(g)|\psi(\mathbf{r}_1,\mathbf{r}_2,\dots,\mathbf{r}_n)\rangle\notag\\
  =&U(g)|\psi(\mathbf{r}_1,\mathbf{r}_2,\dots,\mathbf{r}_n)\rangle=|\widetilde{\psi}(\mathbf{r}_1,\mathbf{r}_2,\dots,\mathbf{r}_n)\rangle
  \label{eq:s_f_assumption}
\end{align}
Pictorially, this condition is shown in Fig.(\ref{fig:s_f_assumption}). 

Note that technically Eq.(\ref{eq:s_f_assumption}) is \emph{not} a general condition for symmetry fractionalization phenomena. For example, let us consider $SG$ to be an on-site $U(1)$ symmetry, and assume that Eq.(\ref{eq:s_f_assumption}) holds for a wavefunction $|\psi(\mathbf{r}_1,\mathbf{r}_2,\dots,\mathbf{r}_n)\rangle$. We can then just add one extra $U(1)$ charge outside the regions that $U_i(g)$ ($i=1,..,n$) act and obtains a new wavefunction $|\overline{\psi}(\mathbf{r}_1,\mathbf{r}_2,\dots,\mathbf{r}_n)\rangle$. It is perfectly fine to imagine the extra charge as if it already exists in the ground state. Physically the local operators that transform quasiparticles: $U_i(g)$ for $|\overline{\psi}\rangle$ should be exactly the same as before, since $|\psi\rangle$ and $|\overline\psi\rangle$ are locally identical around $\mathbf{r}_1,\mathbf{r}_2,\dots,\mathbf{r}_n$. However, clearly Eq.(\ref{eq:s_f_assumption}) is no longer true for $|\overline{\psi}\rangle$, because the global symmetry $U(g)$ picks 
up 
an extra $U(1)$ phase from the added $U(1)$ charge. 

In fact, Eq.(\ref{eq:s_f_assumption}) implicitly assumes that, under a global symmetry transformation, there is no phase ``locally accumulated'' in the ground state wavefunction. But, as demonstrated above, generally there could be such ``locally accumulated'' phases in the ground state, and Eq.(\ref{eq:s_f_assumption}) should be modified up to the ``locally accumulated'' phases outside the blue regions in Fig.\ref{fig:s_f_assumption}. 

\emph{How to sharply define such ``locally accumulated'' phases in general?} The answer to this question is important to provide a general sharp definition of $U_i(g)$. But to answer this question, one needs a tool capable to diagnose wavefunctions locally, which is exactly the power of PEPS. For the moment, let us postpone answering this question in the framework of PEPS, and have some further discussion on symmetry fractionalizations.

 First, fractionalized symmetry transformations are local operators and cannot change the quasiparticle's species (or more precisely, the superselection sector of a quasiparticle). Thus, we can investigate the transformation rules of each anyon species individually. However, anyons do not need to form a representation of $SG$ due to the nontrivial fusion rule. For example, in a $Z_2$ topological ordered phase, two chargons fuse to one trivial particle. We can multiply each chargon in the system by a fixed element in an $IGG'=Z_2=\{1,-1\}$. Clearly, the total phase becomes unity, and physical wavefunction is invariant. Here $IGG'$ is the subgroup of U(1) describing the fusion rule of chargons. Quite generally for a $Z_n$ topological order, $IGG'=Z_n$. 

A PEPS with $IGG=Z_n$ can describe a deconfined phase with a $Z_n$ topological order. We will only consider this case and we do have $IGG'=IGG$. So $IGG$ tells us that when we implement the global symmetry transformation on chargons, it is perfect fine to have a phase ambiguity, if this phase ambiguity is an element in $IGG$. Consequently, a single quasiparticle could form a projective representation of $SG$ with coefficient in $IGG$, which is classified by second cohomology $H^2(SG,IGG)$ (see Appendix \ref{app:proj_rep} for details).
 
Now, let us translate the above discussion into the PEPS language. The main task is to construct the local symmetry transformation operators for a small patch of PEPS with a nontrivial $IGG$. Here we focus on $IGG=Z_2$ case. Without loss of generality, we assume that tensors of the PEPS are all $Z_2$ even. Then we cut a small patch $\mathcal{A}$ from the PEPS. We can view the tensor associated with patch $\mathcal{A}$ as a linear map from boundary virtual legs to physical legs living in the bulk of the patch, which is labeled as $\hat{T}^0_\mathcal{A}$. Here $0$ denotes that there is no quasiparticles inside $\mathcal{A}$. Namely,
\begin{align}
  \hat{T}^0_\mathcal{A}=\sum_{I,V}(T^0_{\mathcal{A}})_{IV}|I\rangle\langle V|
  \label{}
\end{align}
where $|I\rangle$  labels ket states of all physical legs inside $\mathcal{A}$, while $\langle V|$ labels bra states of all boundary virtual legs.

Before studying excitations inside $\mathcal{A}$, we firstly discuss properties of $\hat{T}^0_{\mathcal{A}}$. As a tensor, $\hat{T}^0_\mathcal{A}$ is $Z_2$ even. Namely, action of the nontrivial $Z_2$ element $g$ on the boundary legs of $\hat{T}^0_{\mathcal{A}}$ leaves the tensor invariant. This property implies that $\hat{T}^0_{\mathcal{A}}$, as a linear map, can never be injective. To see this, consider an arbitrary boundary state 
$|V\rangle$, we have 
\begin{align}
  \hat{T}^0_{\mathcal{A}}|V\rangle=\hat{T}^0_{\mathcal{A}}|g\circ V\rangle 
  \label{}
\end{align}
So, the inverse map of $\hat{T}^0_{\mathcal{A}}$ is not well defined. To have a reasonable definition of the inverse map, one observes that an arbitrary boundary state $|V\rangle$ can be rewritten as
\begin{align}
  |V\rangle&=\frac{1}{2}(|V\rangle+|g\circ V\rangle)+\frac{1}{2}(|V\rangle-|g\circ V\rangle)\notag\\
  &=\Pi_{\mathcal{U}}|V\rangle+(1-\Pi_\mathcal{U})|V\rangle
  \label{}
\end{align}
where $\mathcal{U}$ is the $Z_2$ even sector of boundary legs. Namely, $\forall |V\rangle\in\mathcal{U}$, we have $|g\circ V\rangle=|V\rangle$. $\Pi_\mathcal{U}$ is a projection operator which projects a boundary state into $\mathcal{U}$. Under $\hat{T}^0_{\mathcal{A}}$, the second term in the above equation is mapped to zero. For a generic PEPS with $IGG=Z_2$, we can further assume that $\hat{T}^0_{\mathcal{A}}$ is injective on the subspace $\mathcal{U}$ when the patch $\mathcal{A}$ is not too small. This is because the dimension of the physical Hilbert space increases parametrically faster than the dimension of the boundary virtual Hilbert space as the patch size increases. Such a PEPS is named as a $Z_2$ injective PEPS in Ref.\onlinecite{Schuch:2010p2153}. Namely, generically one can find a linear map $(\hat{T}^0_{\mathcal{A}})^{-1}$ from bulk physical legs to boundary virtual legs, such that
\begin{align}
  (\hat{T}^0_{\mathcal{A}})^{-1}\cdot\hat{T}^0_{\mathcal{A}}=\Pi_\mathcal{U}
  \label{}
\end{align}

Next, let us study the case with topological excitations inside patch $\mathcal{A}$. One could create odd number of chargons near the center of the patch $\mathcal{A}$ by modifying $\hat{T}^0_{\mathcal{A}}$ to some $Z_2$ odd tensor $\hat{T}^{e}_{\mathcal{A}}$. Opposite to the previous case, we have
\begin{align}
  \hat{T}^{e}_{\mathcal{A}}|V\rangle=0,\quad\forall |V\rangle\in\mathcal{U}
  \label{}
\end{align}
Generically we can further assume $\hat{T}^{e}_{\mathcal{A}}$ is injective on the $Z_2$ odd sector of boundary legs. Namely, one can construct $(\hat{T}^{e}_{\mathcal{A}})^{-1}$ as linear map from bulk legs to $Z_2$ odd sector of boundary legs, such that
\begin{align}
  (\hat{T}^{e}_{\mathcal{A}})^{-1}\cdot\hat{T}^{e}_\mathcal{A}=\Pi_\mathcal{\overline{U}}
  \label{eq:chargon_tensor_inverse}
\end{align}
where $\Pi_\mathcal{\overline{U}}\equiv1-\Pi_\mathcal{U}$. 

Similarly, one can construct patch tensors with even number chargons inside the patch by modifying $\hat{T}^0_{\mathcal{A}}$ to any other $Z_2$ even and $Z_2$ injective tensors. For example, let us assume $\hat{T}^{\mathbf{1}}_{\mathcal{A}}$ to be such a tensor. Then, one can find it inverse $(\hat{T}^{\mathbf{1}}_{\mathcal{A}})^{-1}$ on the subspace $\mathcal{U}$, such that $(\hat{T}^{\mathbf{1}}_{\mathcal{A}})^{-1}\cdot\hat{T}^{\mathbf{1}}_{\mathcal{A}}=\Pi_\mathcal{U}$.

In the following, we will study the local physical operator acting on small patches for a symmetry $R$. Starting with a PEPS wavefunction $|\Psi\rangle$ with topological excitations inside small patches $\mathcal{A},\mathcal{B},\dots$, while the region outside these patches share the same tensors as the ground state wavefunction $|\Psi_0\rangle$. The action of the symmetry $R$ on $|\Psi\rangle$ is obtained by acting $R$ on all tensors, which is defined in Eq.(\ref{eq:on-site_sym_site_bond},\ref{eq:time_reversal_sym_site_bond},\ref{eq:lattice_sym_site_bond}). Since we try to construct local symmetry operators only on patches $\mathcal{A},\mathcal{B},\dots$, we can apply gauge transformations $W_R$ on all virtual legs in the region \emph{outside} all small patches as well as on the boundaries of all small patches, but leave virtual legs inside small patches untouched. Note that this gauge transformation does \emph{not} modify the $R$-transformed physical wavefunction at all. Because tensors outside small 
patches are the same as 
tensors of ground state, the following relations still hold for them:
\begin{align}
  &T^{\mathrm{s}}=\Theta_RW_RR\circ T^\mathrm{s}\notag\\
  &B_\mathrm{b}=W_RR\circ B_\mathrm{b}
  \label{}
\end{align}
Thus, under the symmetry $R$ together with the gauge transformation $W_R$ defined above, tensors outside patches will be invariant up to an \emph{``locally accumulated''} phase $\prod_{\mathrm{s}\in \mbox{outside}} \Theta_R(\mathrm{s})$. We emphasize that this actually provides the sharp \emph{definition} of the ``locally accumulated'' phases mentioned earlier in this section. As discussed in the previous subsection, $\Theta_R(\mathrm{s})$'s exactly capture the local phases picked up after applying a global symmetry transformation. Without the tool of PEPS, it is actually difficult to sharply define this object.

For tensors inside patches, we have
\begin{align}
  \hat{T}^{R}_{\mathcal{A}}=W_RR\circ \hat{T}_{\mathcal{A}}
  \label{eq:patch_sym}
\end{align}
Here, $\hat{T}_\mathcal{A}$ is the linear map associated with patch $\mathcal{A}$, which is obtained by contraction of all tensors inside $\mathcal{A}$ patch. And $W_R$ in Eq.{\ref{eq:patch_sym}} is defined to \emph{only} act on boundary virtual legs of $\hat{T}_\mathcal{A}$. Note that $\hat{T}_\mathcal{A}$ is either $Z_2$ even or $Z_2$ odd, which corresponds to even number chargons or odd number chargons inside $\mathcal{A}$. Note that we should always choose the patch that is large enough so that all quasiparticles exist in the patch before the transformation keep staying in the patch after the transformation. The above equation can be viewed as the \emph{definition} of $\hat{T}^R_\mathcal{A}$. 

In fact, Eq.(\ref{eq:patch_sym}) is a very general result which is applicable even when the condition of symmetry fractionalizations breaks down. For example, it is possible that certain symmetry transformation interchanges quasiparticle superselection sectors. In the PEPS formulation this happens when $\hat{T}^{R}_{\mathcal{A}}$ and $\hat{T}_{\mathcal{A}}$ describes distinct quasiparticle species, and consequently there is no way to use a local physical operator in $\mathcal{A}$ to send $\hat{T}_{\mathcal{A}}$ to $\hat{T}^{R}_{\mathcal{A}}$. For the kagome example this would never happen. For example, we showed that $W_R$ matrices all commute with the nontrivial $IGG$ element $g=\mathrm{J}$, and therefore the parity of the number of chargons would be the same in $\hat{T}^{R}_{\mathcal{A}}$ and $\hat{T}_{\mathcal{A}}$. But in a symmetric PEPS with a larger $IGG$ (e.g. $IGG=Z_2\times Z_2$), we expect that it is possible that $W_R$ does not commute with a $g\in IGG$. In this case the $R$ may interchange 
quasiparticle species.

Below we only consider the situation that $\hat{T}^{R}_\mathcal{A}$ and $\hat{T}_\mathcal{A}$ support the same superselection sector and consequently share the same $Z_2$ parity. This allows us to construct the fractionalized local \emph{physical} operator $\hat{L}_R^\mathcal{A}$ for the symmetry $R$ acting on patch $\mathcal{A}$ that realizes Eq.(\ref{eq:patch_sym}); namely: 
\begin{align}
  \hat{L}_R^\mathcal{A}\circ\hat{T}_{\mathcal{A}}=W_RR\circ\hat{T}_{\mathcal{A}},
  \label{eq:lantern_operator_def}
\end{align}
at least for those $\hat{T}_{\mathcal{A}}$ describing the relevant low energy states. One should keep in mind that $L_R^\mathcal{A}$ only acts on physical legs, without touching boundary legs; i.e.,
\begin{align}
  \hat{L}_R^\mathcal{A}=\sum_{I,I'}(L_R^\mathcal{A})_{I,I'}|I\rangle\langle I'|.
  \label{}
\end{align}

To obtain the explicit form of this local operator, let us consider a particular tensor $\hat{T}^{e}_\mathcal{A}$, which supports an odd number of chargons in $\mathcal{A}$. We have
\begin{align}
  \hat{T}^{e,R}_\mathcal{A}=[\hat{T}^{e,R}_\mathcal{A}\cdot(\hat{T}^{e}_\mathcal{A})^{-1}]\cdot\hat{T}^{e}_\mathcal{A}
  \label{}
\end{align}
where $\hat{T}^{e,R}_\mathcal{A}\equiv W_RR\circ\hat{T}^{e}_\mathcal{A}$, and $(\hat{T}^{e}_\mathcal{A})^{-1}$ is defined in Eq.(\ref{eq:chargon_tensor_inverse}). In the above equation we assume that both $\hat{T}^e_\mathcal{A}$ and $\hat{T}^{e,R}_\mathcal{A}$ is $Z_2$ odd as well as injective in the $Z_2$ odd subspace of boundary legs, which is expected to be generically true. Note that $[\hat{T}^{e,R}_\mathcal{A}\cdot(\hat{T}^e_\mathcal{A})^{-1}]$ can be viewed as an operator acting only on physical legs. 

To study the transformation rules for a number of chargon excitations, let us consider a finite set $\Lambda$ of tensors: $\Lambda\equiv\{\hat{T}^{(i)}_{\mathcal{A}},i=0,1,\dots\}$ in the patch $\mathcal{A}$. These tensors may describe states with chargon number equal to zero, one, two, etc, and are injective in the corresponding boundary $Z_2$ sectors respectively. But tensors in $\Lambda$ contain \emph{no} fluxon excitations in $\mathcal{A}$. (we will study the symmetry fractionalization of fluxons later in this paper.) We assume that any symmetry transformation as shown in Eq.(\ref{eq:patch_sym}) transform within the linear space spanned by $\Lambda$. 

In addition, we assume the tensors in $\Lambda$ to satisfy $ (\hat{T}^{(j)}_{\mathcal{A}})^{-1}\cdot\hat{T}^{(i)}_{\mathcal{A}}=\mathbf{0}$, $\forall i\neq j$. Physically, this can be achieved by choosing $\Lambda$ so that all tensor states in it can be sharply distinguished from each other by a set of mutually commuting local physical measurements. Mathematically these local physical measurements are Hermitian operators acting near the center of the patch where quasiparticles live. For instance, these measurements could include a measurement of the locations of chargons by inserting small fluxon loops. Then $\{\hat{T}^{(i)}_{\mathcal{A}}\}$ are chosen to be the eigenstates of these measurements with distinct eigenvalues. Since these measurements are locally near the center of the patch, the boundary condition (i.e., the virtual boundary state) will not affect the measurement when the patch is large enough, and the condition $(\hat{T}^{(j)}_{\mathcal{A}})^{-1}\cdot \hat{T}^{(i)}_{\mathcal{A}} =\mathbf{0}$,
 $\forall i\neq j$ is expected to hold.

\begin{figure}
 \includegraphics[width=0.5\textwidth]{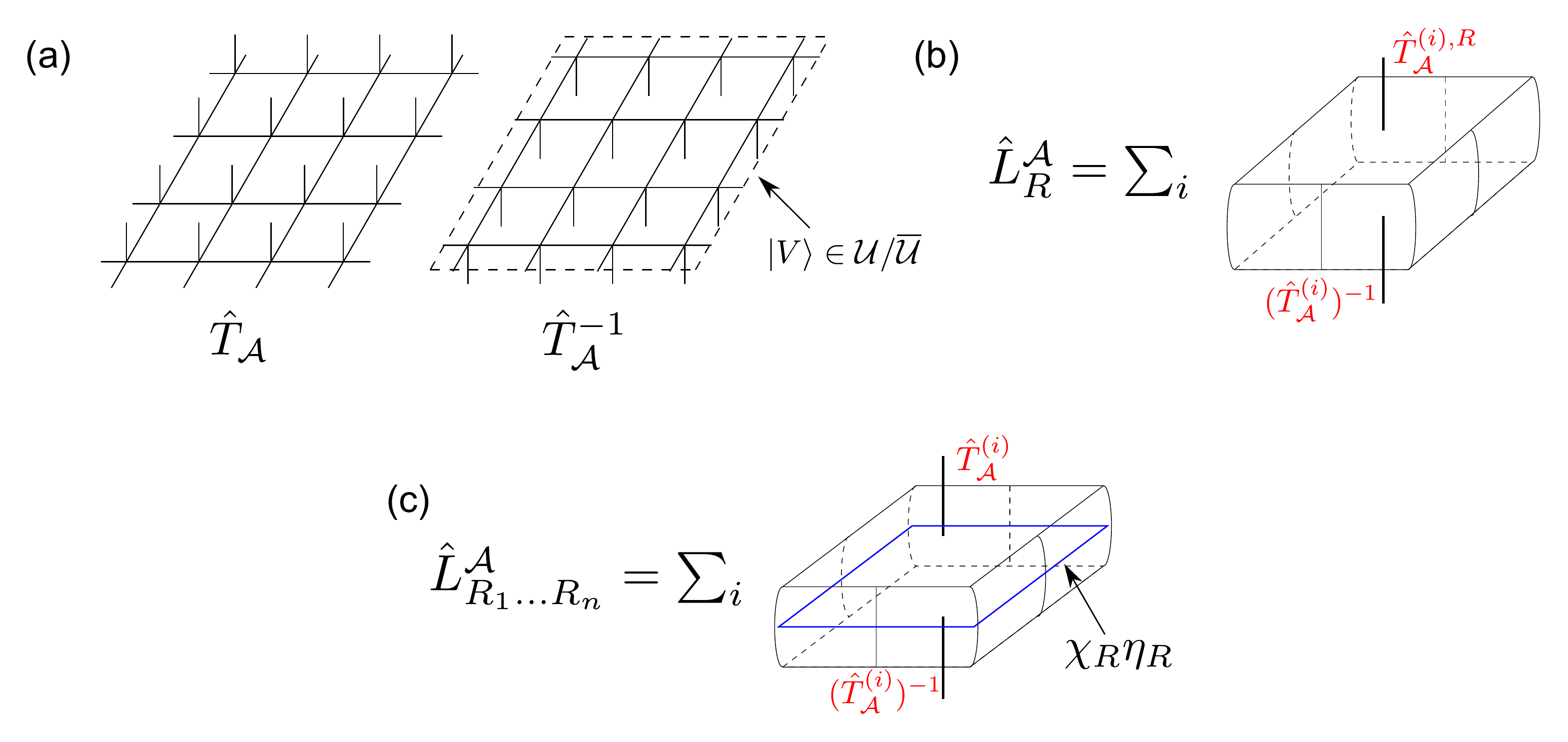}
 \caption{(a): Tensor $\hat{T}_\mathcal{A}$ and its ``generalized inverse'' $\hat{T}_\mathcal{A}^{-1}$ associated with patch $\mathcal{A}$. $\hat{T}_\mathcal{A}$ is obtained by contracting all bond tensors and site tensors inside patch $\mathcal{A}$. As a linear map from boundary legs to bulk legs, $\hat{T}_\mathcal{A}$ is either $Z_2$ even or $Z_2$ odd. (b): The local $R$-symmetry operator on patch $\mathcal{A}$. $\{\hat{T}^{(i)}_\mathcal{A}\}$ is an orthonormal basis, where every state in the basis is either $Z_2$ even or $Z_2$ odd. (c): The local symmetry operator for a series symmetry operations $R_1\dots R_n$, where $R_1\dots R_n=\mathrm{I}$. If $\eta_R$ is nontrivial, action of this operator on $Z_2$ even or $Z_2$ odd tensor gives different phase factor. This indicates symmetry fractionalization of chargons.}
 \label{fig:spinon_lantern_operator}
\end{figure}

We then can construct a local operator to transform states in $\Lambda$ under a symmetry $R$:
\begin{align}
  \hat{L}_R^{\mathcal{A}}=\sum_i [\hat{T}^{(i),R}_{\mathcal{A}}\cdot(\hat{T}^{(i)}_{\mathcal{A}})^{-1}]
  \label{eq:lantern_operator_expression}
\end{align}
as shown in Fig.(\ref{fig:spinon_lantern_operator}b). One can easily verify, $\hat{L}_R^\mathcal{A}$ defined above indeed satisfies Eq.(\ref{eq:lantern_operator_def}) for all states in $\Lambda$. Moreover, such local operators in patches $\mathcal{A},\mathcal{B}...$ satisfy the symmetry fractionalization condition Eq.(\ref{eq:s_f_assumption}) up to the ``locally accumulated'' phase outside these patches $\prod_{\mathrm{s}\in \mbox{outside}} \Theta_R(\mathrm{s})$. 

After the local symmetry operator is defined, we are able to study the symmetry fractionalization of chargons. Consider a relation between symmetry group elements $R_1R_2\dots R_n=\mathrm{e}$, we can construct a local symmetry operators $\hat{L}_{R_1\dots R_n}^\mathcal{A}$ as 
\begin{align}
  \hat{L}_{R_1\dots R_n}^\mathcal{A}\equiv\hat{L}_{R_1}^\mathcal{A}\cdots\hat{L}_{R_n}^\mathcal{A}
  \label{}
\end{align}
By inserting Eq.(\ref{eq:lantern_operator_expression}) into the above equation, we get
\begin{align}
  \hat{L}_{R_1\dots R_n}^\mathcal{A}=\sum_i[(\hat{T}^{(i),R_1\dots R_n}_{\mathcal{A}})\cdot(\hat{T}^{(i)}_{\mathcal{A}})^{-1}]
  \label{}
\end{align}
where
\begin{align}
  \hat{T}^{(i),R_1\dots R_n}_\mathcal{A}&\equiv W_{R_1}R_1\dots W_{R_n}R_n\circ \hat{T}^{(i)}_\mathcal{A}\notag\\
  &=\chi_R\eta_R\circ\hat{T}^{(i)}_\mathcal{A}
  \label{eq:patch_group_identity}
\end{align}
Here, the $Z_2$ element $\eta_R$ and the phase factor $\chi_R$ act on boundary virtual legs, as shown in Fig.(\ref{fig:spinon_lantern_operator}c). The second line of the above equation is obtained by the following fact:
\begin{align}
  \eta_{R}(\mathrm{s},i)\chi_R(\mathrm{s},i)=W_{R_1}(\mathrm{s},i)\dots W_{R_n}(R_{n-1}^{-1}\dots R_1^{-1}(\mathrm{s},i))
  \label{}
\end{align}

When $\eta_R=\mathrm{I}$, the action of $\hat{L}_{R_1\dots R_n}^\mathcal{A}$ on an arbitrary tensor $\hat{T}_\mathcal{A}\in \Lambda$ gives the same phase. When $\eta_R$ is the nontrivial $Z_2$ element, a $Z_2$ odd tensor $\hat{T}^e_\mathcal{A}$ picks up an extra $-1$ comparing to a $Z_2$ even tensor $\hat{T}^{\mathbf{1}}_{\mathcal{A}}$ under the action of $\hat{L}_{R_1\dots R_n}^\mathcal{A}$. This is exactly the phenomena for symmetry fractionalization of chargons: for nontrivial $\eta_R$, under symmetry $R_1\dots R_n$, a single chargon picks up an extra $-1$ comparing to a topologically trivial excitations. 

Note that $\chi_R$ only serves as a global phase, thus does not contribute to the symmetry fractionalization of chargons. It appears in Eq.(\ref{eq:patch_group_identity}) even for the ground state tensor patch. In fact, this result is expected and is consistent with the physical interpretation of $\chi$ discussed in the previous subsection. One way to see this is to repeat the above analysis \emph{only} for the ground states of the 1d spin-1 AKLT model on an open chain, with the patch $\mathcal{A}$ covering one end of the chain. Here one should instead consider an injective matrix project state since the $IGG$ here is trivial. The appearance of $\chi$ in this example can be simply interpreted as the projective representation of the edge states in the AKLT model.

\section{Fluxons and the Decorated PEPS}\label{sec:vison_psg}
In this section, we will construct the decorated PEPS from the original symmetric PEPS with $IGG=Z_2$. The decorated PEPS explicitly captures all $Z_2$ gauge fluctuations, and is a good tool to study properties of fluxons. In particular, we can extract fractional lattice quantum numbers of fluxons by constructing local symmetry operators on a small patch of the decorated PEPS. The result shows that lattice quantum numbers of fluxons is completely determined by the chargon distribution.

\subsection{The decorated PEPS}
Given a symmetric PEPS with $IGG=Z_2$, we can create topological excitations such as chargons and fluxons. As shown in Sec.\ref{subsec:IGG_peps}, creation of chargons is by locally changing site tensors from $Z_2$ even(odd) to $Z_2$ odd(even) and the wavefunction would vanishes if we modify odd number of tensors on a closed manifold. Fluxons are always created in pairs as ends of strings of the nontrivial $Z_2$ action on virtual legs. 

Note that \emph{there is a major difference between chargons and fluxons in this symmetric PEPS language: chargon strings are always ``hidden'' while fluxon strings are explicitly present.} Fluxon strings, which are extended objects, will cause inconvenience as one tries to define local lattice symmetry operations on small patches with fluxons. 

To overcome this inconvenience, we define a decorated PEPS, on which we can create fluxons by changing tensors ``locally'' without creating strings. The decorated PEPS can be viewed as a dual description of the original symmetric PEPS, and they represent the same physical state if one does not consider the boundary effects. In particular, we will show that \emph{fluxon strings are always ``hidden'' while the chargon strings are explicitly present in the decorated PEPS.} For simplicity, we consider the decorated PEPS on the square lattice first, and develop the decoration method. Then we apply the method to the kagome PEPS with $Z_2$ odd site tensors. 

\begin{figure}
\includegraphics[width=0.5\textwidth]{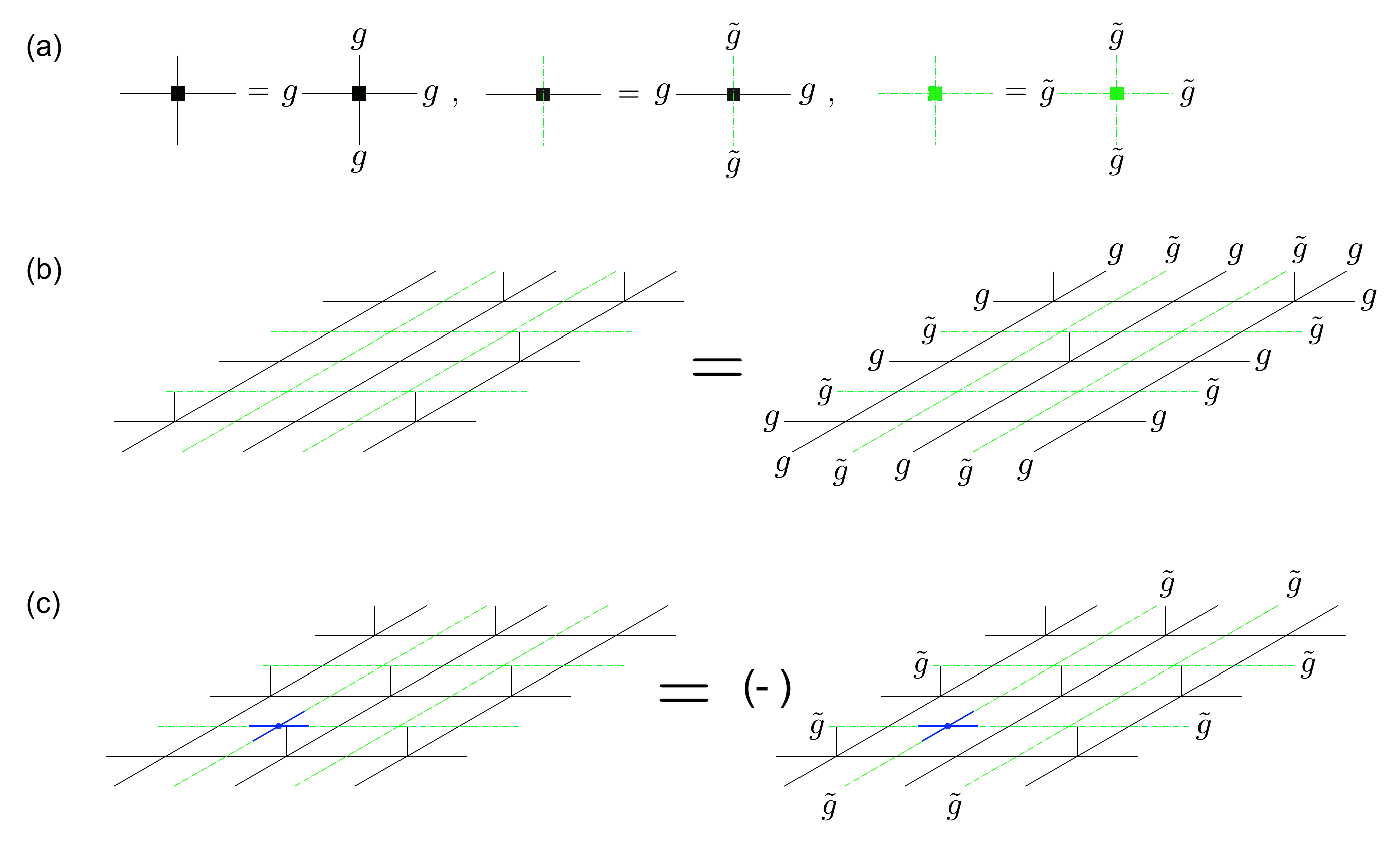}
\caption{(a): We decorate the PEPS by changing bond tensors, adding plaquette tensors, while leaving site tensors invariant. New legs connecting plaquette tensors and bond tensors are depicted by green dashed dotted line. These tensors is invariant under action of $g$ on old virtual legs and $\widetilde{g}$ on new legs. (b): Any patch of PEPS are invariant under action of $g$ and $\widetilde{g}$ on boundary of this patch. The whole PEPS is invariant under $IGG\times\widetilde{IGG}$. (c): By changing a plaquette tensor from $\widetilde{Z_2}$ even to $\widetilde{Z_2}$ odd (blue tensor), one creates a fluxon at the plaquette center. Odd number of fluxons inside a patch can be detected by acting $\widetilde{g}$ loop around the boundary new legs, where one gets an extra minus sign. The physical meaning of $\widetilde{g}$ loop is the Wilson loop of chargons, and the extra minus sign encodes the braiding statistics of chargons and fluxons.}
  \label{fig:decorated_peps}
\end{figure}

Now, let us discuss the method to obtain the decorated PEPS defined on the square lattice, as shown in Fig.(\ref{fig:decorated_peps}). We will first discuss the case where all tensors are $Z_2$ even. The procedure to construct the decorated PEPS is as follows:
\begin{enumerate}
  \item One adds two new virtual legs pointing to plaquette centers for every bond tensor. A new leg has dimension $\widetilde{D}=2$. And we change the bond tensor $B_b$ to $\widetilde{B}_b$ as
    \begin{align}
      \widetilde{B}_b&=B_b\otimes|\widetilde{1},\widetilde{1}\rangle+g\cdot B_b\otimes|-\widetilde{1},-\widetilde{1}\rangle,\notag\\
      \label{eq:decorated_bond_tensor}
    \end{align}
    where $|\pm\widetilde{1}\rangle$ labels the basis for the new virtual leg, and $g$ is the nontrivial $Z_2$ element. Here in Eq.(\ref{eq:decorated_bond_tensor}), \emph{$g$ is defined to act only on one end of the bond tensor $B_b$}. This is shown in the second figure of Fig.\ref{fig:decorated_peps}(a).
  \item All site tensors $T^s$ are the same as those in the undecorated PEPS, as shown in the first figure of Fig.\ref{fig:decorated_peps}(a).
  \item One adds plaquette tensors at all plaquette centers as shown in the third figure of Fig.\ref{fig:decorated_peps}(a). Plaquette tensors connect the modified bond tensors nearby with four new virtual legs. Plaquette tensors are simply superpositions of all (new) virtual states with even numbers of $-\widetilde{1}$'s:
    \begin{align}
      P_\mathrm{c}=|\widetilde{1},\widetilde{1},\widetilde{1},\widetilde{1}\rangle+|\widetilde{1},\widetilde{1},-\widetilde{1},-\widetilde{1}\rangle+\dots
      \label{}
    \end{align}
    where $P_\mathrm{c}$ labels the tensor at a plaquette center $\mathrm{c}$.
\end{enumerate}
Let us visualize the decorated PEPS. Since only configurations with even number of $|-\widetilde{1}\rangle$ contribute to plaquette tensors and bond tensors, we conclude that $|-\widetilde{1}\rangle$ always form loops. Further, every loop configuration contributes equally to the wavefunction. To see this, consider an arbitrary loop configuration $\mathcal{L}$. Then, we can transform the configuration to a PEPS wavefunction defined on the undecorated lattice. The PEPS wavefunction $|\Psi_\mathcal{L}\rangle$ associated with this loop configuration is obtained by modifying the undecorated PEPS $|\Psi\rangle$. For bond tensors intersecting with loops, according to Eq.(\ref{eq:decorated_bond_tensor}), we have 
\begin{align}
  B_b\rightarrow g\cdot B_b
  \label{}
\end{align}
While all other tensors are unchanged. Namely, the action of $g$ forms the same loop configuration $\mathcal{L}$. In fact, $\mathcal{L}$ can be viewed as a fluxon loop in the original PEPS before decoration. Since the tensor obtained by contracting site tensors and bond tensors inside any region is also $Z_2$ even, the action of $g$ on any loops gives the same quantum state. We then conclude that every loop configuration $|\Psi_\mathcal{L}\rangle$ contributes the same wavefunction as $|\Psi\rangle$. The decorated PEPS, which is the linear superposition of all $|\Psi_\mathcal{L}\rangle$, describes the same quantum state as the undecorated one up to normalization (for the moment, let us postpone the discussion on boundary conditions). 

The decorated PEPS still has a $Z_2=\{\mathrm{I},g\}$ invariance with $g$ acting on the old virtual legs only. In addition, the single-sided $g$-action along a loop of old virtual legs is equivalent to flipping all the \emph{new} Ising variables $|\pm \widetilde{1}\rangle\rightarrow \mp\widetilde{1}\rangle$ along the corresponding loop $\mathcal{L}$ of the new virtual legs. 

In the following, we will study fluxon excitations in the decorated PEPS. First, let us point out that for the decorated PEPS, there is an additional $\widetilde{Z_2}$ gauge transformation $\widetilde{IGG}$ leaving all tensors invariant. We define the action $\widetilde{g}$ as
\begin{align}
  \widetilde{g}\circ|\pm\widetilde{1}\rangle=\pm|\pm\widetilde{1}\rangle,
  \label{}
\end{align}
while $\widetilde{g}$ acts trivially on the original virtual legs. The nontrivial element in $\widetilde{IGG}=\widetilde{Z_2}$ is the action of $\widetilde{g}$ on all new virtual legs only. Then, if the action of $\widetilde{g}$ on all new virtual legs of a tensor leaves the tensor invariant, we call this tensor a $\widetilde{Z_2}$ even tensor. Similarly, we can define the $\widetilde{Z_2}$ odd tensor. To create a fluxon living on plaquette center $c$, one can simply change the plaquette tensor $P_c$ from $\widetilde{Z_2}$ even to $\widetilde{Z_2}$ odd. For instance, a plaquette tensor $P_c^m$ which supports one fluxon projects out quantum states with even numbers of $|-\widetilde{1}\rangle$:
\begin{align}
  P_c^{m}=|-\widetilde{1},\widetilde{1},\widetilde{1},\widetilde{1}\rangle+|-\widetilde{1},-\widetilde{1},-\widetilde{1},\widetilde{1}\rangle+\dots
  \label{}
\end{align}
In order to see that $P_c^m$ indeed supports a fluxon, we translate back to the undecorated PEPS. $P_c$ only has configurations with odd numbers of $|-\widetilde{1}\rangle$, Thus, in the undecorated lattice, there are always odd numbers of modified bond tensors $g\cdot B_b$ around the plaquette $c$. In other words, a single fluxon lives at the plaquette $c$.

Fluxons inside a small patch can be detected by acting $\widetilde{g}$ on boundary of that patch, see Fig.(\ref{fig:decorated_peps}c). An odd number of fluxons contributes an additional $(-1)$ due to the $\widetilde{Z_2}$ oddness of the patch tensor. It is natural to interpret the loop of $\widetilde{g}$ (acting only on one end of an involved bond) as the chargon loop which detects fluxons. One can easily show that this is exactly true and the end points of an open $\widetilde{g}$-string are chargons. In the decorated PEPS, the chargon strings are explicit while the fluxon strings are ``hidden''.

\begin{figure}
\includegraphics[width=0.5\textwidth]{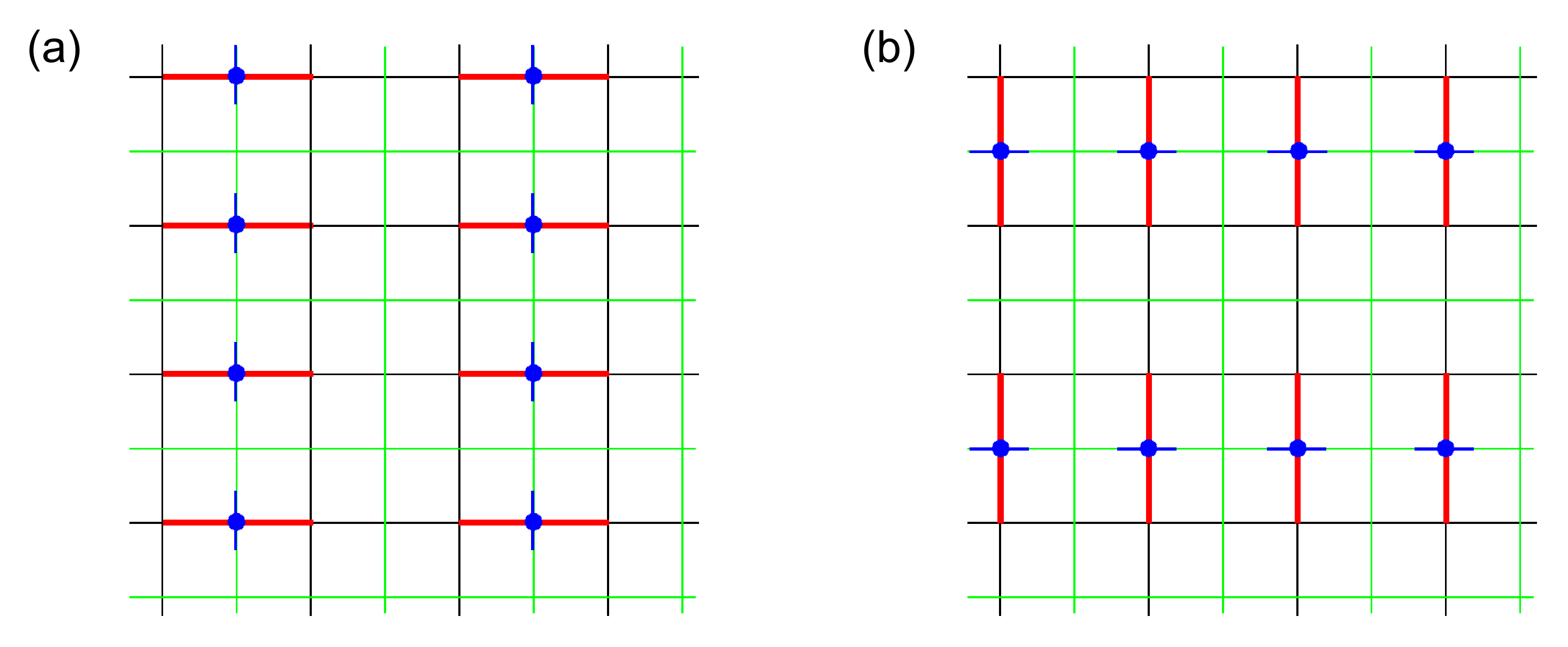}
\caption{(a),(b):The decorated PEPS on square lattice with every site tensor to be $Z_2$ odd. Bond tensors with blue dot and blue line should be modified as $\widetilde{B}'_b=\widetilde{g}\cdot\widetilde{B}_b$ to ensure that every site is the end of some chargon string. One can verify (a) and (b) are gauge equivalent, thus represent the same state.}
  \label{fig:Z2_odd_square_decorated_peps}
\end{figure}

Before we discuss the symmetry fractionalization of fluxons, let us study the case with site tensors being $Z_2$ odd. In this case, if one exactly follows the above decoration procedure, one finds that the similarly constructed $|\Psi_\mathcal{L}\rangle$ wavefunction satisfies:
\begin{align}
  |\Psi_\mathcal{L}\rangle=(-1)^{n_s}|\Psi\rangle
  \label{}
\end{align}
where $n_s$ is the number of sites enclosed by the loop $\mathcal{L}$. To construct a decorated PEPS with in-phase contributions from all $|\Psi_\mathcal{L}\rangle$, we require an additional $-1$ for loops enclosing odd numbers of sites, and need to modify the decoration procedure. As shown in Fig.(\ref{fig:Z2_odd_square_decorated_peps}), we simply modify some bond tensors by the action of $\widetilde{g}$ \emph{only on one of the two new virtual legs}. More precisely, we can choose bond tensors in $x$ direction to be modified for every other column in the following way:
\begin{align}
  \widetilde{B}'_b&=\widetilde{g}\cdot\widetilde{B}_b\notag\\
  &=B_b\otimes|\widetilde{1},\widetilde{1}\rangle-g\cdot B_b\otimes|-\widetilde{1},-\widetilde{1}\rangle
  \label{}
\end{align}
Note that the $\widetilde{g}$ action here is defined to only act on one of the two new virtual legs, and picks up the $(-1)$ for the second term in the second line. One can easily verify that for the modified decorated PEPS, loop configurations enclosing odd numbers of sites contribute the same wavefunction as the undecorated PEPS.

As we pointed out before, $\widetilde{g}$ strings can be interpreted as chargon strings. Further, the $Z_2$ oddness of site tensors can be interpreted as one chargon per site. The action of $\widetilde{g}$ on modified bond tensors in fact creates short ``chargon strings'' connecting background chargons. A loop of $|-\widetilde{1}\rangle$ enclosing an odd number of sites intersects with chargon strings for an odd number of times, and contributes an extra $-1$. Keeping this in mind, one can easily construct other possible decorated PEPS, once ensuring that every background chargon ($Z_2$ odd tensor) is an end of a $\widetilde{g}$ string, as shown in Fig.(\ref{fig:Z2_odd_square_decorated_peps}). We point out that all different decorated PEPS obtained for the same quantum state are gauge equivalent, in the sense that they can be transformed to each other by acting $\widetilde{g}$ on both ends of a collection of virtual legs. 

\begin{figure}
\includegraphics[width=0.45\textwidth]{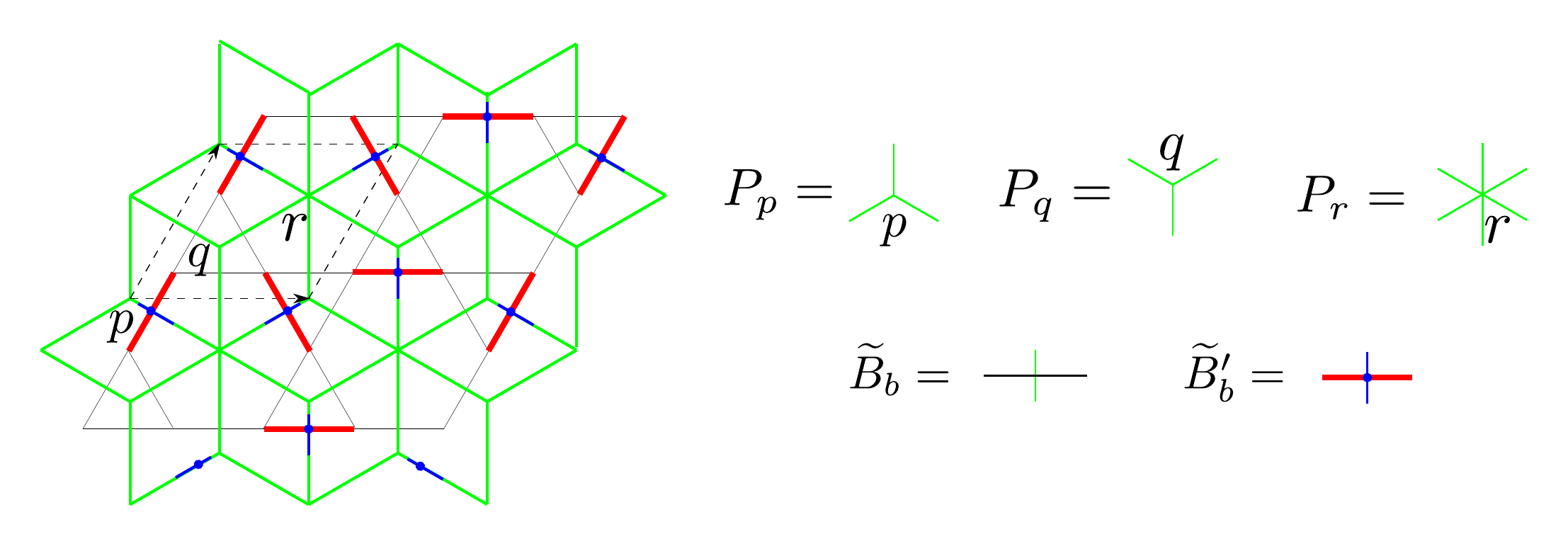}
\caption{The decorated PEPS for the kagome lattice with $Z_2$ odd site tensors. Fluxons live on the dual lattice, which is the dice lattice in this case. There are three plaquette centers in one unit cell, labeled as $p$, $q$ and $r$. Green/blue legs are two dimensional Hilbert space with basis as $|\pm\widetilde{1}\rangle$. Plaquette tensors project out configurations with odd number of $|-\widetilde{1}\rangle$. Bond tensors need to be modified in two different ways according to their color.} 
  \label{fig:kagome_kagome_decorated_peps}
\end{figure}

The above point of view is very useful for us to construct the decorated PEPS in more complicated lattices such as the kagome lattice. In the kagome lattice, plaquette centers form a dice lattice, as shown in Fig.(\ref{fig:kagome_kagome_decorated_peps}). Following the similar procedure as in the square lattice case, we decorate the PEPS by adding plaquette tensors and changing bond tensors. The new virtual legs are 2-dimensional Hilbert space, with basis $|\pm\widetilde{1}\rangle$ . Note that unlike the previous case, there are three kinds of plaquette tensors. Two lie at centers of triangles while one lies at the honeycomb center. These tensors project out configurations with odd numbers of $|-\widetilde{1}\rangle$. Further, due to the $Z_2$ oddness of site tensors, we should ensure that every site tensor in the decorated PEPS is connected with a chargon string. Thus, we can modify bond tensors as 
\begin{align}
  \widetilde{B}_b&=B_b\otimes|\widetilde{1},\widetilde{1}\rangle+g\cdot B_b\otimes(|-\widetilde{1},-\widetilde{1}\rangle)\notag\\
  \widetilde{B}'_b&=B_b\otimes|\widetilde{1},\widetilde{1}\rangle+g\cdot B_b\otimes\tilde{g}\cdot(|-\widetilde{1},-\widetilde{1}\rangle)\notag\\
  &=B_b\otimes|\widetilde{1},\widetilde{1}\rangle-g\cdot B_b\otimes(|-\widetilde{1},-\widetilde{1}\rangle),
  \label{}
\end{align}
where the pattern of bond tensors is shown in Fig.(\ref{fig:kagome_kagome_decorated_peps}).

\subsection{Symmetry fractionalization of fluxons}
In this part, we will develop a general method to extract (fractional) quantum numbers carried by fluxons.

Comparing with the undecorated PEPS, there are more gauge freedoms associated with new virtual legs for the decorated PEPS. We call these new gauge freedoms as $\widetilde{V}$. For the purpose of the discussion in this section, we only need to consider $\widetilde{V}$ gauge transformations such as acting $\tilde g$ on both ends of a number of new virtual legs, which leave the physical wavefunction intact. If the quantum state is invariant under a symmetry $R$, tensors of the decorated PEPS should satisfy the following conditions:
\begin{align}
  T^s&=\Theta_RW_RR\circ T^s\notag\\
  \widetilde{B}_b&=\widetilde{W}_RW_RR\circ\widetilde{B}_b\notag\\
  P_c&=\widetilde{W}_RR\circ P_c
  \label{}
\end{align}
where $W_R$ labels the gauge transformation associated with symmetry $R$ on old virtual legs while $\widetilde{W}_R$ labels that on new legs. Here $W_R$ takes the same value as in the undecorated PEPS. And we will solve $\widetilde{W}_R$ in the following.

Let us firstly consider a simple example: the translation symmetry group generated by $T_1, T_2$ for the decorated PEPS defined on the square lattice. For the case where site tensors are $Z_2$ even, we have $\widetilde{W}_{T_1}$ and $\widetilde{W}_{T_2}$ to be identity. While for the case with site tensors being $Z_2$ odd, $\widetilde{W}_{T_i}$ is nontrivial. If we decorate the PEPS as in Fig.(\ref{fig:Z2_odd_square_decorated_peps}a), we get
\begin{align}
  \widetilde{W}_{T_1}(x,y,i)=\mathrm{\widetilde{J}}^y,\quad \widetilde{W}_{T_2}(x,y,i)=\mathrm{I}
  \label{}
\end{align}
where $i$ labels the four new virtual legs coming out of the plaquette tensor at $(x,y)$, and
$\mathrm{\widetilde{J}}=\bigl(\begin{smallmatrix}
  1&0\\0&-1
\end{smallmatrix}\bigr)$
is the representation of $\widetilde{g}$ on new virtual legs. If the decoration has the form as in Fig.(\ref{fig:Z2_odd_square_decorated_peps}b), we get
\begin{align}
  \widetilde{W}_{T_1}(x,y,i)=\mathrm{I},\quad \widetilde{W}_{T_2}(x,y,i)=\mathrm{\widetilde{J}}^x
  \label{}
\end{align}
As we can see, values of $\widetilde{W}_{T_i}$ depend on the way to decorate the PEPS. However, similar to the undecorated case, we have
\begin{align}
  &\widetilde{W}_{T_2}^{-1}(T_2(x,y,i))\widetilde{W}_{T_1}^{-1}(T_1T_2(x,y,i))\widetilde{W}_{T_2}(T_1T_2(x,y,i))\notag\\
  &\widetilde{W}_{T_1}(T_2^{-1}T_1T_2(x,y,i))=\widetilde{\eta}_{12}
  \label{}
\end{align}
as a gauge invariant quantity. Inserting $\widetilde{W}_{T_i}$ into the above equation, we conclude that $\widetilde{\eta}_{12}=\mathrm{I}$ for the case with site tensors of undecorated PEPS being $Z_2$ even, while $\widetilde{\eta}_{12}=\mathrm{\widetilde{J}}$ for the case with site tensors being $Z_2$ odd. 

Next, we will show that $\widetilde{\eta}_{12}$ is directly related to the translation symmetry fractionalization of fluxons. To see this, let us consider the decorated PEPS with \emph{only} fluxon excitations inside some small patches. (Note that the decorated PEPS is inconvenient to study chargon excitations since chargon-strings are explicit.) Following the similar procedure in Sec.(\ref{subsec:spinon_sym_frac}), one can construct local translation operators for a single patch $\mathcal{\widetilde{A}}$ of the decorated PEPS, labeled as $\hat{L}^{\mathcal{\widetilde{A}}}_{T_i}$, $i=1,2$. Labeling the state associated with $\mathcal{\widetilde{A}}$ as $\hat{T}_\mathcal{\widetilde{A}}$, we get
\begin{align}
  \hat{L}^{\mathcal{\widetilde{A}}}_{T_i}\cdot\hat{T}_\mathcal{\widetilde{A}}=\widetilde{W}_{T_i}W_{T_i}T_i\circ\hat{T}_\mathcal{\widetilde{A}}
  \label{}
\end{align}
By series connecting local symmetry operators, we can define
\begin{align}
  \hat{L}_{T_2^{-1}T_1^{-1}T_2T_1}^{\mathcal{\widetilde{A}}}\equiv(\hat{L}^{\mathcal{\widetilde{A}}}_{T_2})^{-1}\cdot(\hat{L}^{\mathcal{\widetilde{A}}}_{T_1})^{-1}\cdot\hat{L}^{\mathcal{\widetilde{A}}}_{T_2}\cdot\hat{L}^{\mathcal{\widetilde{A}}}_{T_1}
  \label{}
\end{align}
as the local operator associated with $T_2^{-1}T_1^{-1}T_2T_1$. Acting this operator on $\hat{T}_\mathcal{\widetilde{A}}$, we have
\begin{align}
  \hat{L}_{T_2^{-1}T_1^{-1}T_2T_1}^{\mathcal{\widetilde{A}}}\cdot\hat{T}_\mathcal{\widetilde{A}}=\chi_{12}\eta_{12}\widetilde{\eta}_{12}\circ\hat{T}_\mathcal{\widetilde{A}}
  \label{}
\end{align}
For the case with nontrivial $\widetilde{\eta}_{12}$, if there are odd numbers of fluxons inside the patch $\mathcal{\widetilde{A}}$, $\hat{L}_{T_2^{-1}T_1^{-1}T_2T_1}^{\mathcal{\widetilde{A}}}$ will pick up an extra $-1$. This indicates the nontrivial translational symmetry fractionalization of fluxons. Since $\widetilde{\eta}_{12}$ only depends on the $Z_2$ parity of site tensors, we conclude that the translation symmetry fractionalization of fluxons is fully determined by the background chargon distribution. The above argument can be easily generalized to arbitrary lattice symmetry operations.

In the following, we will figure out $\widetilde W_R$ of the decorated PEPS for the kagome lattice example. As we discussed above, this directly implies the symmetry fractionalization pattern for fluxons. One can easily work out symmetry transformation rules directly from the decorated PEPS shown in Fig{\ref{fig:kagome_kagome_decorated_peps}}. The result is listed as follows.
\begin{align}
  &\widetilde W_{T_1}(x,y,\widetilde{s},i)=\widetilde{\mathrm{J}}^y,\notag\\
  &\widetilde W_{T_2}(x,y,\widetilde{s},i)=\mathrm{I}\notag,\\
  &\widetilde W_{C_6}(x,y,p/q,i)=\widetilde{\mathrm{J}}^{xy+\frac{1}{2}x(x+1)+1},\notag\\
  &\widetilde W_{C_6}(x,y,r,i)=\widetilde{\mathrm{J}}^{xy+\frac{1}{2}x(x+1)+x+y},\notag\\
  &\widetilde W_\sigma(x,y,p,i)=\widetilde{\mathrm{J}}^{xy+1},\notag\\
  &\widetilde W_\sigma(x,y,q/r,i)=\widetilde{\mathrm{J}}^{xy},
  \label{eq:kagome_vison_psg}
\end{align}
where $\widetilde s$ can be any one of the plaquette sublattices labeled by $p/q/r$ as shown in Fig.\ref{fig:kagome_kagome_decorated_peps}, and $i$ labels the new virtual legs coming out of the plaquette tensor. These symmetry transformation rules are not gauge invariant. The symmetry fractionalization of fluxons is determined by gauge invariant quantities $\widetilde{\eta}$'s. By replacing $W_R$ in Eq.(\ref{eq:psg_eta12}), (\ref{eq:psg_etaT1C6_etaT2C6}), (\ref{eq:psg_etaC6}), (\ref{eq:psg_etaT1sigma_T2sigma}), (\ref{eq:psg_etasigma}) and (\ref{eq:psg_etasigmaC6}) with $\widetilde{W}_R$ obtained above, we find 
\begin{align}
  &\widetilde{\eta}_{12}=\widetilde{\eta}_{T_1\sigma}=\widetilde{\eta}_{T_2\sigma}=\widetilde{\mathrm{J}},\notag\\
  &\widetilde{\eta}_{T_1C_6}=\widetilde{\eta}_{T_2C_6}=\widetilde{\eta}_{C_6}=\widetilde{\eta}_{\sigma}=\widetilde{\eta}_{\sigma C_6}=\mathrm{I}.
  \label{eq:kagome_vison_sf}
\end{align}

For on-site symmetries such as the spin rotation and the time reversal symmetry, symmetry transformation rules on new virtual legs are trivial. In other words, we expect fluxons constructed here to be spin 0 as well as Kramer singlet.  Our result for symmetry fractionalization of fluxons is consistent with eariler results in Ref.\onlinecite{Huh:2011p94419,Lu:2014p}.

\section{Symmetric PEPS on Torus and Long-Range Order}\label{sec:torus_long_range_order}
In the above discussion, we mainly focus on symmetric PEPS on infinite lattices. In this section, we will consider the symmetric PEPS on a finite torus. Namely, site tensors on the left (up) boundary are connected to sites on the right (down) boundary by bond tensors. Further, we will provide some of our partial understandings on how long-range ordered phases fit into the current symmetric PEPS formulation.

To make the discussion concrete, \emph{we will focus on spin-$\frac{1}{2}$ systems to demonstrate the principle}. We firstly consider the PEPS description of a $Z_2$ spin liquid phase on finite tori in subsection \ref{sec:topo_deg_torus}, and construct the topological degenerate ground state sector. Then we study the finite size effects due to the spinon (vison) condensation in subsection \ref{sec:long_range_order_torus}, which gives rise to the MO (VBS order) in the long range. One result here is that the undecorated PEPS provides the natural basis to represent a MO state, while the decorated PEPS is the natural language to represent a VBS state.

\subsection{Topological degeneracy in the PEPS formulation}\label{sec:topo_deg_torus}
\begin{figure}
\includegraphics[width=0.5\textwidth]{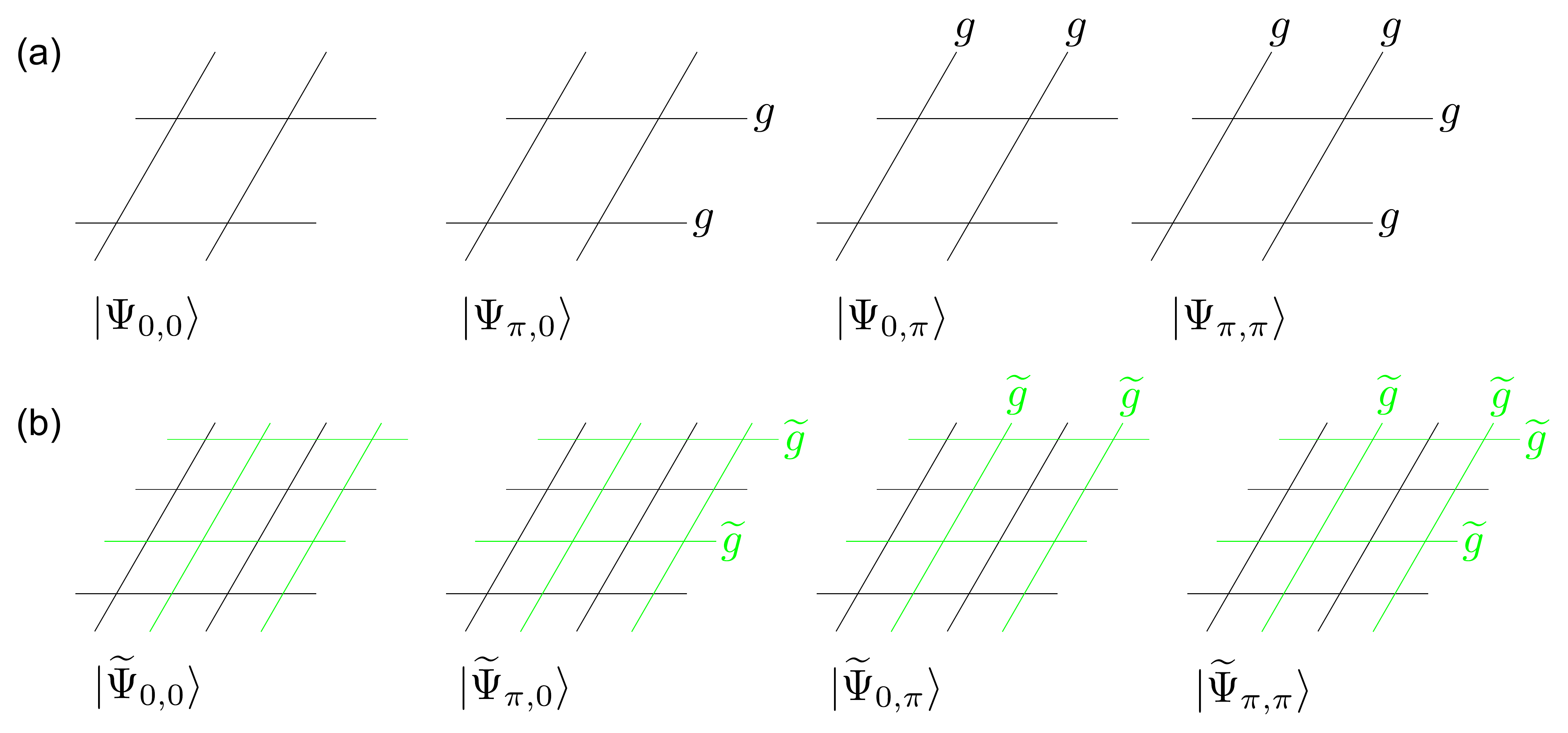}
\caption{Topological degenerate ground states on finite torus. Site tensors at left(up) boundary are connected with those at right(down) boundary . (a) labels $m$-basis while (b) labels $e$-basis.}
  \label{fig:torus_peps}
\end{figure}

It is well known that for the toric code topological ordered phase, the ground state degeneracy (GSD) equals four on torus in the thermodynamic limit. As mentioned in subsection \ref{sec:gauge_dynamics}, in the $Z_2$ invariant PEPS, one can construct these four states on a finite torus by acting the nontrivial $Z_2$ element $g$ on the non-contractible loops of torus, as shown in Fig.(\ref{fig:torus_peps}a).\footnote{Note that in the presence of lattice symmetries, sometimes a finite sample has a geometry which does not match the periodicity of the tensor transformation rules, and the lattice symmetry cannot be respected. For the discussion in this subsection one does not have to worry about the lattice symmetry.} We label these four states as $|\Psi_{0,0}\rangle$, $|\Psi_{\pi,0}\rangle$, $|\Psi_{0,\pi}\rangle$ and $|\Psi_{\pi,\pi}\rangle$. Recall that $g$ strings can be interpreted as flux loop. So these four ground state basis can be visualized as different ways of inserting non-contractible flux loops. We 
call this set of basis for the ground state manifold as the $m$-basis. Note that in general, these four states have different energies on any finite torus. However, if the system is in the deconfined phase, the energy difference between these four states goes to zero in the thermodynamic limit.

Now, let us consider decorated PEPS defined in Sec.\ref{sec:vison_psg}. The decorated PEPS describes the same wavefunction as the undecorated one on the infinite plane. However, this is not true if we consider finite samples on tori. Using the method developed in the last section, we can easily construct the decorated PEPS on a torus based on an undecorated PEPS state $|\Psi_{0,0}\rangle$. Similar to the infinite PEPS case, any configurations with all $|-\widetilde{1}\rangle$ forming loops will contribute to the wavefunction of the decorated PEPS. For a torus sample, we should consider both contractible loops and non-contractible loops of $|-\widetilde{1}\rangle$. First, any configurations with only contractible loops contribute $|\Psi_{0,0}\rangle$ to the decorated wavefunction. Note that for 
configurations with an even number of 
non-contractible loops, one can always decompose them to only contractible loops. For configurations containing an odd number of non-contractible loops in the $x$/$y$/both direction, one gets $\pm|\Psi_{\pi,0}\rangle$/$\pm|\Psi_{0,\pi}\rangle$/$\pm|\Psi_{\pi,\pi}\rangle$. Here, the $\pm$ signs depend on the way of decoration: as pointed out in Sec.(\ref{sec:vison_psg}), to capture the background chargons of original PEPS, some bonds are modified to $\widetilde{B}'_b=\widetilde{g}\cdot\widetilde{B}_b$. When loops of $|-\widetilde{1}\rangle$ intersect with bond $\widetilde{B}'_b$, we will get an extra $-1$. Thus, it is always possible to choose the distribution of $\widetilde{B}'_b$ such that the $\pm$ signs are $+1$. The obtained decorated wavefunction $|\widetilde{\Psi}_{0,0}\rangle$ is then
\begin{align}
  |\widetilde{\Psi}_{0,0}\rangle=|\Psi_{0,0}\rangle+|\Psi_{\pi,0}\rangle+|\Psi_{0,\pi}\rangle+|\Psi_{\pi,\pi}\rangle
  \label{}
\end{align}
up to a normalization factor.

Other three states in the decorated language can be generated from $|\widetilde{\Psi}_{0,0}\rangle$ by threading chargon strings on non-contractible loops of torus, or in other words, by acting nontrivial $\widetilde{Z_2}$ element $\widetilde{g}$ on new virtual legs along non-contractible loops. We label them as $|\widetilde{\Psi}_{\pi,0}\rangle$, $|\widetilde{\Psi}_{0,\pi}\rangle$ and $|\widetilde{\Psi}_{\pi,\pi}\rangle$ respectively, as shown in Fig.(\ref{fig:torus_peps}b). It is straightforward to see that
\begin{align}
  &|\widetilde{\Psi}_{\pi,0}\rangle=|\Psi_{0,0}\rangle+|\Psi_{\pi,0}\rangle-|\Psi_{0,\pi}\rangle-|\Psi_{\pi,\pi}\rangle\notag\\
  &|\widetilde{\Psi}_{0,\pi}\rangle=|\Psi_{0,0}\rangle-|\Psi_{\pi,0}\rangle+|\Psi_{0,\pi}\rangle-|\Psi_{\pi,\pi}\rangle\notag\\
  &|\widetilde{\Psi}_{0,\pi}\rangle=|\Psi_{0,0}\rangle-|\Psi_{\pi,0}\rangle-|\Psi_{0,\pi}\rangle+|\Psi_{\pi,\pi}\rangle
  \label{}
\end{align}
Here, the  basis formed by four $|\widetilde{\Psi}\rangle$ states is named as the $e$-basis. 

As concrete examples, in Appendix \ref{app:quantum_number} we explicitly constructed the four-fold ground state sectors of the $Q_1=Q_2$ QSL and $Q_1=-Q_2$ QSL on an even by odd $(4n+2)$-unit-cell torus sample. Interestingly, on such samples, we find that the ground state sectors of the two QSL share $no$ identical lattice symmetry irreducible representations(irreps); namely, any irrep in the $Q_1=Q_2$ QSL ground state sector is different from any irrep in the $Q_1=-Q_2$ QSL ground state sector. This result is consistent with a recent study based on parton constructions\cite{Qi:2015p100401}.

\subsection{Long-range ordered phases represented by symmetric PEPS}\label{sec:long_range_order_torus}
We claimed that each crude class contains many possible member phases, and these phases are distinguished by long-range physics. Here we mention some concrete examples. In the 32 crude classes on the kagome lattice, we know that each crude class contains one $Z_2$ QSL member phase which has no spontaneous symmetry breaking. Let's consider the crude class which contains the $Q_1=Q_2$ $Z_2$ QSL, firstly constructed using the Schwinger-boson approach\cite{Sachdev:1992p12377}. There are a few known neighboring phases of this QSL in which symmetry is spontaneously broken in different fashions. These symmetry breaking phases and the $Q_1=Q_2$ $Z_2$ QSL  are in the same crude class proposed in this work.

For example, the $120^{\circ}$ $Q=0$ long-range magnetic ordered (MO) state can be obtained if the spinons in the $Q_1=Q_2$ QSL condense in the long range\cite{Sachdev:1992p12377,Wang:2006p174423}. And a valence bond solid(VBS) state with 12-site per unit cell, which breaks the translational symmetry, can be obtained if the visons in the $Q_1=Q_2$ QSL condense in the long range\cite{Huh:2011p94419}. To capture the long-range physics of these symmetry breaking phases, we expect that scaling of bond dimension $D$ and scaling with system sizes need to be performed in our symmetric PEPS methods. For instance, it was shown that generically the entanglement entropy in symmetric states which breaks a continuous symmetry in the long-range contains additive logarithmic corrections\cite{Metlitski:2011p}. To our knowledge, there is no known PEPS construction with a finite $D$ that can be proven to host this behavior. 

Next, let us try to understand the effects of the spinon/boson condensations in the PEPS formulation on finite tori. Note that practical PEPS simulations on a torus sample can be computationally very expensive, so the discussion here is mainly for conceptual purposes. Before the condensation, the system is in a $Z_2$ spin liquid phase with a four-fold ground state sector $\mathcal{H}_{GS}$. After the vison-$m$ condensation, the system is expected to have a multi-fold ground state sector due to the lattice symmetry breaking, and the number of ground states depend on the geometry of the sample and the VBS order pattern. On the other hand, after the spinon-$e$ condensation, there will be gapless Goldstone excitations assuming spin-rotational symmetry. \emph{Exactly how the ground state sector on a torus evolves as the spinon/vison condenses?}

Below we provide our partial answer to this interesting question. Note that, in the ordered phases, for the system to ``know'' the spinon/vison condensation, we always consider torus samples whose sizes are larger than the vison/spinon confinement length scale. For simplicity, we will focus on samples whose geometries are commensurate with the spatial patterns of the long-range orders. We can consider a Ginzburg-Landau theory describing the boson condensation phase transition. In the Ginzburg-Landau language, this commensuration is achieved by choosing the finite lattices together with a proper boundary condition such that the momentum space minima of the condensed bosons are available. 

Here one immediately sees that \emph{the undecorated(decorated) $m$-basis ($e$-basis) is exactly suitable to describe the spinon(vison) condensation}. For instance, the existence of $m$-loops trapped in the torus holes is explicit in the undecorated $m$-basis. Consequently the boundary conditions for the spinon condensation Ginzburg-Landau theory are sharply defined. We conclude that only one of the four states in the $m$-basis is representing the true ground state in the MO phase. The spinon condensation minima in the momentum space are clearly \emph{not} available for the other three states due to the $\frac{\pi}{L}$ momentum shift. Physically, we know that the other three states trap MO vortices in the torus holes. Therefore they have finite excitation energies (but zero excitation energy density) in the MO phase based on simple nonlinear sigma model analysis.

Similarly, the decorated $e$-basis is suitable to describe the VBS order, since the boundary condition for the condensed visons is explicit. Only one of the four states in the $e$-basis corresponds to a true ground state in the VBS phase, while the other three states host VBS domain walls wrapping around the torus holes. Note that the confined spinon-$e$ corresponds to a VBS vortex\cite{Levin:2004p220403} and the $e$-string corresponds to a VBS domain wall in the VBS phase.  These three states are separated from the true ground state by excitation energies proportional to the linear sample size due to the energy cost of the domain wall. 

Note that the decorated PEPS only captures one state in the true ground state sector in the VBS phase. The other ``cat states'' in this symmetry breaking phase are still missing in the current construction. In fact on a torus sample with even by even unit cells, our construction leads to physical states with the center of mass momentum at $\Gamma=(0,0)$ only. We expect that, in order to capture the other cat states at different center of mass momentum, one needs to perform the symmetric PEPS classification on finite tori instead of on the infinite plane.

Next, we discuss some general guiding principles about these symmetry breaking member phases. It is crucial to note that these symmetry breaking patterns in the vicinity of the $Q_1=Q_2$ QSL are \emph{not} arbitrary. In fact, the sharp way to understand how the symmetry breaking patterns arise is exactly to use the Ginzburg-Landau theory --- the golden tool to investigate symmetry breaking. \emph{In order to write down a Ginzburg-Landau theory for the spinon-$e$ (vison-$m$) condensation, the only information that we need to know is how these particles transform under the global symmetry group, which is nothing but the projective symmetry group transformation rules for spinons and visons investigated in Sec.\ref{sec:lantern_operator_peps} and Sec.\ref{sec:vison_psg}.} In the past, this is exactly how the $120^{\circ}$ $Q=0$ MO state and the VBS state with a 12-site unit cell were identified to be the neighboring phases of the $Q_1=Q_2$ $Z_2$ QSL.

Therefore, even though we emphasize that our algorithm can be used to efficiently determine the crude class of the quantum ground states based on short range physics, we still learn sharp constraints on candidate symmetry breaking patterns in the long range physics: these orders must be consistent with the Ginzburg-Landau theory of the given crude class. With a careful scaling with larger system sizes and bond dimension $D$, our algorithm could be practical useful to pin down the possible long range orders in quantum phase diagrams.

Note that the spinon condensation and vison condensation in the 32 $Z_2$ QSL studied here are quite different: The $\eta$ indices are really the symmetry transformation rules for spinons, while the symmetry transformation rules for visons are completely fixed. Consequently, two QSL with different $\eta$ indices are expected to connect to different MO orders after spinon condensations, since the Ginzburg-Landau theories are different. For instance, the $Q_1=-Q_2$ $Z_2$ QSL is connected to the $\sqrt{3}\times\sqrt{3}$ MO\cite{Sachdev:1992p12377}, fundamentally different from the $120^{\circ}$ $Q=0$ order in the vicinity of the  $Q_1=Q_2$ QSL. 

However, the same Ginzburg-Landau analysis indicates that all the 32 $Z_2$ QSL could give rise to the \emph{same} long-range VBS order pattern after vison condensation. There are two possible explanations for this phenomenon: (1) it is possible that the long range VBS orders emerging from different QSL, although sharing the same real space pattern, are still in different quantum phases. (2) it is possible that certain VBS phase could appear in distinct crude classes. \footnote{There is a third possible explanation: in order to describe the VBS phase, the PEPS must be in a class with a larger invariance gauge group than $Z_2$, which has many descendant $IGG=Z_2$ classes. This picture is related to the discussion on the hierarchical structure of the crude classes near the end of this paper. Although conceptually this explanation is self consistent, our current knowledge contains no further evidence supporting it.}

We tend to believe that either scenario could be correct under certain conditions, although we do not have rigorous understandings. It would be very helpful to perform numerical simulations based on the algorithms proposed here in models supporting relevant symmetry breaking phases and see exactly what happens. But we have to leave this as a subject of future investigations. At this moment we only can provide some physical speculations. 

First, the Ginzburg-Landau theory for vison or spinon condensations completely misses $\chi$ indices in the classification, and the long-range real space VBS pattern does not capture the physics described by these indices. Consequently we expect that scenario-(1) is correct if different classes have different $\chi$ indices. In fact, we could compare two classes with $\chi_{\mathcal{T}}=+1$ and $\chi_{\mathcal{T}}=-1$ respectively. The $Z_2$ QSL member phase in the second class is expected to host nontrivial projective representations of $SU(2)\times \mathcal{T}$ on the physical boundary. These boundary degrees of freedom are expected to lead to measurable effects even if the bulk VBS order is established, which is absent in the first class.

Second, we speculate that scenario-(2) could be correct when different classes share the same $\chi$ indices but have different $\eta$ indices, e.g., the $Q_1=Q_2$ QSL and the $Q_1=-Q_2$ QSL. Note that this speculation is not as naive as it appears. In particular, in Appendix \ref{app:quantum_number} we demonstrate that the 4-fold ground state sectors of these two QSL have completely different lattice symmetry irreps on $(4n+2)$-unit-cell samples. Therefore the symmetric PEPS of the two classes in the VBS ordered phase must be describing distinct ground state wavefunctions on such samples. In Appendix \ref{app:vbs} we trace the origin of the lattice quantum number discrepancy of the two PEPS classes in the 12-site VBS phase, and show that it is related to the phase factor due to the quantum fluctuation of valence bonds along a VBS domain wall. However, we expect that this particular phase factor is \emph{not} a universal feature, and the VBS orders in the two classes correspond to the same quantum phase.

\section{Discussion and Conclusions}\label{sec:discussion}

In this paper we attempt to construct generic symmetric ground state wavefunctions for integer or fractionally filled correlated systems using PEPS, under certain assumptions. Here we review the assumptions that we made and discuss the limitations and generalizations of our results. 

We firstly highlight an assumption that is, to our knowledge, due to a more fundamental difficulty. And we currently do not know how to solve this difficulty generally. This assumption is that the on-site symmetry is implemented as the simple tensor product of local representations or projective representations on the virtual legs in PEPS. For instance, this is the origin of the minimal required $Z_2$ $IGG$ in the half-integer spin systems on the kagome lattice. 

This assumption is known to have problems at least in the long range physics when attempting to describe SPT phases. For instance, let's attempt to construct a U(1) charge-conserving Chern insulator using the fermionic version of PEPS (fPEPS)\cite{Barthel:2009p42333,Corboz:2010p165104,Kraus:2010p52338,Pizorn:2010p245110}. Here the exact constructions of free fermion states with a nonzero Chern number using Gaussian fPEPS\cite{Dubail:2013p,Wahl:2013p236805}, in which the virtual legs transform as U(1) representations, are shown to host power law correlation functions in the real space. It has been pointed by Hastings\cite{Hastings:2014p} that for a general U(1) symmetric PEPS with a bounded bond dimension $D$ which is a fully gapped ground state of a local Hamiltonian, the assumption that the virtual legs transform as U(1) representations and the assumption that the PEPS carries nonzero Chern number generically lead to contradictions. 

Another important piece of information can be obtained by understanding the exact PEPS constructions of available short-range correlated SPT ground state wavefunctions of exact solvable models\cite{Williamson:2014p}. In particular, for a finite on-site symmetry group, it is shown that the virtual legs do not form representation (or projective representations) of the symmetry. Instead the virtual degrees of freedom transform in a ``non-on-site'' fashion, which can be described by matrix product operators\cite{Chen:2011p235141,Williamson:2014p}. 

We currently do not know how to generically represent an SPT state in two and higher dimensions with correct long-range physics using PEPS. However, it is possible that our assumption about symmetry representations on virtual legs does not cause problems in capturing the short-range physics of SPT phases under certain conditions. For instance, given an finite size sample, it is possible that the SPT ground state can be accurately approximated with a PEPS after a scaling with respect to bond dimension $D$ is performed. As demonstrated in an example using Gaussian fPEPS in Ref.\onlinecite{Wahl:2013p236805}, the required bond dimension $D$ in practical simulations on intermediate sized sample may not be very large.

Moreover, we speculate that even the short-range physics of a SPT phase may not be captured using the current symmetric PEPS construction. For example, it is known that for inversion symmetric system, the ground state wavefunction of a Chern insulator with an odd Chern number is inversion odd.\cite{Turner:2012p165120} And the inversion quantum number should be completely short-range physics.

We made a second assumption: we study only those symmetric quantum ground states that can be represented by a single tensor network on the infinite lattice. This assumption is made here mainly for technical simplicity rather than fundamental difficulty. Note that this assumption is weaker than the assumption that the ground state sector is composed of one-dimensional representations of the symmetry group on any finite size samples. For instance consider a $Z_2$ QSL studied in this paper with a four-fold ground state sector on tori. When considering a finite size torus, some of them could form multi-dimensional irreducible representations of the space group.

This assumption could be violated in general model simulations. As a trivial example we could consider a ferromagnetic state in an $SU(2)$ symmetric model. In this case the number of degenerate ground states scale linearly as the number of sites, which certainly cannot be represented by one or few PEPS.

As a slightly nontrivial example, we refer to the chiral-spin-charge-Chern liquid (SCCL) in Ref.\onlinecite{Jiang:2014p31040}. The spin dynamics in SCCL is described by a chiral $Z_2$ QSL, which is a $Z_2$ QSL breaking the time reversal symmetry and has nonzero spin-chirality order parameter (e.g., $<\vec S_i\cdot \vec S_j\times \vec S_k>\neq 0$ for three nearby spins $i,j,k$.). This state breaks both time-reversal and mirror reflection symmetries, but leaves the combination of the two respected. In this situation, we found $8=4\times2$ ground states on symmetric torus samples (compatible with the PSG transformations). The factor of $4$ is related to the topological degeneracy of $Z_2$ gauge theory. And the extra factor of $2$ is due to the fact that the time reversal, the mirror reflection and the lattice rotation form nontrivial 2-dimensional irreducible representations. The latter fact dictates that it is impossible to represent such chiral liquids by a single symmetric PEPS, in which case the extra 
factor of $2$ 
degeneracy cannot be captured. 
The simple way to proceed is to instead only consider the combination of the time reversal and the mirror reflection as a symmetry, which allows a description of one of the two time-reversal images using PEPS. The PEPS description of the other state can be obtained by the time-reversal transformation.

We now comment on another fact in our construction. In the half-integer spin systems on the kagome lattice, we show that a spin-singlet symmetric PEPS has an $IGG$ that at least contains a $Z_2$ subgroup. If $IGG=Z_2$ for a PEPS, and if the PEPS is describing a fully gapped QSL, we showed that the topological order is toric-code-like in Sec.\ref{sec:gauge_dynamics}. This remains to be true if we construct some $Z_2$ QSL in the absence of the time-reversal symmetry, using our formulation. However, there are known constructions\cite{Buerschaper:2014p195148,Qi:2014p,Iqbal:2014p115129} of gapped $Z_2$ QSL on the kagome lattice in the absence of the time-reversal symmetry whose topological order is the same as the one in the double-semion model, fundamentally different from toric-code. 

Interestingly, in a PEPS construction of the double semion QSL\cite{Iqbal:2014p115129}, in which spin rotation is still implemented as representations on the virtual legs, the constructed tensors are actually $Z_4$ invariant. Naively, such a state should have a 16-fold degenerate ground state sector on torus, but it was shown that only $4$ of them are linearly independent. 

Next we comment on the connection between our work with previous works. For readers that are familiar with the parton constructions and projective symmetry group analysis of parton wavefunctions\cite{Wen:2002p165113,Wang:2006p174423,Lu:2011p224413}, clearly part of our results can be viewed as generalizations of these analyses into PEPS wavefunctions. In particular, in the kagome half-integer spin $S$ example presented here, every crude class contains a distinct $Z_2$ QSL as a member phase. Part of our results can be viewed as a classification of $Z_2$ QSL on the kagome lattice. Comparing with previous investigations on this topic specifically for $S=1/2$, based on parton constructions\cite{Wang:2006p174423,Lu:2011p224413}, we find that our result captures every phase present in the Schwinger-boson construction\cite{Wang:2006p174423}, and finer than that. Basically the previous PSG analysis of the Schwinger boson construction is related to the $\eta$-indices and $\Theta$-indices in our formulation, while in 
this work $\chi$-indices are revealed. 

However, comparing with the classification based on the Abrikosov-fermion construction of $Z_2$ QSL on the kagome lattice\cite{Lu:2011p224413}, we find that some of them cannot be described in our result. Similar observation was made by Ref.\onlinecite{Lu:2014p} when directly comparing Schwinger-boson and Abrikosov-fermion constructions. We currently do not have a full understanding of the physics behind this phenomenon. But it is worth pointing out that the missing Abrikosov-fermion $Z_2$ QSL are all found to be gapless (at least perturbatively) on the mean-field level\cite{Lu:2011p224413}.

Finally we comment on the hierarchical structure of the crude classes. Sometimes there are physical reasons to believe that the $IGG$ needs to be larger than the minimal required one in order to correctly capture certain quantum phases. The double semion PEPS mentioned above may be viewed as such an example.

A more conventional example in which this is expected to happen is the collinear Neel ordered phase on the square lattice, for which we expect $IGG=U(1)$ in our PEPS construction. In fact, the non-compact CP$^1$ (NCCP$^1$) description\cite{Senthil:2004p1490} for the Neel state signals that the natural gauge dynamics in this state, although Higgs'ed out in the long-range, should be $U(1)$.

One can ask the following question: for instance, suppose we have one parent crude class of symmetric PEPS with a large $IGG_1$, what are the possible descendant crude classes with a smaller $IGG_2\subset IGG_1$? Similar questions were investigated in the context of parton constructions\cite{Lu:2011p24420,Lu:2011p224413}. Generally one expects that there could exist multiple descendant crude classes with $IGG_2$, which eventually gives a hierarchical structure of the crude classes with different $IGG$'s. This hierarchical structure may be useful to understand certain exotic quantum criticalities. For example, two member phases belong to distinct crude classes could be smoothly connected via a critical point described by their parent crude class\footnote{In addition, for the deconfined criticality on the square lattice\cite{Senthil:2004p1490}, we expect that the Neel ordered state is described by a class with $IGG=U(1)$, while the VBS ordered state is described by a class with $IGG=Z_2$.}. 

As one can see from the above discussions, the current work, which is based on the point of view of diagnosing ground state wavefunctions using symmetric PEPS, brings up many open questions and needs future investigations to clarify. In addition, the algorithms proposed here for simulating strongly interacting models need benchmark tests to have a understanding of its practical performance. Nevertheless we believe that separating the short-range part of the physics from the long-range part is a useful idea in investigating quantum phase diagrams of strongly correlated systems. While generally the long-range part is still a difficult task, we expect that the method introduced here can be used to provide sharp information for the short-range physics efficiently.

We thank for help discussions with Fa Wang, Jung-Hoon Han, Hyunyong Lee, Panjin Kim, Xie Chen and Michael Hermele. This work is supported by the Alfred P. Sloan fellowship and National Science Foundation under Grant No. DMR-1151440.

\begin{appendix}

\section{Symmetry group of the kagome lattice}\label{app:sym_group_kagome}
As shown in Fig.(\ref{fig:kagome_lattice}), we label the three lattice sites in each unit cell with sublattice index $\{s=u,v,w\}$. Further, we specify the virtual index $\{i=a,b,c,d\}$ of a given site. We choose Bravais unit vector as $\vec{a}_1=\hat{x}$ and $\vec{a}_2=\frac{1}{2}(\hat{x}+\sqrt{3}\hat{y})$. Thus, we are able to specify the virtual degrees of freedom of site tensors as $(x,y,s,i)$. The symmetry group of such a two-dimensional kagome lattice is generated by the following operations
\begin{align}
  \begin{split}
    T_1 : &\ (x,y,s,i) \to (x+1,y,s,i),\\
    T_2 : &\ (x,y,s,i) \to (x,y+1,s,i),\\
    \sigma : &\ (x,y,u,i) \to (y,x,u,i_{\sigma1}),\\
    &\  (x,y,v,i) \to (y,x,w,i_{\sigma2}),\\
    &\  (x,y,w,i) \to (y,x,v,i_{\sigma2}),\\
    C_6 : &\ (x,y,u,i) \to (-y+1,x+y-1,v,i),\\
    &\  (x,y,v,i) \to (-y,x+y,w,i).\\
    &\  (x,y,w,i) \to (-y+1,x+y,u,i_{C_6}).
  \end{split}
  \label{eq:space_group_generator}
\end{align}
together with time reversal $\mathcal{T}$. Here,
\begin{align}
  \{a_{\sigma1},b_{\sigma1},c_{\sigma1},d_{\sigma1}\}&=\{d,c,b,a\}\notag\\
  \{a_{\sigma2},b_{\sigma2},c_{\sigma2},d_{\sigma2}\}&=\{c,d,a,b\}\notag\\
  \{a_{C_6},b_{C_6},c_{C_6},d_{C_6}\}&=\{b,a,d,c\}\notag\\
  \label{}
\end{align}

The symmetry group of a kagome lattice is defined by the following algebraic relations between its generators:
\begin{align}
  \begin{split}
    T_2^{-1}T_1^{-1}T_2T_1 & =\mathrm{e},\\
    \sigma^{-1}T_1^{-1}\sigma T_2 & =\mathrm{e},\\
    \sigma^{-1}T_2^{-1}\sigma T_1 & =\mathrm{e},\\
    C_6^{-1}T_2^{-1}C_6T_1 & =\mathrm{e},\\
    C_6^{-1}T_2^{-1}T_1C_6T_2 & =\mathrm{e},\\
    \sigma^{-1} C_6\sigma C_6 & =\mathrm{e},\\
    C_6^6=\sigma^2=\mathcal{T}^2 &=\mathrm{e},\\
    g^{-1}\mathcal{T}^{-1}g\mathcal{T}=\mathrm{e},\, \forall g &=T_{1,2},\sigma, C_6
  \end{split}
  \label{eq:space_group_generator_relation}
\end{align}
where $\mathrm{e}$ stands for the identity element in the symmetry group.

Further, consider system with spin rotation symmetry operator $R_{\theta\vec{n}}$, which means spin rotation about axis $\vec{n}$ through angle $\theta$. We mainly consider half-integer spins ($SU(2)$ symmetry) in this paper. The spin rotation symmetry commutes with all lattice symmetries as well as time reversal symmetry:
\begin{align}
  g^{-1}R^{-1}_{\theta\vec{n}}gR_{\theta\vec{n}}&=\mathrm{e},\, \forall g=T_{1,2},\sigma,C_6 , \mathcal{T}\notag\\
  \label{spin_rotation_sym_constraint}
\end{align}

\section{Classification of PEPS wavefunction with $Z_2$ $IGG$ on kagome lattice}\label{app:kagome_PEPS_Z2_PSG}
In this Appendix, we will classify symmetric PEPS wavefunctions defined on kagome lattice with a half-integer spin per site. We first obtain the symmetry transformation rules for all different classes. Then using the symmetry transformation rules, we can solve the constraint Hilbert space for all classes. 

\subsection{Solving symmetry operation on PEPS}
In this part, we will work out symmetry transformation rules for PEPS defined on a infinite kagome lattice with a half-integer spin per site. We will focus on the case with minimal required $IGG$, which equals $Z_2$, as shown in the main text. Further, every site tensor is a $Z_2$ odd tensor as we will see later.

As shown in Sec.(\ref{subsec:classification_kagome_peps}), the representation of $Z_2$ $IGG$ on virtual legs can be set as the same form $\{\mathrm{I},\mathrm{J}\}$, where $\mathrm{J}\equiv\mathrm{I}_{D_1}\oplus(-\mathrm{I}_{D_2})$ with a $\pm1$ ambiguity. The remaining $V$-ambiguity $V(x,y,s,i)$ commute with $\mathrm{J}$. Namely, $V(x,y,s,i)$ is block diagonal with two blocks with blocks' size to be $D_1$ and $D_2$ respectively. Further, for any symmetry $R$, we have proved the associated $W_R$ transformation should also commute with $\mathrm{J}$.

\subsubsection{Implementation of lattice symmetry on PEPS}
For completeness, we copy the calculation for translation symmetry transformation rules done in Sec.(\ref{subsec:classification_kagome_peps}). According to the definition of symmetric PEPS, for $T_i$ $(i=1,2)$ transformation, site tensors and bond tensors satisfy the following condition:
\begin{align}
  &T^{(x,y,s)}=\Theta_{T_i}W_{T_i}T_i\circ T^{(x,y,s)}\notag\\
  &B_{(xysi|x'y's'i')}=W_{T_i}T_i\circ B_{(xysi|x'y's'i')}\notag\\
  \label{eq:Ti_transf_tensor}
\end{align}

From group relation $T_2^{-1}T_1^{-1}T_2T_1=\mathrm{e}$, we have
\begin{align}
  &W_{T_2}^{-1}(T_2(x,y,s,i))W_{T_1}^{-1}(T_1T_2(x,y,s,i))W_{T_2}(T_1T_2(x,y,s,i))\notag\\
  &W_{T_1}(T_2^{-1}T_1T_2(x,y,s,i))=\eta_{12}\chi_{12}(x,y,s,i)\notag\\
  \label{eq:psg_eta12}
\end{align}
as well as
\begin{align}
  &\Theta^*_{T_2}{(T_2(x,y,s))}\Theta^*_{T_1}{(T_1T_2(x,y,s))}\Theta_{T_2}{(T_1T_2(x,y,s))}\notag\\
  &\Theta_{T_1}{(T_2^{-1}T_1T_2(x,y,s))}=\mu_{12}\prod_i\chi^*_{12}(x,y,s,i)
  \label{eq:psg_mu_eta12}
\end{align}
where $\eta_{12}\in\{\mathrm{I},\mathrm{J}\}$, and $\chi_{12}(x,y,s,i)$ is a bond dependent phase. Under $\varepsilon_{T_i}$ ambiguity $W_{T_i}\rightarrow\varepsilon_{T_i}W_{T_i}$, $\Theta_{T_i}\rightarrow\varepsilon_{T_i}\Theta_{T_i}$, we get
\begin{align}
  &\chi_{12}\rightarrow\varepsilon_{T_2}(x,y+1,s,i)\varepsilon_{T_1}^*(x+1,y+1,s,i)\cdot\notag\\
  &\varepsilon_{T_2}(x+1,y+1,s,i)\varepsilon_{T_1}(x+1,y,s,i)\chi_{12}(x,y,s,i)
  \label{}
\end{align}
Thus, we are able to set $\chi_{12}(x,y,s,i)=1$, $\forall(x,y,s,i)$. 

According to Eq.(\ref{eq:gauge_transf_on_psg_coord}) and Eq.(\ref{eq:phase_factor_transf}), using gauge transformation $V(x,y,s,i)$ and phase factor ambiguity $\Phi(x,y,s)$, we get
\begin{align}
  W_{T_2}(x,y,s,i)&\rightarrow V(x,y,s,i)W_{T_2}(x,y,s,i)V^{-1}(x,y-1,s,i)\notag\\
  \Theta_{T_2}{(x,y,s)}&\rightarrow\Theta_{T_2}{(x,y,s)}\Phi{(x,y,s)}\Phi^*{(x,y-1,s)}
  \label{eq:gauge_transf_T2_psg}
\end{align}
Then, we are able to set $W_{T_2}(x,y,s,i)=\mathrm{I}$ as well as $\Theta_{T_2}{(x,y,s,i)}=1$. And we get $T^{(x,y,s)}=T^{(0,y,s)}$. The remaining ambiguity which leaves $W_{T_2}$ and $\Theta_{T_2}$ invariant should satisfy the condition: $V(x,y,s,i)=V(x,0,s,i)$ and $\Phi(x,y,s)=\Phi(x,0,s)$. Any $\varepsilon_{T_2}$ transformation will change $W_{T_2}$, so $\varepsilon_{T_2}$ is fixed to $1$.

Similarly, for $T_1$ transformation, using remaining gauge transformation, we have
\begin{align}
  W_{T_1}(x,y,s,i)&\rightarrow V(x,0,s,i)W_{T_1}(x,y,s,i)V^{-1}(x-1,0,s,i)\notag\\
  \Theta_{T_1}{(x,y,s)}&\rightarrow\Theta_{T_1}{(x,y,s)}\Phi{(x,0,s)}\Phi^*{(x-1,0,s)}
  \label{eq:gauge_transf_T1_psg}
\end{align}
Thus we can set $W_{T_1}(x,0,s,i)=\mathrm{I}$ and $\Theta_{T_1}(x,0,s)=0$. Then $\varepsilon_{T_1}(x,y,s,i)=\varepsilon_{T_1}(x,0,s,i)=1$. Further, according to Eq.(\ref{eq:Ti_transf_tensor}), site tensors are translational invariant under this gauge:
\begin{align}
  T^{(x,y,s)}=T^{(x,0,s)}=T^{s}\doteq T^{(0,0,s)}
  \label{eq:psg_constant_site_translation}
\end{align}
To keep this property, the allowed transformations are only sublattice dependent: $V(x,y,s,i)=V(s,i)$ and $\Phi{(x,y,s)}=\Phi(s)$. 

In the gauge we choose above, we can solve Eq.(\ref{eq:psg_eta12}) as
\begin{align}
  &W_{T_1}(x,y,s,i)=\eta_{12}^y\notag\\
  &W_{T_2}(x,y,s,i)=\mathrm{I}\notag\\
  &\Theta_{T_1}{(x,y,s)}=\mu^y_{12}\notag\\
  &\Theta_{T_2}{(x,y,s)}=1
  \label{eq:translation_symmetry_rule}
\end{align}

Now, let us add $C_6$ rotation symmetry. Under $C_6$ symmetry, tensors will transform as 
\begin{align}
  &T^{(x,y,s)}=\Theta_{C_6}W_{C_6}C_6\circ T^{(x,y,s)}\notag\\
  &B_{(xysi|x'y's'i')}=W_{C_6}C_6\circ B_{(xysi|x'y's'i')}
  \label{eq:C6_transf_tensor}
\end{align}
where
\begin{align}
  C_6\circ(T^{(x,y,u)})^i_{\alpha\beta\gamma\delta}&=(T^{(x+y-1,-x+1,w)})^i_{\beta\alpha\delta\gamma}\notag\\
  C_6\circ(T^{(x,y,v)})^i_{\alpha\beta\gamma\delta}&=(T^{(x+y,-x+1,u)})^i_{\alpha\beta\gamma\delta}\notag\\
  C_6\circ(T^{(x,y,w)})^i_{\alpha\beta\gamma\delta}&=(T^{(x+y,-x,v)})^i_{\alpha\beta\gamma\delta}\notag\\
  \label{}
\end{align}

From group relation $C_6^{-1}T_2^{-1}C_6T_1=C_6^{-1}T_2^{-1}T_1C_6T_2=\mathrm{e}$, we get
\begin{align}
  &W^{-1}_{C_6}(C_6(x,y,s,i))W_{T_2}^{-1}(T_2C_6(x,y,s,i))W_{C_6}(T_2C_6(x,y,s,i))\notag\\
  &W_{T_1}(C_6^{-1}T_2C_6(x,y,s,i))=\eta_{T_1C_6}\chi_{T_1C_6}(x,y,s,i)\notag\\
  &W^{-1}_{C_6}(C_6(x,y,s,i))W_{T_2}^{-1}(T_2C_6(x,y,s,i))W_{T_1}(T_2C_6(x,y,s,i))\notag\\
  &W_{C_6}(C_6T_2(x,y,s,i))W_{T_2}(T_2(x,y,s,i))=\eta_{T_2C_6}\chi_{T_2C_6}(x,y,s,i)\notag\\
  \label{eq:psg_etaT1C6_etaT2C6}
\end{align}
as well as 
\begin{align}
  &\Theta^*_{C_6}{(C_6(x,y,s))}\Theta^*_{T_2}{(T_2C_6(x,y,s))}\Theta_{C_6}{(T_2C_6(x,y,s))}\notag\\
  &\Theta_{T_1}{(C_6^{-1}T_2C_6(x,y,s))}=\mu_{T_1C_6}\prod_i\chi^*_{T_1C_6}(x,y,s,i)\notag\\
  &\Theta^*_{C_6}{(C_6(x,y,s))}\Theta^*_{T_2}{(T_2C_6(x,y,s))}\Theta_{T_1}{(T_2C_6(x,y,s))}\notag\\
  &\Theta_{C_6}{(C_6T_2(x,y,s))}\Theta_{T_2}{(T_2(x,y,s))}=\mu_{T_2C_6}\prod_i\chi^*_{T_2C_6}(x,y,s,i)\notag\\
  \label{eq:psg_theta_etaT1C6_etaT2C6}
\end{align}
Due to the $\eta$-ambiguity, we can always redefine $W_R\rightarrow\eta W_R$ and $\Theta_R\rightarrow\mu\Theta_R$, which has no physics consequence. Thus, by redefining 
\begin{align}
  W_{T_1}\rightarrow\eta_{T_2C_6}W_{T_1},\quad&W_{T_2}\rightarrow\eta_{T_1C_6}\eta_{T_2C_6}W_{T_2},\notag\\
  \Theta_{T_1}\rightarrow\mu_{T_2C_6}\Theta_{T_1},\quad&\Theta_{T_2}\rightarrow\mu_{T_1C_6}\mu_{T_2C_6},
  \label{}
\end{align}
we set the right side of Eq.(\ref{eq:psg_etaT1C6_etaT2C6}) and Eq.(\ref{eq:psg_theta_etaT1C6_etaT2C6}) to be $\mathrm{I}$ and $1$. Then, by performing transformation
\begin{align}
  V(x,y,s,i)&=\eta_{T_1C_6}^y\eta_{T_2C_6}^{(x+y)}\notag\\
  \Phi(x,y,s)&=\mu_{T_1C_6}^y\mu_{T_2C_6}^{(x+y)}\notag\\
  \label{}
\end{align}
$W_{T_i}$ and $\Theta_{T_i}$ are changed back to its original value in Eq.(\ref{eq:translation_symmetry_rule}).

Using $\varepsilon_{C_6}$ ambiguity, we are able to set $\chi_{T_1C_6}(x,y,s,i)=1$ and $\chi_{T_2C_6}(0,y,s,a/b)=1$. The remaining $\varepsilon_{C_6}$ is independent of unit cell coordinate, namely $\varepsilon_{C_6}(x,y,s,i)=\varepsilon_{C_6}(s,i)$. Then, by solving Eq.(\ref{eq:psg_etaT1C6_etaT2C6}) and Eq.(\ref{eq:psg_theta_etaT1C6_etaT2C6}), we get 
\begin{align}
  W_{C_6}(x,y,u,i)&=[\eta_{12}(i)]^{xy+\frac{1}{2}x(x+1)+x+y}w_{C_6}(u,i)\notag\\
  W_{C_6}(x,y,v,i)&=[\eta_{12}(i)]^{xy+\frac{1}{2}x(x+1)+x+y}w_{C_6}(v,i)\notag\\
  W_{C_6}(x,y,w,i)&=[\eta_{12}(i)]^{xy+\frac{1}{2}x(x+1)}w_{C_6}(w,i)\notag\\
  \label{eq:psg_WC6}
\end{align}
as well as
\begin{align}
  &\Theta_{C_6}(x,y,u)=\mu_{12}^{xy+\frac{1}{2}x(x+1)+x+y}\Theta_{C_6}(u)\notag\\
  &\Theta_{C_6}(x,y,u)=\mu_{12}^{xy+\frac{1}{2}x(x+1)+x+y}\Theta_{C_6}(v)\notag\\
  &\Theta_{C_6}(x,y,u)=\mu_{12}^{xy+\frac{1}{2}x(x+1)}\Theta_{C_6}(w)\notag\\
  \label{eq:psg_thetaC6}
\end{align}
where we define $w_R(s,i)\equiv W_R(0,0,s,i)$ and $\Theta_R(s)\equiv\Theta_R(0,0,s)$. Inserting the above result back to Eq.(\ref{eq:psg_etaT1C6_etaT2C6}) and Eq.(\ref{eq:psg_theta_etaT1C6_etaT2C6}), we conclude that all $\chi_{T_2C_6}=1$.

Further, since $C_6^6=\mathrm{e}$, we get
\begin{align}
  &W_{C_6}(x,y,s,i)W_{C_6}(C_6^{-1}(x,y,s,i))W_{C_6}(C_6^{-2}(x,y,s,i))\notag\\
  &W_{C_6}(C_6^{-3}(x,y,s,i))W_{C_6}(C_6^{-4}(x,y,s,i))W_{C_6}(C_6^{-5}(x,y,s,i))\notag\\
  &=\eta_{C_6}\chi_{C_6}(x,y,s,i)
  \label{eq:psg_etaC6}
\end{align}
Using Eq.(\ref{eq:psg_WC6}) and Eq.(\ref{eq:psg_etaC6}), we can simplify the above equation as
\begin{align}
  &w_{C_6}(w,i)w_{C_6}(v,i)w_{C_6}(u,i)w_{C_6}(w,i_{C_6})\cdot\notag\\
  &w_{C_6}(v,i_{C_6})w_{C_6}(u,i_{C_6})=\eta_{12}\eta_{C_6}\chi_{C_6}(x,y,s,i)
  \label{eq:psg_wC6}
\end{align}
So, we conclude that $\chi_{C_6}(i)\equiv\chi_{C_6}(0,0,u,i)=\chi_{C_6}(x,y,s,i)$ and $\chi_{C_6}(i)=\chi_{C_6}(i_{C_6})$. Under remaining $\varepsilon_{C_6}(s,i)$, $\chi_{C_6}(i)$ changes as 
\begin{align}
  \chi_{C_6}(i)\rightarrow\chi_{C_6}(i)\prod_s\varepsilon_{C_6}(s,i)\varepsilon_{C_6}(s,i_{C_6})
  \label{}
\end{align}
By choosing proper $\varepsilon_{C_6}$, we can set $\chi_{C_6}=1$.

Then, by performing unit cell independent gauge transformation $V(s,i)$, $W_{C_6}$ transforms as
\begin{align}
  W_{C_6}(x,y,u,i)&\rightarrow V(u,i)W_{C_6}(x,y,u,i)V^{-1}(w,i_{C_6})\notag\\
  W_{C_6}(x,y,v,i)&\rightarrow V(v,i)W_{C_6}(x,y,v,i)V^{-1}(u,i)\notag\\
  W_{C_6}(x,y,w,i)&\rightarrow V(w,i)W_{C_6}(x,y,w,i)V^{-1}(v,i)\notag\\
  \label{eq:gauge_tranf_C6_psg}
\end{align}
Thus, we can set $w_{C_6}(v,i)=w_{C_6}(w,i)=w_{C_6}(u,a)=w_{C_6}(u,c)=\mathrm{I}$ by choosing proper gauge. Now, $\varepsilon_{C_6}$ is fixed to $1$. And the remaining $V$-ambiguity satisfies $V(s,i)=V(i)=V(i_{C_6})$. We solve Eq.(\ref{eq:psg_wC6}) under this gauge as
\begin{align}
  w_{C_6}(u,b)=w_{C_6}(u,d)=\eta_{12}\eta_{C_6}
  \label{}
\end{align}

Similarly, for phase factor, the corresponding equation reads
\begin{align}
  &\Theta_{C_6}(x,y,s)\Theta_{C_6}(C_6^{-1}(x,y,s))\Theta_{C_6}(C_6^{-2}(x,y,s))\notag\\
  &\Theta_{C_6}(C_6^{-3}(x,y,s))\Theta_{C_6}(C_6^{-4}(x,y,s))\Theta_{C_6}(C_6^{-5}(x,y,s))\notag\\
  &=\mu_{C_6}
  \label{eq:psg_theta_etaC6}
\end{align}
From Eq.(\ref{eq:psg_thetaC6}), we conclude
\begin{align}
  \Theta_{C_6}(u)\Theta_{C_6}(v)\Theta_{C_6}(w)=\pm(\mu_{12}\mu_{C_6})^{\frac{1}{2}}
  \label{eq:psg_thetaC6_uvw}
\end{align}
Due to $\eta$-ambiguity for $C_6$, we can always redefine symmetry transformation rules as
\begin{align}
  W_{C_6}\rightarrow \mathrm{J}\cdot W_{C_6},\quad\Theta_{C_6}\rightarrow-\Theta_{C_6}
  \label{}
\end{align}
to absorb the minus sign in Eq.(\ref{eq:psg_thetaC6_uvw}). Further, $W_{C_6}$ can be transformed back to the original form by performing gauge transformation
\begin{align}
  &V(u/w,a/c)=V(v,b/d)=\mathrm{I},\notag\\
  &V(u/w,b/d)=V(v,a/c)=\eta.
  \label{}
\end{align}
The remaining $V$-ambiguity satisfies $V(x,y,s,i)=V(i)=V(i_{C_6})$.

Under phase transformation $\Phi(s)$, we get
\begin{align}
  \Theta(s)\rightarrow\Theta(s)\Phi(s)\Phi(C_6^{-1}(s))
\end{align}
So, we can set $\Theta(v)=\Theta(w)=1$. Now, $\Phi$-ambiguity is only left with an overall phase factor. Further, according to Eq.(\ref{eq:psg_thetaC6_uvw}) (without minus sign), we have
\begin{align}
  \Theta_{C_6}(u)=(\mu_{12}\mu_{C_6})^{\frac{1}{2}}
  \label{eq:psg_theta_C6u_square}
\end{align}

Notice that in the above gauge, according to Eq.(\ref{eq:C6_transf_tensor}), all site tensor are equal, namely, 
\begin{align}
  T^u=T^v=T^w
  \label{eq:psg_constant_site_rotation}
\end{align}

Now, let us add reflection. For reflection symmetry $\sigma$, we have
\begin{align}
  &T^{(x,y,s)}=\Theta_{\sigma}W_{\sigma}\sigma\circ T^{(x,y,s)}\notag\\
  &B_{(xysi|x'y's'i')}=W_{\sigma}\sigma\circ B_{(xysi|x'y's'i')}
  \label{eq:reflection_transf_tensor}
\end{align}
where
\begin{align}
  \sigma\circ (T^{(x,y,u)})^i_{\alpha\beta\gamma\delta}&=(T^{(y,x,u)})^i_{\delta\gamma\beta\alpha}\notag\\
  \sigma\circ (T^{(x,y,v)})^i_{\alpha\beta\gamma\delta}&=(T^{(y,x,w)})^i_{\gamma\delta\alpha\beta}\notag\\
  \sigma\circ (T^{(x,y,w)})^i_{\alpha\beta\gamma\delta}&=(T^{(y,x,v)})^i_{\gamma\delta\alpha\beta}\notag\\
  \label{}
\end{align}

From group relation $\sigma^{-1}T_1^{-1}\sigma T_2=\mathrm{e}$ and $\sigma^{-1}T_2^{-1}\sigma T_1=\mathrm{e}$, we can list the corresponding equations as
\begin{align}
  &W_{\sigma}^{-1}(\sigma(x,y,s,i))W_{T_2}^{-1}(T_2\sigma(x,y,s,i))W_{\sigma}(T_2\sigma(x,y,s,i))\notag\\
  &W_{T_1}(\sigma^{-1}T_2\sigma(x,y,s,i))=\eta_{\sigma T_1}\chi_{\sigma T_1}(x,y,s,i)\notag\\
  &W_{\sigma}^{-1}(\sigma(x,y,s,i))W_{T_1}^{-1}(T_1\sigma(x,y,s,i))W_{\sigma}(T_1\sigma(x,y,s,i))\notag\\
  &W_{T_2}(\sigma^{-1}T_1\sigma(x,y,s,i))=\eta_{\sigma T_2}\chi_{\sigma T_2}(x,y,s,i)\notag\\
  \label{eq:psg_etaT1sigma_T2sigma}
\end{align}
Using $\varepsilon_{\sigma}$ ambiguity, we are able to set $\chi_{\sigma T_1}(x,y,s,i)=1$ and $\chi_{\sigma T_2}(0,y,s,c/d)=1$, with remaining $\varepsilon_{\sigma}(x,y,s,i)=\varepsilon_{\sigma}(s,i)$. 

Then, we can solve the above equation as
\begin{align}
  W_{\sigma}(x,y,s,i)&=\eta_{\sigma T_1}^y\eta_{\sigma T_2}^x\eta_{12}^{xy}w_{\sigma}(s,i)
  \label{eq:psg_Wsigma}
\end{align}
Inserting the solution back to Eq.(\ref{eq:psg_etaT1sigma_T2sigma}), we get all $\chi_{\sigma T_2}(x,y,s,i)=1$.

Similarly, the equations for phase factor reads
\begin{align}
  &\Theta^*_{\sigma}(\sigma(x,y,s))\Theta^*_{T_2}(T_2\sigma(x,y,s))\Theta_{\sigma}(T_2\sigma(x,y,s))\notag\\
  &\Theta_{T_1}(\sigma^{-1}T_2\sigma(x,y,s))=\mu_{\sigma T_1}\notag\\
  &\Theta^*_{\sigma}(\sigma(x,y,s))\Theta_{T_1}^*(T_1\sigma(x,y,s))\Theta_{\sigma}(T_1\sigma(x,y,s))\notag\\
  &\Theta_{T_2}(\sigma^{-1}T_1\sigma(x,y,s))=\mu_{\sigma T_2}\notag\\
  \label{eq:psg_theta_etaT1sigma_T2sigma}
\end{align}
By solving the above equations, we get
\begin{align}
  \Theta_\sigma(x,y,s)=\mu_{\sigma T_1}^y\mu_{\sigma T_2}^x\mu_{12}^{xy}\Theta_\sigma(s)
  \label{eq:psg_thetasigma}
\end{align}

The equation corresponding to $\sigma^2=\mathrm{e}$ reads
\begin{align}
  W_{\sigma}(x,y,s,i)W_{\sigma}(\sigma(x,y,s,i))=\eta_\sigma\chi_\sigma(x,y,s,i)
  \label{eq:psg_etasigma}
\end{align}
Combine Eq.(\ref{eq:psg_Wsigma}) and Eq.(\ref{eq:psg_etasigma}), we have
\begin{align}
  (\eta_{\sigma T_1}\eta_{\sigma T_2})^{x+y}w_\sigma(s,i)w_\sigma(\sigma(s,i))=\eta_\sigma\chi_\sigma(x,y,s,i)
  \label{eq:psg_wsigmawsigma}
\end{align}
So, we conclude that $\eta_{\sigma T_1}=\eta_{\sigma T_2}$, and $\chi_\sigma(x,y,s,i)=\chi_\sigma(s,i)$. By applying $\sigma$ on both sides of Eq.(\ref{eq:psg_wsigmawsigma}), we get
\begin{align}
  w_\sigma(\sigma(s,i))w_\sigma(s,i)=\eta_\sigma\chi_\sigma(\sigma(s,i))
  \label{eq:psg_wsigmawsigma2}
\end{align}
Since the left side of Eq.(\ref{eq:psg_wsigmawsigma}) and Eq.(\ref{eq:psg_wsigmawsigma2}) are equal, we have $\chi_\sigma(s,i)=\chi_\sigma(\sigma(s,i))$. In particular, we have
\begin{align}
  &\chi_\sigma(w,a/b)=\chi_\sigma(v,c/d)=\chi_\sigma^*(w,a/b)\notag\\
  \label{}
\end{align}
which means 
\begin{align}
  \chi_\sigma(w,a/b)=\pm1
  \label{}
\end{align}

Using the remaining $\varepsilon_\sigma(s,i)$, we can set $\chi_\sigma(u,a/b)=\chi_\sigma(v,a/b)=1$. Then, we are left with $\varepsilon_\sigma(s,i)$ satisfying the following relations:
\begin{align}
  \varepsilon_\sigma(u,a)=\varepsilon_\sigma(v,b),\quad \varepsilon_\sigma(u,b)=\varepsilon_\sigma(v,a)
  \label{}
\end{align}

For group relation $\sigma^{-1}C_6\sigma C_6=\mathrm{e}$, the corresponding equation is
\begin{align}
  &W_\sigma^{-1}(\sigma(x,y,s,i))W_{C_6}(\sigma(x,y,s,i))W_\sigma(C_6^{-1}\sigma(x,y,s,i))\notag\\
  &W_{C_6}(\sigma(x,y,s,i))=\eta_{\sigma C_6}\chi_{\sigma C_6}(x,y,s,i)
  \label{eq:psg_etasigmaC6}
\end{align}
By simplifying the above equation, we have $\eta_{\sigma T_1}=\eta_{12}$, and $\chi_{\sigma C_6}(x,y,s,i)=\chi_{\sigma C_6}(s,i)$, with 
\begin{align}
  &w_\sigma(v,i)w_\sigma(v,i_{\sigma2})=\eta_\sigma\eta_{\sigma C_6}\chi_\sigma(v,i)\chi_{\sigma C_6}(v,i)\notag\\
  &w_\sigma(u,i)w_{C_6}(u,i_{\sigma1})w_\sigma(w,i_{\sigma2})=\eta_\sigma\eta_{\sigma C_6}\eta_{12}\chi_{\sigma C_6}(u,i)\notag\\
  &=\eta_\sigma\eta_{\sigma C_6}\eta_{12}\chi_{\sigma}(w,i_{\sigma2})\chi_{\sigma C_6}(w,i_{\sigma2})
  \label{eq:psg_wsigma_wC6}
\end{align}
Using remaining $\varepsilon_\sigma$ transformation, we are able to set
\begin{align}
  &\chi_{\sigma C_6}(u/v,a/b)=1\notag\\
  &\chi_{\sigma C_6}(w,a)=\chi_\sigma(w,a)=\pm1\notag\\
  &\chi_{\sigma C_6}(w,b)=\chi_\sigma(w,b)=\pm1
  \label{}
\end{align}

Further, by performing the remaining gauge transformation $V(x,y,s,i)=V(i)=V(i_{C_6})$, $W_\sigma$ transforms as
\begin{align}
  W_{\sigma}(x,y,u,i)&\rightarrow V(i)W_{\sigma}(x,y,u,i)V^{-1}(i_{\sigma1})\notag\\
  W_{\sigma}(x,y,v/w,i)&\rightarrow V(i)W_{\sigma}(x,y,v/w,i)V^{-1}(i_{\sigma2})\notag\\
  \label{eq:gauge_transf_sigma_psg}
\end{align}
Then, we can set $w(u,a)=\mathrm{I}$. The only $V$-ambiguity left is a block diagonal independent of sites and legs matrix $V(x,y,s,i)=V$.

According to Eq.(\ref{eq:psg_wsigmawsigma}) and Eq.(\ref{eq:psg_wsigma_wC6}), we can solve the transformation rules for reflection as
\begin{align}
  w_\sigma(u,a)=\mathrm{I},\quad &w_\sigma(u,b)=\chi_\sigma\eta_{\sigma C_6},\notag\\
  w_\sigma(u,c)=\chi_\sigma\eta_\sigma\eta_{\sigma C_6},\quad &w_\sigma(u,d)=\eta_\sigma;\notag\\
  w_\sigma(v,a)=\eta_{C_6}\eta_{\sigma C_6},\quad &w_\sigma(v,b)=\chi_\sigma\eta_{12},\notag\\
  w_\sigma(v,c)=\eta_\sigma\eta_{C_6},\quad &w_\sigma(v,d)=\chi_\sigma\eta_{12}\eta_\sigma\eta_{\sigma C_6};\notag\\
  w_\sigma(w,a)=\chi_\sigma\eta_{C_6},\quad &w_\sigma(w,b)=\eta_{12}\eta_{\sigma C_6},\notag\\
  w_\sigma(w,c)=\eta_\sigma\eta_{C_6}\eta_{\sigma C_6},\quad &w_\sigma(w,d)=\chi_\sigma\eta_{12}\eta_\sigma;
  \label{eq:psg_wsigmauvw}
\end{align}
where $\chi_\sigma\equiv\chi_\sigma(w,a)=\chi_\sigma(w,b)$.

For the phase factor, the equations read
\begin{align}
  [\Theta_\sigma(u)]^2=\Theta_\sigma(v)\Theta_\sigma(w)=\mu_{\sigma}
  \label{eq:psg_thetasigma_thetasigma}
\end{align}
as well as
\begin{align}
  [\Theta_\sigma(v)]^2&=\mu_{\sigma}\mu_{\sigma C_6}\notag\\
  \Theta_\sigma(u)\Theta_\sigma(w)\Theta_{C_6}(u)&=\mu_{12}\mu_{\sigma}\mu_{\sigma C_6}
  \label{eq:psg_thetasigma_thetaC6}
\end{align}
where we have used the fact $\prod_i \chi_\sigma(s,i)=\prod_i\chi_{\sigma C_6}=1$.

According to Eq.(\ref{eq:psg_theta_C6u_square}), Eq.(\ref{eq:psg_thetasigma_thetasigma}) and Eq.(\ref{eq:psg_thetasigma_thetaC6}), we have
\begin{align}
  &[\Theta_\sigma(u)\Theta_\sigma(w)\Theta_{C_6}(u)]^2=(\mu_{12}\mu_{\sigma}\mu_{\sigma C_6})^2=1\notag\\
  &=\mu_\sigma\cdot\mu_\sigma\mu_{\sigma C_6}\cdot\mu_{12}\mu_{C_6}=\mu_{12}\mu_{C_6}\mu_{\sigma C_6}
  \label{}
\end{align}
So, we get $\mu_{\sigma C_6}=\mu_{12}\mu_{C_6}$. In our case, site tensors are $Z_2$ odd, then the relation for $\mu$ causes additional constraint for $\eta$:
\begin{align}
  \eta_{\sigma C_6}=\eta_{12}\eta_{C_6}
  \label{}
\end{align}
From Eq.(\ref{eq:psg_thetasigma_thetasigma}), we get
\begin{align}
  &\Theta_\sigma(u)=\pm(\mu_\sigma)^{\frac{1}{2}}
\end{align}
However, using the $\eta$-ambiguity for $\sigma$, we can absorb the minus sign in above equations by redefining $W_\sigma\rightarrow\eta W_\sigma$ and $\Theta_\sigma\rightarrow\mu\Theta_\sigma$. Further, by performing gauge transformation $V(a/b)=\mathrm{I}$, $V(c/d)=\mathrm{J}$, we can transform $W_\sigma$ to their original forms. 

Then, we get solutions for $\Theta_{\sigma}(s)$ as
\begin{align}
  &\Theta_\sigma(u)=(\mu_\sigma)^{\frac{1}{2}};\notag\\
  &\Theta_\sigma(v)=\mu_{C_6}\Theta_{C_6}(u)\Theta_\sigma(u);\notag\\
  &\Theta_\sigma(w)=\mu_\sigma\mu_{C_6}(\Theta_{C_6}(u)\Theta_\sigma(u))^{-1}.
  \label{eq:psg_solution_theta_oneuc}
\end{align}

Let us summarize the result for lattice symmetry:
\begin{align}
  &W_{T_1}(x,y,s,i)=\eta_{12}^y,\notag\\
  &W_{T_2}(x,y,s,i)=\mathrm{I}\notag\\
  &W_{C_6}(x,y,u,i)=\eta_{12}^{xy+\frac{1}{2}x(x+1)+x+y}w_{C_6}(u,i),\notag\\
  &W_{C_6}(x,y,v,i)=\eta_{12}^{xy+\frac{1}{2}x(x+1)+x+y}\notag\\
  &W_{C_6}(x,y,w,i)=\eta_{12}^{xy+\frac{1}{2}x(x+1)}\notag\\
  &W_\sigma(x,y,s,i)=\eta_{12}^{x+y+xy}w_\sigma(s,i)\notag\\
  \label{eq:psg_solution}
\end{align}
where 
\begin{align}
  &w_{C_6}(u,a)=w_{C_6}(u,c)=\mathrm{I},\notag\\
  &w_{C_6}(u,b)=w_{C_6}(u,d)=\eta_{12}\eta_{C_6},\notag\\
  \label{eq:psg_wC6u}
\end{align}
and $w_\sigma(s,i)$ are given in Eq.(\ref{eq:psg_wsigmauvw}) with additional condition $\eta_{\sigma C_6}=\eta_{12}\eta_{C_6}$.

For phase factor $\Theta_R$, we get
\begin{align}
  &\Theta_{T_1}(x,y,s)=\mu_{12}^y,\notag\\
  &\Theta_{T_2}(x,y,s)=1\notag\\
  &\Theta_{C_6}(x,y,u)=\mu_{12}^{xy+\frac{1}{2}x(x+1)+x+y}\Theta_{C_6}(u),\notag\\
  &\Theta_{C_6}(x,y,v)=\mu_{12}^{xy+\frac{1}{2}x(x+1)+x+y}\notag\\
  &\Theta_{C_6}(x,y,w)=\mu_{12}^{xy+\frac{1}{2}x(x+1)}\notag\\
  &\Theta_\sigma(x,y,s)=\mu_{12}^{x+y+xy}\Theta_\sigma(s)\notag\\
  \label{eq:psg_solution_theta}
\end{align}
where $\Theta_{C_6}(u)$ and $\Theta_{\sigma}(s)$ are given in Eq.(\ref{eq:psg_theta_C6u_square}) and Eq.(\ref{eq:psg_solution_theta_oneuc}).

\subsubsection{Adding time reversal symmetry}

Now, let us consider time reversal symmetry $\mathcal{T}$. The transformation rule for time reversal symmetry is defined on Eq.(\ref{eq:time_reversal_sym_site_bond}). We should keep in mind that time reversal is antiunitary, so for symmetry operation $W_R$, we have $\mathcal{T}W_R\mathcal{T}^{-1}=W_R^*$.

From group relation $T_1^{-1}\mathcal{T}^{-1}T_1\mathcal{T}=T_2^{-1}\mathcal{T}^{-1}T_2\mathcal{T}=\mathrm{e}$, we get
\begin{align}
  &W_{T_1}(x,y,s,i)[W_{\mathcal{T}}^{-1}(x,y,s,i)]^*W_{T_1}^*(x,y,s,i)\notag\\
  &W_{\mathcal{T}}^*(T_1^{-1}(x,y,s,i))=\eta_{T_1\mathcal{T}}\chi_{T_1\mathcal{T}}(T_1^{-1}(x,y,s,i))\notag\\
  &W_{T_2}(x,y,s,i)[W_{\mathcal{T}}^{-1}(x,y,s,i)]^*W_{T_2}^*(x,y,s,i)\notag\\
  &W_{\mathcal{T}}^*(T_2^{-1}(x,y,s,i))=\eta_{T_2\mathcal{T}}\chi_{T_2\mathcal{T}}(T_2^{-1}(x,y,s,i))
  \label{eq:psg_etaT1T_etaT2T}
\end{align}
Similar to previous case, by using $\varepsilon_{\mathcal{T}}$ transformation, we are able to set $\chi_{T_1\mathcal{T}}$ and $\chi_{T_2\mathcal{T}}$ to be identity. The solution for the above equation is
\begin{align}
  W_T(x,y,s,i)=\eta_{T_1\mathcal{T}}^x\eta_{T_2\mathcal{T}}^yw_{\mathcal{T}}(s,i) 
  \label{}
\end{align}
The remaining $\varepsilon_{\mathcal{T}}$ is independent of unit cell coordinate $(x,y)$.

For group relation $\sigma^{-1}\mathcal{T}^{-1}\sigma\mathcal{T}=\mathrm{e}$, we have
\begin{align}
  &(\eta_{T_1\mathcal{T}}\eta_{T_2\mathcal{T}})^{x+y}w_{\sigma}^{-1}(s,i)[w_{\mathcal{T}}^{-1}(s,i)]^*\notag\\
  &w_{\sigma}^*(s,i)w_{\mathcal{T}}^*(\sigma(s,i))=\eta_{\sigma \mathcal{T}}\chi_{\sigma\mathcal{T}}(\sigma(x,y,s,i))
  \label{eq:psg_etasigmaT}
\end{align}
So, we conclude $\chi_{\sigma\mathcal{T}}(x,y,s,i)=\chi_{\sigma\mathcal{T}}(s,i)$ and $\eta_{T_1\mathcal{T}}=\eta_{T_2\mathcal{T}}$. Inserting the solution for $w_{\sigma}(s,i)$ into above equation, we get
\begin{align}
  w_{\mathcal{T}}^{-1}(s,i)w_{\mathcal{T}}(\sigma(s,i))=\eta_{\sigma\mathcal{T}}\chi^*_{\sigma\mathcal{T}}(\sigma(s,i))
  \label{}
\end{align}
Acting $\sigma$ on both side of above equation, we get
\begin{align}
  w_{\mathcal{T}}^{-1}(\sigma(s,i))w_{\mathcal{T}}{(s,i)}=\eta_{\sigma\mathcal{T}}\chi^*_{\sigma\mathcal{T}}(s,i)
  \label{}
\end{align}
Since the left side of above two equations are hermitian conjugate to each other, we conclude that $\chi_{\sigma\mathcal{T}}(s,i)=\chi_{\sigma\mathcal{T}}^*(\sigma(s,i))$.

Let us consider $C_6^{-1}\mathcal{T}^{-1}C_6\mathcal{T}=\mathrm{e}$. The corresponding equation is
\begin{align}
  &W_{C_6}(x,y,s,i)[W_{\mathcal{T}}^{-1}(x,y,s,i)]^*W_{C_6}^*(x,y,s,i)\notag\\
  &W_{\mathcal{T}}^*(C_6^{-1}(x,y,s,i))=\eta_{C_6\mathcal{T}}\chi_{C_6\mathcal{T}}(C_6^{-1}(x,y,s,i))
  \label{eq:psg_etaC6T}
\end{align}
Then, we get $\eta_{T_1\mathcal{T}}=\eta_{T_2\mathcal{T}}=\mathrm{I}$, and $\chi_{C_6\mathcal{T}}(x,y,s,i)=\chi_{C_6\mathcal{T}}(s,i)$. Inserting values of $W_{C_6}$, we have
\begin{align}
  w_{\mathcal{T}}^{-1}(s,i)w_{\mathcal{T}}(C_6^{-1}(s,i))=\eta_{C_6\mathcal{T}}\chi^*_{C_6\mathcal{T}}(C_6^{-1}(s,i))
  \label{}
\end{align}
Under transformation $W_{\mathcal{T}}\rightarrow\varepsilon_{\mathcal{T}}W_{\mathcal{T}}$, $\chi_{C_6\mathcal{T}}$ changes as
\begin{align}
  \chi_{C_6\mathcal{T}}(s,i)\rightarrow\chi_{C_6\mathcal{T}}(s,i)\varepsilon_{\mathcal{T}}(C_6(s,i))\varepsilon_{\mathcal{T}}^*(s,i)
  \label{}
\end{align}
So, we can set all $\chi_{C_6\mathcal{T}}(s,i)=1$, with remaining $\varepsilon_{\mathcal{T}}\equiv\varepsilon_{\mathcal{T}}(s,a/b)=\varepsilon^*_{\mathcal{T}}(s,c/d)$. The above equation is simplified as
\begin{align}
  w_{\mathcal{T}}^{-1}(s,i)w_{\mathcal{T}}(C_6^{-1}(s,i))=\eta_{C_6\mathcal{T}}
  \label{eq:psg_wTwTC6}
\end{align}

Let's try to fix $\chi_{\sigma\mathcal{T}}(s,i)$ by remaining $\varepsilon_{\mathcal{T}}$. We observe that $\chi_{\sigma\mathcal{T}}(s,a/b)\rightarrow\chi_{\sigma\mathcal{T}}(s,a/b)(\varepsilon_\mathcal{T}^*)^2$. So, we can set $\chi_{\sigma\mathcal{T}}(u,a)=1$. Further, we get
\begin{align}
  \chi_{\sigma\mathcal{T}}(v,b)=\chi_{\sigma\mathcal{T}}^*(u,d)=\chi_{\sigma\mathcal{T}}(u,a)
  \label{}
\end{align}
due to the relation $\chi_{\sigma\mathcal{T}}(s,i)=\chi_{\sigma\mathcal{T}}^*(\sigma(s,i))$. So, we have
\begin{align}
  w_{\mathcal{T}}^{-1}(u,a)w_{\mathcal{T}}(u,d)=\eta_{\sigma\mathcal{T}}
  \label{eq:psg_wTuawTud}
\end{align}
From Eq.(\ref{eq:psg_wTwTC6}) and Eq.(\ref{eq:psg_wTuawTud}), we conclude
\begin{align}
  w_{\mathcal{T}}^{-1}(s,i)w_{\mathcal{T}}(\sigma(s,i))=\eta_{\sigma\mathcal{T}}
  \label{eq:psg_wTwTsigma}
\end{align}
Namely, we have $\chi_{\sigma\mathcal{T}}(s,i)=1$.

So once we can determine the value of $w_\mathcal{T}\equiv w_\mathcal{T}(u,a)$, we get the complete the solution of time reversal symmetry $W_\mathcal{T}$ with $W_\mathcal{T}(x,y,s,i)=w_\mathcal{T}(s,i)$. And $w_\mathcal{T}(s,i)$ can be expressed by $w_\mathcal{T}$ as follows:
\begin{align}
  w_\mathcal{T}(u,a)=w_\mathcal{T},\quad &w_\mathcal{T}(u,b)=\eta_{C_6\mathcal{T}}w_\mathcal{T},\notag\\
  w_\mathcal{T}(u,c)=\eta_{\sigma\mathcal{T}}\eta_{C_6\mathcal{T}}w_\mathcal{T},\quad &w_\mathcal{T}(u,d)=\eta_{\sigma\mathcal{T}}w_\mathcal{T};\notag\\
  w_\mathcal{T}(v,a)=\eta_{C_6\mathcal{T}}w_\mathcal{T},\quad &w_\mathcal{T}(v,b)=w_\mathcal{T},\notag\\
  w_\mathcal{T}(v,c)=\eta_{\sigma\mathcal{T}}w_\mathcal{T},\quad &w_\mathcal{T}(v,d)=\eta_{\sigma\mathcal{T}}\eta_{C_6\mathcal{T}}w_\mathcal{T};\notag\\
  w_\mathcal{T}(w,a)=w_\mathcal{T},\quad &w_\mathcal{T}(w,b)=\eta_{C_6\mathcal{T}}w_\mathcal{T},\notag\\
  w_\mathcal{T}(w,c)=\eta_{\sigma\mathcal{T}}\eta_{C_6\mathcal{T}}w_\mathcal{T},\quad &w_\mathcal{T}(w,d)=\eta_{\sigma\mathcal{T}}w_\mathcal{T};
  \label{eq:psg_wTuvw}
\end{align}

Now, let us determine $\Theta_\mathcal{T}$. The equations for $\Theta_\mathcal{T}$ read as
\begin{align}
  \Theta_g^*(x,y,s)\Theta_\mathcal{T}(x,y,s)\Theta_g^*(x,y,s)\Theta_\mathcal{T}^*(g^{-1}(x,y,s))=\mu_{g\mathcal{T}}
  \label{eq:psg_theta_etagT}
\end{align}
where $g$ labels lattice symmetry generators $T_1,T_2,C_6,\sigma$. Further, under action of global phase factor $\Phi$, $\Theta_\mathcal{T}$ changes as 
\begin{align}
  \Theta_\mathcal{T}(x,y,s)\rightarrow\Theta_\mathcal{T}(x,y,s)\Phi^2
  \label{}
\end{align}
Therefore, by choosing proper phase, we can always set $\Theta_\mathcal{T}(u)=1$. Combine with Eq.(\ref{eq:psg_solution_theta}), we are able to solve Eq.(\ref{eq:psg_theta_etagT}). The solution is $\Theta_\mathcal{T}(x,y,s)=\Theta_\mathcal{T}(s)$, where 
\begin{align}
  &\Theta_\mathcal{T}(u)=1,\notag\\
  &\Theta_\mathcal{T}(v)=\mu_{12}\mu_{C_6},\notag\\
  &\Theta_\mathcal{T}(w)=1.\notag\\
  \label{}
\end{align}
And the constraint on $\mu_{g\mathcal{T}}$ is
\begin{align}
  \mu_{\sigma\mathcal{T}}=\mu_\sigma,\quad \mu_{C_6\mathcal{T}}=\mu_{12}\mu_{C_6}
  \label{}
\end{align}
Since site tensors are $Z_2$ odd, we also have constraint on $\eta_{g\mathcal{T}}$ as 
\begin{align}
  \eta_{\sigma\mathcal{T}}=\eta_\sigma,\quad \eta_{C_6\mathcal{T}}=\eta_{12}\eta_{C_6}
  \label{}
\end{align}

Finally, let us consider group relation $\mathcal{T}^2=\mathrm{e}$. For $W_{\mathcal{T}}$, we get
\begin{align}
  w_{\mathcal{T}}(s,i)w^*_\mathcal{T}(s,i)=\eta_\mathcal{T}\chi_\mathcal{T}(x,y,s,i)
  \label{eq:psg_etaT}
\end{align}
where we use the fact that $W_\mathcal{T}(x,y,s,i)=w_\mathcal{T}(s,i)$. Inserting Eq.(\ref{eq:psg_wTuvw}) back to the above equation, we conclude that $\chi_\mathcal{T}\equiv\chi_\mathcal{T}(x,y,s,i)=\pm1$. So, we have
\begin{align}
  w_{\mathcal{T}}(s,i)w^*_\mathcal{T}(s,i)=\eta_\mathcal{T}\chi_\mathcal{T}
  \label{eq:psg_wTwT}
\end{align}
Similarly, for phase factor $\Theta_\mathcal{T}$, we have
\begin{align}
  \Theta_\mathcal{T}(s,i)\Theta_\mathcal{T}^*(s,i)=\mu_\mathcal{T}
  \label{eq:psg_theta_etaT}
\end{align}
We conclude that $\mu_\mathcal{T}=1$.

In our case, a physical leg lives Kramer doublets. Namely, we have $U_\mathcal{T}^2=-1$, where $\mathcal{T}=U_\mathcal{T}K$ is the action of time reversal operator on a physical leg, and $K$ is the complex conjugation. In the following, we will prove that $\eta_\mathcal{T}$ cannot be trivial, and site tensor must be $Z_2$ odd tensor.

To see this, we can act time reversal twice on site tensor $T^s$. Then, we get
\begin{align}
  T^s&=\Theta_\mathcal{T}W_\mathcal{T}T\Theta_\mathcal{T}W_\mathcal{T}\mathcal{T}\circ T^s\notag\\
  &=\prod_i\chi_\mathcal{T}(s,i)\mu_\mathcal{T}\eta_\mathcal{T}\mathcal{T}^2\circ T\notag\\
  &=-\eta_\mathcal{T}\circ T^s
  \label{}
\end{align}
where we use Eq.(\ref{eq:psg_etaT}) and Eq.(\ref{eq:psg_theta_etaT}) to get the second line. And in the third line, we use the fact that $\mu_\mathcal{T}=1$, $\chi_\mathcal{T}(s,i)=\chi_\mathcal{T}=\pm1$, while $\mathcal{T}^2\circ T^s=-T^s$. So, we conclude $\eta_\mathcal{T}\circ T^s=-T^s$, namely, $\eta_\mathcal{T}$ must be nontrivial, and $T^s$ is $Z_2$ odd. 

Let us try to understand the physical meaning of the above statement. We want to construct a PEPS wavefunction on the kagome lattice with every site as a Kramer doublet, and preserving all lattice symmetries as well as time reversal symmetry. However, the above proof tells us that, we are forced to introduce $Z_2$ gauge structure due to the nontrivialness of $\eta_\mathcal{T}$. The $Z_2$ gauge structure leads to either a spin liquid phase or a symmetry breaking phase in the long range physics. In other words, we can never be able to write a trivial symmetric wavefunction with Kramer doublets on physical sites with no ground state degeneracy in this formulation! The above argument can be viewed as manifestation of Hastings-Oshikawa-Lieb-Schultz-Mattis theorem on PEPS.

\subsubsection{Adding spin rotation symmetry}
At last, let us consider the spin rotation symmetry. The action of a group element of the spin rotation symmetry on site tensors is defined as
\begin{align}
  U_{\theta\vec{n}}\circ T^s\doteq(\ee^{\ii\theta\vec{n}\cdot\vec{S}})_{ij}(T^s)^j_{\alpha\beta\gamma\delta}
  \label{}
\end{align}
where $\vec{S}$ labels physical spins. In our case, the system is formed by half-integer spins. 
For PEPS invariant under spin rotation symmetry, we have
\begin{align}
  &T^s=\Theta_{\theta\vec{n}}W_{\theta\vec{n}}U_{\theta\vec{n}}\circ T^s\notag\\
  &B_b=W_{\theta\vec{n}}\circ B_b
  \label{}
\end{align}
Here, $W_{\theta\vec{n}}$ and $\Theta_{\theta\vec{n}}$ are projective representations of spin rotation symmetry. To see this, let us consider the group multiplication relation:
\begin{align}
  U_{\theta_1\vec{n}_1}\cdot U_{\theta_2\vec{n}_2}=U_{\theta_3\vec{n}_3}
  \label{}
\end{align}
The corresponding equation on virtual legs reads
\begin{align}
  &W_{\theta_1\vec{n}_1}(x,y,s,i)\cdot W_{\theta_2\vec{n}_2}(x,y,s,i)=\notag\\
  &\chi_{\theta_1\vec{n_1},\theta_2\vec{n}_2}\eta_{\theta_1\vec{n_1},\theta_2\vec{n}_2}\cdot W_{\theta_3\vec{n}_3}(x,y,s,i)
  \label{}
\end{align}
The above equation implies $W_{\theta\vec{n}}(x,y,s,i)$ form a projective representation of $SU(2)$ symmetry, with coefficient $\UU(1)\times Z_2$. However, it is well known that $SU(2)$ group has no nontrivial projective symmetry. Thus, we can always set $\chi$'s and $\eta$'s to be trivial ones by group extension ambiguities. Further, we have $\Theta_{\theta\vec{n}}$ always equal to 1, since $SU(2)$ has no nontrivial 1D representation.

The on-site spin rotation symmetry are commute with all lattice symmetry. Namely, we have $g^{-1}U^{-1}_{\theta\vec{n}}gU_{\theta\vec{n}}=\mathrm{e}$, where $g=T_1,T_2,C_6,\sigma$. So, for $W_g$ and $W_{\theta\vec{n}}$. The corresponding equations for $W_R$ are:
\begin{align}
  &W_g^{-1}(x,y,s,i)W_{\theta\vec{n}}^{-1}(x,y,s,i)W_g(x,y,s,i)\notag\\
  &W_{\theta\vec{n}}(g^{-1}(x,y,s,i))=\eta_{g,\theta\vec{n}}\chi_{g,\theta\vec{n}}(x,y,s,i)
  \label{eq:psg_etagspin}
\end{align}
According to the above solution, we have $W_g(x,y,s,i)=\mathrm{I}/\mathrm{J}$, which always commute with $W_{\theta\vec{n}}$. So, we get
\begin{align}
  W_{\theta\vec{n}}^{-1}(x,y,s,i)W_{\theta\vec{n}}(g^{-1}(x,y,s,i))=\eta_{g,\theta\vec{n}}\chi_{g,\theta\vec{n}}(x,y,s,i)
  \label{}
\end{align}
One can prove that $W_{\theta\vec{n}}^{-1}(x,y,s,i)W_{\theta\vec{n}}(g^{-1}(x,y,s,i))$ form a 1D representation of $SU(2)$ symmetry. Thus, $\eta_{g,\theta\vec{n}}$ and $\chi_{g,\theta\vec{n}}$ can be set to trivial. We have
\begin{align}
  W_{\theta\vec{n}}(x,y,s,i)=w_{\theta\vec{n}}
  \label{eq:psg_Wspin_xysi}
\end{align}
Namely, the representation of spin rotation symmetry shares the same form on virtual legs.

Now, we consider the relation:
\begin{align}
  W_{\theta=2\pi}(x,y,s,i)=\eta_{\theta=2\pi}\chi_{\theta=2\pi}(x,y,s,i)
  \label{}
\end{align}
Using the fact that $W_{\theta=2\pi}(x,y,s,i)=w_{\theta=2\pi}$, which is independent of sites and virtual legs, we conclude that $\chi_{\theta=2\pi}(x,y,s,i)=\chi_{\theta=2\pi}=\pm1$. 

At last, we have the relation $U_{\theta\vec{n}}^{-1}\mathcal{T}^{-1}U_{\theta\vec{n}}\mathcal{T}=\mathrm{e}$, which is the equivalent way to say that a spin reverses its direction under time reversal symmetry. Then, for transformation rules on virtual legs, we get
\begin{align}
  &W^{-1}_{\theta\vec{n}}(x,y,s,i)[W_{\mathcal{T}}^{-1}(x,y,s,i)]^{*}W_{\theta\vec{n}}^*(x,y,s,i)\notag\\
  &W_{\mathcal{T}}^*(x,y,s,i)=\eta_{\mathcal{T},\theta\vec{n}}\chi_{\mathcal{T},\theta\vec{n}}(x,y,s,i)
  \label{eq:psg_etaTspin}
\end{align}
We can easily conclude that $\eta_{\mathcal{T},\theta\vec{n}}=\mathrm{I}$, and $\chi_{\mathcal{T},\theta\vec{n}}(x,y,s,i)=1$. The result is reasonable, since we also expect spins living on virtual legs reverses direction under time reversal action. Then, we have
\begin{align}
  w^{-1}_{\theta\vec{n}}[w_{\mathcal{T}}^{-1}(s,i)]^*w_{\theta\vec{n}}^*w_{\mathcal{T}}^*(s,i)=\mathrm{I}
  \label{eq:psg_wspinwT}
\end{align}

In our case, physical legs are spin doublets, so $U_{\theta=2\pi}=-\mathrm{I}$. Using similar proof as in Kramer doublet case, we are able to show that $\eta_{\theta=2\pi}$ must be nontrivial and site tensors must be $Z_2$ odd. So, we conclude $\eta_{\theta=2\pi}=\eta_{\mathcal{T}}=\mathrm{J}$. Further, it is straightforward to see that one can always redefine $\mathrm{J}$ such that $\chi_{\theta=2\pi}=1$. WLOG, we assume $\mathrm{J}=\mathrm{I}_{D_1}\oplus(-\mathrm{I}_{D_2})$, where $D_1+D_2=D$ is dimension of a virtual leg.

Now, let us fix the form of $w_{\theta\vec{n}}$ and $w_{\mathcal{T}}$ using the remaining $V$-ambiguity, which is leg-independent block diagonal matrix. Under gauge transformation $V$, we have
\begin{align}
  w_{\theta\vec{n}}&\rightarrow Vw_{\theta\vec{n}}V^{-1}\notag\\
  w_{\mathcal{T}}&\rightarrow Vw_\mathcal{T}[V^{-1}]^*
  \label{}
\end{align}
So, first, we are able to set the (reducible) representation of spin rotation symmetry on virtual legs as
\begin{align}
  w_{\theta\vec{n}}=\bigoplus_{i=1}^{M}(\mathrm{I}_{n_i}\otimes\ee^{\ii\theta\vec{n}\cdot\vec{S_i}})
  \label{eq:psg_wspin}
\end{align}
where $\vec{S_i}$ labels spin quantum number while $n_i$ is the extra degeneracy associated with spin $\vec{S_i}$. In other words, a virtual leg is formed by $n_i$ number of spin $S_i$, where $i=1,2,\dots,M$. The dimension for spin ${S_i}$ is $n_i(2S_i+1)$, and we get the total dimension of a virtual leg $D=\sum_{i=1}^{M}n_i(2S_i+1)$. Further, we can arrange the order of $S_i$, such that $S_i$ is integer (half-integer) for $i<=m_1$ ($i>m_1$), and we have $S_1<\dots<S_{m_1}$ as well as $S_{m_1+1}<\dots<S_{M}$. Apparently, $D_1=\sum_{i=1}^{m_1}n_i(2S_i+1)$ and $D_2=\sum_{i=m_1+1}^Mn_i(2S_i+1)$. 

After fixing the form of $w_{\theta\vec{n}}$, we still left with overall gauge transformation $V=\bigoplus_i(\widetilde{V}_{S_i}\otimes\mathrm{I}_{2S_i+1})$, where $\widetilde{V}_{S_i}$ is arbitrary $n_i$ dimensional invertible matrix. We will use the remaining gauge degree of freedom to fix the representation of time reversal symmetry on virtual legs $w_{\mathcal{T}}(s,i)$. Particularly, let us focus on $w_\mathcal{T}\equiv w_{\mathcal{T}}(u,a)$, since representations on other legs can be generated using Eq.(\ref{eq:psg_wTuvw}).

According to Eq.(\ref{eq:psg_wspinwT}) and Eq.(\ref{eq:psg_wspin}), time reversal reverses spin direction. So the most general form of $w_{\mathcal{T}}$ reads: 
\begin{align}
  w_{\mathcal{T}}=\bigoplus_{i=1}^M(\widetilde{w}_{\mathcal{T}}^{S_i}\otimes\ee^{\ii\pi S_i^y})
  \label{eq:psg_wT}
\end{align}
Here, $\widetilde{w}_{\mathcal{T}}^{S_i}$ is $n_i$-dimensional invertible matrix. Further, according to Eq.(\ref{eq:psg_wTwT}), we have
\begin{align}
  \bigoplus_{i=1}^M\left([\widetilde{w}_{\mathcal{T}}^{S_i}\cdot(\widetilde{w}_{\mathcal{T}}^{S_i})^*]\otimes[\ee^{\ii\pi S_i^y}\cdot\ee^{-\ii\pi (S_i^y)^*}]\right)=\eta_{\mathcal{T}}\chi_{\mathcal{T}}
  \label{}
\end{align}
where $\ee^{\ii\pi S_i^y}\cdot\ee^{-\ii\pi (S_i^y)^*}=\mathrm{I}(-\mathrm{I})$ for integer (half-integer) spin. Note that we focus on the case where $\eta_{\mathcal{T}}=\mathrm{J}$, so we have
\begin{align}
\widetilde{w}_{\mathcal{T}}^{S_i}(\widetilde{w}_{\mathcal{T}}^{S_i})^*=\chi_{\mathcal{T}}
  \label{}
\end{align}

Then, under gauge transformation $V=\bigoplus_i(\widetilde{V}_{S_i}\otimes\mathrm{I}_{2S_i+1})$, 
\begin{align}
  \widetilde{w}_{\mathcal{T}}^{S_i}\rightarrow\widetilde{V}_{S_i}\widetilde{w}_{\mathcal{T}}^{S_i}[\widetilde{V}^{-1}_{S_i}]^{*}
  \label{}
\end{align} 
If two matrices are related by above transformation, then they are consimilar to each other. The canonical form for matrix under consimilarity has already been studied in the mathematical literature Ref.\onlinecite{Hong:1988p143}. In the following, we will give results for the two cases $\chi_\mathcal{T}=\pm1$. 

First, consider $\chi_\mathcal{T}=1$. Then the $n_i$ dimensional extra degeneracy space should accommodate representation of Kramer singlet. By choosing proper basis, we are able to obtain 
\begin{align}
  \widetilde{w}_\mathcal{T}^{S_i}=\mathrm{I}_{n_i}
  \label{eq:psg_tildewT_sym}
\end{align}
Then, the remaining gauge transformation must satisfy 
\begin{align}
  \widetilde{V}_{S_i}=\widetilde{V}_{S_i}^*
  \label{}
\end{align}
Namely, $\widetilde{V}_{S_i}$ is real matrix.

Then, we consider the case where $\chi_\mathcal{T}=-1$. Then the $n_i$ dimensional space are Kramer doublets, so $n_i$ should be even number in this case. By choosing proper basis, the canonical form for $\widetilde{w}_\mathcal{T}^{S_i}$ is 
\begin{align}
  \widetilde{w}^{S_i}_{\mathcal{T}}=\Omega_{n_i}
  \label{eq:psg_tildewT_antisym}
\end{align}
where $\Omega_{n_i}=\ii\sigma_y\otimes \mathrm{I}_{n_i/2}$. After choosing the basis for canonical $\widetilde{w}^{S_i}_{\mathcal{T}}$, we are left with gauge transformation $V=\bigoplus_i(\widetilde{V}_{S_i}\otimes\mathrm{I}_{S_i(S_i+1)})$ that satisfies 
\begin{align}
  \widetilde{V}_{S_i}\cdot\Omega_{n_i}=\Omega_{n_i}\cdot\widetilde{V}_{S_i}^*
  \label{}
\end{align}

Now, let us summarize the result. For a fully symmetric wavefunction on kagome lattice with a half-integer spin per site, there are at most $2^3\times2^2=32$ classes in the above framework. The different classes are distinguished by parameter $\eta$'s and $\chi$'s. Here, $\eta_{12},\eta_{C_6},\eta_\sigma\in\{\mathrm{I},\mathrm{J}\}$, while other $\eta$'s are fixed. And $\chi_\sigma,\chi_\mathcal{T}=\pm 1$.

\subsection{Construction of PEPS state for different classes}
In this part, we will use the symmetry transformation rules obtained above as constraint to determine the Hilbert space for a symmetric PEPS state for all classes. We will present the general framework first, and then work out possible forms of bond tensors and site tensors separately.

\subsubsection{Genenal framework}
In the following, we will set up the framework to get the constraint Hilbert space for symmetric PEPS.

Let us start by reviewing Hilbert space of PEPS without any symmetry. We require every virtual leg of site tensors is isomorphism to a $D$ dimensional Hilbert space $\mathbb{V}$. Virtual legs of bond tensors are all isomorphism to $\bar{\mathbb{V}}$, which is the dual space of $\mathbb{V}$. Further, every physical leg is isomorphism a $d$-dimensional Hilbert space $\mathbb{U}$.

The Hilbert space of a single bond tensor $\mathbb{V}_B$ and a site tensor $\mathbb{V}_T$ have the following tensor product structure:
\begin{align}
  &\mathbb{V}_B\cong\bar{\mathbb{V}}\otimes\bar{\mathbb{V}}\notag\\
  &\mathbb{V}_T\cong\mathbb{U}\otimes\mathbb{V}\otimes\mathbb{V}\otimes\mathbb{V}\otimes\mathbb{V}\notag\\
  \label{eq:psg_bond_site_hilbert}
\end{align}

Then, let us add the spin rotation symmetry. For a physical leg, there lives a half-integer spin $S_0$, so we have 
\begin{align}
  \mathbb{U}\cong \mathbb{V}_{S_0}
  \label{eq:psg_physical_hilbert}
\end{align}
where $\mathbb{V}_{S_0}$ accommodates a irreducible representation of $SU(2)$ with dimension $2S_0+1$. 

Let us consider virtual legs of site tensors. Notice that all virtual legs of site tensors are related by lattice group transformations. So, in the presence of lattice symmetries, all virtual legs of site tensors share the same representation (can be reducible) for spin rotation up to isomorphism. In particular, as we show in the previous subsection, we can always choose a proper basis, such that the spin operators have the same form on all virtual legs of site tensors, as in Eq.(\ref{eq:psg_Wspin_xysi}). Further, we can decompose $\mathbb{V}=\mathbb{V}_1\oplus\mathbb{V}_2$, where $\mathbb{V}_1$ with dimension $D_1$ denotes integer spin representations, and $\mathbb{V}_2$ ($\bar{\mathbb{V}}_2$) with dimension $D_2$ accommodates half integer spin representations. 

We can further decompose $\mathbb{V}$ to $SU(2)$ irreducible representations as 
\begin{align}
  \mathbb{V}\cong\bigoplus_{i=1}^M(\mathbb{D}_{S_i}\otimes\mathbb{V}_{S_i})
  \label{eq:psg_local_hilbert_structure}
\end{align}
Here $\mathbb{D}_{S_i}$ is an $n_i$ dimensional space that labels the extra degeneracy for spin-$S_i$. According to the decomposition, the orthonormal basis of $\mathbb{V}$ can be chosen as 
\begin{align}
  |S_i,t_\alpha,m_\beta\rangle\equiv|S_i,t_\alpha\rangle\otimes|S_i,m_\beta\rangle
  \label{}
\end{align}
where $|S_i,m_\beta\rangle\in\mathbb{V}_{S_i}$ labels an eigenstate of $\vec{S}^2$ and $S_z$, while $|S_i,t_\alpha\rangle\in\mathbb{D}_i$ labels basis in the extra degenerate space. Under this basis, the spin rotatio operator shares the form as in Eq.(\ref{eq:psg_wspin}).

Similarly, virtual legs of bond tensors can be decomposed as
\begin{align}
  \bar{\mathbb{V}}\cong\bigoplus_{i=1}^M(\bar{\mathbb{D}}_{S_i}\otimes\bar{\mathbb{V}}_{S_i})
  \label{eq:psg_dual_local_hilbert_structure}
\end{align}
with basis as
\begin{align}
  \langle S_i,t_\alpha,m_\beta|\equiv\langle S_i,t_\alpha|\otimes\langle S_i,m_\beta|
  \label{}
\end{align}
We point out that the nontrivial $Z_2$ $IGG$ element is in fact $2\pi$ spin rotation on all virtual legs. In other words, we get $Z_2$ group $\{\mathrm{I},g\}$ with a trivial representation on $\mathbb{V}_1$ ($\bar{\mathbb{V}}_1$) and a nontrivial representation on $\mathbb{V}_2$ ($\bar{\mathbb{V}}_2$).

After establishing the structure of a single virtual leg as in Eq.(\ref{eq:psg_local_hilbert_structure}), we are able to get the structure of $\mathbb{V}_B$ and $\mathbb{V}_T$ according to Eq.(\ref{eq:psg_bond_site_hilbert}). $\mathbb{V}_B$ and $\mathbb{V}_T$ are formed by tensor product of $SU(2)$ representation, which can be decomposed to direct sums of irreps of $SU(2)$ by Clebsch-Gordan coefficients. Further, we require bond states and site states to be spin singlets, which gives extra constraint to the possible Hilbert space.

Now, let us add lattice symmetries. There are two kinds of constraint caused by lattice symmetries. First, a lattice symmetry may act as a linear mapping between the Hilbert space of different bonds and sites. In other words, if a single bond/site tensor is fixed, one can use transformation rules of lattice symmetries to generate other symmetry related tensors. Second, a lattice symmetry may also be a self-mapping (automorphism) on the Hilbert space of a single bond/site. In this case, the possible Hilbert space of a single bond/site tensor will be further constraint by lattice symmetry transformation rules. 

At last, due to time reversal symmetry, we require all tensors to be Kramer singlets.

In the following, we will apply the method developed above to solve the possible Hilbert space for the symmetric PEPS wavefunction of all classes.

\subsubsection{Constraint on bond tensors}
Let us consider bond tensors first. A bond tensor can be viewed as a matrix with dimension $D\times D$. Let us define $B_b=B_{(xysi|x'y's'i')}$ as the bond tensor connecting two virtual legs $(x,y,s,i)$ and $(x',y',s',i')$. It is obvious that $B_{(xysi|x'y's'i')}=B_{(x'y's'i'|xysi)}^\mathrm{t}$. 

Under the action of spin rotation symmetry $w_{\theta\vec{n}}=\bigoplus_{i=1}^{M}(\mathrm{I}_{n_i}\otimes\ee^{\ii\theta\vec{n}\cdot\vec{S_i}})$, the bond tensor is a spin singlet in the sense
\begin{align}
  B_b=w_{\theta\vec{n}}^*\cdot B_b\cdot w_{\theta\vec{n}}^{-1}
  \label{eq:psg_bond_constraint_spin}
\end{align}
Then, we can explicitly write bond tensor $B_b$ as a block diagonal matrix according to the spin quantum number of virtual legs as
\begin{align}
  B_b=\bigoplus_{i=1}^{M}\left( \widetilde{B}_b^{S_i}\otimes K_{S_i} \right)
  \label{eq:psg_bond_constraint_spin_state}
\end{align}
where $\widetilde{B}_b^{S_i}$ is an $n_i$ dimensional matrix, and $K_{S_i}$ is a $(2S_i+1)$ dimensional matrix labeling singlet state. More precisely, the quantum state $\hat{K}_{S_i}\equiv\langle S_i,m_\alpha,S_i,m_\beta|(K_{S_i})_{\alpha\beta}$ is a singlet state under $S_{tot}=S\otimes\mathrm{I}+\mathrm{I}\otimes S$. Here, $m_\alpha=-S_i+\alpha-1$ labels the quantum number of $S_i^z$. Namely, we have 
\begin{align}
  \langle S,m_\alpha,S,m_\beta|(K_S)_{\alpha\beta}S_{tot}^2=0
  \label{}
\end{align}
Using Clebsch-Gordan(CG) coefficients, we get
\begin{align}
  (K_{S})_{\alpha\beta}=\ee^{\ii\phi_S}(-1)^{S-m_\alpha}\delta_{m_\alpha,-m_\beta}
  \label{}
\end{align}
where $\phi_S$ is an indefinite phase. We can absorb the phase factor to $\widetilde{B}_b^{S}$, thus $K_S$ is always real. For example, we have $K_{S=0}=1$, $K_{S=\frac{1}{2}}=\ii\sigma_y$.

From another point of view, the Hilbert space of a bond tensor $\mathbb{V}_B$ is the tensor product of two virtual legs $\bar{\mathbb{V}}$, where $\bar{\mathbb{V}}$ is decomposed as Eq.(\ref{eq:psg_dual_local_hilbert_structure}). So, we can decompose $\mathbb{V}_B$ as
\begin{align}
  \mathbb{V}_B&\cong\bigoplus_{i,j}\left((\bar{\mathbb{D}}_{S_i}\otimes\bar{\mathbb{D}}_{S_j})\otimes(\bar{\mathbb{V}}_{S_i}\otimes\bar{\mathbb{V}}_{S_j}) \right)\notag\\
  &\cong\bigoplus_{i,j,k}\left(\bar{\mathbb{D}}_{S_i}\otimes\bar{\mathbb{D}}_{S_j}\otimes\bar{\mathbb{V}}_{S_iS_j}^{S_k}\otimes\bar{\mathbb{V}}_{S_k} \right)\notag\\
  \label{eq:psg_bond_hilbert_decomposition}
\end{align}
where $\bar{\mathbb{V}}^{S_k}_{S_iS_j}$ is the ``fusion space'', which means different ways to fuse spin $S_i$ and $S_j$ to spin $S_k$. According to representation theory of $SU(2)$, $\bar{\mathbb{V}}^{S_k}_{S_iS_j}$ is isomorphic to $\mathbb{C}$ if $|S_i-S_j|\leq S_k\leq S_i+S_j$. Otherwise, $\bar{\mathbb{V}}^{S_k}_{S_iS_j}$ vanishes. Since we only focus on spin singlet bond states $S_k=0$, so we conclude the possible Hilbert space of a bond tensor should be
\begin{align}
  \bar{\mathbb{V}}_B^{S=0}\cong\bigoplus_i (\bar{\mathbb{D}}_{S_i}\otimes\bar{\mathbb{D}}_{S_i}\otimes\bar{\mathbb{V}}_{S_iS_i}^{S=0}\otimes\bar{\mathbb{V}}_{S=0})
  \label{eq:psg_bond_singlet_hilbert}
\end{align}
where we use the fact $\bar{\mathbb{V}}_{S_iS_j}^{S=0}$ vanishes if $S_i\neq S_j$. Then $\hat{B}_b\in\bar{\mathbb{V}}_B^{S=0}$ can be decomposed to $\hat{\widetilde{B}}_b^{S_i}\in\bar{\mathbb{D}}_{S_i}\otimes\bar{\mathbb{D}}_{S_i}$ and $\hat{K}_{S_i}\in\bar{\mathbb{V}}_{S_iS_i}^{S=0}\otimes\bar{\mathbb{V}}_{S=0}$. Namely, we have
\begin{align}
  \hat{B}_b=\sum_{i;\alpha_1,\alpha_2;\beta_1,\beta_2}\langle S_i,t_{\alpha_1},m_{\alpha_2}; S_i,t_{\beta_1},m_{\beta_2}|(\tilde{B}_b^{S_i})_{\alpha_1\beta_1}(K_{S_i})_{\alpha_2,\beta_2}
  \label{}
\end{align}
The above equation is just another way to express Eq.(\ref{eq:psg_bond_constraint_spin_state}). Notice that we use $\hat{B}_b$ to denote the quantum state associated with matrix (tensor) $B_b$.

Let us add the lattice symmetry. Given a single bond tensor $B_{b_0}$, we can generate all other bond tensors by using the relation $R^{-1}B_{b_0}=R^{-1}W_RR\circ B_{b_0}$, where $R$ is some lattice symmetry here. The explicitly expression is 
\begin{align}
  &B_{(R(xysi)|R(x'y's'i'))}=\notag\\
  &W_R^*(R(x,y,s,i))\cdot B_{(xysi|x'y's'i')}\cdot W_R^{-1}(R(x',y',s',i'))
  \label{}
\end{align}
It is obvious that we can generate all bond tensors if we consider the group generated by $T_1, T_2$ and $C_6$.

Further, reflection $\sigma$ will provide extra constraint on the Hilbert space of a single bond tensor. Let us consider $B_{(vd|wb)}\equiv B_{(00vd|00wb)}$. It is straightforward to see that 
\begin{align}
  &B_{(vd|wb)}=w_\sigma^*(v,d)\cdot B^{\mathrm{t}}_{(vd|wb)}\cdot w^{-1}_\sigma(w,b)
  \label{eq:psg_bond_tensor_constraint_reflection}
\end{align}
According to Eq.(\ref{eq:psg_wsigmauvw}), we have $w_{\sigma}(v,d)=\chi_\sigma\eta_{12}\eta_\sigma\eta_{\sigma C_6}$ and $w_\sigma(w,b)=\eta_{12}\eta_{\sigma C_6}$. Then, we get
\begin{align}
  B_{(vd|wb)}=\chi_\sigma\eta_{\sigma}\cdot B_{(vd|wb)}^\mathrm{t}.
  \label{eq:psg_bond_constraint_reflection_simplified}
\end{align}
Namely, for any block of $B_{(vd|wb)}$, it is either symmetric or antisymmetric, depending on values of $\chi_\sigma$ and $\eta_\sigma$. One can easily verify that the above result is not limited to bond $B_{(vd|wb)}$. In fact, it is true for all bond tensors.
 
Note that for integer (half-integer) spin $S$, $K_S$ is symmetric (antisymmetric). So, we conclude matrix $\widetilde{B}_{b}^{S_i}$ must be either symmetric or antisymmetric depending on values of $\chi_\sigma$ and $\eta_\sigma$. In particular, we can write
\begin{align}
  \eta_R=\bigoplus_i\,(\mu_R)^{2S_i}\cdot\mathrm{I}_{n_i(2S_i+1)}
  \label{eq:psg_blocked_eta}
\end{align}
where $\mu_R=1 (-1)$ for $\eta_R=\mathrm{I} (\mathrm{J})$.
Since we also have $K_{S_i}^\mathrm{t}=(-1)^{2S_i}K_{S_i}$, we conclude
\begin{align}
  \widetilde{B}_b^{S_i}=(-\mu_\sigma)^{2S_i}\chi_\sigma(\widetilde{B}_b^{S_i})^{\mathrm{t}}
  \label{eq:psg_bond_constraint_spin_reflection}
\end{align}

Finally, let us consider time reversal symmetry. Bond tensor should be a Kramer singlet in the sense
\begin{align}
  B_{(s_1i_1|s_2i_2)}^*=W_{\mathcal{T}}(s_1,i_1)\cdot B_{(s_1i_1|s_2i_2)}\cdot W^\mathrm{t}_\mathcal{T}(s_2,i_2)
  \label{eq:psg_bond_constraint_T}
\end{align}
By inserting Eq.(\ref{eq:psg_wTuvw}), we conclude, for any bond $B_b$, we have
\begin{align}
  B_b^*=\eta_{\sigma}w_\mathcal{T}\cdot B_b\cdot w_\mathcal{T}^\mathrm{t}
  \label{eq:psg_bond_constraint_T_simplified}
\end{align}
Further, using Eq.(\ref{eq:psg_wT}) and Eq.(\ref{eq:psg_bond_constraint_spin_state}), we get
\begin{align}
  (\widetilde{B}_b^{S_i})^*=(\mu_\sigma)^{2S_i}\widetilde{w}^{S_i}_\mathcal{T}\cdot \widetilde{B}_b^{S_i}\cdot (\widetilde{w}^{S_i}_\mathcal{T})^\mathrm{t}
  \label{eq:psg_bond_constraint_spinT}
\end{align}
where we use the fact that $K_S$ is invariant under time reversal symmetry: 
\begin{align}
  \ee^{\ii\pi S^y}\cdot K_S\cdot(\ee^{\ii\pi S^y})^\mathrm{t}=K_S^*
  \label{}
\end{align}
Here we have $\widetilde{w}_\mathcal{T}^{S_i}=\mathrm{I}_{n_i}$ for $\chi_\mathcal{T}=1$ while $\widetilde{w}_\mathcal{T}^{S_i}=\Omega_{n_i}$ for $\chi_\mathcal{T}=-1$.

So, to summarize, the constraint on the Hilbert space of a single bond tensor is determined by parameter $\chi_\sigma$, $\eta_\sigma$ and $\chi_\mathcal{T}$. We will list the constraint case by case.
\begin{enumerate}
  \item $\chi_\mathcal{T}=1$\\
    In this case, time reversal symmetry reads as
    \begin{align}
      w_{\mathcal{T}}=\bigoplus_{i=1}^M(\mathrm{I}_{n_i}\otimes\ee^{\ii\pi S_i^y})
      \label{}
    \end{align}
    So, according to Eq.(\ref{eq:psg_bond_constraint_spinT}), 
    \begin{align}
      (\widetilde{B}_b^{S_i})^*=(\mu_\sigma)^{2S_i}\widetilde{B}_b^{S_i}
      \label{}
    \end{align}
    Further, we are left with remaining global gauge transformation
    \begin{align}
      V=\bigoplus_i(\widetilde{V}_{S_i}\otimes\mathrm{I}_{2S_i+1})
      \label{}
    \end{align}
    where $\widetilde{V}_{S_i}$ is matrix defined on $\mathbb{R}$. Under gauge transformation $V$, we get
    \begin{align}
      \widetilde{B}_b^{S_i}\rightarrow\widetilde{V}_{n_i}\cdot\widetilde{B}_b^{n_i}\cdot\widetilde{V}_{n_i}^\mathrm{t}
      \label{}
    \end{align}

  \begin{enumerate}
  \item $\eta_\sigma=\mathrm{I}$, $\chi_\sigma=1$

    In this case, we get $\widetilde{B}_b^{S_i}$ is real symmetric for integer $S_i$, while real antisymmetric for half-integer $S_i$.

  \item $\eta_\sigma=\mathrm{I}$, $\chi_\sigma=-1$

    In this case, we get $\widetilde{B}_b^{S_i}$ is real antisymmetric for integer $S_i$, while real symmetric for half-integer $S_i$.

  \item $\eta_\sigma=\mathrm{J}$, $\chi_\sigma=1$

    In this case, we get $\widetilde{B}_b^{S_i}$ is real symmetric for integer $S_i$, while imaginary symmetric for half-integer $S_i$.

  \item $\eta_\sigma=\mathrm{J}$, $\chi_\sigma=-1$

    In this case, we get $\widetilde{B}_b^{S_i}$ is real antisymmetric for integer $S_i$, while imaginary antisymmetric for half-integer $S_i$.

  \end{enumerate}

  By using the remaining gauge transformation $V$, we can set the bond tensor to a maximal entangled states. Namely, if $\widetilde{B}_b^{S_i}$ is real symmetric, the canonical form is
  \begin{align}
    \widetilde{B}_b^{S_i}=\mathrm{Diag}(1,\dots,1,-1,\dots,-1)
    \label{}
  \end{align}
  where the number of $\pm$ sign $n_{i\pm}$ is not fixed. After doing this, we still left with gauge transformation $\widetilde{V}_{n_i}\in\mathrm{O}(n_{i+})\otimes\mathrm{O}(n_{i-})$. Here, we point out that different $n_{i\pm}$ does not lead to new classes. If we do not require bond tensors to be maximal entangled, different $n_{i\pm}$ can be connected adiabatically by continuously tuning the entries of bond tensors.

  If $\widetilde{B}_b^{S_i}$ is real antisymmetric, the canonical form is
  \begin{align}
    \widetilde{B}_b^{S_i}=\Omega_{n_i/2}
    \label{}
  \end{align}
  where $\Omega\equiv\ii\sigma_y\otimes\mathrm{I}_{n_i/2}$. The remaining gauge transformation satisfies $\Omega_{n_i/2}=\widetilde{V}_{n_i}\cdot\Omega_{n_i/2}\cdot\widetilde{V}_{n_i}^\mathrm{t}$.

  If $\widetilde{B}_b^{S_i}$ is imaginary, the canonical form is similar as the real case, except that all entries are replaced by $\pm\ii$.

  \item $\chi_\mathcal{T}=-1$

    In this case, the time reversal symmetry reads as
    \begin{align}
      w_{\mathcal{T}}=\bigoplus_{i=1}^M(\Omega_{n_i}\otimes\ee^{\ii\pi S_i^y})
      \label{}
    \end{align}
    So, we get the constraint on the bond tensor to be
    \begin{align}
      (\widetilde{B}_b^{S_i})^*=(\mu_\sigma)^{2S_i}\Omega_{n_i}\cdot\widetilde{B}_b^{S_i}\cdot\Omega_{n_i}^{-1}
      \label{}
    \end{align}
    Then, depending on values of $\mu_\sigma$ and $S_i$, either $\widetilde{B}_b^{S_i}$ or $\ii\widetilde{B}_b^{S_i}$ is quaternion matrix. In this case, the remaining gauge transformation is block diagonal matrix $V$, which reads
    \begin{align}
      V=\bigoplus_i(\widetilde{V}_{S_i}\otimes\mathrm{I}_{2S_i+1})
      \label{}
    \end{align}
    where $\widetilde{V}_{S_i}$ satisfies
    \begin{align}
      \widetilde{V}_{S_i}\cdot\Omega_{n_i}=\Omega_{n_i}\cdot\widetilde{V}_{S_i}^*
      \label{}
    \end{align}

    Under the gauge transformation $V$, we have
    \begin{align}
      \widetilde{B}_b^{S_i}\rightarrow\widetilde{V}_{n_i}\cdot\widetilde{B}_b^{S_i}\cdot\widetilde{V}_{n_i}^\mathrm{t}
      \label{}
    \end{align}

  \begin{enumerate}
  \item $\eta_\sigma=\mathrm{I}$, $\chi_\sigma=1$

    In this case, we get $\widetilde{B}_b^{S_i}$ is quaternion symmetric for integer $S_i$, while quaternion antisymmetric for half-integer $S_i$.

  \item $\eta_\sigma=\mathrm{I}$, $\chi_\sigma=-1$

    In this case, we get $\widetilde{B}_b^{S_i}$ is quaternion antisymmetric for integer $S_i$, while quaternion symmetric for half-integer $S_i$.

  \item $\eta_\sigma=\mathrm{J}$, $\chi_\sigma=1$

    In this case, we get $\widetilde{B}_b^{S_i}$ is quaternion symmetric for integer $S_i$, while $\ii\widetilde{B}_b^{S_i}$ is quaternion symmetric for half-integer $S_i$.

  \item $\eta_\sigma=\mathrm{J}$, $\chi_\sigma=-1$

    In this case, we get $\widetilde{B}_b^{S_i}$ is quaternion antisymmetric for integer $S_i$, while $\ii\widetilde{B}_b^{S_i}$ is quaternion antisymmetric for half-integer $S_i$.

  \end{enumerate}
\end{enumerate}

\subsubsection{Constraint on site tensors}
The Hilbert space of a site tensor $\mathbb{V}_T$ is defined in Eq.(\ref{eq:psg_bond_site_hilbert}). In the presence of $SU(2)$ symmetry, $\mathbb{V}_T$ can be decomposed as
\begin{align}
  \mathbb{V}_T\cong&\mathbb{U}\otimes\mathbb{V}\otimes\mathbb{V}\otimes\mathbb{V}\otimes\mathbb{V}\notag\\
  \cong&\bigoplus_{i_a,i_b,i_c,i_d}(\mathbb{D}_{S_{i_a}S_{i_b}S_{i_c}S_{i_d}}\otimes\mathbb{V}_{S_0}\otimes\notag\\
  &\mathbb{V}_{S_{i_a}}\otimes\mathbb{V}_{S_{i_b}}\otimes\mathbb{V}_{S_{i_c}}\otimes\mathbb{V}_{S_{i_d}})\notag\\
  \cong&\bigoplus_{i_a,i_b,i_c,i_d,k}(\mathbb{D}_{S_{i_a}S_{i_b}S_{i_c}S_{i_d}}\otimes\mathbb{V}_{S_0S_{i_a}S_{i_b}S_{i_c}S_{i_d}}^{S_k}\otimes\mathbb{V}_{S_k})
  \label{eq:psg_site_spin_hilbert}
\end{align}
where
\begin{align}
  \mathbb{D}_{S_{i_a}S_{i_b}S_{i_c}S_{i_d}}\equiv\mathbb{D}_{S_{i_a}}\otimes\mathbb{D}_{S_{i_b}}\otimes\mathbb{D}_{S_{i_c}}\otimes\mathbb{D}_{S_{i_d}}
  \label{}
\end{align}
labels the extra degenerate space associated with spins $S_{i_a}, S_{i_b}, S_{i_c}, S_{i_d}$ on four virtual legs. The basis of $\mathbb{D}_{S_{i_a}S_{i_b}S_{i_c}S_{i_d}}$ is labeled as
\begin{align}
  |S_{i_a},t_{\alpha}\rangle\otimes|S_{i_b},t_{\beta}\rangle\otimes|S_{i_c},t_{\gamma}\rangle\otimes|S_{i_d},t_{\delta}\rangle
  \label{}
\end{align}
$V_{S_0S_{i_a}S_{i_b}S_{i_c}S_{i_d}}^{S_k}$ is the fusion space, which denotes different ways to fuse spin $S_0, S_{i_a},S_{i_b},S_{i_c},S_{i_d}$ to spin $S_k$. The complicated fusion rules with six spins can be obtained by the fusion rules with only three spins as
\begin{align}
  \mathbb{V}_{S_0S_{i_a}S_{i_b}S_{i_c}S_{i_d}}^{S_k}\cong\bigoplus_{\alpha,\beta,\gamma}\mathbb{V}_{S_0S_{i_a}}^{S_\alpha}\otimes\mathbb{V}_{S_\alpha S_{i_b}}^{S_\beta}\otimes\mathbb{V}_{S_\beta S_{i_c}}^{S_\gamma}\otimes\mathbb{V}_{S_\gamma S_{i_d}}^{S_k}
  \label{eq:psg_fusion_space_decompose}
\end{align}
Since site tensors are $SU(2)$ singlet, we should focus on the Hilbert space with $S=0$:
\begin{align}
  \mathbb{V}_T^{S=0}\cong\bigoplus_{i_a,i_b,i_c,i_d}(\mathbb{D}_{S_{i_a}S_{i_b}S_{i_c}S_{i_d}}\otimes\mathbb{V}_{S_0S_{i_a}S_{i_b}S_{i_c}S_{i_d}}^{S=0}\otimes\mathbb{V}_{S=0})
  \label{}
\end{align}
The basis for space $\mathbb{V}_{S_0S_{i_a}S_{i_b}S_{i_c}S_{i_d}}^{S=0}\otimes\mathbb{V}_{S=0}$ can be expressed as
\begin{align}
  &\hat{K}^l_{S_0S_{i_a}S_{i_b}S_{i_c}S_{i_d}}\equiv (K^l_{S_0S_{i_a}S_{i_b}S_{i_c}S_{i_d}})_{\alpha\beta\gamma\delta}^{j}\times\notag\\
  &|S_0,m_j\rangle\otimes|S_{i_a},m_{\alpha}\rangle\otimes|S_{i_b},m_{\beta}\rangle\otimes|S_{i_c},m_{\gamma}\rangle\otimes|S_{i_d},m_{\delta}\rangle
  \label{}
\end{align}
where $\hat{K}^l_{S_0S_{i_a}S_{i_b}S_{i_c}S_{i_d}}$ labels orthogonal singlet states for different $l$.

Then, in terms of the tensor representation, we decompose the site tensor $T^s$ as
\begin{align}
  T^s=\bigoplus_{i_a,i_b,i_c,i_d,l}(\widetilde{T}^l_{S_{i_a}S_{i_b}S_{i_c}S_{i_d}}\otimes K^l_{S_0S_{i_a}S_{i_b}S_{i_c}S_{i_d}})
  \label{eq:psg_site_tensor_blocked}
\end{align}
where the state
\begin{align}
  \sum_{\alpha,\beta,\gamma,\delta}(\widetilde{T}^l_{S_{i_a}S_{i_b}S_{i_c}S_{i_d}})_{\alpha\beta\gamma\delta}&|S_{i_a},t_{\alpha}\rangle\otimes|S_{i_b},t_{\beta}\rangle\notag\\
  \otimes&|S_{i_c},t_{\gamma}\rangle\otimes|S_{i_d},t_{\delta}\rangle
  \label{}
\end{align}
is an arbitrary state lives in the extra degenerate space $\mathbb{D}_{S_{i_a}S_{i_b}S_{i_c}S_{i_d}}$.

Due to the representation theory of $SU(2)$, $K_{S_0S_{i_a}S_{i_b}S_{i_c}S_{i_d}}$ does not vanish only if there are even number of half integer spins for $S_0,\dots,S_{i_d}$. Since $S_0$ is a half-integer spin, we conclude that there should always be odd number of half integer spins living on virtual legs. For a site tensor, there are four virtual legs, so we get two different cases:
\begin{enumerate}
  \item Only one virtual leg are a half integer spin, while other three are integer spins;
  \item Three virtual legs are half integer spins, while the remaining one is an integer spin.
\end{enumerate}

We now consider the constraint from lattice symmetry. Remember that in the presence of translation and rotation, we can always choose a gauge such that all site tensors share the same form, as shown in Eq.(\ref{eq:psg_constant_site_translation}) and Eq.(\ref{eq:psg_constant_site_rotation}). Then, in the following, we only need to focus on a single site tensor.

We figure out lattice symmetries that maps site tensor $T^u$ to itself as follows
\begin{align}
  \sigma\circ(T^u)^i_{\alpha\beta\gamma\delta}&=(T^u)^i_{\delta\gamma\beta\alpha}\notag\\
  T_1T_2C_6^3\circ(T^u)^i_{\alpha\beta\gamma\delta}&=C_6\circ(T^u)^i_{\alpha\beta\gamma\delta}=(T^u)^i_{\beta\alpha\delta\gamma}
  \label{eq:psg_site_automorphism_lattice_symmetry}
\end{align}
Besides, combining reflection $\sigma$ and rotation $C_6$, we get
\begin{align}
  \sigma C_6\circ(T^u)^i_{\alpha\beta\gamma\delta}&=(T^u)^i_{\gamma\delta\alpha\beta}\notag\\
  \label{eq:psg_site_automorphism_lattice_symmetry2}
\end{align}
In the following, we will solve the constraint from above symmetry operations. Further, we can prove that the whole site tensor can be generated by lattice symmetries once we fix quantum states in the Hilbert space satisfying the two situations below:
\begin{enumerate}
  \item $S_{i_a}$ is a half integer spin, while other three are integer spins;
  \item $S_{i_a}$ is an integer spin, while other three are half integer spins.
\end{enumerate}

To see this, let us first consider reflection symmetry $\sigma$. Under the action of $\sigma$, for the decomposed parts of the site tensor, we have
\begin{align}
  &\sigma\circ(\widetilde{T}^l_{S_{i_a}S_{i_b}S_{i_c}S_{i_d}})_{\alpha\beta\gamma\delta}=(\widetilde{T}^l_{S_{i_a}S_{i_b}S_{i_c}S_{i_d}})_{\delta\gamma\beta\alpha}\notag\\
  &\sigma\circ(K_{S_0S_{i_a}S_{i_b}S_{i_c}S_{i_d}}^l)_{\alpha\beta\gamma\delta}^j=(K^l_{S_0S_{i_a}S_{i_b}S_{i_c}S_{i_d}})^j_{\delta\gamma\beta\alpha}
  \label{eq:psg_sigma_block_site}
\end{align}
It is obvious that we can choose $K_{S_0S_{i_a}S_{i_b}S_{i_c}S_{i_d}}$ to be either symmetric or antisymmetric under the permutation of $\{S_{i_a},S_{i_b},S_{i_c},S_{i_d}\}$: 
\begin{align}
  (K_{S_0\mathbf{P}(S_{i_a}S_{i_b}S_{i_c}S_{i_d})})^j_{\mathbf{P}(\alpha\beta\gamma\delta)}=\pm(K_{S_0S_{i_a}S_{i_b}S_{i_c}S_{i_d}})^j_{\alpha\beta\gamma\delta}
  \label{eq:psg_site_K_permutation_symmetric}
\end{align}
where $\mathbf{P}$ is any permutation. The $\pm$ sign depends on the definition of $K$. Particularly, we have
\begin{align}
  \sigma\circ(K_{S_0S_{i_a}S_{i_b}S_{i_c}S_{i_d}})_{\alpha\beta\gamma\delta}^j=\pm(K_{S_0S_{i_d}S_{i_c}S_{i_b}S_{i_a}})^j_{\alpha\beta\gamma\delta}
  \label{}
\end{align}
For the two cases we consider here, $S_{i_a}$ is always different from spins of other three virtual legs. So, $K_{S_0S_{i_a}S_{i_b}S_{i_c}S_{i_d}}$ and $K_{S_0S_{i_d}S_{i_c}S_{i_b}S_{i_a}}$ are always independent tensors. Thus, we can absorb minus sign in the above equation by redefining $K_{S_0S_{i_d}S_{i_c}S_{i_b}S_{i_a}}$. Then, we get
\begin{align}
  \sigma\circ T^l&=\bigoplus_{i_a,i_b,i_c,i_d,l}(\sigma\circ\widetilde{T}^l_{S_{i_a}S_{i_b}S_{i_c}S_{i_d}}\otimes K^l_{S_0S_{i_d}S_{i_c}S_{i_b}S_{i_a}})
  \label{eq:psg_blocked_site_sigma}
\end{align}
Remember that the site tensor is symmetric under $\sigma$, so we have
\begin{align}
  T^u&=\Theta_\sigma W_\sigma\sigma\circ T^u\notag\\
  &=\Theta_\sigma W_\sigma\bigoplus_{i_a,i_b,i_c,i_d,l}\left(\sigma\circ\widetilde{T}^l_{S_{i_a}S_{i_b}S_{i_c}S_{i_d}}\otimes K^l_{S_0S_{i_d}S_{i_c}S_{i_b}S_{i_a}}\right)
  \label{}
\end{align}
As shown in the last subsection, we always choose the basis such that $W_R(x,y,s)\in\{\mathrm{I},\mathrm{J}\}$ for any lattice symmetry $R$. So, we can always define the action of $W_\sigma$ trivially on $K^l_{S_0S_{i_a}S_{i_b}S_{i_c}S_{i_d}}$. Namely, we can decompose $W_R(x,y,s,i)$ as
\begin{align}
  W_R(x,y,s,i)=\bigoplus_i\left( \widetilde{W}_R^{S_i}(x,y,s,i)\otimes\mathrm{I}_{2S_i+1} \right)
  \label{}
\end{align}
Then, from the above analysis, we conclude 
\begin{align}
  \widetilde{T}^l_{S_{i_d}S_{i_c}S_{i_b}S_{i_a}}=\Theta_\sigma \widetilde{W}_\sigma\sigma\circ\widetilde{T}^l_{S_{i_a}S_{i_b}S_{i_c}S_{i_d}}
  \label{}
\end{align}
Writing the above equation explicitly, we get
\begin{align}
  [\widetilde{T}^l_{S_{i_d}S_{i_c}S_{i_b}S_{i_a}}]_{\alpha\beta\gamma\delta}=&\Theta_\sigma(u)[\widetilde{w}^{S_{i_a}}_\sigma]_{\alpha\alpha'}[\widetilde{w}^{S_{i_b}}_\sigma]_{\beta\beta'}[\widetilde{w}^{S_{i_c}}_\sigma]_{\gamma\gamma'}\notag\\
  &[\widetilde{w}^{S_{i_d}}_\sigma]_{\delta\delta'}[\widetilde{T}^l_{S_{i_a}S_{i_b}S_{i_c}S_{i_d}}]_{\delta'\gamma'\beta'\alpha'}
  \label{}
\end{align}
According to Eq.(\ref{eq:psg_wsigmauvw}), we have
\begin{align}
  &\widetilde{w}^{S}_\sigma(u,a)=\mathrm{I},\quad \widetilde{w}^{S}_\sigma(u,b)=\chi_\sigma(\mu_{12}\mu_{C_6})^{2S},\notag\\
  &\widetilde{w}^{S}_\sigma(u,c)=\chi_\sigma(\mu_{12}\mu_{C_6}\mu_\sigma)^{2S},\quad \widetilde{w}^{S}_\sigma(u,d)=(\mu_{\sigma})^{2S}.\notag\\
  \label{eq:psg_blocked_wsigma}
\end{align}
Then, we can simplify the constraint as
\begin{align}
  [\widetilde{T}^l_{S_{i_d}S_{i_c}S_{i_b}S_{i_a}}]_{\alpha\beta\gamma\delta}=&\Theta_\sigma(u)\cdot(\mu_\sigma)^{2S_{i_c}+2S_{i_d}}(\mu_{12}\mu_{C_6})^{2S_{i_b}+2S_{i_d}}\notag\\
  &\times[\widetilde{T}^l_{S_{i_a}S_{i_b}S_{i_c}S_{i_d}}]_{\delta\gamma\beta\alpha}
  \label{}
\end{align}
Since $S_{i_b}$, $S_{i_c}$ and $S_{i_d}$ are all integer spins or all half-integer spins, the above equation reads
\begin{align}
  [\widetilde{T}^l_{S_{i_d}S_{i_c}S_{i_b}S_{i_a}}]_{\alpha\beta\gamma\delta}=\Theta_\sigma(u)[\widetilde{T}^l_{S_{i_a}S_{i_b}S_{i_c}S_{i_d}}]_{\delta\gamma\beta\alpha}
  \label{eq:psg_blocked_site_sigma_simplified}
\end{align}
where $\Theta_\sigma(u)=(\mu_\sigma)^{\frac{1}{2}}$.

We consider the constraint by the rotation symmetry now. Similarly, $\widetilde{T}^l_{S_{i_b}S_{i_a}S_{i_d}S_{i_c}}$ and $\widetilde{T}^l_{S_{i_c}S_{i_d}S_{i_b}S_{i_a}}$ can be obtained from $\widetilde{T}^l_{S_{i_a}S_{i_b}S_{i_c}S_{i_d}}$ and $\widetilde{T}^l_{S_{i_d}S_{i_c}S_{i_b}S_{i_a}}$ by rotaion symmetry: 
\begin{align}
  &\widetilde{T}^l_{S_{i_b}S_{i_a}S_{i_d}S_{i_c}}=\Theta_{C_6}\widetilde{W}_{C_6}C_6\circ\widetilde{T}^l_{S_{i_a}S_{i_b}S_{i_c}S_{i_d}}\notag\\
  &\widetilde{T}^l_{S_{i_c}S_{i_d}S_{i_a}S_{i_b}}=\Theta_{C_6}\widetilde{W}_{C_6}C_6\circ\widetilde{T}^l_{S_{i_d}S_{i_c}S_{i_b}S_{i_a}}
  \label{}
\end{align}
By inserting $w_{C_6}(u,i)$ defined in Eq.(\ref{eq:psg_wC6u}), we get
\begin{align}
  &[\widetilde{T}^l_{S_{i_b}S_{i_a}S_{i_d}S_{i_c}}]_{\alpha\beta\gamma\delta}=\Theta_{C_6}(u)[\widetilde{T}^l_{S_{i_a}S_{i_b}S_{i_c}S_{i_d}}]_{\beta\alpha\delta\gamma}\notag\\
  &[\widetilde{T}^l_{S_{i_c}S_{i_d}S_{i_a}S_{i_b}}]_{\alpha\beta\gamma\delta}=\mu_{12}\mu_{C_6}\Theta_{C_6}(u)\Theta_\sigma(u)[\widetilde{T}^l_{S_{i_a}S_{i_b}S_{i_c}S_{i_d}}]_{\gamma\delta\alpha\beta}
  \label{eq:psg_blocked_site_C6_simplified}
\end{align}
where $\Theta_{C_6}(u)=(\mu_{12}\mu_{C_6})^{\frac{1}{2}}$.

Thus, once we know tensors $\widetilde{T}_{S_{i_a}S_{i_b}S_{i_c}S_{i_d}}^l$ with $S_{i_a}$ to be a half-integer/integer spin and $S_{i_b},S_{i_c},S_{i_d}$ to be integer/half-integer spins, by the above lattice symmetries, we are able to generate tensors $\widetilde{T}^l_{S_{i_a}S_{i_b}S_{i_c}S_{i_d}}$ which satisfy one virtual leg to be a half-integer/integer spin and other three virtual legs to be integer/half-integer spins.

At last, we add time reversal symmetry. The constraint of time reversal symmetry reads $T^s=\Theta_\mathcal{T}W_\mathcal{T}\mathcal{T}\circ T^s$. Since $K^l_{S_0S_{i_a}S_{i_b}S_{i_c}S_{i_d}}$ is a Kramer singlet state and real, we have 
\begin{align}
  &[K^l_{S_0S_{i_a}S_{i_b}S_{i_c}S_{i_d}}]^j_{\alpha\beta\gamma\delta}=[\ee^{\ii\pi S^y_0}]_{jj'}[\ee^{\ii\pi S^y_{i_a}}]_{\alpha\alpha'}[\ee^{\ii\pi S^y_{i_b}}]_{\beta\beta'}\times\notag\\
  &[\ee^{\ii\pi S^y_{i_c}}]_{\gamma\gamma'}[\ee^{\ii\pi S^y_{i_d}}]_{\delta\delta'}[K^{l\,*}_{S_0S_{i_a}S_{i_b}S_{i_c}S_{i_d}}]^{j'}_{\alpha'\beta'\gamma'\delta'};\notag\\
  \label{}
\end{align}
Then, according to Eq.(\ref{eq:psg_wT}), the constraint on $\widetilde{T}^l_{S_{i_a}S_{i_b}S_{i_c}S_{i_d}}$ reads
\begin{align}
  &[\widetilde{T}^l_{S_{i_a}S_{i_b}S_{i_c}S_{i_d}}]_{\alpha\beta\gamma\delta}=[\widetilde{w}^{S_{i_a}}_\mathcal{T}(u,a)]_{\alpha\alpha'}[\widetilde{w}^{S_{i_b}}_\mathcal{T}(u,b)]_{\beta\beta'}\times\notag\\
  &[\widetilde{w}^{S_{i_c}}_\mathcal{T}(u,c)]_{\gamma\gamma'}[\widetilde{w}^{S_{i_d}}_\mathcal{T}(u,d)]_{\delta\delta'}[\widetilde{T}^{l\,*}_{S_{i_a}S_{i_b}S_{i_c}S_{i_d}}]_{\alpha'\beta'\gamma'\delta'}\notag\\
  \label{eq:psg_blocked_site_T}
\end{align}
where according to Eq.(\ref{eq:psg_wTuvw}), we obtain $\widetilde{w}_\mathcal{T}^S(u,i)$ as
\begin{align}
  \widetilde{w}^{S}_\mathcal{T}(u,a)=\widetilde{w}^{S}_\mathcal{T},\quad &\widetilde{w}^{S}_\mathcal{T}(u,b)=(\mu_{12}\mu_{C_6})^{2S}\widetilde{w}^{S}_\mathcal{T},\notag\\
  \widetilde{w}^{S}_\mathcal{T}(u,c)=(\mu_{12}\mu_{C_6}\mu_{\sigma})^{2S}\widetilde{w}^{S}_\mathcal{T},\quad &\widetilde{w}^{S}_\mathcal{T}(u,d)=(\mu_{\sigma})^{2S}\widetilde{w}^{S}_\mathcal{T};\notag\\
  \label{eq:psg_blocked_wT}
\end{align}
And $\widetilde{w}_\mathcal{T}^S$ depends on $\chi_\mathcal{T}$. Remember that we will focus on the case where $S_{i_b}$, $S_{i_c}$ and $S_{i_d}$ must be all integer or half integer spins. Then, by inserting Eq.(\ref{eq:psg_blocked_wT}) back to Eq.(\ref{eq:psg_blocked_site_T}), we get
\begin{align}
  &[\widetilde{T}^l_{S_{i_a}S_{i_b}S_{i_c}S_{i_d}}]_{\alpha\beta\gamma\delta}=[\widetilde{w}^{S_{i_a}}_\mathcal{T}]_{\alpha\alpha'}[\widetilde{w}^{S_{i_b}}_\mathcal{T}]_{\beta\beta'}\times\notag\\
  &[\widetilde{w}^{S_{i_c}}_\mathcal{T}]_{\gamma\gamma'}[\widetilde{w}^{S_{i_d}}_\mathcal{T}]_{\delta\delta'}[\widetilde{T}^{l\,*}_{S_{i_a}S_{i_b}S_{i_c}S_{i_d}}]_{\alpha'\beta'\gamma'\delta'}\notag\\
  \label{eq:psg_blocked_site_T_simplified}
\end{align}

When $\chi_\mathcal{T}=1$, we have $\widetilde{w}_\mathcal{T}^{S_i}=\mathrm{I}_{n_i}$, then Eq.(\ref{eq:psg_blocked_site_T_simplified}) is simplified as
\begin{align}
  \widetilde{T}^l_{S_{i_a}S_{i_b}S_{i_c}S_{i_d}}=\widetilde{T}^{l\,*}_{S_{i_a}S_{i_b}S_{i_c}S_{i_d}}
  \label{eq:psg_blocked_site_chiT=1}
\end{align}
So, $\widetilde{T}^l_{S_{i_a}S_{i_b}S_{i_c}S_{i_d}}$ is real tensor.

When $\chi_\mathcal{T}=-1$, we have $\widetilde{w}_\mathcal{T}^{S_i}=\Omega_{n_i}$, then Eq.(\ref{eq:psg_blocked_site_T_simplified}) becomes
\begin{align}
  &[\widetilde{T}^l_{S_{i_a}S_{i_b}S_{i_c}S_{i_d}}]_{\alpha\beta\gamma\delta}=[\Omega_{n_{i_a}}]_{\alpha\alpha'}[\Omega_{n_{i_b}}]_{\beta\beta'}\times\notag\\
  &[\Omega_{n_{i_c}}]_{\gamma\gamma'}[\Omega_{n_{i_d}}]_{\delta\delta'}[\widetilde{T}^{l\,*}_{S_{i_a}S_{i_b}S_{i_c}S_{i_d}}]_{\alpha'\beta'\gamma'\delta'}
  \label{eq:psg_blocked_site_chiT=-1}
\end{align}

\subsubsection{Examples}\label{subsubapp:examples_spin0_spin1/2}
Now, let us focus on a special case where the physical spin $S_0=\frac{1}{2}$ and there are only spin-0 and spin-$\frac{1}{2}$ living on virtual legs. Then, to obtain site tensors, according the above analysis, we restrict ourselves to two subspace of $\mathbb{V}_T^{S=0}$:

{\bf{1.}} \,$S_{i_a}=\frac{1}{2}$ and $S_{i_b}=S_{i_c}=S_{i_d}=0$. Then the corresponding Hilbert space $\mathbb{H}_0$ is 
\begin{align}
  \mathbb{H}_0=\mathbb{D}_{\frac{1}{2}}\otimes(\mathbb{D}_{0})^{3}\otimes\mathbb{V}_{\frac{1}{2}\frac{1}{2}000}^0\otimes\mathbb{V}_{0}
  \label{}
\end{align}
where fusion space $\mathbb{V}_{\frac{1}{2}\frac{1}{2}000}^{0}\cong\mathbb{V}_{\frac{1}{2}\frac{1}{2}}^0$ with dimension one. The singlet state $K_0\doteq K_{\frac{1}{2}\frac{1}{2}000}$ is 
\begin{align}
  (K_0)^j_{\alpha\beta\gamma\delta}=(\ii\sigma^y)_{j\alpha}
  \label{}
\end{align}
since $\beta=\gamma=\delta\equiv1$ in this case. One can easily verify that $K_0$ is invariant under time reversal operator:
\begin{align}
  (K_0^*)^j_{\alpha\beta\gamma\delta}=(\ii\sigma^y)_{jj'}(\ii\sigma^y)_{\alpha\alpha'}\cdot (K_0^*)_{\alpha'\beta\gamma\delta}^{j'}
  \label{}
\end{align}

Then, quantum state in $\mathbb{H}_0$ can be expressed by the tensor form as
\begin{align}
  \widetilde{T}_0\otimes K_0
  \label{}
\end{align}
where $\widetilde{T}_0$ denotes an arbitrary quantum state in $\mathbb{D}_{\frac{1}{2}}\otimes(\mathbb{D}_{0})^3$.

{\bf{2.}} \,$S_{i_a}=0$ and  $S_{i_b}=S_{i_c}=S_{i_d}=\frac{1}{2}$. The the corresponding Hilbert space $\mathbb{H}_1$ is 
\begin{align}
  \mathbb{H}_1=\mathbb{D}_{0}\otimes(\mathbb{D}_{\frac{1}{2}})^{3}\otimes\mathbb{V}_{\frac{1}{2}0\frac{1}{2}\frac{1}{2}\frac{1}{2}}^0\otimes\mathbb{V}_{0}
  \label{}
\end{align}
Using representation theory of $SU(2)$, we get
\begin{align}
  \mathbb{V}_{\frac{1}{2}0\frac{1}{2}\frac{1}{2}\frac{1}{2}}^0\cong&(\mathbb{V}_{\frac{1}{2}\frac{1}{2}}^0\otimes\mathbb{V}_{\frac{1}{2}0}^{\frac{1}{2}}\otimes\mathbb{V}_{\frac{1}{2}\frac{1}{2}}^0)\oplus\notag\\
  &(\mathbb{V}_{\frac{1}{2}\frac{1}{2}}^0\otimes\mathbb{V}_{\frac{1}{2}1}^{\frac{1}{2}}\otimes\mathbb{V}_{\frac{1}{2}\frac{1}{2}}^1)
  \label{}
\end{align}
So, $\mathbb{V}_{\frac{1}{2}0\frac{1}{2}\frac{1}{2}\frac{1}{2}}^0$ has dimension 2. And we choose the two basis in $\mathbb{V}_{\frac{1}{2}0\frac{1}{2}\frac{1}{2}\frac{1}{2}}^0\otimes\mathbb{V}_0$ as
\begin{align}
  &(K_1)^j_{\alpha\beta\gamma\delta}=(\ii\sigma^y)_{j\beta}(\ii\sigma^y)_{\gamma\delta}\notag\\
  &(K_2)^j_{\alpha\beta\gamma\delta}=\sum_{\mu\nu}(\ii\sigma^y)_{j\nu}C^{\frac{1}{2}m_\nu}_{\frac{1}{2}m_\beta1m_\mu}C^{1m_\mu}_{\frac{1}{2}m_\gamma\frac{1}{2}m_\delta}\notag\\
  \label{}
\end{align}
where $\alpha\equiv1$ in this case. Here, $C_{S_1m_1S_2m_2}^{Jm_J}\doteq\langle S_1m_1S_2m_2|Jm_J\rangle$ is the CG coefficient. And $m_i=-S-1+i$ is the $S_z$ quantum number. Simialr as the previous case, $K_1$ and $K_2$ are also chosen to be invariant under time reversal operator:
\begin{align}
  (K_{1(2)})_{\alpha\beta\gamma\delta}^j=(\ii\sigma^y)_{jj'}(\ii\sigma^y)_{\beta\beta'}(\ii\sigma^y)_{\gamma\gamma'}(\ii\sigma^y)_{\delta\delta'}(K^*_{1(2)})_{\alpha\beta'\gamma'\delta'}^{j'}
  \label{}
\end{align}

Then quantum state in $\mathbb{H}_1$ can be expressed by the tensor form as
\begin{align}
  \widetilde{T}_1\otimes K_1\oplus\widetilde{T}_2\otimes K_2
  \label{}
\end{align}
where $\widetilde{T}_1$, $\widetilde{T}_2$ are tensor representation of arbitrary states in $\mathbb{D}_0\otimes(\mathbb{D}_{\frac{1}{2}})^3$.

We can explicitly write down $K_i$ as a quantum state by introducing basis $|0\rangle$ for spin-$0$ and $|\uparrow\rangle, |\downarrow\rangle$ for spin-$\frac{1}{2}$:
\begin{align}
  \hat{K}_0=&|\uparrow\rangle\otimes|\downarrow000\rangle-|\downarrow\rangle\otimes|\uparrow000\rangle\notag\\
  \hat{K}_1=&|\uparrow\rangle\otimes(|0\downarrow\uparrow\downarrow\rangle-|0\downarrow\downarrow\uparrow\rangle)-\notag\\
  &|\downarrow\rangle\otimes(|0\uparrow\uparrow\downarrow\rangle-|0\uparrow\downarrow\uparrow\rangle)\notag\\
  \hat{K}_2=&|\uparrow\rangle\otimes(2|0\uparrow\downarrow\downarrow\rangle-|0\downarrow\uparrow\downarrow\rangle-|0\downarrow\downarrow\uparrow\rangle)+\notag\\
  &|\downarrow\rangle\otimes(2|0\downarrow\uparrow\uparrow\rangle-|0\uparrow\uparrow\downarrow\rangle-|0\uparrow\downarrow\uparrow\rangle)\notag\\
  \label{eq:K0_1_2_form}
\end{align}
where we define $\hat{K}_i=(K_i)^j_{\alpha\beta\gamma\delta}|m_j\rangle\otimes|m_\alpha m_\beta m_\gamma m_\delta\rangle$.

Now, let us consider case $D=3$ virtual legs $\mathbb{V}\cong0\oplus\frac{1}{2}$. Then, There is no extra degeneracy of spins. So, according to constraint from bond tensors, only classes with $\eta_{\sigma}=\mathrm{J}$, $\chi_\sigma=1$ and $\chi_\mathcal{T}=1$ can be realized. Other classes require even dimensional extra degenerate spaces, since bond tensors of those classes are either antisymmetric or symplectic in the extra degenerate spaces. Thus, when $D=3$, we only left with $\eta_{12}$, $\eta_{C_6}$, and the number of classes can be realized is $2^2=4$.

Given a bond tensor $B_{b_0}$, we can fix it as a maximal entangled state with the following form
\begin{align}
  B_{b_0}=
\begin{pmatrix}
  \pm1 & 0 & 0 \\
  0 & 0 & -\mathrm{i} \\
  0 & \mathrm{i} & 0 \\
\end{pmatrix}
\label{eq:bond_tensor_D3}
\end{align}
Other bonds are all related to $B_{b_0}$ by translation and rotation symmetry, and can be generated as 
\begin{align}
  B_{R(b)}=R^{-1}W_RR\circ B_{b_0}
  \label{}
\end{align}
where $R=T_1^{n_1}T_2^{n_2}C_6^{n_{C_6}}$ with $n_1,n_2,n_{C_6}\in\mathbb{Z}$.

For site tensors, they all share the same form. The spin singlet state $\hat{K}_{S_0S_{i_a}S_{i_b}S_{i_c}S_{i_d}}$ is fixed as Eq.(\ref{eq:K0_1_2_form}). So we can express the site tensor using the quantum state representation as
\begin{align}
  \hat{T}^s=&\{\hat{K}_0+\hat{K}_{12}(p_1,p_2)\}+\Theta_{C_6}(u)\{a\leftrightarrow b,c\leftrightarrow d\}+\Theta_{\sigma}(u)\cdot\notag\\
  &\{a\leftrightarrow d,b\leftrightarrow c\}+\mu_{12}\mu_{C_6}\Theta_{C_6}(u)\Theta_{\sigma}(u)\{a\leftrightarrow c,b\leftrightarrow d\}\notag\\
  \label{eq:site_tensor_D3}
\end{align}
where we have
\begin{align}
  \hat{K}_{12}=&a_1\hat{K}_1+a_2\hat{K}_2\notag\\
  =&p_1\cdot(|\uparrow\rangle\otimes|0\downarrow\uparrow\downarrow\rangle+|\downarrow\rangle\otimes|0\uparrow\downarrow\uparrow\rangle)+\notag\\
  &p_2\cdot(|\uparrow\rangle\otimes|0\downarrow\downarrow\uparrow\rangle+|\downarrow\rangle\otimes|0\uparrow\uparrow\downarrow\rangle)-\notag\\
  &(p_1+p_2)\cdot(|\uparrow\rangle\otimes|0\uparrow\downarrow\downarrow\rangle+|\downarrow\rangle\otimes|0\downarrow\uparrow\uparrow\rangle
  \label{eq:K_12_form}
\end{align}
where we define $p_1\equiv a_1-a_2$ and $p_2\equiv-a_1-a_2$ as the two tunable parameters.

Now, let us consider virtual legs $\mathbb{V}\cong0\oplus0\oplus\frac{1}{2}\oplus\frac{1}{2}$ with $D=6$. In this case, there are extra two dimensional degeneracy spaces for both spin-$0$ and spin-$\frac{1}{2}$. We believe all 32 classes can be realized in this case. However, here we will focus on the 4 classes realized in $D=3$ case.

Fixing $\eta_\sigma=\mathrm{J}$, $\chi_\sigma=1$ and $\chi_\mathcal{T}=1$, the bond tensor $B_{b_0}$ now reads
\begin{align}
  B_{b_0}=
\begin{pmatrix}
  \pm1&0\\
  0&\pm1
\end{pmatrix}
\oplus
\begin{pmatrix}
  \pm1&0\\
  0&\pm1
\end{pmatrix}
\otimes
\begin{pmatrix}
  0&\ii\\
  -\ii&0
\end{pmatrix}
\label{eq:bond_tensor_D6}
\end{align}
Other bonds can be generated by translation and reflection symmetry as discussed above.

For the site tensor, we have
\begin{align}
  \hat{T}^s=&\{\hat{\widetilde{T}}_0\otimes\hat{K}_0+\hat{\widetilde{T}}_1\otimes\hat{K}_1+\hat{\widetilde{T}}_2\otimes\hat{K}_2\}\notag\\
  &+\Theta_{C_6}(u)\{a\leftrightarrow b,c\leftrightarrow d\}+\Theta_{\sigma}(u)\{a\leftrightarrow d,b\leftrightarrow c\}\notag\\
  &+\mu_{12}\mu_{C_6}\Theta_{C_6}(u)\Theta_{\sigma}(u)\{a\leftrightarrow c,b\leftrightarrow d\}
  \label{eq:site_tensor_D6}
\end{align}
where $\hat{\widetilde{T}}_i$ labels a quantum state in extra degenerate space, which has dimension $2^4=16$. Further, the transformation rules of $\widetilde{T}$'s are given in Eq.(\ref{eq:psg_blocked_site_sigma_simplified}) and Eq.(\ref{eq:psg_blocked_site_C6_simplified}). So, there are three $\hat{\widetilde{T}}$\,'s ($\hat{\widetilde{T}}_0$, $\hat{\widetilde{T}}_1$ and $\hat{\widetilde{T}}_2$) serving as tunable parameters. Then the tunable parameters in $D=6$ case should be $16\times3-1=47$, where the additional $-1$ comes from the fact that the norm of the wavefunction has no physical consequence.

\section{Projective representation, group extension and second cohomology}\label{app:proj_rep}
In this appendix, we will introduce mathematical tools for symmetry fractionalization, including projective representation, group extension as well as the second cohomology. Readers may refer Ref.\onlinecite{Essin:2013p104406} for more details.

Consider a group $G$ with elements $g\in G$. We call $\Gamma(g)$ a projective representation of $G$ with coefficient $A$, where $A$ is an Abelian group, if 
\begin{align}
  \Gamma(g_1)\Gamma(g_2)=\omega(g_1,g_2)\Gamma(g_1g_2)
  \label{}
\end{align}
Here $\omega$ is a map, which is defined as $\omega:G\times G\rightarrow A$. According to associativity of matrix product, we get
\begin{align}
  \Gamma(g_1)\Gamma(g_2)\Gamma(g_3)&=\omega(g_1,g_2)\omega(g_1g_2,g_3)\Gamma(g_1g_2g_3)\notag\\
  &=\omega(g_1,g_2g_3)\,{}^{g_1}\!\omega(g_2,g_3)\Gamma(g_1g_2g_3).
  \label{}
\end{align}
where appearance of ${}^{g_1}\!\omega(g_2,g_3)$ comes from commutation of $g_1$ and $\omega(g_2,g_3)$, which indicates action of $G$ on coefficient $A$ may be nontrivial. Further, we require the action of $\forall g\in G$ on Abelian group $A$ should be an automorphism of $A$. Then the associativity constraint for $\omega$ is
\begin{align}
  \omega(g_1,g_2)\omega(g_1g_2,g_3)=\omega(g_1,g_2g_3)\,{}^{g_1}\!\omega(g_2,g_3).
  \label{eq:factor_system_associativity}
\end{align}
Any function $\omega$ satisfy the associativity constraint is called a factor set.

If $\omega_a$ and $\omega_b$ are both factor sets, then $\omega_{ab}=\omega_a\omega_b$ is also a factor set, where $(\omega_{a}\omega_b)(g_1,g_2)\equiv\omega_{a}(g_1,g_2)\omega_b(g_1,g_2)$. The product of factor sets is associated with tensor product of projective representations: if $\omega_a,\omega_b$ are factor sets of $\Gamma_a,\Gamma_b$, respectively, then $\omega_{ab}$ is factor set of the tensor product representation $\Gamma_a\otimes\Gamma_b$.

Now, let us define the equivalent class for factor sets. Suppose we allow a redefinition of the $\Gamma$'s by
\begin{align}
  \Gamma'(g)=\lambda(g)\Gamma(g)
  \label{eq:proj_rep_redefine}
\end{align}
where $\lambda$ defined as $\lambda:G\rightarrow A$. This induces transformation of the factor set:
\begin{align}
  \omega'(g_1,g_2)=\lambda(g_1)\,{}^{g_1}\!\lambda(g_2)\lambda(g_1g_2)^{-1}\omega(g_1,g_2)
  \label{eq:factor_set_redefine}
\end{align}
where $\omega'$ is also a factor set. Two factor sets $\omega$ and $\omega'$ are said to be equivalent if they are related by the above equation for some $\lambda$, and we write as $\omega\sim\omega'$. We group all equivalent $\omega$ as a class, and define equivalence class by $c(\omega)$. Then, one can easily verify that the equivalent classes form an Abelian group with product defined by
\begin{align}
  c(\omega_1)c(\omega_2)=c(\omega_1\omega_2)
  \label{}
\end{align}
The Abelian group of factor set equivalence classes is isomorphic to the cohomology group $H^2(G,A)$. We can view it as definition of the second group cohomology. Any factor set $\omega$ is named as cocycle, which is classified by $Z^2(G,A)$, while $\lambda$ is called coboundary, classified by $B^2(G,A)$. Then, we have
\begin{align}
  H^2(G,A)=Z^2(G,A)/B^2(G,A)
  \label{}
\end{align}

We point out here, the definition of $H^2(G,A)$ depends on the action of $G$ on $A$. In mathematics language, $A$ is a $G$-module, which is equivalent to say that $\forall g\in G$ may have nontrivial action on $A$ with the action to be automorphism of $A$, rather than just an Abelian group. For example, a trivial module just means $A$ is invariant under $G$:
\begin{align}
  {}^g\!a=a,\quad \forall g\in G, a\in A
  \label{}
\end{align}
One should always fix a $G$-module $A$, and then classify projective representation with coefficient $A$. However, in our case, $IGG=Z_2$, the automorphism of $Z_2$ only contains trivial one. 

We now put projective representation aside and turn to discussion about group extension. Assume group $E$ has a normal subgroup $A$. Then, we can define $G$ as quotient group
\begin{align}
  G=E/A
  \label{}
\end{align}
with associated homomorphism $\pi:E\rightarrow G$. Then, it is natural to define $G$-module $A$: given $g\in G$, the action of $g$ on $A$ is characterized by 
\begin{align}
  {}^g\!a=\widetilde{g}a\widetilde{g}^{-1}
  \label{}
\end{align}
where we choose $\widetilde{g}$ so that $\pi(\widetilde{g})=g$. $E$ is called an extension of a group $G$ by $G$-module $A$. In particular, $A$ is central in $E$ if and only if the $G$-action is trivial. In this case, the extension is called a central extension.

Now, let us discuss about the relation between group extension and projective representation. Roughly speaking, the equivalent class of projective representation has one-to-one correspondance with group extension of $G$ by $G$-module $A$. Namely, group extension $E$ is also classified by 2-cohomology $H^2(G,A)$. Projective representation $\Gamma$ can be viewed as a map $\Gamma:G\rightarrow E$ such that $\pi\circ\Gamma=\mathrm{id}_G$. Then, factor set $\omega$ is naturally induced by $\Gamma$, and automatically satisfies associativity constraint Eq.(\ref{eq:factor_set_redefine}). Notice that the choice of $\Gamma$ is far from unique, and we can always redefine $\Gamma$ as shown in Eq.(\ref{eq:proj_rep_redefine}).

In the following, we will develop a general method to solve the inequivalent projective representations for discrete group $G$ with $G$-module $A$, and the corresponding extended group is $E$. Particularly, we will focus on $G$ as the symmetry group of kagome PEPS defined in Appendix \ref{app:sym_group_kagome}, while $A=IGG=Z_2$ is a trivial $G$-module. Spin-rotation symmetry will also be discussed.

Let us first set up the general framework. $G$ is defined by generators $\{T_1,T_2,C_6,\sigma,\mathcal{T}\}$ as well as the relation between these generators, as shown in Eq.(\ref{eq:space_group_generator_relation}). In other words, $\forall g\in G$, there is one integer set $\{n_1,n_2,n_{C_6},n_\sigma\}$ such that
\begin{align}
  g=T_1^{n_1}T_2^{n_2}C_6^{n_{C_6}}\sigma^{n_\sigma}
  \label{}
\end{align}

As discussed before, projective representation can be constructed from $E$ by map $\Gamma$. Let us first choose the gauge such that $\Gamma(1)=1$. Then, this implies $\omega(1,1)=\omega(g,1)=\omega(1,g)=1,\forall g\in G$. Let us consider a particular relation between generators as 
\begin{align}
  T_2^{-1}T_1^{-1}T_2T_1=\mathrm{e}
  \label{}
\end{align}
Lifting this relation to $E$ by $\Gamma$, we have
\begin{align}
  \Gamma(T_2)^{-1}\Gamma(T_1)^{-1}\Gamma(T_2)\Gamma(T_1)=\eta_{12}
  \label{}
\end{align}
where $\eta_{12}\in Z_2$. Similarly, for all relations, we obtain a set of $\eta$'s.

These $\eta$'s are closely related to classification of projective representation. However, there are some issues arise. First, the $\eta$'s are not, in general, one-to-one correspondance with cohomology classes. There may be some redundancy in this description. Second, some choices of $\eta$'s may be inconsistent and not give a legitimate factor set. To see this, let us solve the classification of projective representations for the kagome lattice symmetry group completely. Conditions for group relations are as follows
\begin{align}
  &\Gamma(T_2)^{-1}\Gamma(T_1)^{-1}\Gamma(T_2)\Gamma(T_1)=\eta_{12},\notag\\
  &\Gamma(\sigma)^{-1}\Gamma(T_1)^{-1}\Gamma(\sigma) \Gamma(T_2)  =\eta_{\sigma T_2},\notag\\
  &\Gamma(\sigma)^{-1}\Gamma(T_2)^{-1}\Gamma(\sigma) \Gamma(T_1)  =\eta_{\sigma T_1},\notag\\
  &\Gamma(C_6)^{-1}\Gamma(T_2)^{-1}\Gamma(C_6)\Gamma(T_1)  =\eta_{C_6T_1},\notag\\
  &\Gamma(C_6)^{-1}\Gamma(T_2)^{-1}\Gamma(T_1)\Gamma(C_6)\Gamma(T_2)  =\eta_{C_6 T_2},\notag\\
  &\Gamma(\sigma)^{-1} \Gamma(C_6)\Gamma(\sigma) \Gamma(C_6)  =\eta_{\sigma C_6},\notag\\
  &\Gamma(C_6)^6=\eta_{C_6},\notag\\
  &\Gamma(\sigma)^2=\eta_{\sigma},\notag\\
  &\Gamma(\mathcal{T})^2 =\eta_{\mathcal{T}},\notag\\
  &\Gamma(g)^{-1}\Gamma(\mathcal{T})^{-1}\Gamma(g)\Gamma(\mathcal{T})=\eta_{g\mathcal{T}},\, \forall g =T_{1,2},\sigma, C_6
  \label{eq:projective_space_group_generator_relation}
\end{align}
where we get 13 $\eta$'s. One may expect the number of cohomology classes should be $2^{13}$, however, as we will see later, there is lots redundancy.

Now, let's try to eliminate those redundant parameters by choosing gauge of $\Gamma(g)$. By doing gauge transformation $\Gamma(T_1)\rightarrow\eta_{C_6T_2}\Gamma(T_1)$ and $T_2\rightarrow\eta_{C_6T_2}\eta_{\sigma T_1}T_2$, we are able to set $\eta_{C_6T_2}=\eta_{\sigma T_1}=\mathrm{I}$. Notice, other $\eta$'s may also change, however, we can always absorb the change by redefining other $\eta$'s.

Then, we have
\begin{align}
  \Gamma(T_2)=\Gamma(\sigma)\Gamma(T_1)\Gamma(\sigma)^{-1}
  \label{}
\end{align}
By applying the above equation, we get
\begin{align}
  \eta_{\sigma T_2}&=\Gamma(\sigma)^{-1}\Gamma(T_1)^{-1}\Gamma(\sigma)\Gamma(T_2)\notag\\
  &=\Gamma(\sigma)^{-1}\Gamma(T_1)^{-1}\Gamma(\sigma)\Gamma(\sigma)\Gamma(T_1)\Gamma(\sigma)^{-1}=\mathrm{I}.
  \label{}
\end{align}
as well as
\begin{align}
  \eta_{T_2\mathcal{T}}&=\Gamma(T_2)^{-1}\Gamma(\mathcal{T})^{-1}\Gamma(T_2)\Gamma(\mathcal{T})\notag\\
  &=\Gamma(\sigma)\Gamma(T_1)^{-1}\Gamma(\sigma)^{-1}\Gamma(\mathcal{T})^{-1}\Gamma(\sigma)\Gamma(T_1)\Gamma(\sigma)^{-1}\Gamma(\mathcal{T})\notag\\
  &=\eta_{\sigma\mathcal{T}}\Gamma(\sigma)\Gamma(T_1)\Gamma(\mathcal{T})^{-1}\Gamma(T_1)\Gamma(\sigma)^{-1}\Gamma(\mathcal{T})\notag\\
  &=(\eta_{\sigma\mathcal{T}})^2\Gamma(\mathcal{T})\Gamma(T_1)^{-1}\Gamma(\mathcal{T})^{-1}\Gamma(T_1)=\eta_{T_1\mathcal{T}}.
  \label{}
\end{align}

After above calculation, we are left with 9 free tunable $Z_2$ parameters,
\begin{align}
  \{\eta_{12},\eta_{C_6 T_1},\eta_{\sigma C_6},\eta_{\sigma},\eta_{C_6},\eta_{\mathcal{T}},\eta_{T_1\mathcal{T}}\eta_{\sigma\mathcal{T}},\eta_{C_6\mathcal{T}}\}
  \label{}
\end{align}
So, we expect $H^2(G,Z_2)=2^9$.

Now, let's adding spin rotation symmetry $R_s(\theta\vec{n})$. We have
\begin{align}
  \Gamma(R_s(2\pi))&=\eta_{\theta=2\pi},\notag\\
  \Gamma(R_s(\theta\vec{n}))\Gamma(g)&=\eta_{g,\theta\vec{n}}\Gamma(g)\Gamma(R_s(\theta\vec{n}))
  \label{}
\end{align}
where $g\in {T_1,T_2,C_6,\sigma,\mathcal{T}}$. As argued in Ref.\onlinecite{Essin:2013p104406}, one can always set $\eta_{g,\theta\vec{n}}=\mathrm{I}$. Thus, by including spin rotation symmetry, we get an extra parameter $\eta_{\theta=2\pi}$. So the number of cohomology classes becomes $2^{10}$.

\section{Distinguishing different classes by lattice quantum numbers}\label{app:quantum_number}
It has been shown that for system defined on a torus, lattice quantum numbers can be served as useful tools to distinguish different phases\cite{Varney:2011p241105,Jiang:2014p31040,Qi:2015p100401,Zaletel:2015p}. Here, we will show lattice quantum numbers are also very useful to distinguish different classes.

First, let us set the framework to extract quantum numbers of symmetric PEPS wavefunction $|\Psi\rangle$ defined on a torus. Similar to the PEPS on a infinite plane, tensors of the torus PEPS wavefunction $|\Psi\rangle$ satisfy
\begin{align}
  T^{(x,y,s)}=\Theta_RW_RR\circ T^{(x,y,s)}\notag\\
  B_{(x,y,b)}=W_RR\circ B_{(x,y,b)}
  \label{}
\end{align}
where $R$ is the global symmetry operator, while $W_R$ acts on virtual legs as a gauge transformation. Then, the global quantum number of $R$ is simply product of all $\Theta_R$. Namely, we get
\begin{align}
  R|\Psi\rangle=\prod_{x,y,s}\Theta_R(x,y,s)|\Psi\rangle
  \label{eq:quant_num}
\end{align}

Let us focus on two particular classes:
\begin{enumerate}
  \item Class-\rm{I} is labeled by $\eta_{12}=\eta_{C_6}=\mathrm{I}$, $\eta_\sigma=\mathrm{J}$ and $\chi_\sigma=\chi_\mathcal{T}=1$. $Q_1=Q_2$ spin liquid phase\cite{Sachdev:1992p12377} belongs to this class.
  \item Class-\rm{II} is labeled by $\eta_{12}=\mathrm{I}$, $\eta_{C_6}=\eta_\sigma=\mathrm{J}$ and $\chi_\sigma=\chi_\mathcal{T}=1$. $Q_1=-Q_2$ spin liquid phase\cite{Sachdev:1992p12377} belongs to this class.
\end{enumerate}
To distinguish quantum numbers of these two classes, let us consider systems on a torus with $(4n+2)$ unit cells with even number of unit cells in $T_1$ direction and odd number of unit cells in $T_2$ direction. Notice that wavefunctions defined on this system explicitly break $C_6$ rotation symmetry. However, it still preserves inversion symmetry $R_\pi=C_6^3$. In the following, we will show that the ground state manifold of these two classes defined on systems with $(4n+2)$ unit cells form distinct representations of symmetry group generated by translation $T_1$, $T_2$ and inversion $R_\pi$.

Let us first list symmetry transformation rules for infinite PEPS. For translation, we can choose a proper gauge such that $W_{T_1}(x,y,s,i)=W_{T_2}(x,y,s,i)=\mathrm{I}$ as well as $\Theta_{T_1}(x,y,s)=\Theta_{T_2}(x,y,s)=1$ for both two classes. The symmetry transformation rule of reflection $R_\pi$ can be generated by $C_6$ rotation as  
\begin{align}
  W_{R_\pi}(x,y,s,i)=&W_{C_6}(x,y,s,i)W_{C_6}(C_6^{-1}(x,y,s,i))\cdot\notag\\
  &W_{C_6}(C_6^{-2}(x,y,s,i))\notag\\
  \Theta_{R_\pi}(x,y,s)=&\Theta_{C_6}(x,y,s)\Theta_{C_6}(C_6^{-1}(x,y,s))\cdot\notag\\
  &\Theta_{C_6}(C_6^{-2}(x,y,s))
  \label{}
\end{align}
Thus, for infinite PEPS in Class-I, we have
\begin{align}
  W_{R_\pi}(x,y,s,i)&=\mathrm{I},\notag\\
  \Theta_{R_\pi}(x,y,s)&=1.
  \label{}
\end{align}
And for Class-II, we have
\begin{align}
  W_{R_\pi}(x,y,s,a/c)&=\mathrm{I},\notag\\
  W_{R_\pi}(x,y,s,b/d)&=\mathrm{J},\notag\\
  \Theta_{R_\pi}(x,y,s)&=\ii.
  \label{}
\end{align}

Now, let us turn to PEPS on a torus with $(4n+2)$ unit cells. For these two classes, one can construct a symmetric wavefunction $|\Psi_{0,0}\rangle$ with the symmetry transformation rules defined the same as infinite PEPS, since the transformation rules of $T_1$, $T_2$ and $R_\pi$ on virtual legs of both classes are compatible with the system size. Symmetry quantum numbers of these states can be calculated using Eq.(\ref{eq:quant_num}), where the result is listed in the first columns in Table \ref{tab:quant_num}.

\begin{table}
  \caption{Translation and inversion quantum numbers for topological degenerate ground states on $(4n+2)$-uc lattice (even by odd) samples}
  \center{(a): Quantum number for Class-I}
  \begin{tabular}{|c|c|c|c|c|}
    \hline
    Sym. & $|\Psi_{0,0}\rangle$ & $|\Psi_{\pi,0}\rangle$ & $|\Psi_{0,\pi}\rangle$ & $|\Psi_{\pi,\pi}\rangle$\\\hline
    $T_1$ & 1 & -1 & 1 & -1\\\hline
    $T_2$ & 1 & 1 & 1 & 1\\\hline
    $R_\pi$ & 1 & -1 & 1 & -1 \\\hline
  \end{tabular}\\
  \center{(b): Quantum number for Class-II}
  \begin{tabular}{|c|c|c|c|c|}
    \hline
    Sym. & $|\Psi_{0,0}\rangle$ & $|\Psi_{\pi,0}\rangle$ & $|\Psi_{0,\pi}\rangle$ & $|\Psi_{\pi,\pi}\rangle$\\\hline
    $T_1$ & 1 & -1 & 1 & -1\\\hline
    $T_2$ & 1 & 1 & 1 & 1\\\hline
    $R_\pi$ & -1 & 1 & -1 & 1 \\\hline
  \end{tabular}\\
  \label{tab:quant_num}
\end{table}

As discussed in Sec.\ref{sec:vison_psg}, other bases of ground state manifold is obtained by inserting non-contractible flux loops, labeled by $|\Psi_{\pi,0}\rangle$, $|\Psi_{0,\pi}\rangle$ and $|\Psi_{\pi,\pi}\rangle$. $W_R$ and $\Theta_R$ change their values after the loop insertion, but it is easy to extract the quantum numbers of these states. For example, let us consider $T_1$ quantum number of $|\Psi_{\pi,0}\rangle$. $T_1$ will move the non-contractible $g$ loop with one lattice spacing, leading to a new wavefunction $|\Psi'_{\pi,0}\rangle$. One can easily figure out that the PEPS wavefunction $|\Psi'_{0,\pi}\rangle$ is related to $|\Psi_{0,\pi}\rangle$ by a $Z_2$ gauge transformation on the column sandwiched by $g$ loops of $|\Psi_{0,\pi}\rangle$ and $|\Psi'_{\pi,0}\rangle$, plus the $T_1$ symmetry transformation rule of $|\Psi_{0,0}\rangle$. Since there are odd number of sites per column, and site tensors are $Z_2$ odd, the single column $Z_2$ gauge transformation contributes an extra $-1$ to $T_1$ 
quantum number. Following similar strategy, one can obtain the representation of symmetry group on the whole ground state manifold for both classes. We list the result in Table \ref{tab:quant_num}.

As seen in Table \ref{tab:quant_num}, ground state manifolds of the two classes have distinct representations. So, these two classes can be distinguished by lattice quantum numbers.

\section{Valence bond solid phase}\label{app:vbs}
As we mentioned before, a single class includes many different phases, which can only be distinguished by finite size scaling. One may ask, is it possible that some particular phase can be described by different classes? We think that the answer is yes, and we will give an example in the following.

Let us focus on Class-I and Class-II discussed in Appendix \ref{app:quantum_number}. As we argued before, the ground state spaces of these two corresponding QSL have different lattice quantum numbers on systems with $(4n+2)$ unit cells. 

Now, let us consider the valence bond solid (VBS) phases. The VBS pattern can be obtained by Landau-Ginzburg theory of visons. The effective Lagrangian of visons is constrained by visons' transformation rules under symmetries, which are calculated in Sec.\ref{sec:vison_psg}. It turns out that the transformation rules only depend on spinon distributions, which are the same for all classes. Thus, we expect both Class-I and Class-II give the same VBS order pattern after vison condensation. This seems to contradict with the quantum number discrepancy mentioned before.

The first observation is that for those samples where two classes have different lattice quantum numbers, it is impossible to write a compatible VBS order with the lattice size. In other words, there are always domain wall configurations on those samples. 

\begin{figure}
  \includegraphics[width=0.35\textwidth]{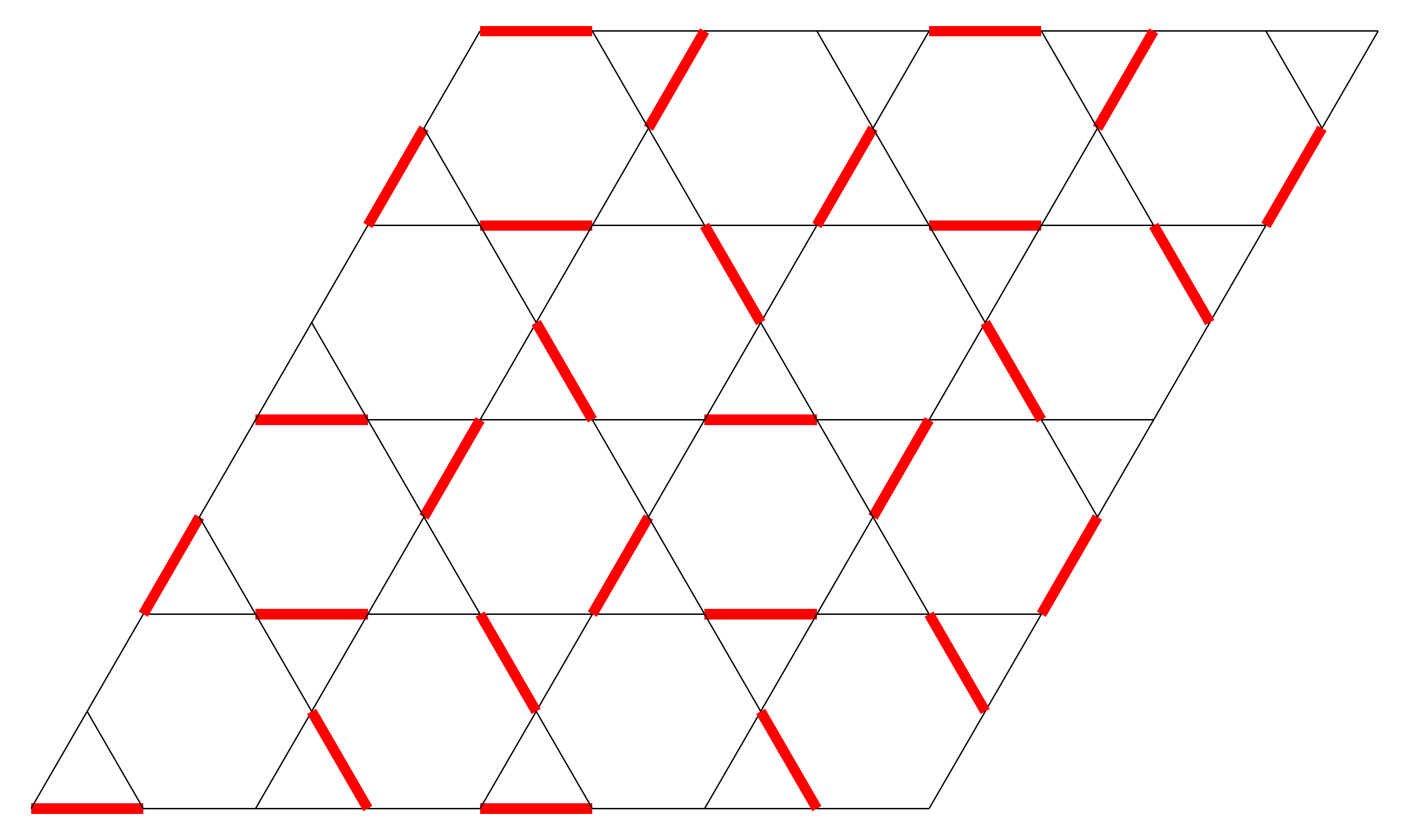}
  \caption{Visulization of the 12-site valence bond solid state as coverings of spin singlets (red thick bonds) on the kagome lattice.}
  \label{fig:12-site_VBS}
\end{figure}

\begin{figure}
  \includegraphics[width=0.5\textwidth]{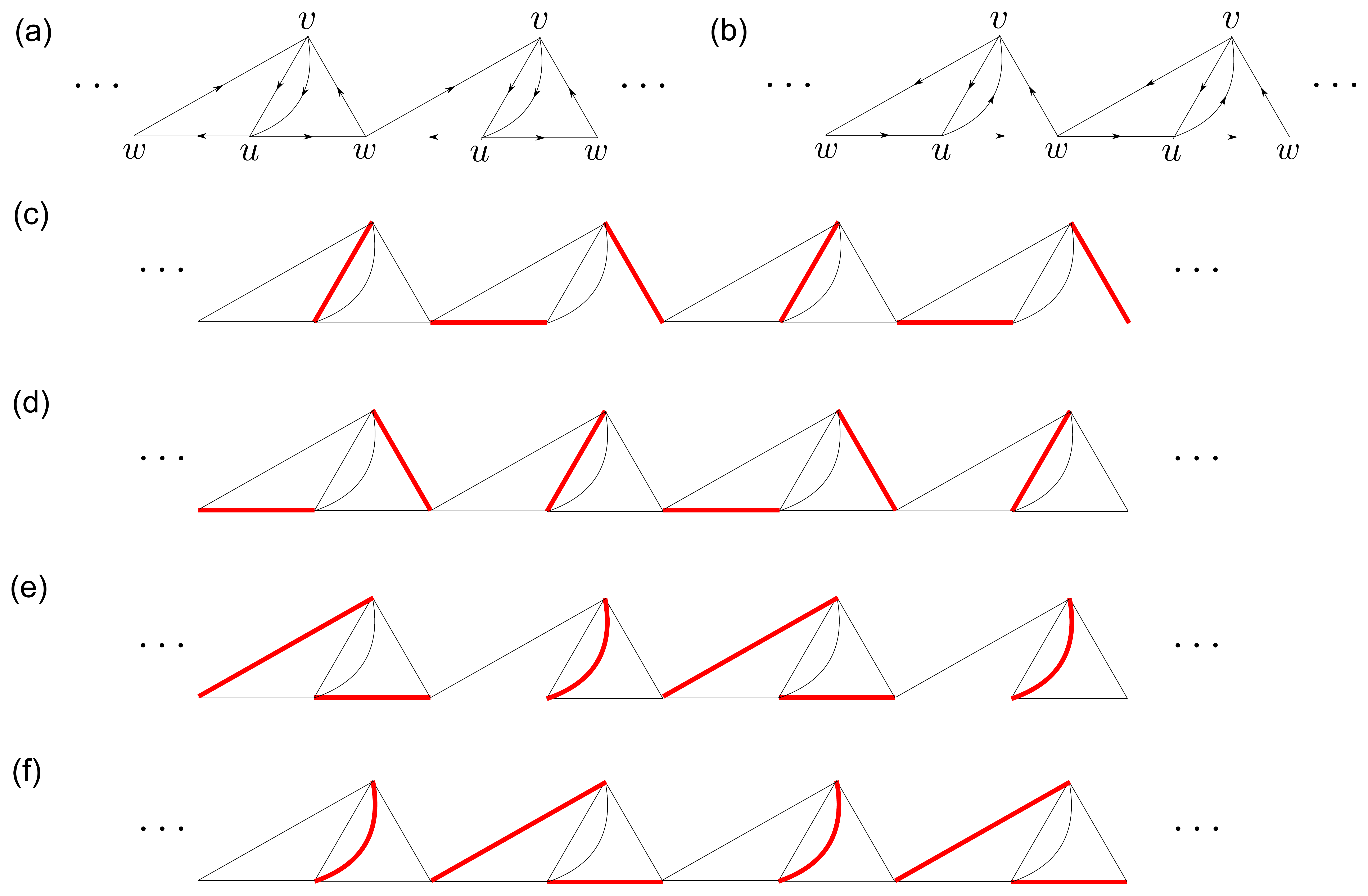}
  \caption{(a),(b): 1D chain in $T_1$ direction. The periodic boundary condition is imposed, so every site shares four bonds. If two sites connected by a bond form a spin singlet, then we require the direction of the singlet is along the arrow on the bond. Arrows on (a) is consistent with Class-I, while arrows on (b) is consistent with Class-II. (c),(d),(e): Four possible domain wall configurations connected by symmetry $T_2$ and $R_\pi$. The direction of spin singlets (red thick bonds) follows arrows in (a).}
  \label{fig:kagome_chain_domain_wall}
\end{figure}

To see this, let us consider a particular 12-site VBS order pattern shown in Fig.(\ref{fig:12-site_VBS}) which is compatible with vison symmetry transformation rules\cite{Huh:2011p94419}. As shown in Appendix \ref{app:quantum_number}, Class-I and Class-II have different lattice quantum numbers on a torus with $(4n+2)$ unit cells with even number of unit cells in $T_1$ direction and odd in $T_2$ direction. It is straight forward to see that on those samples, one can never avoid domain walls. In the following , we will show the different lattice quantum numbers are actually caused by different quantum fluctuations along the domain wall.

Let us focus on the simplest case where the sample is a chain along the $T_1$ direction with even number of unit cells, but with only one unit cell along the $T_2$ direction, as shown in Fig.(\ref{fig:kagome_chain_domain_wall}). Notice that the periodic boundary condition is imposed. Then, every site is connected with four bonds. Arrows on Fig.(\ref{fig:kagome_chain_domain_wall}a,b) denotes the direction of singlet bonds of Class-I and Class-II respectively. The direction of single bonds (red ones) on Fig.(\ref{fig:kagome_chain_domain_wall}c-f) follows the convention in Fig.(\ref{fig:kagome_chain_domain_wall}). We label these four domain wall states as $|\phi_1\rangle$, $|\phi_2\rangle$, $|\phi_3\rangle$ and $|\phi_4\rangle$ respectively. In the following, we will show how these four simple VBS configurations give states with different quantum numbers. 

On the Hilbert space spanned by these four basis, the representation of $T_1$, $T_2$ and $R_\pi$ reads
\begin{align}
  T_1=
  \begin{pmatrix}
    0 & 1 & 0 & 0 \\
    1 & 0 & 0 & 0 \\
    0 & 0 & 0 & 1 \\
    0 & 0 & 0 & 1
  \end{pmatrix},
  T_2=
  \begin{pmatrix}
    1 & 0 & 0 & 0 \\
    0 & 1 & 0 & 0 \\
    0 & 0 & 1 & 0 \\
    0 & 0 & 0 & 1
  \end{pmatrix},
  R_\pi=
  \begin{pmatrix}
    0 & 0 & 1 & 0 \\
    0 & 0 & 0 & 1 \\
    1 & 0 & 0 & 0 \\
    0 & 1 & 0 & 0
  \end{pmatrix}
  \label{}
\end{align}
Then, eigenstates and eigenvalues for these three matrix are as following
\begin{center}
  \begin{tabular}{|c|c|c|c|}
    \hline
    Eigenstates & $T_1$ & $T_2$ & $R_\pi$ \\\hline
    $|\phi_1\rangle+|\phi_2\rangle+|\phi_3\rangle+|\phi_4\rangle$ & $1$ & $1$ & $1$ \\\hline
    $|\phi_1\rangle+|\phi_2\rangle-|\phi_3\rangle-|\phi_4\rangle$ & $1$ & $1$ & $-1$ \\\hline
    $|\phi_1\rangle-|\phi_2\rangle+|\phi_3\rangle-|\phi_4\rangle$ & $-1$ & $1$ & $1$ \\\hline
    $|\phi_1\rangle-|\phi_2\rangle-|\phi_3\rangle+|\phi_4\rangle$ & $-1$ & $1$ & $-1$ \\\hline
  \end{tabular}\\
  \label{tab:1D_chain_quantum_number}
\end{center}
So different superpositions of VBS configurations give different quantum numbers. The above picture is similar to spin liquid phases (RVB states). Different spin liquid phases are distinguished by relative phases of different configurations of bond coverings. These relative phase factors may result in different quantum numbers on some finite size sample. However, unlike spin liquid phases, fluctuation along the VBS domain wall is essentially 1D physics. This remains true after considering samples with more unit cells along the $T_2$ direction. In one dimension, in the thermodynamic limit, the system will be pinned to a particular VBS configuration, and the information of phase factors are lost. So we believe different ways of fluctuations along the VBS domain wall will not give new phases.

Now, let us check the correctness of the fluctuation-along-domain-wall picture by studying the possible superposition of these four states in Class-I and Class-II respectively. By comparing directions of singlet bonds, We conclude that for Class-I, the possible superposition of these four states is $(|\phi_1\rangle+|\phi_2\rangle+|\phi_3\rangle+|\phi_4\rangle)$. Further, we observe $|\phi_1\rangle-|\phi_2\rangle-|\phi_3\rangle-|\phi_4\rangle$ is also consistent with Class-I with a non-contractible flux loops in $T_2$ direction. While for Class-II, there are two cases. For chains with $(4n+2)$ uc, the possible superposition of these four states is $(-|\phi_1\rangle-|\phi_2\rangle+|\phi_3\rangle+|\phi_4\rangle)$ as well as $|\phi_1\rangle-|\phi_2\rangle+|\phi_3\rangle-|\phi_4\rangle$ (with no-ncontractible flux loop in $T_2$ direction). For chains with $4n$ uc, we get the same result as Class-I. Thus, for systems with $(4n+2)$ sites, the two classes always have different quantum numbers. The above observation 
is 
consistent with the quantum numbers of Class-I and Class-II obtained in the previous appendix.

\section{An example on the square lattice}\label{app:square_C4}
As a pedagogical example, here we present the classification of symmetric PEPS with $IGG=Z_2$ for systems on the square lattice with a half-integer spin per site, in the presence of lattice translation, lattice $C_4$ rotation and spin $SU(2)$ rotation symmetries.

\begin{figure}
  \includegraphics[width=0.2\textwidth]{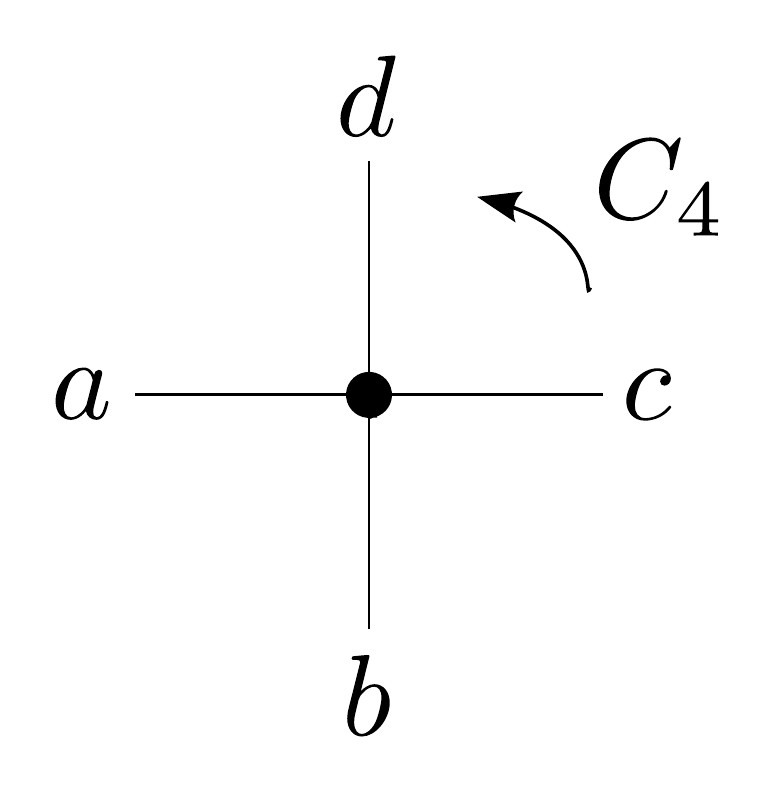}
  \caption{A site on the square lattice with four virtual legs labeled as $a,b,c,d$.}
  \label{fig:square_lattice}
\end{figure}

The lattice symmetry group is generated by $T_1,T_2,C_4$, which transform the virtual leg labeled by $(x,y,i)$ as:
\begin{align}
 T_1(x,y,i)&=(x+1,y,i)\notag\\
 T_2(x,y,i)&=(x,y+1,i)\notag\\
 C_4(x,y,i)&=(-y,x,C_4(i)),
\end{align}
where $i=a,b,c,d$ labels the four virtual legs on a site tensor at $(x,y)$ in a counter-clockwise fashion, as shown in Fig.(\ref{fig:square_lattice}). The counter-clockwise $C_4$ rotates the legs as:
\begin{align}
 C_4(a)=b,\;\;C_4(b)=c,\;\;C_4(c)=d,\;\;C_4(d)=a.
\end{align}

These generators satisfy the following identities which define the space group:
\begin{align}
  [1]:&& T_1T_2T_1^{-1}T_2^{-1}&=\mathrm{e}\notag\\
  [2]:&& C_4^{-1}T_1C_4T_2&=\mathrm{e}\notag\\
  [3]:&& C_4^{-1}T_2C_4T_1^{-1}&=\mathrm{e}\notag\\
  [4]:&& C_4^4&=\mathrm{e}\notag
\end{align}
And the on-site physical spin rotation by an angle $\theta$ around the spin axis $\vec n$: $U_{\theta\vec n}$, which forms a half-integer spin irrep of the $SU(2)$, commutes with all lattice symmetries:
\begin{align}
  g^{-1}\cdot U_{\theta\vec n}\cdot g\cdot U_{\theta\vec n}^{-1}&=\mathrm{e}, &\forall g=T_1,T_2,C_4\label{eq:square_sym_g_SU(2)}
\end{align}
Below we solve the implementations of these symmetries on PEPS with $IGG=Z_2=\{\mathrm{I},\mathrm{J}\}$, step by step. As discussed in Sec.\ref{subsec:classification_general_framework}, each symmetry element $R$ is associated with its own $\eta$-ambiguity and $\epsilon$-ambiguity. And the PEPS itself has an $\Phi$-ambiguity and a $V$-ambiguity, a statement unrelated to specific symmetry elements.

\textbf{(1)} Choose the virtual basis such that $\mathrm{J}(x,y,i)=\mathrm{J}=\mathrm{I}_{D_1}\oplus(-\mathrm{I}_{D_2})$ is a site and leg independent diagonal matrix, according to the discussion in \ref{subsec:classification_kagome_peps}. This determines $\mathrm{J}$ up to an overall $\pm1$ sign, which we will use in step-(10). (Note that we have not attached any physical meanings for the $D_1$ and $D_2$ sectors yet.) All the remaining $V$-ambiguity matrices (could be site and leg dependent) as well as all the $W_R$ matrices (could be site and leg dependent) commute with $\mathrm{J}$, so they are block diagonal and act within the $D_1$ and $D_2$ subspaces.

\textbf{(2)} Consider identity-[2]. Applying Eq.(\ref{eq:psg_equation_general}) to this identity:
\begin{align}
 &W_{C_4}^{-1}(-y,x,C_4(i))W_{T_1}(-y,x,C_4(i))W_{C_4}(-y-1,x,C_4(i))\notag\\
 &\cdot W_{T_2}(x,y+1,i)=\eta_{[2]}\chi_{[2]}(x,y,i)
\end{align}
Since $W_{T_1}$ appears in this equation only once, we can always use the $\eta_{T_1}$ ambiguity: $W_{T_1}\rightarrow \mathrm{J}W_{T_1}$ to tune $\eta_{[2]}=\mathrm{I}$. This fixes the relative $\eta$-ambiguity for $W_{T_1}$,$W_{T_2}$, and still leaves an overall $\eta$-ambiguity $W_{T_1},W_{T_2}\rightarrow \mathrm{J}W_{T_1},\mathrm{J}W_{T_2}$.

\textbf{(3)} Using the remaining $V$-ambiguity to transform $W_{T_2}$, according to Eq.(\ref{eq:gauge_transf_on_psg_coord}):
\begin{align}
 W_{T_2}(x,y,i)\rightarrow V(x,y,i) W_{T_2}(x,y,i) V^{-1}(x,y-1,i).
\end{align}
So we can set $W_{T_2}(x,y,i)=\mathrm{I}$. There is \emph{no} $\epsilon_{T_2}$ ambiguity left. The remaining $V$-ambiguity satisfies $V(x,y,i)=V(x,0,i)$.

Next, we use this remaining $V$-ambiguity to transform $W_{T_1}$ along the row of sites at $y=0$:
\begin{align}
 W_{T_1}(x,0,i)\rightarrow V(x,0,i) W_{T_1}(x,0,i) V^{-1}(x-1,0,i),
\end{align}
so that $W_{T_1}(x,0,i)=\mathrm{I}$. Now the remaining $V$-ambiguity is site independent but could be leg-dependent: $V(x,y,i)=V(i)$. The remaining $\epsilon_{T_1}$ ambiguity satisfy $\epsilon_{T_1}(x,0,i)=1$.

Consider identity-[1]. Applying Eq.(\ref{eq:psg_equation_general}) to this identity:
\begin{align}
 &W_{T_1}(x,y,i)W_{T_2}(x-1,y,i)W_{T_1}^{-1}(x,y-1,i)\notag\\
 &\cdot W_{T_2}^{-1}(x,y,i)=\eta_{12}\chi_{12}(x,y,i)
\end{align}
The remaining $\eta$-ambiguity for $W_{T_1},W_{T_2}$ \emph{cannot} tune away $\eta_{12}$. Using results obtained previously in the step, we have
\begin{align}
 W_{T_1}(x,y,i)W_{T_1}^{-1}(x,y-1,i)=\eta_{12}\chi_{12}(x,y,i)\label{eq:square_T1}
\end{align}
We then use the remaining $\epsilon_{T_1}$ ambiguity to transform $\chi_{12}$, according to Eq.(\ref{eq:epsilon_transf_chi}):
\begin{align}
 \chi_{12}(x,y,i)\rightarrow\chi_{12}(x,y,i)\epsilon_{T_1}(x,y,i)\epsilon_{T_1}^*(x,y-1,i).
\end{align}
So we can set $\chi_{12}(x,y,i)=1$. After this Eq.(\ref{eq:square_T1}) leads to $W_{T_1}(x,y,i)=\eta_{12}^yW_{T_1}(x,0,i)=\eta_{12}^y$, and there is no $\epsilon_{T_1}$ ambiguity left. In this gauge all site tensors are the same: $T^{(x,y)}=T^{(0,0)}\equiv T$, while bond tensors are spatial dependent if the $\eta_{12}$ index is nontrivial.

Next we study $\Theta_{T_1}$ and $\Theta_{T_2}$. First we use the $\Phi$-ambiguity, which we have not used before, to transform $\Theta_{T_2}$. According to Eq.(\ref{eq:phase_factor_transf}):
\begin{align}
 \Theta_{T_2}(x,y)\rightarrow\Theta_{T_2}(x,y)\Phi(x,y)\Phi^*(x,y-1),
\end{align}
so we can set $\Theta_{T_2}(x,y)=1$, and the remaining $\Phi$-ambiguity satisfies $\Phi(x,y)=\Phi(x,0)$.

We then use the remaining $\Phi$-ambiguity to transform $\Theta_{T_1}$ along the row of sites at $y=0$:
\begin{align}
 \Theta_{T_1}(x,0)\rightarrow\Theta_{T_1}(x,0)\Phi(x,0)\Phi^*(x-1,0),
\end{align}
so we can set $\Theta_{T_1}(x,0)=1$. The remaining $\Phi$-ambiguity is only a site-independent overall phase: $\Phi(x,y)=\Phi(0,0)\equiv\Phi$. Since we will not study time-reversal symmetry here, it turns out that this overall phase ambiguity is not useful.

Applying Eq.(\ref{eq:psg_equation_general_phase}) to the identity-[1], we have
\begin{align}
 &\Theta_{T_1}(x,y)\Theta_{T_2}(x-1,y)\Theta_{T_1}^*(x,y-1)\Theta_{T_2}^*(x,y)\notag\\
 &=\mu_{12}\prod_{i}\chi_{12}^*(x,y,i),
\end{align}
where $\mu_{12}$ is the site-independent phase factor obtained when applying $\eta_{12}$ on a site tensor, as mentioned in \ref{subsec:classification_kagome_peps}. Using the results obtained so far, we find $\Theta_{T_1}(x,y)=\mu_{12}^y\Theta_{T_1}(x,0)=\mu_{12}^y$.

We can summarize the results obtained in step-(3):
\begin{align}
W_{T_1}(x,y,i)&=\eta_{12}^y\notag\\
W_{T_2}(x,y,i)&=\mathrm{I}\notag\\
\Theta_{T_1}(x,y)&=\mu_{12}^y\notag\\
\Theta_{T_2}(x,y)&=1.
\end{align}
The remaining $\Phi$-ambiguity is an overall phase. The remaining $V(x,y,i)=V(i)$. There is no remaining $\epsilon_{T_1},\epsilon_{T_2}$ ambiguities left. There is a remaining overall $\eta$-ambiguity for $W_{T_1},W_{T_2}$.

\textbf{(4)} Consider identity-[2] again. Based on the discussion in step-(2), we have:
\begin{align}
 &W_{C_4}^{-1}(-y,x,C_4(i))W_{T_1}(-y,x,C_4(i))W_{C_4}(-y-1,x,C_4(i))\notag\\
 &\cdot W_{T_2}(x,y+1,i)=\chi_{[2]}(x,y,i).
\end{align}
Plugging in the results in step-(3), we obtain:
\begin{align}
 W_{C_4}^{-1}(-y,x,C_4(i))W_{C_4}(-y-1,x,C_4(i))=\eta_{12}^x\chi_{[2]}(x,y,i).\label{eq:square_W_C4}
\end{align}

Now we use the $\epsilon_{C_4}$ ambiguity. Applying Eq.(\ref{eq:epsilon_transf_chi}), this transforms $\chi_{[2]}$ as:
\begin{align}
 &\chi_{[2]}(x,y,i)\rightarrow\notag\\
 &\chi_{[2]}(x,y,i)\epsilon_{C_4}(-y-1,x,C_4(i))\epsilon_{C_4}^*(-y,x,C_4(i)).
\end{align}
So we can set $\chi_{[2]}(x,y,i)=1$, and the remaining $\epsilon_{C_4}$-ambiguity satisfies $\epsilon_{C_4}(x,y,i)=\epsilon_{C_4}(0,y,i)$. After this, Eq.(\ref{eq:square_W_C4}) leads to $W_{C_4}(x,y,i)=\eta_{12}^{xy}W_{C_4}(0,y,i)$.

\textbf{(5)} Consider identity-[3]. Applying Eq.(\ref{eq:psg_equation_general}):
\begin{align}
 &W_{C_4}^{-1}(-y,x,C_4(i))\cdot\mathrm{I}\cdot W_{C_4}(-y,x-1,C_4(i))W_{T_1}^{-1}(x,y,i)\notag\\
 &=\eta_{C_4T}\chi_{C_4T}(x,y,i)
\end{align}
Using results in step-(3) and step-(4), we have
\begin{align}
 W_{C_4}^{-1}(0,x,C_4(i))W_{C_4}(0,x-1,C_4(i))=\eta_{C_4T}\chi_{C_4T}(x,y,i),
\end{align}
so we know $\chi_{C_4T}(x,y,i)=\chi_{C_4T}(x,0,i)$, independent of $y$.

we then can use the remaining $\epsilon_{C_4}$-ambiguity from step-(4) to transform $\chi_{C_4T}$:
\begin{align}
 &\chi_{C_4T}(x,0,i)\rightarrow \notag\\
 &\chi_{C_4T}(x,0,i)\epsilon_{C_4}^*(0,x,C_4(i))\epsilon_{C_4}(0,x-1,C_4(i)).
\end{align}
So we can set $\chi_{C_4T}(x,0,i)=1$ and the remaining $\epsilon_{C_4}$-ambiguity is site independent: $\epsilon_{C_4}(x,y,i)= \epsilon_{C_4}(i)$.

After this, the site-dependence of $W_{C_4}$ is solved: $W_{C_4}(0,y,i)=\eta_{C_4T}^yW_{C_4}(0,0,i)$. Together with results in step-(4):
\begin{align}
 W_{C_4}(x,y,i)=\eta_{12}^{xy}\eta_{C_4T}^yW_{C_4}(i),
\end{align}
where we defined $W_{C_4}(i)\equiv W_{C_4}(0,0,i)$.

So far we have not used the $\eta$-ambiguity for $W_{C_4}$. The remaining $\epsilon_{C_4}(x,y,i)=\epsilon_{C_4}(i)$, and there is still a remaining $V(x,y,i)=V(i)$ ambiguity.

 \textbf{(6)} Consider $\Theta_{C_4}$. Applying Eq.(\ref{eq:psg_equation_general_phase}) to identity-[2], together with results in step-(2,4):
\begin{align}
 \Theta_{C_4}^{*}(-y,x)\Theta_{T_1}(-y,x)\Theta_{C_4}(-y-1,x)\Theta_{T_2}(x,y)=1.
\end{align}
Pluggin in results in step-(3), this leads to:
\begin{align}
 \Theta_{C_4}(x,y)=\mu_{12}^{xy}\Theta_{C_4}(0,y).\label{eq:square_Theta_C4_1}
\end{align}

Similarly, we apply Eq.(\ref{eq:psg_equation_general_phase}) to identity-[3]:
\begin{align}
 \Theta_{C_4}^*(-y,x)\Theta_{C_4}(-y,x-1)\mu_{12}^y=\mu_{C_4T}\prod_i\chi_{C_4T}^*(x,y,i),
\end{align}
where we used results in step-(3). Plugging in $\chi_{C_4T}=1$, which has been obtained in step-(5), and use Eq.(\ref{eq:square_Theta_C4_1}), the site-dependence of $\Theta_{C_4}$ is solved:
\begin{align}
 \Theta_{C_4}(x,y)=\mu_{C_4T}^y\mu_{12}^{xy}\Theta_{C_4},
\end{align}
where we introduced $\Theta_{C_4}\equiv\Theta_{C_4}(0,0)$.

\textbf{(7)} Consider identity-[4]. Applying Eq.(\ref{eq:psg_equation_general}) and the site-dependence of $W_{C_4}$ obtained in step-(5), we have:
\begin{align}
 &W_{C_4}(i)W_{C_4}(C_4^3(i))W_{C_4}(C_4^2(i))W_{C_4}(C_4(i))\notag\\
 &=\eta_{C_4}\chi_{C_4}(i),\label{eq:square_C4_4}
\end{align}
where $\chi_{C_4}(i)\equiv\chi_{C_4}(x,y,i)$ since the above relation dictates $\chi_{C_4}$ to be site independent. 

Applying Eq.(\ref{eq:psg_equation_general_phase}) and the site-dependence of $\Theta_{C_4}$ obtained in step(6), we find:
\begin{align}
 \Theta_{C_4}^4=\mu_{C_4}\prod_i\chi_{C_4}^*(i).
\end{align}
But due to the definition of the $\chi-group$, we know that $\chi_{C_4}(a)=\chi_{C_4}^*(c)$ and  $\chi_{C_4}(b)=\chi_{C_4}^*(d)$, so $\prod_i\chi_{C_4}^*(i)=1$. 

Consequently $\Theta_{C_4}=(\mu_{C_4})^{\frac{1}{4}}$. Naively there would be four allowed roots for a given choice of $\eta_{C_4}$. However, we have not used the $\eta$-ambiguity for $W_{C_4}$. Under the $\eta$-ambiguity transformation: $W_{C_4}\rightarrow \mathrm{J}\cdot W_{C_4}$, clearly $\Theta_{C_4}$ transforms as: $\Theta_{C_4}\rightarrow \mu_{\mathrm{J}}\Theta_{C_4}$.

We will prove that $\mu_{\mathrm{J}}=-1$ in step-(10); i.e., every site tensor is $Z_2$-odd. Thus, we can use the $\eta$-ambiguity for $W_{C_4}$ to tune away the sign in $\Theta_{C_4}$, and there are only two independent values for the root. We made the following choice:
\begin{align}
 \Theta_{C_4}&=1 \mbox{ or } i, \mbox{ if }\eta_{C_4}=\mathrm{I};\notag\\
 \Theta_{C_4}&=e^{i\frac{\pi}{4}} \mbox{ or } e^{-i\frac{\pi}{4}}, \mbox{ if }\eta_{C_4}=\mathrm{J}.\label{eq:theta_C4}
\end{align}
After this, there is no remaining $\eta$-ambiguity for $W_{C_4}$.

Next, we can use the remaining $V(i)$ ambiguity to transform $W_{C_4}(i)$:
\begin{align}
 W_{C_4}(i)\rightarrow V(i)W_{C_4}(i)V^{-1}(C_4^{-1}(i)).
\end{align}
So we can set $W_{C_4}(b)=W_{C_4}(c)=W_{C_4}(d)=\mathrm{I}$. This leaves \emph{no} remaining $\epsilon_{C_4}$-ambiguity, because $\epsilon_{C_4}(a)=\epsilon_{C_4}^*(c)$ as required in the definition of the $\chi-group$. The remaining $V$-ambiguity is site and leg independent: $V(x,y,i)=V$. 

Coming back to Eq.(\ref{eq:square_C4_4}), in the current gauge we have:
\begin{align}
 W_{C_4}(a)=\eta_{C_4}\chi_{C_4}(i), \;\;\forall i=a,b,c,d.
\end{align}
Consequently $\chi_{C_4}(i)\equiv\chi_{C_4}$ is leg independent, and $\chi_{C_4}=\pm1$ since e.g., $\chi_{C_4}(a)=\chi^*_{C_4}(c)$. 

One may worry that if $\eta_{C_4}=\mathrm{J}$, we simply have $W_{C_4}(a)=\chi_{C_4}\mathrm{J}$, and the $\pm1$ sign here may be tuned away by redefining the $\mathrm{J}$ element (recall that there is such sign freedom as mentioned in step-(1)). But for the moment let us not use this sign freedom in the definition of $\mathrm{J}$ because it will be used later in step-(10). Consequently after step-(10), the $\chi_{C_4}$ index here cannot be tuned away.

\textbf{(8)} Consider the on-site $SU(2)$ symmetry. We can apply Eq.(\ref{eq:psg_equation_general}) for a group identity in the multiplication table of $SU(2)$:
\begin{align}
 [\theta_2\vec n_2]\cdot [\theta_1\vec n_1]=[\theta_3\vec n_3],
\end{align}
and obtain:
\begin{align}
 &W_{\theta_3\vec n_3}^{-1}(x,y,i)W_{\theta_2\vec n_2}(x,y,i)W_{\theta_1\vec n_1}(x,y,i)\notag\\
 &=\eta_{[\theta_2\vec n_2],[\theta_1\vec n_1]}\chi_{[\theta_2\vec n_2],[\theta_1\vec n_1]}(x,y,i)\label{eq:psg_SU(2)}
\end{align}
Let us focus on a single virtual leg $(x,y,i)$, and consider all the possible $[\theta_2\vec n_2],[\theta_1\vec n_1]$. One can then immediately see that both  $\eta_{[\theta_2\vec n_2],[\theta_1\vec n_1]}$ and $\chi_{[\theta_2\vec n_2],[\theta_1\vec n_1]}(x,y,i)$ must satisfy 2-cocycle conditions as a function of $[\theta_2\vec n_2]$ and $[\theta_1\vec n_1]$ (see Appendix \ref{app:proj_rep} for detailed discussions). Namely, for a fixed $(x,y,i)$:
\begin{align}
 \eta_{[\theta_2\vec n_2],[\theta_1\vec n_1]}& \in H^2(SU(2),Z_2)=Z_1,\notag\\
 \chi_{[\theta_2\vec n_2],[\theta_1\vec n_1]}(x,y,i)& \in H^2(SU(2),U(1))=Z_1
\end{align}

Because both 2-cohomology groups are trivial, we find both $\eta_{[\theta_2\vec n_2],[\theta_1\vec n_1]}$ and $\chi_{[\theta_2\vec n_2],[\theta_1\vec n_1]}(x,y,i)$ are 2-coboundaries. Consequently, one can use the $\eta$-ambiguities for $W_{\theta\vec n}$ and the $\epsilon_{\theta\vec n}(x,y,i)$-ambiguities to set $\eta_{[\theta_2\vec n_2],[\theta_1\vec n_1]}=\mathrm{I}$ and $\chi_{[\theta_2\vec n_2],[\theta_1\vec n_1]}(x,y,i)=1$. After this, Eq.(\ref{eq:psg_SU(2)}) simply means that $W_{\theta\vec n}(x,y,i)$ forms a representation of $SU(2)$, $\forall (x,y,i)$: 
\begin{align}
 W_{\theta_3\vec n_3}^{-1}(x,y,i)W_{\theta_2\vec n_2}(x,y,i)W_{\theta_1\vec n_1}(x,y,i)=\mathrm{I}.\label{eq:su(2)_rep}
\end{align}

\textbf{(9)} Study the site and leg dependence of $W_{\theta\vec n}(x,y,i)$. For any space group symmetry element $g$, we have the group identity:
\begin{align}
 g^{-1}\cdot[\theta\vec n]\cdot g\cdot [\theta\vec n]^{-1}=e.
\end{align}
Applying Eq.(\ref{eq:psg_equation_general})  to this identity, we find:
\begin{align}
 &W_g^{-1}(g(x,y,i))W_{\theta\vec n}(g(x,y,i))W_g(g(x,y,i))W_{\theta\vec n}^{-1}(x,y,i)\notag\\
 &=\eta_{g,[\theta \vec n]}\chi_{g,[\theta \vec n]}(x,y,i)
\end{align}
Because we have already determined the form of $W_g(x,y,i)$ in step-(3,5,7), and $W_g(x,y,i)$ can only be a power of $\mathrm{J}$ up to a factor, we conclude that $W_g(x,y,i)$ commutes with $W_{\theta\vec n}(x,y,i)$. So the above equation reduces to:
\begin{align}
 W_{\theta\vec n}(g(x,y,i))W_{\theta\vec n}^{-1}(x,y,i)=\eta_{g,[\theta \vec n]}\chi_{g,[\theta \vec n]}(x,y,i).\label{eq:su(2)_site_dep}
\end{align}
Next, one can plug in $[\theta_2\vec n_2]\cdot [\theta_1\vec n_1]=[\theta_3\vec n_3]$ on the RHS of this equation and apply Eq.(\ref{eq:su(2)_rep}). One then concludes that, for any fixed $(x,y,i)$, $\eta_{g,[\theta \vec n]}$ ($\chi_{g,[\theta \vec n]}(x,y,i)$)  must be a representation of $SU(2)$ in $Z_2$ (U(1)). But such representations must be trivial. Therefore,  $\eta_{g,[\theta \vec n]}=\mathrm{I}$ and $\chi_{g,[\theta \vec n]}(x,y,i)=1$, $\forall (x,y,i)$. Eq.(\ref{eq:su(2)_site_dep}) dictates that $W_{\theta\vec n}$ must be site and leg independent:
\begin{align}
 W_{\theta\vec n}(x,y,i)=W_{\theta\vec n},\;\;\forall(x,y,i).
\end{align}

In addition, we can consider $\Theta_{\theta\vec n}(x,y)$. Due to the definition of $\Theta$, the following is true for any site tensor $T^{(x,y)}$:
\begin{align}
 T^{(x,y)}=\Theta_{\theta\vec n}W_{\theta\vec n}U_{\theta\vec n}\circ T^{(x,y)}, \label{eq:su(2)_on_site}
\end{align}
However, since $U_{\theta\vec n}$ is the $SU(2)$ representation on the physical leg and we already showed that $W_{\theta\vec n}$ also form a representation of $SU(2)$, $\Theta_{\theta\vec n}$ must be a representation of $SU(2)$ in U(1), which again must be trivial. So $\Theta_{\theta\vec n}(x,y)=1$.

\textbf{(10)} Finally, consider the $\theta=2\pi$ $SU(2)$ rotation. Eq.(\ref{eq:su(2)_on_site}) for this particular case becomes:
\begin{align}
 T^{(x,y)}=-W_{\theta=2\pi}\circ T^{(x,y)},\label{eq:2pi}
\end{align}
where we use the fact that the physical spin is half-integer and $\Theta_{\theta=2\pi}=1$ obtained in step-(9). This means that $W_{\theta=2\pi}$, a site and leg independent transformation, must be an element in $\overline{IGG}$ and thus can be written as:
\begin{align}
 W_{\theta=2\pi}=\eta_{\theta=2\pi}\chi_{\theta=2\pi}(x,y,i),
\end{align}
we see that $\chi_{\theta=2\pi}(x,y,i)\equiv \chi_{\theta=2\pi}$ is site and leg independent. Due to the definition of the $\chi-group$, this limits $\chi_{\theta=2\pi}=\pm1$. Plugging the above equation back in Eq.(\ref{eq:2pi}), one has:
\begin{align}
 -\mu_{\theta=2\pi}\prod_i\chi_{\theta=2\pi}=-\mu_{\theta=2\pi}=1
\end{align}
So $\mu_{\theta=2\pi}=-1$. This dictates that $\eta_{\theta=2\pi}=\mathrm{J}$ and every site tensor must be $Z_2$ odd: $\mu_{\mathrm{J}}=-1$.

We then have:
\begin{align}
 W_{\theta=2\pi}=\chi_{\theta=2\pi}\mathrm{J}
\end{align}
Note that we still have a sign ambiguity in the definition of $\mathrm{J}$, as mentioned in step-(1). We now use this sign ambiguity to set $\chi_{\theta=2\pi}=1$, and $W_{\theta=2\pi}=\mathrm{J}$.

\textbf{(11)} We still have the remaining $V(x,y,i)=V$ ambiguity. Now we use this ambiguity to tranform the site and leg independent $W_{\theta\vec n}$ to the standard form:
\begin{align}
 W_{\theta\vec n}=\oplus_{i=1}^M (\mathrm{I}_{n_i}\otimes e^{i\theta\vec n\cdot \vec S_i});
\end{align}
namely each virtual leg is a direct sum of $n_i$ number of $\vec S_i$ with different spin representation $\vec S_i$. Here we are only left with the $V$-ambiguity that is a direct sum of the similarity transformations acting in the $\mathrm{I}_{n_i}$ spaces.

\textbf{Summary: } We find in the presence of translational symmetry, $C_4$ symmetry, spin-rotational symmetry, the $IGG=Z_2$ symmetric PEPS for a half-integer spin system on the square lattice are classified by the following three sets of algebraic data:
\begin{enumerate}
 \item $\eta_{12},\eta_{C_4T},\eta_{C_4}\in IGG=\{\mathrm{I},\mathrm{J}\}$.
 \item $\chi_{C_4}$ which can be $\pm 1$.
 \item $\Theta_{C_4}$ which can choose values as defined in Eq.(\ref{eq:theta_C4}).
\end{enumerate}
Since each index can choose two values, there are $2^5=32$ classes. These indices completely determine the transformation rules of the site and bond tensors as:
\begin{align}
 W_{\theta\vec n}(x,y,i)&=W_{\theta\vec n}=\oplus_{i=1}^M (\mathrm{I}_{n_i}\otimes e^{i\theta\vec n\cdot \vec S_i}),\notag\\
 \mathrm{J}&=W_{\theta=2\pi},\notag\\
 W_{T_1}(x,y,i)&=\eta_{12}^y,\notag\\
 W_{T_2}(x,y,i)&=\mathrm{I},\notag\\
 W_{C_4}(x,y,a)&=\chi_{C_4}\eta_{12}^{xy}\eta_{C_4T}^y\eta_{C_4}\notag\\
 W_{C_4}(x,y,b/c/d)&=\eta_{12}^{xy}\eta_{C_4T}^y,\notag\\
\end{align}
and:
\begin{align}
 \Theta_{\theta\vec n}(x,y)&=1\notag\\
 \Theta_{T_1}(x,y)&=\mu_{12}^y,\notag\\
 \Theta_{T_2}(x,y)&=1,\notag\\
 \Theta_{C_4}(x,y)&=\mu_{C_4T}^y\mu_{12}^{xy}\Theta_{C_4}.
\end{align}
Here $\mu_{12}=1$($-1$) if $\eta_{12}=\mathrm{I}$ ($\mathrm{J}$), and similarly $\mu_{C_4T}=1$($-1$) if $\eta_{C_4T}=\mathrm{I}$ ($\mathrm{J}$). $\Theta_{C_4}=1$ or $i$($\Theta_{C_4}=e^{i\pi/4}$ or $e^{-i\pi/4}$) if $\eta_{C_4}=\mathrm{I}$ ($\mathrm{J}$).

After the physical half-integer spin is specified, e.g. $S=1/2$ or $S=3/2$, we know the transformation rules for both physical and virtual legs. One can thus determine the generic form of the symmetric tensor network for each class and use it for numerical simulation as discussed in Sec.\ref{sec:minimization}.

\end{appendix}

\bibliography{symmetric_peps_algorithm}

\end{document}